\documentclass[11pt,a4paper,english,german,american]{book}
\usepackage[T1]{fontenc}
\usepackage[latin1]{inputenc}
\usepackage{amsmath}
\usepackage{babel}
\usepackage{color}
\usepackage{graphics}
\usepackage{amssymb}

\makeatletter

\providecommand{\LyX}{L\kern-.1667em\lower.25em\hbox{Y}\kern-.125emX\@}
\let\SF@@footnote\footnote
\def\footnote{\ifx\protect\@typeset@protect
    \expandafter\SF@@footnote
  \else
    \expandafter\SF@gobble@opt
  \fi
}
\expandafter\def\csname SF@gobble@opt \endcsname{\@ifnextchar[
  \SF@gobble@twobracket
  \@gobble
}
\edef\SF@gobble@opt{\noexpand\protect
  \expandafter\noexpand\csname SF@gobble@opt \endcsname}
\def\SF@gobble@twobracket[#1]#2{}

 \newenvironment{lyxlist}[1]
   {\begin{list}{}
     {\settowidth{\labelwidth}{#1}
      \setlength{\leftmargin}{\labelwidth}
      \addtolength{\leftmargin}{\labelsep}
      }}
   {\end{list}}
 \newenvironment{lyxcode}
   {\begin{list}{}{
     \setlength{\rightmargin}{\leftmargin}
     \raggedright
     \setlength{\itemsep}{0pt}
     \setlength{\parsep}{0pt}
     \normalfont\ttfamily}%
    \item[]}
   {\end{list}}

\usepackage{feynarts}
\usepackage{amssymb}

\usepackage{color}

\usepackage{hyperref}

\newcommand{\cL}{\mathcal{L}}

\newcommand{\cD}{\mathcal{D}}
\newcommand{\cN}{\mathcal{N}}

\newcommand{\bk}{\mathbf{k}}
\newcommand{\bx}{\mathbf{x}}

\newcommand{\rem}[1]{}

\newcommand{\diag}{\text{diag}}

\newcommand{\dd}{\text{d}}

\newcommand{\slD}{\!\not\!\!D}
\newcommand{\slp}{\!\!\not\!p}
\newcommand{\slP}{\!\not\!\!P}
\newcommand{\slpd}{\!\!\not\!\partial}

\newcommand{\dsOne}{1\hspace{-0.243em}\text{l}}

\newcommand{\bad}{\blacktriangleleft}

\newcommand{\be}{\begin{equation}}
\newcommand{\ee}{\end{equation}}
\newcommand{\bea}{\begin{eqnarray}}
\newcommand{\eea}{\end{eqnarray}}

\newcommand{\0}{\over }
\newcommand{\2}{{1\over2}}
\newcommand{\4}{{1\over4}}

\def\0{\over } \def\2{{1\over2}} \def\4{{1\over4}}
\def\5{\hat } \def\6{\partial }

\def\g{g_{\rm eff}}

\newcommand{\geff}{g_{\rm eff}}
\newcommand{\vac}{{\rm vac}}
\newcommand{\muMS}{\bar\mu_{\rm MS}}
\newcommand{\LO}{{\rm LO}}
\newcommand{\NLO}{{\rm NLO}}
\newcommand{\im}{{\rm Im}}
\newcommand{\re}{{\rm Re}}
\newcommand{\qmax}{q_{\rm max}}

\newcommand{\ds}{\displaystyle}

\newcommand{\red}[1]{\color{red} #1 \color{black}}
\newcommand{\blue}[1]{\color{blue} #1 \color{black}}
\newcommand{\green}[1]{\color{green} #1 \color{black}}

\newcommand{\is}{{\color{white}\-}}

\newcommand{\tmpbibtex}{}

\newcommand{\non}{\mbox{\scriptsize non-}}

\newcommand{\DR}{{\rm DR}}
\newcommand{\slQ}{{\not \hspace{-0.745mm} Q}}
\newcommand{\FACm}{\mbox{FAC-m}}
\newcommand{\FACg}{\mbox{FAC-g}}

\newcommand{\clearemptydoublepage}{\newpage{\pagestyle{empty}\cleardoublepage}}

\newcommand{\rmi}[1]{{\mbox{\scriptsize #1}}}

\makeatother
\begin{document}


\newcommand{\bra}[1]{\left\langle #1 \right| }
 
\newcommand{\ket}[1]{\left| #1 \right\rangle }

\newcommand{\braket}[2]{\left\langle #1 \textrm{ }|\textrm{ }#2 \right\rangle }
\renewcommand{\braket}[2]{\left\langle #1 \right| \! \! \left. #2 \right\rangle }

\thispagestyle{empty}


\numberwithin{equation}{section}

\vspace{0cm}
{\centering \includegraphics{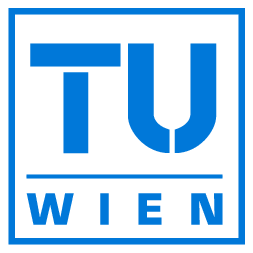} \par}
\vspace{0.3cm}

\bigskip{}
{\centering {\Large DISSERTATION}\Large \par}
\vfill{}

{\centering {\huge Quantum Corrections to Thermodynamic~Properties
in the Large~\( N_{f} \)~Limit of the Quark~Gluon~Plasma }\huge \par}
\vfill{}

\selectlanguage{german}
{\centering ausgeführt zum Zwecke der Erlangung des akademischen Grades
eines\par}

{\centering Doktors der technischen Wissenschaften\par}

\bigskip{}
{\centering unter der Anleitung von\par}
\bigskip{}

{\centering {\large Ao. Univ.-Prof. Dr. Anton Rebhan}\large \par}

{\centering Institutsnr: E 136\par}

{\centering Institut für Theoretische Physik\par}
\bigskip{}

{\centering eingereicht an der Technischen Universität Wien\par}

{\centering Fakultät für technische Naturwissenschaften und Informatik\par}

~

\bigskip{}
{\centering von\par}
\bigskip{}

~

{\centering \textbf{\large Dipl.-Ing. Andreas Ipp}\large \par}
\bigskip{}

{\centering Matr. Nr: 9271057\par}

{\centering Krotenbachgasse 27\par}

{\centering A-2345 Brunn am Gebirge\par}

{\centering Austria, Europe\par}
\vfill{}

Wien, 4. November 2003\hfill{}\underbar{~~~~~~~~~~~~~~~~~~~~~~~~~~~~~~~~~~~~}

\newpage
\clearemptydoublepage

\thispagestyle{plain}

\setcounter{page}{1}

\renewcommand{\thepage}{\roman{page}}
\selectlanguage{american}

\selectlanguage{german}
\section*{Deutsche Kurzfassung}

\rem{Oct 9, 2003}Das theoretische Studium des Quark-Gluon Plasmas
gewinnt immer mehr an Bedeutung seit Teilchenbeschleuniger wie das
SPS, der RHIC, oder der sich im Bau befindliche LHC die erforderlichen
hohen Energiedichten in Schwerionenkollisionen erreichen, die es erlauben,
diesen neuen Materiezustand experimentell zu untersuchen. Einfache
Anwendungen der Quantenfeldtheorie im Rahmen einer störungstheoretischen
Entwicklung nach der Kopplungskonstante versagen bei hohen Temperaturen,
und trotz eifriger Bemühungen, die Situation in den Griff zu bekommen,
haben wir quantitative theoretische Aussagen über den Phasenübergang
nur von Gittersimulationen. Diese wiederum versagen für ein Quark-Gluon
Plasma bei hohem chemischem Potential und niedrigen Temperaturen,
wie man es im Kern von dichten Sternen vermutet. Large-\( N_{f} \)-QCD
- das ist Quantenchromodynamik (QCD) mit einer großen Zahl von Quark-Sorten
(number of quark flavors - \( N_{f} \)) - erlaubt es, Wechselwirkungseffekte
von thermodynamischen Größen wie dem thermischen Druck oder der Entropie
exakt in der effektiven Kopplung \( \geff ^{2}\propto g^{2}N_{f} \)
für alle Temperaturen \( T \) und chemische Potentiale \( \mu _{q} \)
zu berechnen. Dies macht Large \( N_{f} \) QCD zu einem idealen Testwerkzeug
für verschiedene Näherungsmethoden.

In der vorliegenden Arbeit präsentieren wir das exakte Large-\( N_{f} \)
Resultat für den thermischen Wechselwirkungsdruck in der kompletten
\( T \)-\( \mu _{q} \)-Ebene in einem Bereich, in dem der Einfluss
durch den Landau-Pol numerisch vernachlässigt werden kann. Für kleine
Werte der Kopplung vergleichen wir unser Resultat mit existierenden
störungstheoretischen Ergebnissen in der Literatur, einschließlich
der aktuellen Berechnung des Drucks durch Vuorinen für endliche Temperatur
und chemisches Potential sowie einer älteren Rechnung von Freeman
und McLerran für verschwindende Temperatur und hohes chemisches Potential.
Unsere numerische Genauigkeit erlaubt uns, existierende störungstheoretische
Koeffizienten zu verifizieren und zum Teil sogar zu verbessern, und
auch störungstheoretische Koeffizienten zur sechsten Ordnung in der
Kopplung numerisch zu bestimmen, die analytisch bislang noch nicht
berechnet wurden. Für verschwindendes chemisches Potential berechnen
wir lineare und nicht-lineare Quarkzahl-Suszepti\-bili\-täten. Wir
zeigen, dass das moderate Skalierungsverhalten, das durch die Quarkzahl-Suszeptibilitäten
nahegelegt wird, ziemlich abrupt bei \( \mu _{q}\gtrsim \pi T \)
zusammenbricht, aber dass dieser nicht-pertubative Effekt in \( \mu _{q} \)
immer noch in guter Näherung durch die Ergebnisse von Vuorinen bei
kleinen Kopplungen und endlichem \( T \) beschrieben wird. Nur für
\( T\ll \mu _{q} \) versagt auch dieser Zugang, und wir kommen in
den Bereich der sogenannten Non-Fermi-Flüssigkeit, die im Gegensatz
zur klassischen Fermi-Flüssigkeit von langreichweitigen, quasistatischen
transversalen Eichbosonen dominiert wird. In diesem Limes können wir
nicht nur den bereits bekannten führenden \( T\ln T^{-1} \) Beitrag
zur spezifischen Wärme vervollständigen, sondern auch über die führende
Ordnung eine störungstheoretische Reihe mit anomalen gebrochene Potenzen
\( T^{(3+2n)/3} \), die durch dynamische Abschirmung verursacht werden,
angeben. Wir berechnen deren Koeffizienten analytisch bis zur Ordnung
\( T^{7/3} \) und finden, dass diese tatsächlich das führende anomale
Verhalten der vollen QED und QCD bestimmen (also bei endlichem \( N_{f} \)).

~
\selectlanguage{american}

\newpage

\section*{Abstract}

\rem{Oct 9, 2003}The theoretical study of the quark gluon plasma gains
increasing interest as particle accelerators like the SPS, RHIC, or
the currently built LHC will reach sufficiently high energy densities
in heavy ion collisions that allow us to probe this new state of matter
experimentally. Straightforward application of quantum field theory
at high temperatures fails in a perturbative expansion in the coupling
constant, and despite some effort during the last decades to improve
the situation, so far quantitative theoretical knowledge about the
phase transition merely comes from lattice simulations. Lattice simulations
on the other hand fail for a deconfined quark-gluon plasma at large
quark chemical potential and small temperatures which is expected
to be found in the core of dense stars. Large \( N_{f} \) QCD - that
is quantum chromodynamics with large number of quark flavors - allows
one to calculate thermodynamic properties like the interaction contribution
of thermal pressure or entropy exactly in the effective coupling \( \geff ^{2}\propto g^{2}N_{f} \)
for all temperatures \( T \) and chemical potentials \( \mu _{q} \).
This makes large \( N_{f} \) QCD an ideal testing ground for various
approximation methods.

In this work we present the exact large \( N_{f} \) NLO calculation
of the thermal interaction pressure in the whole \( T \)-\( \mu _{q} \)-plane
where the presence of the Landau pole is negligible numerically. For
small values of the coupling we compare our results to existing perturbative
results in the literature, in particular the recent calculation by
Vuorinen for finite temperature and chemical potential or an older
calculation by Freedman and McLerran for zero temperature and high
chemical potential. Our numerical accuracy allows us to verify and
even improve some of the existing perturbative coefficients, and to
predict new coefficients to the sixth order in the coupling numerically
that have not been calculated analytically yet. For larger couplings
we determine where perturbation theory ceases to be applicable. At
zero chemical potential we calculate linear and non-linear quark number
susceptibilities. We show that the moderate scaling behavior suggested
by the quark number susceptibilities breaks down rather abruptly at
\( \mu _{q}\gtrsim \pi T \), but that this non-perturbative effect
in \( \mu _{q} \) can still be reproduced well by the calculation
by Vuorinen for small couplings and finite \( T \). Only for \( T\ll \mu _{q} \)
also this approach breaks down and we enter the range of a so-called
non-Fermi liquid, which in contrast to a classical Fermi liquid is
dominated by long-range quasi-static transverse gauge-boson interactions.
In this limit, we complete the previously known leading \( T\ln T^{-1} \)
contribution to the specific heat, and also go beyond this order
to find a series involving anomalous fractional powers \( T^{(3+2n)/3} \)
caused by dynamical screening. We calculate their coefficients analytically
up to order \( T^{7/3} \) and find that these contributions indeed
determine the leading anomalous contribution in full QED and QCD (i.e.~at
finite \( N_{f} \)).

~

\newpage

\cleardoublepage

~
\vfill{}

\vspace{0.3cm}
{\centering \raisebox{-4ex}{\resizebox*{7cm}{!}{\includegraphics{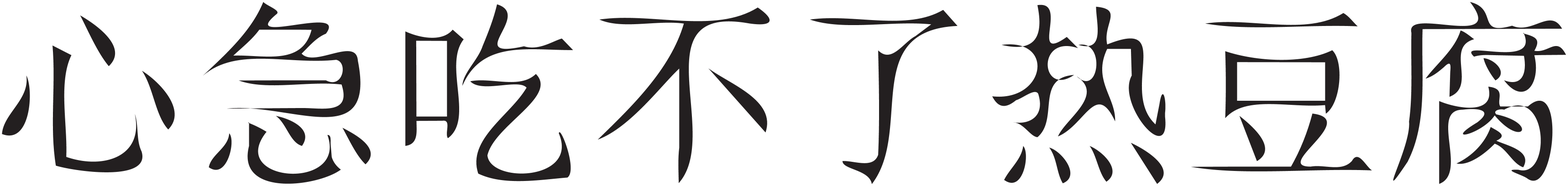}} }%
\footnote{\( \mathbf{x}\bar{\mathbf{i}}\mathbf{n}\, \mathbf{j}\acute{\mathbf{i}}\, \mathbf{ch}\bar{\mathbf{i}}\, \mathbf{b}\grave{\mathbf{u}}\, \mathbf{li}\breve{\mathbf{a}}\mathbf{o}\, \mathbf{r}\grave{\mathbf{e}}\, \mathbf{d}\grave{\mathbf{o}}\mathbf{u}\, \mathbf{fu} \):
Those of impatient heart can not eat hot tofu (bean curd).
(\textit{traditional Chinese saying})
}\par}
\vspace{0.3cm}

\vfill{}
~

\newpage
~

\vfill{}
\noindent Web-Version 3: 13.~Mai~2004 \\
\noindent Version 2: 10.~Nov.~2003 \\
(Original-Version: 4.~Nov.~2003)

\newpage

\tableofcontents{}

\newpage



\setcounter{page}{1}

\renewcommand{\thepage}{\arabic{page}}

{\raggedright \par} 

{\raggedright \par}

\chapter*{Prolog}

\addcontentsline{toc}{chapter}{Prolog}

\subsubsection*{LHC/CERN in Swiss, 2007%
\footnote{Data in this prolog are taken from the \rem{Large Hadron Collider
(LHC) }LHC design performance website 2003 \cite{LHC:2003:cern.ch}.
}}

\textsl{SWOOOOSH....!!!}

{}``Oh my god! What was that?'' - {}``They are trying to smash
us!'' - {}``Oh mum! I wanna get out!''. A bunch of hysterically
screaming lead ions flying almost at the speed of light
can barely fight the 8.36 Tesla magnetic fields (some 100,000 times
the strength of Earth's magnetic field) that keep them on spiraling
orbits in the Large Hadron Collider (LHC).

\textsl{SWOOOOSH....!!!}

{}``Aahhh!'' - {}``That was VERY close!!! I can't believe they
are doing this to us!''. With only 27 km collider circumference,
the ion bunches approach themselves some
11,000 times per second.

\textsl{SWOOOOSH...!!!}

{}``Ok, that's it! Let us oooouut!!'' - {}``They already ripped
off all electrons from me! What else do
they wanna do?''. The darkness in the collider is tantalizing, as
are the 1.9 K (some -270\ensuremath{°}C) coldness from the 5,000 surrounding
super-conducting magnet coils.

\textsl{SWOOOOSH...!!!}

{}``I'm cold!!!'' - {}``Brrrr... I'm freezing!!!''. {}``You'll
feel warmer soon'', Old uncle
Joe tried to soothe them. In his billions of years of life he had
gone through many phases. So far he had survived everything.

\textsl{SWOOOOSH...!!!}

{}``What the..!'' - {}``They are not... no!! Don't tell me they
want to smash us into...'' - {}``..into the Quark Gluon Plasma!!!''

\textsl{SWOOO -} KABOOM!!!!

~

Let me not go into too much detail - it's a horrible view: 
atoms are smashed head on head, their inner protons
and neutrons are torn apart, even the formerly strictly confined up-
and down-quarks end in a disgusting soup of unconfined quarks and
gluons, whining and wincing only as shock fronts penetrate them. But
this is science: Scientists have to take the loss of some lead atoms
for the sake of knowledge about the Quark Gluon Plasma (QGP).

\clearemptydoublepage

\chapter{Introduction}

\section{The experiment}

The study of the quark-gluon plasma in heavy ion collision experiments
is one of the focal points of contemporary high energy physics. Historically,
scientists smashed elementary particles since the 1930's with Earnest
Lawrence's invention of the cyclotron, which improved earlier attempts
with linear accelerators tremendously. In a cyclotron, charged particles
from electron or ion sources are accelerated on a circular orbit 
before they hit a target. Even higher energies can
be reached with synchrotrons, first constructed at the General Electric
Research Laboratory for the University of California, Berkeley, in
1949. Contrary to the continuous beam of a cyclotron, the synchrotron
works with beam pulses which ride on electro-magnetic waves. \rem{\cite{Hoult:2002:cern.ch}}

Modern particle accelerators like the Super Proton Synchrotron (SPS\rem{,
in operation since 1976}) at CERN or the Relativistic Heavy Ion Collider
(RHIC) at Brookhaven National Laboratory are based on this concept
of the synchrotron, relying on several steps of particle acceleration.
For example, heavy ions at RHIC are produced in the Tandem Van de
Graaff by static electricity. \rem{(Some experiments use beams of
protons which are produced in the Linac - Linear Accelerator). }These
particles are then carried to the Booster synchrotron where they are
accelerated to 37\% the speed of light, before they are transfered
to the Alternating Gradient Synchrotron \rem{(AGS) }where they reach
99.7\% the speed of light. From there they will be injected \rem{by
the ATR (AGS-To-RHIC) transfer line }into the first RHIC ring in clockwise
direction or into the second RHIC ring in counter-clockwise direction,
where particles will be collided into one another in one of the six
interaction points of the ring \cite{RHIC:2003:bnl.gov}.

It is interesting to note that for proton-anti-proton collisions like
in the LHC one only needs one collider ring instead of two, as positively
charged protons and negatively charged anti-protons can travel clockwise
and counter-clockwise in the same magnetic field. However, the LHC
also has two collider rings to also allow for lead-lead collisions,
proton-lead collisions, proton-oxygen collisions, or collisions of
protons with other light ions \cite{Jowett:2003dc}. A summary of
accelerators that are producing or will produce the quark-gluon plasma
is given in table \ref{tabAccelerators}. Besides hadron colliders
(using ions or protons) which will probe the quark-gluon plasma, future
collider plans also include lepton colliders as CERN's proposed electron
linear accelerator CLIC (Compact Linear Collider), or muon colliders.
Since leptons are fundamental particles (contrary to hadrons which
consist of quarks and gluons), these colliders are especially apt
for precision measurements of particle properties. \cite{Hoult:2002:cern.ch}
\begin{table}
{\centering \begin{tabular}{|c|c|c|c|c|c|}
\hline 
\selectlanguage{english}
Name
\selectlanguage{american}&
\selectlanguage{english}
Location
\selectlanguage{american}&
\selectlanguage{english}
Year
\selectlanguage{american}&
\selectlanguage{english}
Energy
\selectlanguage{american}&
\selectlanguage{english}
Beam
\selectlanguage{american}&
\selectlanguage{english}
Orbit
\selectlanguage{american}\\
\hline
\hline 
\selectlanguage{english}
SPS
\selectlanguage{american}&
\selectlanguage{english}
CERN
\selectlanguage{american}&
\selectlanguage{english}
1994
\selectlanguage{american}&
\selectlanguage{english}
160 GeV
\selectlanguage{american}&
\selectlanguage{english}
lead ions
\selectlanguage{american}&
\selectlanguage{english}
6.9 km
\selectlanguage{american}\\
\hline 
\selectlanguage{english}
RHIC
\selectlanguage{american}&
\selectlanguage{english}
Brookhaven
\selectlanguage{american}&
\selectlanguage{english}
2000
\selectlanguage{american}&
\selectlanguage{english}
200 GeV
\selectlanguage{american}&
\selectlanguage{english}
gold ions
\selectlanguage{american}&
\selectlanguage{english}
3.8 km
\selectlanguage{american}\\
\hline 
\selectlanguage{english}
LHC
\selectlanguage{american}&
\selectlanguage{english}
CERN
\selectlanguage{american}&
\selectlanguage{english}
2007
\selectlanguage{american}&
\selectlanguage{english}
7 TeV
\selectlanguage{american}&
\selectlanguage{english}
proton, lead
\selectlanguage{american}&
\selectlanguage{english}
26.7 km
\selectlanguage{american}\\
\hline
\end{tabular}\par}

\caption{Particle accelerators that produce or will produce the quark gluon
plasma (QGP). The year marks the date of starting operation for search
of the QGP, center-of-mass energy is given per nucleon for SPS and
RHIC. The abbreviations of the names mean Super Proton Synchrotron
(SPS) \cite{SPS:2003:cern.ch}, Relativistic Heavy Ion Collider (RHIC)
\cite{RHIC:2003:bnl.gov}, and Large Hadron Collider (LHC) \cite{LHC:2003:cern.ch}\rem{,
and Very Large Hadron Collider (VLHC) stages 1 and 2 \cite{VLHC:2003:vlhc.org}}.\label{tabAccelerators}}

\end{table}

The press releases about evidence for the production of the quark-gluon
plasma in heavy ion collisions at the SPS were first announced in
2000 \cite{Heinz:2000:cern.ch}. Only indirect observations are possible
since the QGP formed in the initial stage is quickly turned into a
system of hadrons - a process called {}``hadronization''. Detection
of single quarks is not possible due to color confinement, a property
of strong interactions at low energies, according to which quarks
must always combine to color-neutral hadrons before being able to
travel to the detector. A number of experiments was necessary before
scientists dared to announce the discovery of the QGP: The theoretical
analysis of the measured hadron abundances resembles a state of \char`\"{}chemical
equilibrium\char`\"{} at a temperature of about 170 MeV which marks
the quark-hadron transition. In particular, there is an observed enhancement
of hadrons containing strange quarks by a factor of 2 to 15 (depending
on the hadron), relative to proton-induced collision\rem{ (in experiments
labelled WA97, NA49, and NA50)}, and an observed suppression of the
charmonium states \( J/\psi  \) and \( \psi ' \)\rem{ (NA50)} \cite{Santos:2003qa}
which contain charm quarks with masses of about 1.2 GeV, much higher
than the transition temperature of about 170 MeV. Direct observations
of the QGP via electromagnetic radiation at the SPS\rem{ (WA98, NA45,
NA50)} are difficult due to high backgrounds from other sources. Higher
energies at RHIC and the LHC may allow for better observation of the
plasma radiation and enable detailed studies of the early thermalization
processes and dynamical evolution of the quark gluon plasma. These
experiments might also help to study the order of the phase transition
and to locate the tricritical point within the QCD phase diagram,
where the first-order transition changes into a crossover behavior.

\section{QCD Phase Diagram}

\rem{11. Oct 2003}Depending on temperature and particle density, matter
exists in various phases. Just like water can exist in the familiar
phases of solid, liquid, or gaseous state with phase transitions between
them, so will it enter a new phase of a plasma (ionized gas) at high
enough temperature to break up chemical bonds and to ionize hydrogen
and oxygen atoms. If the temperature still increases,
matter undergoes another phase transition where quarks and gluons,
which are the constituents of protons and neutrons, leave their nuclear
confinement to form a quark gluon plasma. On the lower end of the
temperature scale, matter may exist in the phase of a Bose-Einstein
condensate, superfluids, or superconductors.

The fundamental theory describing quarks and gluons is the theory
of strong interactions, quantum chromodynamics (QCD). Unlike 
electromagnetism, which
becomes stronger as the mutual distance decreases, strong interaction
gets weaker. As a consequence, quarks, the fundamental particles of
QCD, are confined in packages of three (called {}``baryons'', like
protons and neutrons) or two ({}``mesons'', e.g. pions) at low temperatures.
Any attempt to separate confined quarks rather produces a new quark-antiquark
pair from the gluon field in-between that combines with the separated
quarks so that no single quark or net color charge can be observed.

\begin{figure}
\resizebox*{1\textwidth}{!}{\includegraphics{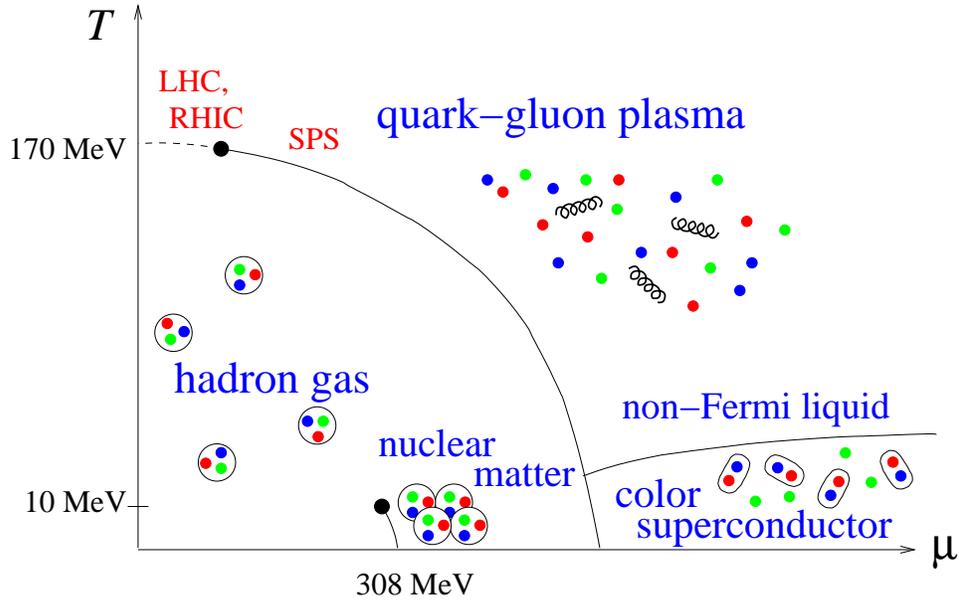}}

\caption{Schematic QCD phase diagram of strongly interacting matter. \label{fig_qcdphasediagram}}
\end{figure}
In figure \ref{fig_qcdphasediagram} we see a schematic view of the
QCD phase diagram. Strongly interacting matter can exist in three
distinct phases, depending on temperature \( T \) and chemical potential
\( \mu  \): the hadronic phase (confined to hadrons as a hadron gas
or nuclear matter), the quark-gluon plasma, and the color-superconducting
quark-matter. Our world is located on the lower line of the diagram
around the phase transition of 308 to 313 MeV, in the form of hadron
droplets (nuclei). Temperature \( T \) is given in units of MeV which
can be converted to Kelvin using the Boltzmann constant \( k_{B}=1.3807\times 10^{-23}\textrm{J}/\textrm{K}=8.617\times 10^{-5}\textrm{eV}/\textrm{K} \).
For a nice day of 27\ensuremath{°}C \rem{(80.6~F) }we get \( 27^{\circ }\textrm{C}\approx300 \, \textrm{K}\approx25 .9\, \textrm{meV} \).
The core of our sun has a temperature of about \rem{Sun fact sheet:
Model values at center of Sun: Central pressure: 2.477 x 10\textasciicircum{}11
bar, Central temperature: 1.571 x 10\textasciicircum{}7 K, Central
density: 1.622 x 10\textasciicircum{}5 kg/m\textasciicircum{}3. Temperature
on surface: 5880 K=5606.85 \ensuremath{°}C }\( 1.6\times10 ^{7}\textrm{K}\approx1 .35\, \textrm{keV} \).
To get to the quark-gluon plasma we still need to increase temperature
by five orders of magnitude to at least \( 170\, \textrm{MeV} \).

\rem{12. Okt 2003}The other axis of the diagram shows the quark chemical
potential \( \mu  \)\rem{Information: www.wikipedia.org}. The term
{}``chemical potential'' may be misleading since we deal with temperatures
and densities far from ordinary chemistry, but it denotes a useful
quantity that is applicable to any thermodynamic system: The chemical
potential is the change in the energy of the system when an additional
constituent particle is introduced, with the entropy and volume held
fixed (or equivalently the change in the Helmholtz free energy with
temperature and volume held fixed). For systems containing different
species of particles, there is a separate chemical potential associated
with each species. If the particles can be transformed into one another,
particle species with higher chemical potential will transform into
a species with lower chemical potential, releasing heat. Therefore
in an equilibrium state of several species, the chemical potentials
of all species must equal each other. At zero temperature, (infinite)
nuclear matter has a ground state at a quark chemical potential of
\( \mu =308\, \textrm{ MeV} \). This is the nucleon mass \( m_{N}\approx 939\, \textrm{MeV} \)
minus the binding energy 16 MeV divided by three for the average energy
per quark or quark chemical potential. In our everyday life we find
nuclear matter density only in atomic nuclei - the droplets in which
quarks are confined\rem{ - and zero quark number density outside of
the nuclei}. Only if we increase the pressure of the system so that
all droplets start touching and overlapping each other, we can build
a phase of infinite nuclear matter. Up to this point, the pressure
of the system at zero temperature is zero, because any compression
just means decrease of the void space between the droplets. If the
density is increased beyond the point where the droplets start to
touch, pressure will start to increase, too. This is the so-called
liquid-gas phase transition between the hadron gas and nuclear matter.

\rem{The regions in figure ??1?? show in each case the energetically
most favorable phase and the phase transitions in between them. }The
pressure always changes continuously if we change from one phase to
another, but if its first derivative with respect to \( T \) is discontinuous,
we call this a phase transition of first order. If the first derivative
of the pressure is continuous, but its second derivative is discontinuous,
we have a second order phase transition. Finally, if the pressure
is continuous to all orders, but there is a rapid change in pressure,
we can talk of a crossover. Usually, points of phase transition in
the diagram form a line that starts at one of the axis of the diagram
and may terminate at a critical point. Also the first-order liquid-gas
phase transition has a critical point for a temperature of about 10
MeV where the transition becomes second order. Above this temperature
one cannot distinguish between the gaseous and the liquid phase.

Below a temperature of 170 MeV and chemical potential of about 350
MeV, quarks stay confined in hadrons. If we further increase temperature
and/or quark density, we arrive at the quark-hadron phase transition
at which quarks cease to be bound in individual nuclei and become
either a deconfined quark-gluon plasma or a color superconductor.
For a range of medium quark chemical potential, the phase transition
is of first order, but for higher temperatures and smaller chemical
potentials there is a critical point below which we only have a crossover,
as suggested by lattice calculations. Particle detectors like SPS,
RHIC or LHC probe will probe the quark-gluon plasma in a range around
this tricritical point. This is also the path that our universe took
in early times, some 20-30 \( \mu \textrm{s} \) (\( \approx 170\, \textrm{MeV} \))
after the big-bang, starting with high temperature and a quark chemical
potential of about \( 10^{-6}\textrm{MeV} \) passing the quark-gluon
phase transition down to our universe temperature of 2.725 K \cite{Bennett:2003bz}
and an average quark chemical potential of about 313.6 MeV \cite{Fromerth:2002wb}.

For small temperatures and high chemical potential we enter the phase
of a color superconductor \cite{Rajagopal:2000wf,Alford:2001dt}.
This phase has been studied extensively during the last decade, and
its theoretical foundations include concepts taken from the theory
of ordinary superconductors: In ordinary superconductors, electrons
are bound together by phonon interactions into Cooper pairs. Color
superconductivity occurs because in the anti-triplet channel there
is an attractive interaction between two quarks at the Fermi surface
\cite{Bailin:1984bm,Barrois:1977xd}\rem{??ref 36, 37 Rischke, p14}.
These quarks then condense in the new ground state of the system by
forming Cooper pairs. Depending on \( T \), color superconductivity
can again be classified in different phases: If only up and down flavors
are involved, we have a 2-flavor color superconductor (2SC), for three
flavors including the strange quark we have a color-flavor-locked
(CFL) phase. Other phases studied involve a color-spin-locked (CSL)
phase and a polar phase. Matter may further occur in the form of crystalline
LOFF (named after Larkin, Ovchinnikov, Fulde and Ferrell) where Cooper
pairs with nonzero total momentum are favored \cite{Alford:2000ze}.
\rem{{[}Crystalline Color Superconductor hep-ph/008208. (LOFF){]}}Color
superconductivity may exist up to a temperature of 6 to 60 MeV \cite{Rischke:2003mt},
above which we enter the quark-gluon-plasma phase. 

For small to medium temperature ranges, the QGP may be described as
a non-Fermi liquid. It is called {}``Non-Fermi'', because its behavior
cannot entirely be described by the classical (Landau-)Fermi liquid
theory of non-interacting cold fermions. This is due to transverse
gauge bosons (transverse gluons) which are not screened at small temperatures
due to the properties of dynamical screening. We have long-range,
quasi-static interactions which give anomalous contributions to thermodynamic
quantities like the specific heat. Such non-Fermi liquid behavior
might have influence on astrophysical calculations, for example the
cooling rate of proto-neutron stars \cite{Iwamoto:1980eb}. Non-Fermi
liquid behavior also appears in solid state physics experiments, for
example in the specific heat of the \( \textrm{YbRh}_{2}\textrm{Si}_{2} \)
crystal \cite{Custers:2003aa,Trovarelli:1999aa}.

\section{Outline}

This thesis is organized as follows: 

In chapter 2 we will introduce large \( N_{f} \) QCD where the number
of quark flavors \( N_{f} \) is sent to infinity while the effective
coupling \( \geff \propto g^{2}N_{f} \) stays of order 1. We will
see that this theory is exactly solvable in the effective coupling
at next-to-leading order of the \( 1/N_{f} \) expansion. Also, the
scale dependence is completely solvable to this order. Therefore this
theory is particularly suitable for testing other perturbative and
non-perturbative approaches against this limit. In contrast to lattice
gauge theory, large \( N_{f} \) can be readily extended to finite
\( \mu  \). 

In chapter 3 we will explore the non-Fermi-liquid regime of the large
\( N_{f} \) theory. We will derive the pressure and the specific
heat perturbatively and show that beyond the leading logarithm there
are anomalous fractional powers and we calculate their coefficients
up to order \( T^{7/3} \) in the specific heat. We will see that
fractional powers appear from the transverse gluon propagator whose
calculation is carried out in careful detail, but both, longitudinal
and transverse contributions to the pressure are needed to complete
the leading logarithm of the specific heat.

The final chapter 4 gives a summary and an outlook.

\tmpbibtex


\rem{ ========= Large Nf CHAPTER ========== }

\chapter{Large \protect\( N_{f}\protect \)}

\section{Introduction to Large \protect\( N_{f}\protect \) QCD}

\subsection{Why Large \protect\( N_{f}\protect \)?}

Large \( N_{f} \) QCD is quantum chromodynamics (QCD) with the number
of quark flavors \( N_{f} \) assumed to be large (\( N_{f}\rightarrow \infty  \))
and the strong coupling \( \alpha _{s}=g^{2}/(4\pi )\rightarrow 0 \)
assumed to be small so that the effective coupling (which we will
define later) \( \geff ^{2}\sim g^{2}N_{f}\sim \alpha _{s}N_{f}\sim O(1) \)
stays of order 1. Also, the number of colors \( N_{c} \) stays small
\( N_{c}\sim O(1) \). 

Why do we want to study \( N_{f}\rightarrow \infty  \)? After all,
in our everyday world we only experience two quark flavors, that are
the {}``up'' and {}``down'' quarks with masses of the order of
1-10 MeV. For the quark gluon phase transition at a temperature of
about 170 MeV, only one more quark flavor, the {}``strange'' quark
with a mass of about 80-155 MeV \cite{Hagiwara:2002fs}, will play
a considerable role in the thermal state. Even if we consider the
other known quark types {}``charm'', {}``bottom'' (also called
{}``beauty''), and {}``top'' (which is barely called {}``truth''
anymore) with masses beyond a GeV, we end up with a total of \( N_{f}\leq 6 \)
quark flavors. Clearly, the aim is not to realistically model QCD,
but to study the theory in a well defined limit that turns out to
be exactly solvable up to next-to-leading order in an \( 1/N_{f} \)
expansion while containing essential physics of the full theory.

Ideally, we would like to solve full QCD, but so far only approximate
numerical results from lattice simulations are available (at least
for small \( \mu /T \)), but with not completely understood systematic
uncertainties. For temperatures far above the intrinsic scale of QCD,
\( T\gg \Lambda _{\rm {QCD}} \), we expect the coupling \( g \)
to get small \cite{Kap:FTFT}. Due to this property of strong interaction,
called asymptotic freedom, we can expand the theory in terms of small
\( g \). It turns out that a strict perturbative expansion of thermodynamic
parameters like the pressure in terms of the small coupling \( g \)
would only work for ridiculously high temperatures and fails to work
for the interesting region of temperature values slightly above the
phase transition \( T_{c} \)\rem{\( T_{c}\approx 170\, \textrm{MeV} \)}
or the QCD scale \( \Lambda _{\rm {QCD}} \)\rem{\( \Lambda _{\rm {QCD}}\approx 200\! \! \sim \! \! 250\, \textrm{MeV} \)}:
Successive orders in perturbation theory show very poor convergence.
But this is not the only problem: Hot QCD has an intrinsically nonperturbative
scale in the magnetostatic sector at wave numbers \( k\sim g^{2}T \).
For the pressure this means that perturbation theory ceases to work
at order \( g^{6}T^{4} \) \cite{Linde:1980ts,Gross:1981br}. For
other observables, this might even mean stricter restrictions to an
expansion in \( g \): Nonperturbative corrections are suppressed
only by a power of \( g^{2} \) for photon emissivity \cite{Arnold:2001ba},
a power of \( g \) for the Debye screening length \cite{Rebhan:1993az,Arnold:1995bh},
or even not suppressed at all but appear at leading order in the case
of the baryon number violation rate \cite{Arnold:1997dy,Bodeker:1998hm}.
But even in theories without the problem of a nonperturbative magnetostatic
scale like in hot QED, the poor convergence of the perturbative series
remains \cite{Arnold:1994ps,Arnold:1995eb,Coriano:1994re,Parwani:1994je,Parwani:1995xi}.

There have been attempts to improve the situation by reorganizations
or partial resummations of the perturbative expansions of hot QCD.
In scalar models where one already meets similar difficulties of poor
convergence \cite{Parwani:1995zz}, Karsch et al. \cite{Karsch:1997gj}
proposed to keep a screening mass unexpanded at any given order of
the loop expansion and to fix this mass by a stationarity principle,
which technically means subtracting a mass term from the bare Lagrangian
and adding it to the interaction part. This successful approach was
extended to QCD by Andersen et al. \cite{Andersen:1999va,Andersen:1999sf,Andersen:1999fw,Andersen:2002ey,Andersen:2002er,Baier:1999db}
by replacing the simple mass term by the gauge-invariant hard-thermal-lop
(HTL) action \cite{Braaten:1992gm,Frenkel:1992ts}, a method which
they termed HTL perturbation theory. Another approximation method
is based on \( \Phi  \)-derivable approximations \cite{Baym:1962,Vanderheyden:1998ph}
for 2PI skeleton diagrams advocated by Blaizot et al. \cite{Blaizot:1999ip,Blaizot:2000fc,Blaizot:2001vr,Blaizot:2003tw,Peshier:2000hx}.
Here, the starting point is an expression of the thermodynamic potential
in terms of dressed propagators, where bare propagators are functionals
of the full propagators that have to satisfy a stationarity condition.
Truncation in this scheme does not happen in terms of the coupling
\( g \), but one resums diagrams to a given loop order of the 2PI
skeleton diagrams.

All of these approaches show improved convergence properties, but
it is difficult to tell to what extent these resummations can predict
the correct behavior of full QCD. Apart from the necessary condition
that these theories should coincide with perturbative QCD at very
small couplings (corresponding to temperatures far beyond the scope
of current collision experiments), only lattice calculations \cite{Boyd:1996bx,Okamoto:1999hi}
could give truly independent indications as to which resummation scheme
to trust. Lattice theory on the other hand works more like a {}``black
box'' for theorists and provides only limited insight into the actual
underlying physics, like screening effects or collective modes. One
has therefore been looking for simpler theories that could be solved
exactly, so that other resummation schemes could be tested within
that simpler framework. One such theory is large-\( N \) scalar field
theory with \( \phi ^{4} \) interaction \cite{Drummond:1997cw,Bodeker:1998an}
with an exactly solvable \( N\rightarrow \infty  \) limit. Another
limit that has been studied in the past is the large \( N \) limit
of a scalar field theory in 6 dimensions with cubic interactions \cite{Bodeker:1998an}
that mimics QED with \( N \) flavors. It is richer than large \( N \)
\( \phi ^{4} \)-theory in that self-energies vary with momentum and
wave-function renormalization is needed just like in full QCD. Yet,
this theory involves instabilities which prevents one from direct
comparisons to QCD. A more appealing test bed is large \( N_{f} \)
QCD.

Besides being exactly solvable up to next-to-leading (NLO) order in
\( 1/N_{f} \), large-\( N_{f} \) QCD contains the same complicated
frequency and momentum dependent gauge field screening and damping
as full QCD and its perturbative series shows similarly poor behavior
as full QCD. One should keep in mind though that since large \( N_{f} \)
QCD only resembles part of the full QCD theory, we do not expect large
\( N_{f} \) to give any useful predictions for a small number of
flavors. In fact, with a running coupling that leads to a Landau pole
(like in QED) and excludes the possibility of confinement, large \( N_{f} \)
at NLO is quite different from full QCD. Also, the leading order contributions
to full QCD in an expansion in the coupling \( g \) are numerically
dominated by subleading powers of \( 1/N_{f} \) for small \( N_{f} \),
as we will see in the next section. Yet, large \( N_{f} \) can be
regarded as an important test theory, and any resummation scheme had
better get close to the exact NLO result in the large \( N_{f} \)
limit. On the other hand, even if a theory turns out to reproduce
the right large \( N_{f} \) limit, that does not guarantee that it
will correctly predict the full QCD result. After all, large \( N_{f} \)
is a test theory, but contains a lot of the physics of full QCD, so
if we can do this test, let's better do it.

\subsection{Comparison to strict perturbative expansion in \protect\( g\protect \)}

\begin{figure}
\selectlanguage{english}
{\centering Perturbative calculations of QCD at high \( T/\mu  \)\\
\[
\begin{array}{cc}
\ds P=\frac{8\pi ^{2}}{45}T^{4}\left\{ \red {1}+\frac{21}{32}N_{f}\right.  & \rm {Planck\, 1900}\\
\ds -\frac{15}{4}(1+\frac{5}{12}\red {N_{f}})\blue {\frac{\alpha _{s}}{\pi }} & \rm {Shuryak/Chin\, 1978}\\
\ds +30(1+\frac{\red {N_{f}}}{6})^{\frac{3}{2}}\blue {(\frac{\alpha _{s}}{\pi })^{\frac{3}{2}}} & \rm {Kapusta\, 1979}\\
+\frac{135}{2}(1+\frac{N_{f}}{6})\blue {\ln \frac{\alpha _{s}}{\pi }(1+\frac{N_{f}}{6})}\cdot \blue {(\frac{\alpha _{s}}{\pi })^{2}} & \rm {Toimela\, 1983}\\
+\left\{ 237.2+15.97N_{f}-0.413\red {N_{f}^{2}}\right. \qquad \qquad  & \rm {Arnold\&Zhai\, 1995}\\
\left. -\frac{165}{8}(1+\frac{5}{12}\red {N_{f}})(1-\frac{2}{33}\red {N_{f}})\green {\ln \frac{\bar{\mu }}{2\pi T}}\right\} \blue {(\frac{\alpha _{s}}{\pi })^{2}} & \\
+(1+\frac{\red {N_{f}}}{6})^{\frac{1}{2}}\left\{ -799.2-21.96N_{f}-1.93\red {N_{f}^{2}}\right. \quad  & \rm {Zhai\&Kastening\, 1995}\\
\left. +\frac{495}{2}(1+\frac{\red {N_{f}}}{6})(1-\frac{2}{33}\red {N_{f}})\green {\ln \frac{\bar{\mu }}{2\pi T}}\right\} \blue {(\frac{\alpha _{s}}{\pi })^{\frac{5}{2}}} & \rm {Braaten\&Nieto\, 1996}\\
+\left\{ 1139.8+65.89N_{f}+7.653N_{f}^{2}\right. \qquad \qquad  & \rm {\raisebox {-0.8ex}{Kajantie,\, Laine,}}\\
\left. -\frac{1485}{2}(1+\frac{N_{f}}{6})(1-\frac{2}{33}N_{f})\green {\ln \frac{\bar{\mu }}{2\pi T}}\right\} \blue {\ln \frac{\pi }{\alpha _{s}}(\frac{\alpha _{s}}{\pi })^{3}} & \begin{array}{c}
\rm {Rummukainen\, \&}\\
\rm {Schr\ddot{o}der\, 2002}
\end{array}\\
\left. +\left\{ \red {?!}+?N_{f}+?N_{f}^{2}+\red {C(\bar{\mu })N_{f}^{3}}\right\} \blue {(\frac{\alpha _{s}}{\pi })^{3}}+O(\frac{\alpha _{s}}{\pi })^{\frac{7}{2}}\right\}  & 
\end{array}\]
\par}

\noindent {*} extension to \( \mu _{q}\neq 0:\, \rm {Vuorinen\, 2003} \)\\
{*} \( \red {C(\bar{\mu })}=0.040...\textrm{for }\bar{\mu }=\pi T \)
~~(A.I.\&A.Rebhan 2003, see section \ref{section_pressurezeromu})\\
{*} \( \red {?!} \): completely non-perturbative (Linde 1980; Gross,
Pisarski \& Yaffe 1980)

\caption{Known coefficients for the pressure of perturbative QCD by 2003.
The coefficient marked as \protect\( \red {?!}\protect \) denotes
the breakdown of perturbation theory in the order \protect\( \alpha _{s}^{3}\protect \).
Only a few coefficients of this series \rem{(marked with red \protect\( \red {N_{f}}\protect \)) }contribute
to the NLO contribution of the large-\protect\( N_{f}\protect \)
limit, namely those of the form \rem{\protect\( \blue {\alpha _{s}^{\, }}\! \! {}_{\, }^{n}\red {N_{f}^{\, }}\! \! {}_{\, }^{n}\protect \)}\protect\( \blue {\alpha _{s}^{n}}\red {N_{f}^{n}}\protect \).
Particularly, terms containing a logarithm of the coupling \protect\( \blue {\ln \alpha _{s}}\protect \)
do not survive the large-\protect\( N_{f}\protect \) limit. The large-\protect\( N_{f}\protect \)
coefficient \protect\( \red {C(\bar{\mu })}\protect \) of order \protect\( N_{f}^{3}\alpha _{s}^{3}\protect \)
has been extracted numerically from the exact NLO result.\label{fig:perturbativeQCD}}
\selectlanguage{american}
\end{figure}

It is instructive to see which terms of perturbation theory survive
in the large \( N_{f} \) limit. Strict perturbation theory has a
long tradition. An overview over the known coefficients is given in
figure \ref{fig:perturbativeQCD}. The leading order result to the
pressure of perturbative QCD is just the Stefan-Boltzmann pressure
of non-interacting quarks and gluons. The first non-trivial contributions
to this result proportional to \( \alpha _{s} \) were calculated
independently by Shuryak and Chin in 1978 \cite{Shuryak:1978ut,Chin:1978gj}
by calculating one- and two-loop vacuum graphs. The next order already
turned out to be non-analytic in \( \alpha _{s} \) and required the
resummation of an infinite subset of diagrams. This was done by Kapusta
to order \( \alpha _{s}^{3/2} \) \cite{Kapusta:1979fh} and Toimela
to order \( \alpha _{s}^{2}\ln \alpha _{s} \) \cite{Toimela:1983hv}.
Subsequent orders demanded new tools for evaluating sum-integrals
by Arnold, Zhai and Kastening \cite{Arnold:1994ps,Arnold:1995eb,Zhai:1995ac},
or the introduction of dimensional reduction \cite{Ginsparg:1980ef,Appelquist:1981vg,Kajantie:1996dw}
by Braaten and Nieto \cite{Braaten:1996jr}. The last perturbative
series coefficient that has be calculated so far for full QCD is the
\( \alpha _{s}^{3}\ln \alpha _{s} \) coefficient by Kajantie, Laine,
Rummukainen, and Schröder \cite{Kajantie:2002wa,Kajantie:2003ax}.
The order \( \alpha _{s}^{3} \) coefficient is completely non-perturbative
as pointed out by Linde \cite{Linde:1980ts} and Gross, Pisarski,
and Yaffe \cite{Gross:1981br}. Recently the perturbative result has
been extended to finite chemical potential by Vuorinen \cite{Vuorinen:2003fs}. 

In this perturbative series the effect of large \( N_{f} \) is straightforward
to be seen: Since we keep the quantity \( \alpha _{s}N_{f}\sim O(1) \)
of order one while we send \( N_{f}\rightarrow \infty  \) and \( \alpha _{s}\rightarrow 0 \),
the leading order (LO) contribution to the pressure is just given
by the contribution proportional to \( N_{f} \): \( P=7\pi ^{2}N_{f}T^{4}/60 \).
Actually, this contribution is infinitely large for \( N_{f}\rightarrow \infty  \),
but we can always subtract this known leading order contribution to
obtain a finite correction at next-to-leading order (NLO). From figure
\ref{fig:perturbativeQCD} we see that now all orders in the coupling
\( \alpha _{s} \) contribute to the large \( N_{f} \) NLO result
which are of the form \( \alpha _{s}^{p}N_{f}^{p} \) with \( p=0,\, 1,\, \frac{3}{2},\, 2,\, \frac{5}{2} \).
A graphical comparison of the partial sums of the series against the
exact large \( N_{f} \) NLO calculation will be presented below in
figure \ref{figurePnlo}.\rem{ Note that in the large \( N_{f} \)
limit it is also possible to numerically extract the analytically
yet unknown coefficient of \( \alpha _{s}^{3}N_{f}^{3} \) which turns
out to be \( C(\bar{\mu })\approx 0.040 \) for \( \bar{\mu }=\pi T \).}

\subsection{Outline}

The pressure in large \( N_{f} \) QCD was first calculated by Moore
for zero chemical potential \cite{Moore:2002md}%
\footnote{\selectlanguage{english}
The first published version contained an unfortunate coding error
which was revealed and corrected by an independent calculation involving
the author of this thesis \cite{Ipp:2003zr}.
\selectlanguage{american}
} and successively extended to finite chemical potential \cite{Ipp:2003jy}.
In the following sections we will present the exact large \( N_{f} \)
result for the thermal pressure and some derived quantities thereof
like the quark number susceptibilities or the entropy. The theory
inevitably contains a Landau pole. We will study its effect on the
result and show for which range of the coupling the effect of the
Landau pole is negligible. At zero chemical potential we compare the
exact large \( N_{f} \) result of the pressure and the quark number
susceptibilities with known results from thermal perturbation theory
\cite{Arnold:1995eb,Zhai:1995ac,Hart:2000ha} obtained at small chemical
potential where dimensional reduction \cite{Ginsparg:1980ef,Braaten:1995na,Kajantie:1996dw}
is applicable. We verify the recent three-loop result of Vuorinen
\cite{Vuorinen:2002ue} on quark number susceptibilities numerically
in the large \( N_{f} \) limit as well as a numerical coefficient
in the pressure at zero temperature obtained long ago by Freedman
and McLerran \cite{Freedman:1977dm,Baluni:1978ms}. For small values
of the coupling \( g \), our numerical accuracy allows us to extract
a number of perturbative coefficients at order \( g^{6} \) that are
not yet known from analytical calculations. 

It is remarkable that large \( N_{f} \) can be safely calculated
all the way down from large temperatures and zero chemical potential
to large chemical potential and zero temperature for effective couplings
\( \geff ^{2}\lesssim 20 \), which is a main advantage compared to
lattice gauge theory. For a long time, the so-called {}``sign problem''
prohibited calculations of fermions with finite chemical potential
on the lattice. Only recently there has been important progress regarding
calculation of thermodynamic quantities within lattice gauge theory
\cite{Fodor:2001au,Fodor:2002km,deForcrand:2002ci,Allton:2002zi,D'Elia:2002gd,Gavai:2003mf}.
Using the large \( N_{f} \) limit, we can test the scaling behavior
noticed in lattice calculations by Fodor, Katz, and Szabo \cite{Fodor:2002km}
and find that it breaks down rather abruptly at \( \mu _{q}\gtrsim \pi T \).
We can also test the range of applicability of dimensional reduction
by comparison to perturbative calculations at finite temperature and
chemical potential by Vuorinen \cite{Vuorinen:2003fs}, where we study
the effect of the choice of the renormalization scale. At small temperatures
\( T\ll \mu  \) we enter the region of a non-Fermi-liquid behavior,
which will be studied in detail in the next chapter.

Further discussions on the large \( N_{f} \) limit can be found in
the literature by Peshier \cite{Peshier:2002fm} on the quasiparticle
picture \cite{Peshier:1996ty,Levai:1997yx,Peshier:1999ww}, who points
out the large differences between large and small \( N_{f} \) QCD
with respect to the different strong coupling behavior and questions
the immediate significance of a comparison within large \( N_{f} \)
NLO. A detailed discussion of the HTL-quasiparticle picture of QCD
with respect to large \( N_{f} \) and its implications on 2PI \( \phi  \)-derivable
approximations \cite{Blaizot:1999ip} is given by Rebhan \cite{Rebhan:2003fj}.

\section{Pressure at Large-\protect\( N_{f}\protect \) QCD}

\begin{figure}
\selectlanguage{english}
{\centering \resizebox*{1\textwidth}{!}{\includegraphics{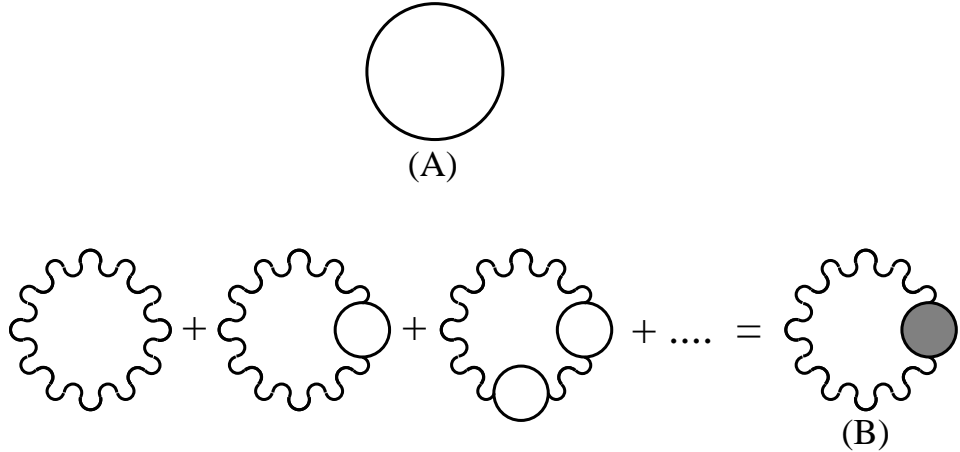}} \par}

\caption{Diagrams contributing to the large \protect\( N_{f}\protect \) limit.
At leading order (LO) of the order \protect\( O(N_{f})\protect \)
there is only the fermion loop (A) contributing. At next-to-leading
order (NLO) of order \protect\( O(1)\protect \) there is an infinite
number of diagrams contributing (B) which can be resummed by the Schwinger-Dyson
method.\label{fig:largenfdiagrams}}
\selectlanguage{american}
\end{figure}

In the following we want to calculate the thermal pressure in the
large \( N_{f} \) limit. As we already mentioned there are indeed
two quantities that are involved in the limit: The number of flavors
\( N_{f} \) and the coupling \( g \). By forming the limit \( N_{f}\rightarrow \infty  \)
we actually keep the effective coupling \( \geff ^{2}\sim g^{2}N_{f} \)
of order 1. In this sense, a suppression by a factor \( 1/N_{f} \)
is equivalent to a suppression by a factor \( g^{2} \). We will see
how this provides a very strong ordering principle for diagrams in
the large \( N_{f} \) expansion. There are no special assumptions
on the number of colors. We will calculate the pressure at next-to-leading
order (NLO) where the theory can be solved exactly, that is to all
orders in the effective coupling, apart from possible difficulties
introduced by a Landau pole that we will discuss later. We will follow
the original approach of references \cite{Moore:2002md,Ipp:2003zr}
and define the effective coupling as \begin{equation}
\g ^{2}\equiv \frac{g^{2}N_{f}C_{\rm F}d_{\rm F}}{d_{\rm A}}=\left\{ \begin{array}{cc}
\displaystyle \frac{g^{2}N_{f}}{2}\, , & {\textrm{QCD}}\, ,\\
 & \\
g^{2}N_{f}\, , & {\textrm{QED}}\, .\\

\end{array}\right. 
\end{equation}
where \( C_{\rm F} \) is the quadratic Casimir of the representation
containing the fermions and \( d_{\rm F} \) and \( d_{\rm A} \)
are the dimensions of the fermionic and adjoint representations. This
allows us to treat the two cases of massless QCD and ultrarelativistic
QED the theory at the same time. Since the coupling runs in these
theories, we need to know the \( \beta  \) function. The one-loop
\( \beta  \) function for QCD, that is a \( SU(N) \) gauge theory
with \( N_{f} \) fermions in the fundamental representation, is given
by

\begin{equation}
\label{oneloopbetafunction}
2g\frac{\mu dg}{d\mu }=\frac{\mu d(g^{2})}{d\mu }=\beta (g^{2})=-\frac{2g^{4}}{(4\pi )^{2}}\left( \frac{11}{3}N-\frac{2}{3}N_{f}\right) .
\end{equation}
For sufficiently small \( N_{f} \), that is \( N_{f}<\frac{11}{2}N \),
the \( \beta  \) function is negative, which means that QCD is asymptotically
free. In the limit of \( N_{f}\rightarrow \infty  \) we see that
this behavior changes as the sign of \( \beta (g) \) gets positive
and large \( N_{f} \) therefore is no more an asymptotically free
theory. The \( \beta  \) function then becomes\begin{equation}
\beta (\geff ^{2})=\frac{\mu d\geff ^{2}}{d\mu }=\frac{(\geff ^{2})^{2}}{6\pi ^{2}}.
\end{equation}
 This relation turns out to be exact at leading order in \( 1/N_{f} \),
even if we add higher loop beta function contributions (see for example
reference \cite{vanRitbergen:1997va}) to the one-loop result (\ref{oneloopbetafunction})
because they are diagrammatically suppressed at least by a factor
of \( 1/N_{f}\propto g^{2} \). We can solve the renormalization scale
dependence exactly, giving \begin{equation}
\label{gscal}
\frac{1}{g_{{\textrm{eff}}}^{2}(\mu )}=\frac{1}{g_{{\textrm{eff}}}^{2}(\mu ')}+\frac{\ln (\mu '/\mu )}{6\pi ^{2}}\, .
\end{equation}
This theory contains a Landau pole of the order of \( \Lambda _{\textrm{L }}\sim \mu \exp (6\pi ^{2}/\geff ^{2}) \).
Following reference \cite{Moore:2002md} we define the Landau scale
\( \Lambda _{\textrm{L}} \) such that the vacuum gauge field propagator
diverges at \( Q^{2}=\Lambda _{\textrm{L}}^{2} \). The vacuum gauge
field propagator is defined as \( D_{\vac }^{-1}=q^{2}-q_{0}^{2}+\Pi _{\vac } \)
with \( \Pi _{\vac } \) from equation (\ref{selfenergy_vacuum_1})
below, which leads to \begin{equation}
\label{LLandau}
\Lambda _{\textrm{L}}=\bar{\mu }_{\textrm{MS}}e^{5/6}e^{6\pi ^{2}/\g ^{2}(\bar{\mu }_{\textrm{MS}})}\, .
\end{equation}
 This Landau singularity means that the exact definition of the theory
is ambiguous unless the UV completion is specified. But that should
not hinder us from doing finite temperature calculations since the
ambiguity for the thermal pressure at NLO is suppressed by a factor
\( ({\textrm{max}(\textrm{T},\mu )}/\Lambda _{\textrm{L}})^{4} \).
As long as we stay well below the Landau pole with temperature \( T \)
and chemical potential \( \mu  \), we can expect to get results which
are not afflicted by the UV incompleteness of the large \( N_{f} \)
theory.

The pressure can be calculated by computing the diagrams contributing
to the free energy, which is the trace of the sum of all 1PI vacuum
bubble graphs. At leading order in the \( 1/N_{f} \) expansion we
only have one diagram - that is the bare fermion loop in figure \ref{fig:largenfdiagrams}.
To obtain the thermal pressure, we subtract off the vacuum part of
the diagrams. The leading order pressure is thus given by\begin{equation}
P_{\LO }=N_{f}\textrm{Tr}\left( T\sum _{n,odd}\int \frac{d^{3}\boldsymbol {q}}{(2\pi )^{3}}\ln \slQ -\int \frac{d^{4}Q}{(2\pi )^{4}}\ln \slQ \right) .
\end{equation}
It is well-known how to calculate this contribution for finite temperature
and chemical potential and the pressure is given by

\begin{equation}
\label{largenfP_LO}
P_{\LO }=NN_{f}\left( {7\pi ^{2}T^{4}\0180 }+{\mu ^{2}T^{2}\06 }+{\mu ^{4}\012 \pi ^{2}}\right) .
\end{equation}
The leading order contribution in the pressure is thus of order \( N_{f} \)
and strictly speaking infinitely large in the limit of \( N_{f}\rightarrow \infty  \).
Of course, we can always think of this as the leading contribution
to an expansion in terms of \( 1/N_{f} \) with large but finite \( N_{f} \). 

The next-to-leading order (NLO) of this expansion is proportional
to \( N_{f}^{0}=1 \). We can calculate it from the gauge boson loop
with an arbitrary number of fermion loop insertions (plus corresponding
counterterm insertions), which are all of order \( N_{f}^{0} \).
For each fermion loop insertion we get a factor \( N_{f} \) and a
factor \( g^{2} \) from the two vertices, thus just our effective
coupling \( g^{2}N_{f}\propto \geff ^{2}\sim O(1) \). Any additional
bosonic or ghost insertion would give factors of \( g^{2} \) or higher,
without introducing new fermion loop factors of \( N_{f} \), and
are thus suppressed by at least a factor of \( g^{2}\propto 1/N_{f} \).
Therefore this gauge boson loop is all there is to NLO. It can be
resummed by the standard Schwinger-Dyson resummation. Its contribution
to the pressure is (again subtracting the vacuum part)\begin{eqnarray}
P_{\NLO } & = & -\frac{1}{2}\textrm{Tr}\left( T\sum _{n,even}\int \frac{d^{3}\boldsymbol {q}}{(2\pi )^{3}}\ln ([D_{0}^{-1}]^{\mu \nu }(Q)+\Pi _{\rm th}^{\mu \nu }(Q))\right. \nonumber \\
 &  & \qquad \qquad \left. -\int \frac{d^{4}Q}{(2\pi )^{4}}\ln ([D_{0}^{-1}]^{\mu \nu }(Q)+\Pi _{\vac }^{\mu \nu }(Q))\right) \label{PNLOstartingpoint} 
\end{eqnarray}
where the trace runs over group and Lorentz indices. The vacuum part
of the gauge-boson self energy is given by \begin{equation}
\label{selfenergy_vacuum_1}
\Pi _{{\textrm{vac}}}^{\mu \nu }(Q)=-\frac{g_{{\textrm{eff}}}^{2}}{12\pi ^{2}}\left( \eta ^{\mu \nu }Q^{2}-Q^{\mu }Q^{\nu }\right) \left( \ln \frac{Q^{2}}{\bar{\mu }^{2}}-\frac{5}{3}\right) ,
\end{equation}
 and the one-loop bosonic self-energy can be calculated via \begin{eqnarray}
\widetilde{\Pi }_{G}(Q) & \equiv  & \eta _{\mu \nu }\Pi ^{\mu \nu }(Q)-[{\rm vac}]\, \, =\, \, 2\Pi _{T}(Q)+\Pi _{L}(Q),\label{selfenergy_G} \\
\widetilde{\Pi }_{H}(Q) & \equiv  & \Pi ^{00}(Q)-[{\rm vac}]\, \, =\, \, \frac{q^{2}}{Q^{2}}\Pi _{L}(Q)\label{selfenergy_H} 
\end{eqnarray}
 where \( \widetilde{\Pi }_{G} \) and \( \widetilde{\Pi }_{H} \)
will be calculated in the next section in equations (\ref{WeldonRealG}),
(\ref{WeldonRealH}), and (\ref{imaginaryselfenergy1}). These cannot
be given in closed form except for their imaginary parts \cite{Ipp:2003zr},
but are represented by one-dimensional integrals involving the fermionic
distribution function \( n_{f} \). Using this decomposition of the
self energy, the trace of the resummed propagator can be rewritten
as\begin{equation}
\textrm{Tr}\ln ([D_{0}^{-1}]^{\mu \nu }(Q)+\Pi _{\rm th}^{\mu \nu }(Q))\! =\! 2\ln (Q^{2}\! +\Pi _{T}+\Pi _{\vac })+\ln (Q^{2}+\Pi _{L}+\Pi _{\vac })-\ln (Q^{2})
\end{equation}
if we choose an appropriate gauge in which \( D_{0}^{-1} \) factorizes
in the same way as \( \Pi  \), as for example in the Feynman gauge
\( [D_{0}^{-1}]^{\mu \nu }(Q)=\eta ^{\mu \nu }Q^{2} \). Any gauge
dependence drops out to our order of interest, since we only regard
the difference between thermal and vacuum contributions. We can now
perform the sum over Matsubara frequencies in the usual way of replacing
it by a contour integral that is deformed appropriately. The result
is then given by \cite{Moore:2002md}\begin{eqnarray}
\frac{P_{\rm {NLO}}}{N_{g}} & \! \! \! \! \! =\! \!  & \! \! \! -\int \! \frac{d^{3}q}{(2\pi )^{3}}\int _{0}^{\infty }\! \frac{dq_{0}}{\pi }\left[ 2\left( [n_{b}+\frac{1}{2}]\textrm{Im}\ln (q^{2}-q_{0}^{2}+\Pi _{T}+\Pi _{\vac })\right. \right. \label{PNLO} \\
 & \! \! \! \! \!  & \qquad \qquad \qquad \qquad \quad \left. -\frac{1}{2}\textrm{Im}\ln (q^{2}-q_{0}^{2}+\Pi _{\vac })\right) \nonumber \\
 & \! \! \! \! + & \! \! \! \! \! \! \left. \left( [n_{b}+\frac{1}{2}]\textrm{Im}\ln (\frac{q^{2}-q_{0}^{2}+\Pi _{L}+\Pi _{\vac }}{q^{2}-q_{0}^{2}})-\frac{1}{2}\textrm{Im}\ln (\frac{q^{2}-q_{0}^{2}+\Pi _{\vac }}{q^{2}-q_{0}^{2}})\right) \right] \, \, \, \, \nonumber 
\end{eqnarray}
 with the bosonic distribution function \( n_{b}(\omega )=1/(e^{\omega /T}-1) \)
coming directly from the sum over even frequencies \( q_{0}=2n\pi T \).
When evaluating the integrals above exactly by numerical means, we
can safely integrate parts proportional to \( n_{b} \) in Minkowski
space, since those are exponentially ultraviolet safe. For terms without
\( n_{b} \) more care is required. We will refer to the former contributions
as {}``\( n_{b} \)''-parts and the latter as {}``non-\( n_{b} \)''
contributions. The {}``non-\( n_{b} \)'' contributions are potentially
logarithmically divergent, unless a Euclidean invariant cutoff is
introduced \cite{Moore:2002md}. We will discuss this point in more
detail in section \ref{section_landaupoleambiguity}.

For non-vanishing chemical potential \( \mu  \) we use the fermionic
distribution function \begin{equation}
\label{fermionicdistributionfunction}
n_{f}(k,T,\mu )=\frac{1}{2}\left( \frac{1}{e^{(k-\mu )/T}+1}+\frac{1}{e^{(k+\mu )/T}+1}\right) 
\end{equation}
 which enters via the gauge boson self-energy expressions \( \Pi _{T} \)
and \( \Pi _{L} \).

\section{Leading-order gauge boson self-energy\label{section_gaugebosonselfenergy}}

\rem{Fr, 22. Nov 2002}The leading-order gauge boson self-energy in
the large \( N_{f} \) limit is diagrammatically represented by a
gauge boson propagator containing one fermion loop insertion. In the
imaginary time formalism it can be written as\begin{eqnarray}
\widetilde{\pi }\Pi ^{\mu \nu }(i\omega _{n},\mathbf{p}) & \! =\!  & 2N_{f}\widetilde{g}^{2}g^{2}\int \frac{d^{3}k}{(2\pi )^{3}}\int _{-\infty }^{\infty }\frac{dk_{0}}{2\pi }\int _{-\infty }^{\infty }\frac{dq_{0}}{2\pi }\rho _{0}(k_{0},\mathbf{k})\rho _{0}(q_{0},\mathbf{q})\qquad \nonumber \\
 & \! \! \!  & \times \frac{n(k_{0})-n(q_{0})}{k_{0}-q_{0}-i\omega _{n}}\left( k^{\mu }q^{\nu }+q^{\mu }k^{\nu }-g^{\mu \nu }k^{\alpha }q_{\alpha }+g^{\mu \nu }m^{2}\right) 
\end{eqnarray}
with \( \mathbf{q}\equiv \mathbf{k}-\mathbf{p} \). The constants
\( \widetilde{g} \) and \( \widetilde{\pi } \) can be adjusted to
fit various author's conventions (see also the conventions used in
appendix \ref{section_qcdfeynmanrules}). We follow Weldon's conventions
\cite{Weldon:1982aq} by setting \( \widetilde{g}^{2}/\widetilde{\pi }=-1 \).
The fermionic Boltzmann-factor is given by \( n(k_{0})=1/(e^{k_{0}/T}+1) \)
and the spectral function is given by \( \rho _{0}(k_{0},\mathbf{k})=2\pi \epsilon (k_{0})\delta (k_{0}^{2}-\mathbf{k}^{2}-m^{2})=\frac{\pi }{\varepsilon _{k}}(\delta (k_{0}-\varepsilon _{k})-\delta (k_{0}+\varepsilon _{k})) \)
with \( \epsilon (k_{0})=k_{0}/|k_{0}| \) and \( \varepsilon _{k}\equiv \sqrt{\mathbf{k}^{2}+m^{2}} \).
We can find a spectral form of the self energy by using\begin{equation}
\label{spectralform1}
\Pi _{\mu \nu }(z,\mathbf{p})=\int _{-\infty }^{\infty }\frac{dp_{0}}{2\pi }\frac{\pi _{\mu \nu }(p_{0},p)}{z-p_{0}}
\end{equation}
 which is valid for any complex \( z \) \cite[eq (3.1.36)]{Landsman:1987uw}.
The spectral density is then given by

\begin{eqnarray}
\widetilde{\pi }\pi ^{\mu \nu }(p_{0},\mathbf{p}) & = & -2N_{f}\widetilde{g}^{2}g^{2}\int \frac{d^{3}k}{(2\pi )^{3}}\int _{-\infty }^{\infty }\frac{dk_{0}}{2\pi }\int _{-\infty }^{\infty }\frac{dq_{0}}{2\pi }\rho _{0}(k_{0},\mathbf{k})\rho _{0}(q_{0},\mathbf{q})\nonumber \\
 &  & \times \left( n(k_{0})-n(q_{0})\right) 2\pi \delta (p_{0}-k_{0}+q_{0})I^{\mu \nu }\label{spectraldensity1} 
\end{eqnarray}
with \[
I^{\mu \nu }=\left( k^{\mu }q^{\nu }+q^{\mu }k^{\nu }-g^{\mu \nu }k^{\alpha }q_{\alpha }+g^{\mu \nu }m^{2}\right) .\]
The spectral density is manifestly real which will simplify our calculations.
Since the spectral density will play an important role, we will try
to push its calculation as far as we can. After integrating over \( q_{0} \),
we are left with with an expression depending only on \( k^{\mu }=(k_{0},\mathbf{k}) \)
and \( p^{\mu }=(p_{0},\mathbf{p}) \). A transformation of variables
\( k^{\mu }\rightarrow -k'^{\mu }+p^{\mu } \) in the part proportional
to \( n(q_{0})=n(k_{0}-p_{0}) \) leaves all other parts invariant
up to minus signs. It basically replaces \( n(k_{0}-p_{0}) \) by
\( -n(-k'_{0})=n(k'_{0})-1 \) so that we can replace \( \left( n(k_{0})-n(q_{0})\right)  \)
by \( \left( 2n(k_{0})-1\right)  \) in equation (\ref{spectraldensity1}).
If we use \( z=\cos \theta  \) as the angle between \( \mathbf{pk}=pkz \),
we can express\begin{equation}
\rho _{0}(k_{0}-p_{0},\mathbf{k}-\mathbf{p})=2\pi \epsilon (k_{0}-p_{0})\frac{1}{2kp}\delta (z-\frac{k^{2}+p^{2}-(k_{0}-p_{0})^{2}}{2kp})
\end{equation}
 and integrate over \( z \) in \( \int \frac{d^{3}k}{(2\pi )^{3}}=\frac{1}{(2\pi )^{2}}\int _{-1}^{1}dz\int _{0}^{\infty }k^{2}dk \).
Analogous to Weldon \cite{Weldon:1982aq} we can then define\begin{eqnarray}
I_{G} & \equiv  & g^{\mu \nu }I_{\mu \nu }=-k_{0}^{2}+k^{2}+p_{0}^{2}-p^{2},\\
I_{H} & \equiv  & u^{\mu }u^{\nu }I_{\mu \nu }=\frac{1}{2}\left( k^{2}-p^{2}+(k_{0}-p_{0})(3k_{0}-p_{0})\right) \nonumber 
\end{eqnarray}
 with \( u^{\mu }=(1,0,0,0) \) and integrate over \( k_{0} \). This
is the last integral that we can do analytically. The spectral density
turns out to be antisymmetric in \( k_{0} \) meaning \( \pi (k_{0},\mathbf{k})=-\pi (-k_{0},\mathbf{k}). \)
The final result for the spectral density of the self-energy can then
be written as a sum of two parts that only differ by the sign of \( p_{0} \)
\begin{equation}
\label{spectraldensityantisym1}
\pi _{X}(p_{0},\mathbf{p})=\pi _{X}^{+}(p_{0},p)-\pi _{X}^{+}(-p_{0},p),
\end{equation}
 \begin{eqnarray}
\pi _{X}^{+}(p_{0},p) & = & -\frac{\widetilde{g}^{2}}{\widetilde{\pi }}\frac{g^{2}N_{f}}{2\pi p}\int _{0}^{\infty }dk\left( n(k)-\frac{1}{2}\right) \bar{I}_{X}\\
 &  & \times \epsilon (k-p_{0})\theta (|k-p|\leq |k-p_{0}|\leq |k+p|)\nonumber 
\end{eqnarray}
with \( X \) = \( G \) or \( H \) as in\begin{eqnarray}
\bar{I}_{G}\, \equiv \, I_{G}(k_{0}=k) & = & p_{0}^{2}-p^{2},\\
\bar{I}_{H}\, \equiv \, I_{H}(k_{0}=k) & = & \frac{1}{2}(2k+p-p_{0})(2k-p-p_{0}).\nonumber 
\end{eqnarray}
The \( \theta  \)-function stems from the integration over \( z \)
with \( -1\leq z\leq 1 \) and its usage here means \( \theta (\textrm{true expression})=1 \)
and \( \theta (\textrm{false expression})=0 \).

\subsection*{Self-energy in the complex energy plane}

Starting from the spectral form (\ref{spectralform1}) we can calculate
the self-energy in Minkowski or in Euclidean metric or anywhere between
in the complex plane. To compare our results to the real-time formalism
we start with the self-energy with Feynman prescription\begin{equation}
\widetilde{\Pi }_{F}(p_{0},\mathbf{p})\equiv \Pi (p_{0}+ip_{0}\varepsilon ,\mathbf{p}).
\end{equation}
Separating \( \frac{1}{x+i\varepsilon }=\frac{\textrm{P}}{x}-i\pi \delta (x) \)
with \( \textrm{P} \) denoting the principal value, and knowing that
\( \pi (k_{0},\mathbf{k}) \) is a real function, we can easily separate
this equation into real and imaginary parts\begin{equation}
\label{selfenergyfeynmanreal}
\textrm{Re}\widetilde{\Pi }_{F}(p_{0},\mathbf{p})=\int _{-\infty }^{\infty }\frac{dp'_{0}}{2\pi }\pi (p'_{0},p)\frac{\textrm{P}}{p_{0}-p'_{0}},
\end{equation}
\begin{equation}
\textrm{Im}\widetilde{\Pi }_{F}(p_{0},\mathbf{p})=-\frac{1}{2}\epsilon (p_{0})\pi (p_{0},p).
\end{equation}
In the general case we can write \begin{eqnarray}
\Pi (a+ib,\mathbf{p}) & = & \int _{-\infty }^{\infty }\frac{dp_{0}}{2\pi }\pi (p_{0},p)\frac{1}{a+ib-p_{0}}\nonumber \\
 & = & \int \frac{dp_{0}}{2\pi }\pi (p_{0},p)\left( \frac{a-p_{0}}{(a-p_{0})^{2}+b^{2}}-i\frac{b}{(a-p_{0})^{2}+b^{2}}\right) \qquad \qquad 
\end{eqnarray}
where we separated real- and imaginary parts. By using the antisymmetry
property of the spectral density (\ref{spectraldensityantisym1})
we can finally write\begin{eqnarray}
\Pi (a+ib,\mathbf{p}) & = & \int \frac{dp_{0}}{2\pi }\pi ^{+}(p_{0},p)\left( \frac{a-p_{0}}{(a-p_{0})^{2}+b^{2}}-\frac{a+p_{0}}{(a+p_{0})^{2}+b^{2}}\right) \nonumber \\
 & - & \! \! \! \! \! i\! \int \frac{dp_{0}}{2\pi }\pi ^{+}(p_{0},p)\left( \frac{b}{(a-p_{0})^{2}+b^{2}}-\frac{b}{(a+p_{0})^{2}+b^{2}}\right) \qquad \quad \label{selfenergycomplex2} 
\end{eqnarray}
where all integrals to be evaluated are real. Particularly for the
case \( a\rightarrow 0 \) and \( b=\omega \neq 0 \) we recover the
result for the Euclidean metric \begin{equation}
\widetilde{\Pi }_{Eucl}(i\omega ,\mathbf{p})\equiv \Pi (i\omega +\varepsilon ,\mathbf{p}),
\end{equation}
\begin{eqnarray}
\textrm{Re}\widetilde{\Pi }_{Eucl}(i\omega ,\mathbf{p}) & = & \int _{-\infty }^{\infty }\frac{dp_{0}}{2\pi }\pi ^{+}(p_{0},p)\frac{-2p_{0}}{\omega ^{2}+p_{0}^{2}}\nonumber \\
 & = & \int _{-\infty }^{\infty }\frac{dp_{0}}{2\pi }\pi (p_{0},p)\frac{-p_{0}}{\omega ^{2}+p_{0}^{2}},\label{selfenergyeuclidreal} 
\end{eqnarray}
\begin{equation}
\label{selfenergyeuclidimaginary}
\textrm{Im}\widetilde{\Pi }_{Eucl}(i\omega ,\mathbf{p})=0.
\end{equation}

\subsubsection{Minkowski space results}

Plugging our self-energy (\ref{spectraldensityantisym1}) in the expressions
(\ref{selfenergyfeynmanreal}) above we get the Minkowski space results.
First we note that we can write the integral over \( p_{0} \) that
we have to deal with in a simple way since \( p \) and \( k \) are
positive: \[
\int _{-\infty }^{\infty }dp_{0}\epsilon (k-p_{0})\theta (|k-p|\leq |k-p_{0}|\leq |k+p|)f(p_{0})\, \, \, \, \, \, \, \, \, \, \]
\begin{eqnarray}
 & = & \theta (k-p)\left( \int _{-p}^{p}dp_{0}f(p_{0})-\int _{2k-p}^{2k+p}dp_{0}f(p_{0})\right) \nonumber \\
 &  & +\theta (p-k)\left( \int _{-p}^{2k-p}dp_{0}f(p_{0})-\int _{p}^{2k+p}dp_{0}f(p_{0})\right) \nonumber \\
 & = & \int _{-p}^{p}dp_{0}f(p_{0})-\int _{2k-p}^{2k+p}dp_{0}f(p_{0}).\label{integralsignepsilon1} 
\end{eqnarray}
The resulting integral for the real part can then be written as\begin{eqnarray}
\textrm{Re}\widetilde{\Pi }_{X}(p_{0},\mathbf{p}) & \! \! =\! \!  & -\frac{\widetilde{g}^{2}}{\widetilde{\pi }}\frac{g^{2}N_{f}}{2\pi p}\int _{0}^{\infty }dk\left( n(k)-\frac{1}{2}\right) \\
 & \! \! \! \!  & \! \! \times \left[ \int _{-p}^{p}dp'_{0}\! -\! \int _{2k-p}^{2k+p}dp'_{0}\right] \left( \frac{\textrm{P}}{p_{0}-p'_{0}}-\frac{\textrm{P}}{p_{0}+p'_{0}}\right) \bar{I}_{X}(k,p'_{0},p).\nonumber 
\end{eqnarray}
Using\[
\int dp'_{0}\left( \frac{\textrm{P}}{p_{0}-p'_{0}}-\frac{\textrm{P}}{p_{0}+p'_{0}}\right) \bar{I}_{G}(k,p'_{0},p)\qquad \qquad \qquad \qquad \qquad \qquad \]
\begin{equation}
=-p'{}_{0}^{2}-(p_{0}^{2}-p^{2})\left( \log (|p'_{0}-p_{0}|)+\log (|p'_{0}+p_{0}|)\right) 
\end{equation}
and\[
\int dp'_{0}\left( \frac{\textrm{P}}{p_{0}-p'_{0}}-\frac{\textrm{P}}{p_{0}+p'_{0}}\right) \bar{I}_{H}(k,p'_{0},p)\qquad \qquad \qquad \qquad \qquad \qquad \]
\begin{eqnarray}
 & = & \frac{1}{2}\left[ 8kp'_{0}-p'{}_{0}^{2}-(2k-p-p_{0})(2k+p-p_{0})\log (|p'_{0}-p_{0}|)\right. \nonumber \\
 &  & \left. -(2k-p+p_{0})(2k+p+p_{0})\log (|p'_{0}+p_{0}|)\right] 
\end{eqnarray}
we finally arrive at the real part of the self-energy that is stated
in Weldon's paper \cite{Weldon:1982aq}\begin{eqnarray}
\textrm{Re}\widetilde{\Pi }_{G}(p_{0},\mathbf{p}) & = & \left( -\frac{\widetilde{g}^{2}}{\widetilde{\pi }}\right) \frac{g^{2}N_{f}}{2\pi ^{2}}\int _{0}^{\infty }dk\left( n(k)-\frac{1}{2}\right) \nonumber \\
 &  & \times \left[ 4k+\frac{p_{0}^{2}-p^{2}}{2p}\log \left| \frac{2k+p_{0}+p}{2k+p_{0}-p}\frac{2k-p_{0}+p}{2k-p_{0}-p}\right| \right] \qquad \label{WeldonRealG} 
\end{eqnarray}
and\begin{eqnarray}
\textrm{Re}\widetilde{\Pi }_{H}(p_{0},\mathbf{p}) & = & \left( -\frac{\widetilde{g}^{2}}{\widetilde{\pi }}\right) \frac{g^{2}N_{f}}{2\pi ^{2}}\int _{0}^{\infty }dk\left( n(k)-\frac{1}{2}\right) \nonumber \\
 & \times  & \! \! \left[ 2k\left( 1-\frac{p_{0}}{p}\log \left| \frac{p_{0}+p}{p_{0}-p}\right| \right) \right. \nonumber \\
 &  & \left. +\frac{(2k+p_{0}+p)(2k+p_{0}-p)}{4p}\log \left| \frac{2k+p_{0}+p}{2k+p_{0}-p}\right| \right. \nonumber \\
 &  & \left. -\frac{(2k-p_{0}-p)(2k-p_{0}+p)}{4p}\log \left| \frac{2k-p_{0}-p}{2k-p_{0}+p}\right| \right] .\qquad \quad \, \, \label{WeldonRealH} 
\end{eqnarray}
Note that Weldon already subtracted the \( T=0 \) vacuum part and
replaced \( n(k)-\frac{1}{2} \) by \( n(k) \). For the imaginary
part no further calculation is necessary and we can simply relate
the imaginary part of the self-energy to its spectral density (\ref{spectraldensityantisym1})\begin{equation}
\textrm{Im}\widetilde{\Pi }_{X}(p_{0},\mathbf{p})=-\frac{1}{2}\epsilon (p_{0})\left( \pi _{X}^{+}(p_{0},p)-\pi _{X}^{+}(-p_{0},p)\right) .
\end{equation}
We are left with integrals over \( k \) that can be solved analytically.
We rewrite the integral over \( k \) for the sign- and the \( \theta  \)-function
as\[
\int _{-\infty }^{\infty }dk\epsilon (k-p_{0})\theta (|k-p|\leq |k-p_{0}|\leq |k+p|)f(k)\, \, \, \, \, \, \, \, \, \, \]
\begin{equation}
=\theta (p^{2}-p_{0}^{2})\int _{\frac{p_{0}+p}{2}}^{\infty }f(k)dk-\theta (p_{0}-p)\int _{\frac{p_{0}-p}{2}}^{\frac{p_{0}+p}{2}}f(k)dk.
\end{equation}
The function \( f(k) \) is one of the following \( f(k)=n(k), \)
\( kn(k), \) or \( k^{2}n(k) \) if we already subtract the vacuum
part from \( n(k)-\frac{1}{2} \). This leaves us with the following
integrals\begin{eqnarray}
F_{1}(x) & \equiv  & \int _{x}^{\infty }n(k)dk=-x+T\log (e^{x/T}+1),\nonumber \\
F_{2}(x) & \equiv  & \int _{x}^{\infty }kn(k)dk=\frac{\pi ^{2}T^{2}}{6}-\frac{x^{2}}{2}+xT\log (e^{x/T}+1)+T^{2}\textrm{Li}_{2}(-e^{x/T}),\nonumber \\
F_{3}(x) & \equiv  & \int _{x}^{\infty }k^{2}n(k)dk=-\frac{x^{3}}{3}+x^{2}T\log (e^{x/T}+1)\nonumber \\
 &  & \qquad \qquad \qquad \quad \, \, +2xT^{2}\textrm{Li}_{2}(-e^{x/T})-2T^{3}\textrm{Li}_{3}(-e^{x/T})\label{functionsF123finiteT} 
\end{eqnarray}
with \( \textrm{Li}_{n}(x) \) being the polylogarithm function. The
functions \( F_{i}(x) \) have the property that \( F_{i}(x\rightarrow \infty )\rightarrow 0 \)
and \( F_{i}(x>0,T\rightarrow 0)\rightarrow 0 \). The thermal contribution
to the imaginary part of the self-energy can then be written as\begin{eqnarray}
\textrm{Im}\widetilde{\Pi }_{X}(p_{0},\mathbf{p}) & \! \! =\! \!  & -\frac{1}{2}\epsilon (p_{0})\left( -\frac{\widetilde{g}^{2}}{\widetilde{\pi }}\right) \frac{g^{2}N_{f}}{2\pi p}\nonumber \\
 & \! \!  & \times \left[ \theta (p^{2}-p_{0}^{2})\left( F^{+}_{X}(\frac{p_{0}+p}{2})-F^{-}_{X}(\frac{-p_{0}+p}{2})\right) \right. \nonumber \\
 & \! \!  & \quad \left. +\theta (p_{0}-p)\left( F^{+}_{X}(\frac{p_{0}+p}{2})-F^{+}_{X}(\frac{p_{0}-p}{2})\right) \right. \nonumber \\
 & \! \!  & \quad \left. -\theta (-p_{0}-p)\left( F^{-}_{X}(\frac{-p_{0}+p}{2})-F^{-}_{X}(\frac{-p_{0}-p}{2})\right) \right] \qquad \quad \label{imaginaryselfenergy1} 
\end{eqnarray}
with \( X=G \) or \( H \) and \begin{equation}
F^{\pm }_{G}(x)=\int _{x}^{\infty }n(k)\bar{I}_{G}dk=(p_{0}^{2}-p^{2})F_{1}(x),
\end{equation}
\begin{equation}
F^{\pm }_{H}(x)=\int _{x}^{\infty }n(k)\bar{I}_{H}(k,\pm p_{0},p)dk=\frac{p_{0}^{2}-p^{2}}{2}F_{1}(x)\mp 2p_{0}F_{2}(x)+2F_{3}(x).
\end{equation}
Using symmetric and antisymmetric versions of these functions \( F_{X}^{S}\equiv (F_{X}^{+}+F_{X}^{-})/2 \)
and \( F_{X}^{A}\equiv (F_{X}^{+}-F_{X}^{-})/2 \) we can rewrite
equation (\ref{imaginaryselfenergy1}) as\begin{eqnarray}
\textrm{Im}\widetilde{\Pi }_{X}(p_{0},\mathbf{p}) & = & -\frac{1}{2}\epsilon (p_{0})\left( -\frac{\widetilde{g}^{2}}{\widetilde{\pi }}\right) \frac{g^{2}N_{f}}{2\pi p}\\
 &  & \times \left[ F^{S}_{X}(\frac{\left| p_{0}+p\right| }{2})-F^{S}_{X}(\frac{\left| p_{0}-p\right| }{2})\right. \nonumber \\
 &  & \qquad \left. +\epsilon (p_{0}+p)F^{A}_{X}(\frac{\left| p_{0}+p\right| }{2})-\epsilon (p_{0}-p)F^{A}_{X}(\frac{\left| p_{0}-p\right| }{2})\right] .\nonumber 
\end{eqnarray}
This way of writing is especially convenient for calculating the \( G \)-part
of the self-energy, since \( F_{G}^{A}=0 \) vanishes.

\subsubsection{Euclidean space results}

To obtain the results in Euclidean space we start from equation (\ref{selfenergyeuclidreal})
with \( \pi ^{+} \). As in the case for the Minkowski metric, we
can simplify the integral over \( p_{0} \) with sign- and \( \theta  \)-function
according to equation (\ref{integralsignepsilon1}) and we are left
with real integrals of the basic form\begin{equation}
\int dp_{0}\frac{-2p_{0}}{\omega ^{2}+p_{0}^{2}}\bar{I}_{G}(k,p_{0},p)=-p_{0}^{2}+(\omega ^{2}+p_{0}^{2})\log (\omega ^{2}+p_{0}^{2})
\end{equation}
and\begin{eqnarray}
\int dp_{0}\frac{-2p_{0}}{\omega ^{2}+p_{0}^{2}}\bar{I}_{H}(k,p_{0},p) & = & \frac{1}{2}p_{0}(8k-p_{0})-4k\omega \arctan \left( \frac{p_{0}}{\omega }\right) \nonumber \\
 &  & -\frac{1}{2}(4k^{2}-p^{2}-\omega ^{2})\log (\omega ^{2}+p_{0}^{2}).\qquad 
\end{eqnarray}
For \( \omega ^{2}+p_{0}^{2}>0 \) there is no danger of running into
singularities. Using the integral limits for \( p_{0} \) from \( -p \)
to \( p \) and \( 2k-p \) to \( 2k+p \) as in (\ref{integralsignepsilon1})
we can write the self-energy in the Euclidean metric as\begin{eqnarray}
\textrm{Re}\widetilde{\Pi }_{G}(i\omega ,\mathbf{p}) & = & \left( -\frac{\widetilde{g}^{2}}{\widetilde{\pi }}\right) \frac{g^{2}N_{f}}{2\pi ^{2}}\int _{0}^{\infty }dk\left( n(k)-\frac{1}{2}\right) \label{selfenergy_Euclid_RePiG} \\
 &  & \times \left( 4k+\frac{\omega ^{2}+p^{2}}{2p}\log \frac{\omega ^{2}+(2k-p)^{2}}{\omega ^{2}+(2k+p)^{2}}\right) ,\qquad \qquad \nonumber 
\end{eqnarray}
\begin{eqnarray}
\textrm{Re}\widetilde{\Pi }_{H}(i\omega ,\mathbf{p}) & = & \left( -\frac{\widetilde{g}^{2}}{\widetilde{\pi }}\right) \frac{g^{2}N_{f}}{2\pi ^{2}}\int _{0}^{\infty }dk\left( n(k)-\frac{1}{2}\right) \label{selfenergy_Euclid_RePiH} \\
 &  & \times \left[ 2k+\frac{\omega ^{2}+p^{2}-4k^{2}}{4p}\log \frac{\omega ^{2}+(2k-p)^{2}}{\omega ^{2}+(2k+p)^{2}}\right. \nonumber \\
 &  & \, \, \left. -\frac{2k\omega }{p}\left( \arctan \frac{2k-p}{\omega }+2\arctan \frac{p}{\omega }-\arctan \frac{2k+p}{\omega }\right) \right] .\nonumber 
\end{eqnarray}
This is in principle the result given by reference \cite{Moore:2002md}
up to the appearance of the arc tangents in the last line. The expression
given there can be fully recovered by using \( \arctan (z)=\frac{1}{2}i\log (1-iz)-\frac{1}{2}i\log (1+iz) \)
for complex \( z \) and gathering the sum of logarithms into one
logarithmic expression \begin{equation}
\arctan \frac{2k-p}{\omega }+2\arctan \frac{p}{\omega }-\arctan \frac{2k+p}{\omega }\simeq -\frac{i}{2}\log \frac{1+\frac{4k^{2}}{(\omega -ip)^{2}}}{1+\frac{4k^{2}}{(\omega +ip)^{2}}}.
\end{equation}
However this expression should be used with care. If one takes the
principal branch of the logarithmic expression on the r.h.s., one
obtains results that might differ from the arc tangent expression
by some multiple of \( \pi  \). Solving for the branch-cut, that
is where the expression inside the logarithm equals \( -1 \), one
finds that expression for real \( \omega , \) \( p, \) and \( k \)
can be written as\begin{equation}
\sum \arctan =-\frac{i}{2}\log \frac{1+\frac{4k^{2}}{(\omega -ip)^{2}}}{1+\frac{4k^{2}}{(\omega +ip)^{2}}}+\pi \epsilon (p\omega )\theta (p^{2}-\omega ^{2})\theta (|k|-\frac{p^{2}+\omega ^{2}}{2\sqrt{p^{2}-\omega ^{2}}})
\end{equation}
where one should take the principal value of the \( \arctan  \) and
\( \log  \) expressions involved. As shown in equation (\ref{selfenergyeuclidimaginary})
the imaginary part of the self-energy in the Euclidean space vanishes.

\subsubsection{Complex Result}

The result in the complex plane allows us to link results in Minkowski
space and Euclidean space by great arcs. The safest way to obtain
the result in the complex plane is to start with equation (\ref{selfenergycomplex2}),
where all integrals involved are real, and the result for the self-energy
is separated into real and imaginary part. Starting point are the
following indefinite integrals\begin{eqnarray}
F_{G}^{Re}(p_{0}) & \equiv  & \int dp_{0}\left( \frac{a-p_{0}}{(a-p_{0})^{2}+b^{2}}-\frac{a+p_{0}}{(a+p_{0})^{2}+b^{2}}\right) \bar{I}_{G}\\
 & = & -p_{0}^{2}-2ab\left( \arctan \frac{a-p_{0}}{b}+\arctan \frac{a+p_{0}}{b}\right) \nonumber \\
 &  & -\frac{1}{2}(a^{2}-b^{2}+p^{2})\left( \log (b^{2}+(a-p_{0})^{2})+\log (b^{2}+(a+p_{0})^{2})\right) ,\nonumber 
\end{eqnarray}
\begin{eqnarray}
F_{G}^{Im}(p_{0}) & \equiv  & \int dp_{0}\left( \frac{b}{(a-p_{0})^{2}+b^{2}}-\frac{b}{(a+p_{0})^{2}+b^{2}}\right) \bar{I}_{G}\\
 & = & -(a^{2}-b^{2}-p^{2})\left( \arctan \frac{a-p_{0}}{b}+\arctan \frac{a+p_{0}}{b}\right) \nonumber \\
 &  & +ab\left( \log (b^{2}+(a-p_{0})^{2})+\log (b^{2}+(a+p_{0})^{2})\right) .\nonumber 
\end{eqnarray}
Analogous results for \( F_{H}^{Re} \) and \( F_{H}^{Im} \) can
be easily derived with any symbolic integrator. We combine these integrals
according to equation (\ref{integralsignepsilon1}) to form the result
in general complex plane (\( X=G\textrm{ or }H \)):\begin{eqnarray}
\textrm{Re}\Pi _{X}(a+ib,\mathbf{p}) & \! \! =\! \!  & \left( -\frac{\widetilde{g}^{2}}{\widetilde{\pi }}\right) \frac{g^{2}N_{f}}{2\pi ^{2}}\int _{0}^{\infty }dk\left( n(k)-\frac{1}{2}\right) \frac{1}{2p}\\
 & \! \! \!  & \times \left( F_{X}^{Re}(p)-F_{X}^{Re}(-p)-F_{X}^{Re}(2k+p)+F_{X}^{Re}(2k-p)\right) ,\nonumber 
\end{eqnarray}
\begin{eqnarray}
\textrm{Im}\Pi _{X}(a+ib,\mathbf{p}) & \! \! =\! \!  & -\left( -\frac{\widetilde{g}^{2}}{\widetilde{\pi }}\right) \frac{g^{2}N_{f}}{2\pi ^{2}}\int _{0}^{\infty }dk\left( n(k)-\frac{1}{2}\right) \frac{1}{2p}\\
 & \! \! \!  & \! \! \! \times \left( F_{X}^{Im}(p)-F_{X}^{Im}(-p)-F_{X}^{Im}(2k+p)+F_{X}^{Im}(2k-p)\right) .\nonumber 
\end{eqnarray}
For \( b\neq 0 \) the expressions in the logarithms are positive,
and the arc tangent functions stay finite, so that there is no danger
of crossing a branch cut or encountering other singularities. Therefore
these expressions are safe to be used in calculating the arc between
Minkowski and Euclidean space.

Again, to subtract the vacuum part, just replace \( \left( n(k)-\frac{1}{2}\right)  \)
by \( n(k) \).

\section{Landau pole ambiguity\label{section_landaupoleambiguity}}

In the pressure formula (\ref{PNLO}) that we intend to integrate,
we had two different kinds of contributions that we called {}``\( n_{b} \)''
and {}``non-\( n_{b} \)''. We already mentioned that the former
are safe to be calculated in Minkowski space: for large \( q_{0} \)
the bosonic distribution function suppresses the integrand exponentially
by \( O(\exp (-q_{0}/T)) \), while for large \( q \) but moderate
\( q_{0} \), the thermal self-energy has an exponentially small imaginary
part \( \im \Pi \sim \exp (-(q-q_{0})/2T) \) as can be seen from
the exact result for the imaginary part of the self energy in Minkowski
space (\ref{imaginaryselfenergy1}), knowing that the functions \( F_{i} \)
from (\ref{functionsF123finiteT}) vanish exponentially as their argument
increases. For the {}``non-\( n_{b} \)'' parts in (\ref{PNLO})
the situation is more subtle.

The {}``non-\( n_{b} \)'' parts are not exponentially suppressed
for large \( q_{0} \). Also, in the region where \( q_{0}\sim q \),
the self-energy functions are only suppressed by \( 1/q \) so that
they might cause UV divergencies that should be studied. It is best
to go into Euclidean space and study the relevant terms of the pressure
function (\ref{PNLO}). In the limit of small \( T^{2}/Q^{2} \),
the logarithms involved can be expanded, giving\begin{eqnarray}
2\ln \lefteqn {\frac{Q^{2}+\Pi _{\vac }+\Pi _{T}}{Q^{2}+\Pi _{\vac }}+\ln \frac{Q^{2}+\Pi _{\vac }+\Pi _{L}}{Q^{2}+\Pi _{\vac }}=} &  & \nonumber \\
 & = & \frac{2\Pi _{T}+\Pi _{L}}{Q^{2}+\Pi _{\vac }}-\frac{2\Pi _{T}^{2}+\Pi _{L}^{2}}{2(Q^{2}+\Pi _{\vac })^{2}}+...\label{logarithmseriesexpansion} 
\end{eqnarray}
The functions \( \Pi _{T} \) and \( \Pi _{L} \) can be expanded
using (\ref{selfenergy_G}) and (\ref{selfenergy_H}) and the expressions
for \( \textrm{Re}\widetilde{\Pi }_{G}(i\omega ,\mathbf{p}) \) (\ref{selfenergy_Euclid_RePiG})
and \( \textrm{Re}\widetilde{\Pi }_{H}(i\omega ,\mathbf{p}) \) (\ref{selfenergy_Euclid_RePiH}).
One can expand the two functions in the limit \( T^{2}\ll Q^{2} \)
by series expansion of the expression within the brackets in terms
of small \( k \) and then performing the \( k \) integration for
each order of \( k \) separately. The result is given by \cite{Moore:2002md}
\begin{equation}
\left. \frac{\textrm{Re}\widetilde{\Pi }_{G}(i\omega ,\mathbf{p})}{2\geff ^{2}}\right| _{Q\gg T}\! \! \! \! \! \! =\frac{7\pi ^{2}\left( 3\omega ^{2}\! -q^{2}\right) }{45(Q^{2})^{2}}T^{4}\! -\frac{248\pi ^{4}(5\omega ^{4}\! -10\omega ^{2}q^{2}\! +q^{4})}{315(Q^{2})^{4}}T^{6}\! +O(T^{8}),
\end{equation}
\begin{equation}
\left. \frac{\textrm{Re}\widetilde{\Pi }_{H}(i\omega ,\mathbf{p})}{2\geff ^{2}}\right| _{Q\gg T}\! \! =\frac{7\pi ^{2}q^{2}}{45(Q^{2})^{2}}T^{4}-\frac{248\pi ^{4}q^{2}(5\omega ^{2}-q^{2})}{945(Q^{2})^{4}}T^{6}+O(T^{8}).
\end{equation}
The integration over the first term of the series (\ref{logarithmseriesexpansion})
with the vacuum self-energy from (\ref{selfenergy_vacuum_1}) then
gives\begin{equation}
\label{expansionintegral4dimensional}
\frac{1}{2}\int \! \frac{d^{4}Q}{(2\pi )^{4}}\frac{2\Pi _{T}\! +\Pi _{L}}{Q^{2}+\Pi _{\vac }}\simeq \frac{7\pi ^{2}T^{4}\geff ^{2}}{45}\! \! \int \! \frac{d^{4}Q}{(2\pi )^{4}}\frac{3\omega ^{2}\! -q^{2}}{(Q^{2})^{3}}\frac{1}{1\! -\geff ^{2}\ln (Q^{2}\! /\muMS ^{2})/12\pi ^{2}}.
\end{equation}
By simple power counting we see that this integral is potentially
logarithmicly divergent. However, if we perform the angular integral
over the 4-sphere first, we see that the contribution \( \int d\Omega _{4}(3\omega ^{2}-q^{2})=0 \)
vanishes for any distance \( |Q| \). Since the rest of (\ref{expansionintegral4dimensional})
is a function of \( Q^{2} \) only, any regularization or cutoff which
respects Euclidean invariance will be UV well behaved. It is important
to average over the Euclidean four spheres first, before integrating
over the radius \( |Q| \). If we first perform one of the integrations
\( \omega  \) or \( q \) first, and then integrate over the other
variable without respecting Euclidean invariance, we might get spurious
logarithmic divergencies. Of course, as long as we choose the upper
integration limits such that in the end we integrate over the complete
4-sphere, nothing should happen (except for numerical cancellations
to become more demanding). Integrating over a 4-dimensional cylinder
in the \( \omega /q \) space, or cutting out a four-dimensional cube
might bring in potentially logarithmic contributions for a large cutoff
\( Q \). But numerically it is easiest to integrate over the surface
of the 4-sphere for a fixed \( Q \), and then integrate over the
radius up to the cutoff. 

Introducing a cutoff might cause gauge fixing dependence. For Lorentz
and Coulomb gauges this turns out not to be the case because the self-energy
is gauge invariant as large \( N_{f} \) is basically an Abelian theory
up to NLO. By the very same procedure as above, it can furthermore
be shown that the introduction of an Euclidean invariant cutoff can
be applied safely also to finite chemical potential. 

As in references \cite{Moore:2002md,Ipp:2003zr} we apply a cutoff
and stop the \( d^{4}Q \) integration at \( Q^{2}=a\Lambda _{\textrm{L}}^{2} \),
varying the value of \( a \) between 1/4 and 1/2 to estimate the
irreducible ambiguity. 

In reference \cite{Moore:2002md}, the {}``\( n_{b} \)'' terms
from equation (\ref{PNLO}) were calculated in Minkowski space. Terms
without \( n_{b} \) were computed along a complex frequency contour
which ran up the Minkowski axis to \( \omega _{\textrm{max}}<\Lambda _{\textrm{Landau}}\sqrt{a} \)
for some \( a<1 \), then along the great arc to Euclidean space,
and back down to \( q_{0}=\sqrt{q^{2}_{max}-q^{2}} \); finally, a
Euclidean integration of the \( n_{b} \) free term was performed
over 4-spheres in Euclidean space up to \( Q^{2}<\Lambda ^{2}_{\textrm{Landau}}a \).
In reference \cite{Ipp:2003zr} a simpler way of calculating the integral
was pursued: all pieces linear in \( n_{b} \) were calculated in
Minkowski space, and all terms without \( n_{b} \) were calculated
in Euclidean space. By actually performing calculations in both ways,
we had a rather non-trivial numerical check on the result which finally
helped to reveal a coding error in the original result published \cite{Ipp:2003zr,Moore:2002md}.
In the numerical implementation both ways turned out to agree within
numerical errors of about \( 10^{-5} \). The term after the vanishing
leading term in equation (\ref{logarithmseriesexpansion}) is of the
order \( (\Pi _{T}^{2}+\Pi _{L}^{2})^{2}/(Q^{2}+\Pi _{\vac })^{2}\sim T^{8} \)
so that the ambiguity introduced by the Landau pole is of the order
\( O(T^{8}/\Lambda _{\textrm{Landau}}^{4}) \).

\section{Results and discussion}

\begin{figure}
\selectlanguage{english}
{\centering \includegraphics{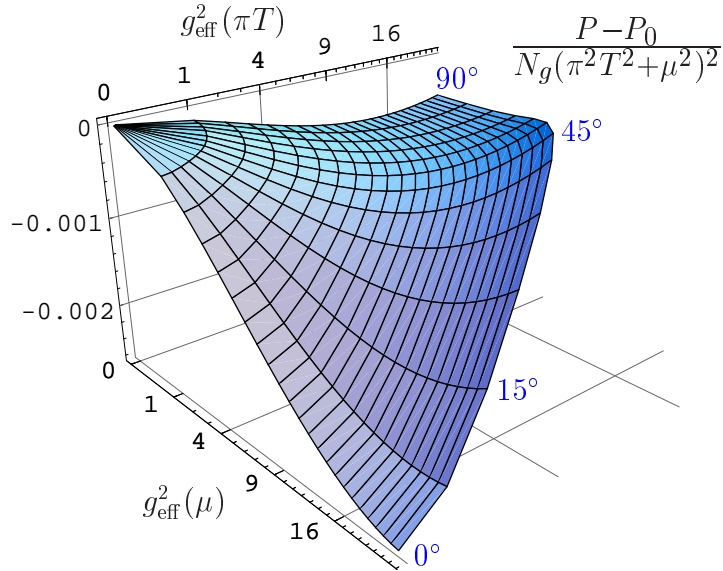} \par}

\caption{Exact result for the large-\(N_{f}\) interaction pressure \(P-P_{0}\) normalized
to \(N_{g}(\pi^{2}T^{2}+\mu^{2})^{2}\) as a function of \(\g^{2}(\bar{\mu}_{\textrm{MS}})\)
with \(\bar{\mu}_{\textrm{MS}}^{2}=\pi^{2}T^{2}+\mu^{2}\), which is the radial
coordinate, and \(\phi=\arctan{\pi T\0\mu}\).\label{fig:3d}}
\selectlanguage{american}
\end{figure}

In Fig.~\ref{fig:3d} we display our exact results%
\footnote{\selectlanguage{english}
Tabulated results are available on-line at \href{http://hep.itp.tuwien.ac.at/~ipp/data/}{\texttt{http://\is hep.itp.tuwien.ac.at/\is \~{}ipp/\is data/}}.
\selectlanguage{american}
} for the interaction pressure \( P-P_{0}\propto N_{f}^{0} \), where
the ideal-gas limit \begin{equation}
\label{largenfP0}
P_{0}=NN_{f}\left( {7\pi ^{2}T^{4}\0180 }+{\mu ^{2}T^{2}\06 }+{\mu ^{4}\012 \pi ^{2}}\right) +N_{g}{\pi ^{2}T^{4}\045 }
\end{equation}
 has been subtracted, for the entire \( \mu  \)-\( T \) plane (but
reasonably below the scale Landau pole). For this we introduce an
angle \( \phi =\arctan {\pi T\0 \mu } \) and encode the magnitudes
\( T/\Lambda _{\textrm{L}} \) and \( \mu /\Lambda _{\textrm{L}} \)
through the running coupling \( \g ^{2}(\bar{\mu }_{\textrm{MS}}) \)
with \( \bar{\mu }_{\textrm{MS}}^{2}=\pi ^{2}T^{2}+\mu ^{2} \) according
to (\ref{LLandau}).

We found that the ambiguity arising from the presence of a Landau
pole reaches the percent level for \( \g ^{2}\gtrapprox 28 \), where
\( \Lambda _{\textrm{L}}/\sqrt{\pi ^{2}T^{2}+\mu ^{2}}\lessapprox 19 \).
At larger coupling (corresponding to larger \( T \) and/or \( \mu  \)),
this ambiguity grows rapidly and will be shown in the two-dimensional
plots below by a (tiny) red area.

In the following we shall compare the exact large-\( N_{f} \) result
with known results from perturbation theory at high temperature and
small chemical potential, where dimensional reduction is an effective
organizing principle, and with results at zero temperature, where
dimensional reduction does not apply. We also investigate to what
extent quark number susceptibilities at vanishing chemical potential
determine the behavior at larger chemical potential.

\subsection{Comparison to dimensional reduction}

Dimensional reduction is a method of expressing the static properties
of a (3+1)-dimensional field theory at high temperature in terms of
an effective field theory in 3 space dimensions \cite{Gross:1981br,Appelquist:1981vg}.
In the imaginary time formalism, all nonstatic modes involve the temperature
scale, so integrating them out to obtain an effective field theory
on scales \( \ll T \) leaves only the static modes with zero Matsubara
frequency. The dimensionally reduced theory involves the color-electric
screening scale \( gT \) and a dimensionful coupling \( g^{2}T \)
which is also the scale of color-magnetic screening. The perturbative
result for the thermal pressure of hot gauge theories with fermions
has been obtained by using this approach in reference \cite{Braaten:1996jr}
at zero chemical potential to order \( g^{5} \). The analytic result
of dimensional reduction through order \( g^{5} \) with complete
analytic dependence on arbitrary \( T \) and \( \mu  \) was calculated
by \cite{Vuorinen:2003fs} and in the large \( N_{f} \) limit this
reduces to\begin{eqnarray}
\left. \frac{P-P_{0}}{N_{g}}\right| _{\rm {DR}} & = & -\left( \frac{5}{9}T^{4}+\frac{2}{\pi ^{2}}\mu ^{2}T^{2}+\frac{\mu ^{4}}{\pi ^{4}}\right) \frac{\geff ^{2}}{32}\nonumber \\
 &  & +T\frac{1}{12\pi }m_{E}^{3}+\frac{\geff ^{4}}{(48\pi )^{2}}\bar{\alpha }_{\rmi {E3}}+O(\geff ^{6}T^{4})
\end{eqnarray}
where the effective field theory parameter \( m_{E}^{2} \) is given
by\begin{equation}
\label{mE2new}
\frac{m_{E}^{2}}{T^{2}}=\left( 1+\frac{3\mu ^{2}}{\pi ^{2}T^{2}}\right) \left[ \frac{\geff ^{2}}{3}-\frac{\geff ^{4}}{(6\pi )^{2}}\left( 2\ln \frac{\muMS }{4\pi T}-1-\aleph (z)\right) \right] +O(\geff ^{6})
\end{equation}
and the coefficient of order \( g^{4} \) is analytic in \( T \)
and \( \mu  \) and given by\begin{eqnarray}
\bar{\alpha }_{\rmi {E3}} & = & 12\left\{ \frac{5}{9}T^{4}+\frac{2}{\pi ^{2}}\mu ^{2}T^{2}+\frac{\mu ^{4}}{\pi ^{4}}\right\} \ln \frac{\muMS }{4\pi T}\nonumber \\
 &  & +4T^{4}\left\{ \frac{1}{12}+\gamma -\frac{16}{15}\frac{\zeta '(-3)}{\zeta (-3)}-\frac{8}{3}\frac{\zeta '(-1)}{\zeta (-1)}\right\} \nonumber \\
 &  & +\frac{\mu ^{2}T^{2}}{\pi ^{2}}\left\{ 14+24\gamma -32\frac{\zeta '(-1)}{\zeta (-1)}\right\} +\frac{\mu ^{4}}{\pi ^{4}}(43+36\gamma )\nonumber \\
 &  & -96T^{4}\left\{ 3\aleph (3,1)+8\aleph (3,z)+3\aleph (3,2z)-2\aleph (1,z)\right\} \nonumber \\
 &  & +\frac{48i\mu T^{3}}{\pi }\left\{ \aleph (0,z)-12\aleph (2,z)-12\aleph (2,2z)\right\} \nonumber \\
 &  & +\frac{96\mu ^{2}T^{2}}{\pi ^{2}}\left\{ 4\aleph (1,z)+3\aleph (1,2z)\right\} +\frac{144i\mu ^{3}T}{\pi ^{3}}\aleph (0,z)\label{coefficient_aE3_largeNf} 
\end{eqnarray}
with the definitions from \cite{Vuorinen:2003fs}\begin{eqnarray}
z & \equiv  & \frac{1}{2}-i\frac{\mu }{2\pi T},\\
\zeta '(x,y) & \equiv  & \partial _{x}\zeta (x,y),\\
\aleph (n,w) & \equiv  & \zeta' (-n,w)+(-1)^{n+1}\zeta '(-n,w^{*}),\\
\aleph (w) & \equiv  & \Psi (w)+\Psi (w^{*}).
\end{eqnarray}
Here, \( n \) is assumed to be a non-negative integer and \( w \)
a general complex number. \( \zeta  \) is the Riemann zeta function
and \( \Psi  \) is the digamma function \( \Psi (w)\equiv \Gamma '(w)/\Gamma (w) \).
Note that despite the appearance of complex quantities in (\ref{coefficient_aE3_largeNf})
the coefficient \( \bar{\alpha }_{\rmi {E3}} \) is real as it has
to be.

For later usage we also present the effective field theory parameter
of the coupling from dimensional reduction\begin{equation}
\geff ^{2}{}_{E}=T\left[ \geff ^{2}+\frac{\geff ^{4}}{3(2\pi )^{2}}\left( 2\ln \frac{\muMS }{4\pi T}-\aleph (z)\right) \right] +O(\geff ^{6}).
\end{equation}

The expansion of the pressure in terms of small \( \mu /T \) is given
by\begin{eqnarray}
{P-P_{0}\0N _{g}}\Big |_{T\gg \mu } & \! \! \! \! \! = & \! \! -\left[ {5\09 }T^{4}+{2\0 \pi ^{2}}\mu ^{2}T^{2}+{1\0 \pi ^{4}}\mu ^{4}\right] {\g ^{2}\032 }\nonumber \label{pintdr} \\
 & \! \! + & \! \! \! {1\012 \pi }Tm_{E}^{3}+\biggl [\left( {20\03 }T^{4}+{24\0 \pi ^{2}}T^{2}\mu ^{2}\right) \ln {\bar{\mu }_{\textrm{MS}}\04 \pi T}\nonumber \\
 & \! \! + & \! \! \! \left( {1\03 }-{88\05 }\ln 2+4\gamma -{8\03 }{\zeta' (-3)\0 \zeta (-3)}+{16\03 }{\zeta' (-1)\0 \zeta (-1)}\right) T^{4}\nonumber \\
 & \! \! - & \! \! \! {26+32\ln 2-24\gamma \0 \pi ^{2}}T^{2}\mu ^{2}\nonumber \\
 & \! \! + & \! \! \! {12\mu ^{4}\0 \pi ^{4}}\! \left[ \ln {\bar{\mu }_{\textrm{MS}}\04 \pi T}\! +\! \gamma \! +\! C_{4}\right] +\ldots \biggr ]{\g ^{4}\0 (48\pi )^{2}}+O(\g ^{6}T^{4}),\qquad \quad \label{pressureDRorder5} 
\end{eqnarray}
where the terms \( \propto \g ^{4} \) and involving \( \mu  \) have
been first obtained by Vuorinen \cite{Vuorinen:2002ue}. The contribution
to order \( g^{5} \) arises from the NLO correction to the effective-field-theory
parameter \( m_{E}^{2} \), which at finite \( \mu  \) was first
computed in reference \cite{Hart:2000ha}. An alternative way of writing
(\ref{mE2new}) is given by\begin{equation}
\label{mE2}
{m_{E}^{2}\0T ^{2}}=\left( {1\03 }+{\mu ^{2}\0 \pi ^{2}T^{2}}\right) \g ^{2}\left\{ 1-{\g ^{2}\06 \pi ^{2}}\left[ \ln {\bar{\mu }_{\textrm{MS}}\0e ^{1/2-\gamma }\pi T}+{1\02 }{\mathcal{D}}({\mu \0 \pi T})\right] \right\} +O(\g ^{6})
\end{equation}
 with the function \begin{equation}
\mathcal{D}(x)=-2\gamma -4\ln 2-2\, \textrm{Re}\, \Psi (\frac{1}{2}(1+ix)).
\end{equation}
 For small \( x \) this function can be expanded as \begin{equation}
\mathcal{D}(x)=4\sum _{n=1}^{\infty }(-1)^{n}\left( 1-\frac{1}{2^{2n+1}}\right) \zeta (2n+1)\, x^{2n}\, ,
\end{equation}
 with a radius of convergence of 1, which corresponds to \( \mu =\pi T \).
The only nonanalytic terms in \( \g ^{2} \) to the pressure in the
dimensionally reduced effective field theory at large \( N_{f} \)
stem from the {}``plasmon term'' \( \propto m_{E}^{3} \). As can
be seen from the expansion in figure \ref{fig:perturbativeQCD} of
the perturbative QCD pressure, the logarithms \( \ln (g) \) do not
survive the large \( N_{f} \) limit.

\begin{figure}
{\centering \includegraphics{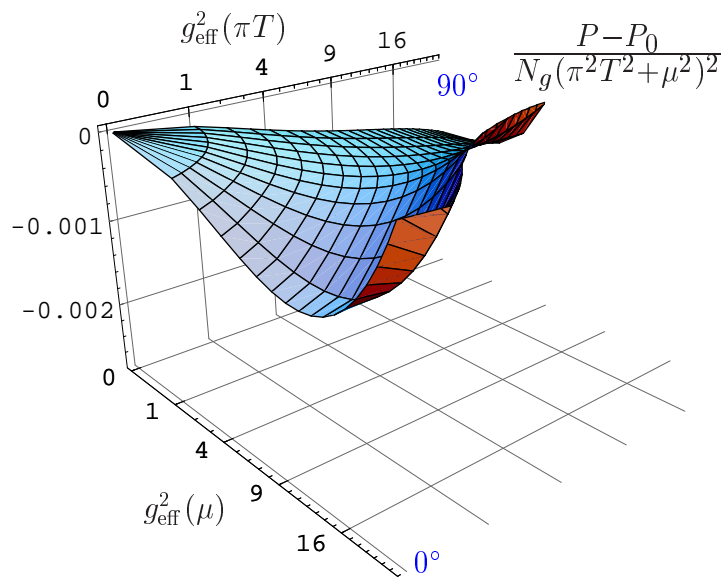} \par}

\caption{Dimensional reduction (DR) result of the pressure \protect\( (P_{\DR }-P_{0})/(N_{g}(\pi ^{2}T^{2}+\mu ^{2})^{2})\protect \)
in the large \protect\( N_{f}\protect \) limit in the same range
as in figure \ref{fig:3d}. The renormalization scale is chosen as
\protect\( \muMS =\sqrt{(\pi T)^{2}+\mu ^{2}}\protect \) for \protect\( \pi T=\muMS \sin \phi \protect \)
and \protect\( \mu =\muMS \cos \phi \protect \). \label{fig:3D_DR}}
\end{figure}
In figure \ref{fig:3D_DR} we plot the result obtained from dimensional
reduction by Vuorinen \cite{Vuorinen:2003fs} in the large \( N_{f} \)
limit. In this plot we choose a naive renormalization scale of \( \muMS =\sqrt{(\pi T)^{2}+\mu ^{2}} \).
By comparison to figure \ref{fig:3d} we see that the small coupling
range is well reproduced for couplings \( \geff ^{2}\lesssim 9 \)
for smaller \( \mu /T \) but deviates considerably for larger \( \mu /T \).
This is partly due to the choice of the renormalization scale which
apparently is not the best. The exact result is independent of the
choice of the initial renormalization scale. More precisely the scale
dependence of the exact result enters only via the coupling \( \geff (\muMS ) \)
and results can be easily converted from one renormalization scale
to another by applying the renormalization scale dependence relation
(\ref{gscal}). In the perturbative result this holds true only up
to the order of perturbative accuracy. Numerically, due to truncation
to a certain order in the coupling, the pressure not only depends
on the renormalization scale indirectly via the coupling \( \geff (\muMS ) \),
but also directly, since higher order contributions that would fix
the renormalization scale dependence are not taken into account. The
result will therefore in general differ whether we started from a
renormalization scale \( \bar{\mu }_{1} \) and rescaled the resulting
pressure according to (\ref{gscal}) to a renormalization scale \( \bar{\mu }_{1}\rightarrow \bar{\mu }_{2} \),
or if we directly calculated the perturbative result with the latter
renormalization scale \( \bar{\mu }_{2} \).\rem{ Again, for the exact
result this does not make any difference, and we can calculate the
pressure for the exact NLO contribution in any renormalization scale
we want, since we can always rescale it exactly to another renormalization
scale by equation (\ref{gscal}).} Therefore, for perturbative results
there is this additional question of which renormalization scale to
choose.

A possible optimization is provided by the FAC (fastest apparent convergence)
scale, which is derived by demanding that the effective-field-theory
parameter \( m_{E}^{2} \) from (\ref{mE2}) vanishes to order \( \geff ^{4} \)
and all higher orders. This means solving \( \ln {\bar{\mu }_{\textrm{MS}}\0e ^{1/2-\gamma }\pi T}+{1\02 }{\mathcal{D}}({\mu \0 \pi T}) \)
so that we obtain \begin{equation}
\label{FACm}
\muMS (\FACm )=\pi T\exp \left[ \frac{1}{2}-\gamma _{E}-\frac{1}{2}\mathcal{D}(\frac{\mu }{\pi T})\right] .
\end{equation}
For \( \mu =0 \) this expression reduces to \( \muMS (\FACm )=\pi Te^{\frac{1}{2}-\gamma _{E}} \)
while for \( T=0 \) it is given by \( \muMS (\FACm )=2\mu e^{\frac{1}{2}} \).

\begin{figure}
{\centering \resizebox*{12cm}{!}{\includegraphics{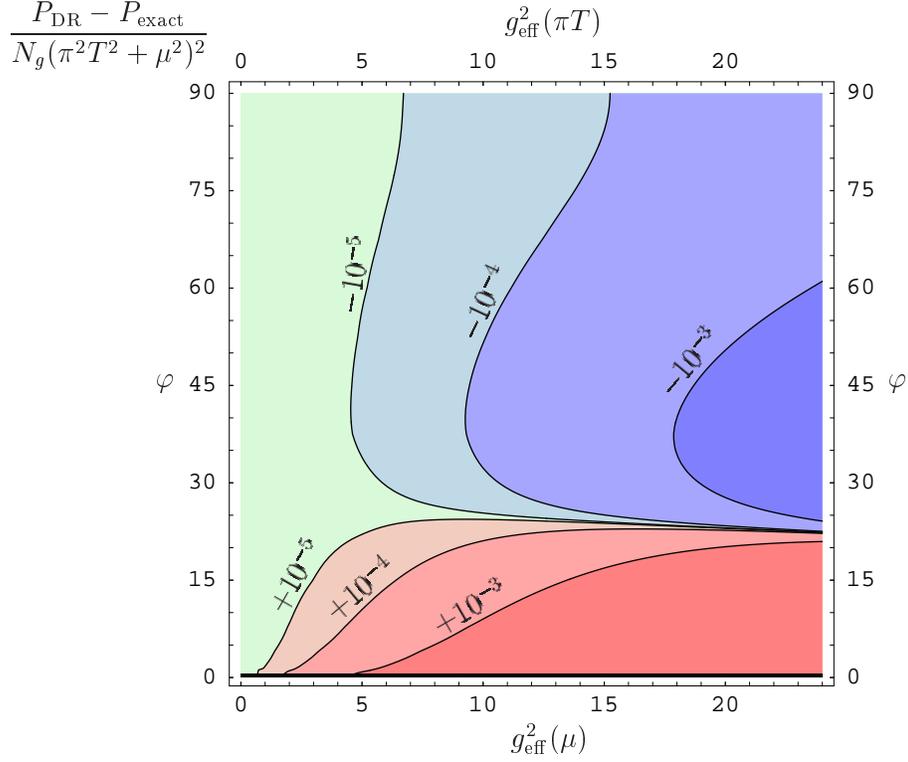}} \par}

\caption{Difference between the pressure obtained from dimensional reduction
to the exact large \protect\( N_{f}\protect \) result for the FAC
(fastest apparent convergence) renormalization scale \protect\( \muMS (\FACm )=\pi T\exp (\frac{1}{2}-\gamma _{E}-\frac{1}{2}\mathcal{D}(\frac{\mu }{\pi T}))\protect \)
for different values of \protect\( \phi =\arctan \frac{\pi T}{\mu }\protect \)
and \protect\( \geff ^{2}\protect \), normalized to \protect\( N_{g}(\pi ^{2}T^{2}+\mu ^{2})^{2}\protect \).
The result of the dimensionally reduced pressure is rescaled by the
exact renormalization scale dependence (\ref{gscal}) to the scale
\protect\( \muMS =\sqrt{(\pi T)^{2}+\mu ^{2}}\protect \) to allow
for unified comparison. \label{fig:contourFACm}}
\end{figure}
In figure \ref{fig:contourFACm} we see the difference between the
pressure obtained from dimensional reduction to the exact large \( N_{f} \)
result for this \( \FACm  \) renormalization scale \cite{Ipp:2003yz}.
As in the three-dimensional plots we vary temperature \( T \) and
chemical potential \( \mu  \) depending on the angle \( \phi =\arctan \frac{\pi T}{\mu } \).
The result is given for values of the coupling \( 0\leq \geff ^{2}\leq 24 \)
where the ambiguity introduced by the Landau pole is negligible numerically.
In the plot we rescaled the dimensionally reduced pressure by the
exact renormalization scale dependence (\ref{gscal}) to the scale
\( \muMS =\sqrt{(\pi T)^{2}+\mu ^{2}} \) to allow for unified comparison.
The lines enclose those areas where the difference is less than \( \pm 10^{-5} \),
\( \pm 10^{-4} \), or \( \pm 10^{-3} \) in units of the normalized
pressure. \( 10^{-3} \) is about the size of the NLO results, and
thus marks the breakdown of the agreement between the two approaches
(\( 100\% \) deviation). Notice that there is always a factor 10
of increase of the inaccuracy between neighboring lines of the contour
plot. We have very good agreement for \( \geff ^{2}\lesssim 7 \),
but this value slowly decreases as the ratio \( \mu /T \) is increased.
Only for small angles \( \phi \lesssim 20^{\circ } \) the behavior
changes considerably and the ranges of good agreement shrink faster
with decreasing \( \mu /T \). For the FAC-m scale dimensional reduction
gives lower values of pressure than the exact large \( N_{f} \) result
for \( \phi \gtrsim 25^{\circ } \) and higher predictions for smaller
\( \mu /T \). In between there is a transition from negative to positive
differences which of course is of no real relevance to the error ranges
in the plot.

\begin{figure}
{\centering \resizebox*{12cm}{!}{\includegraphics{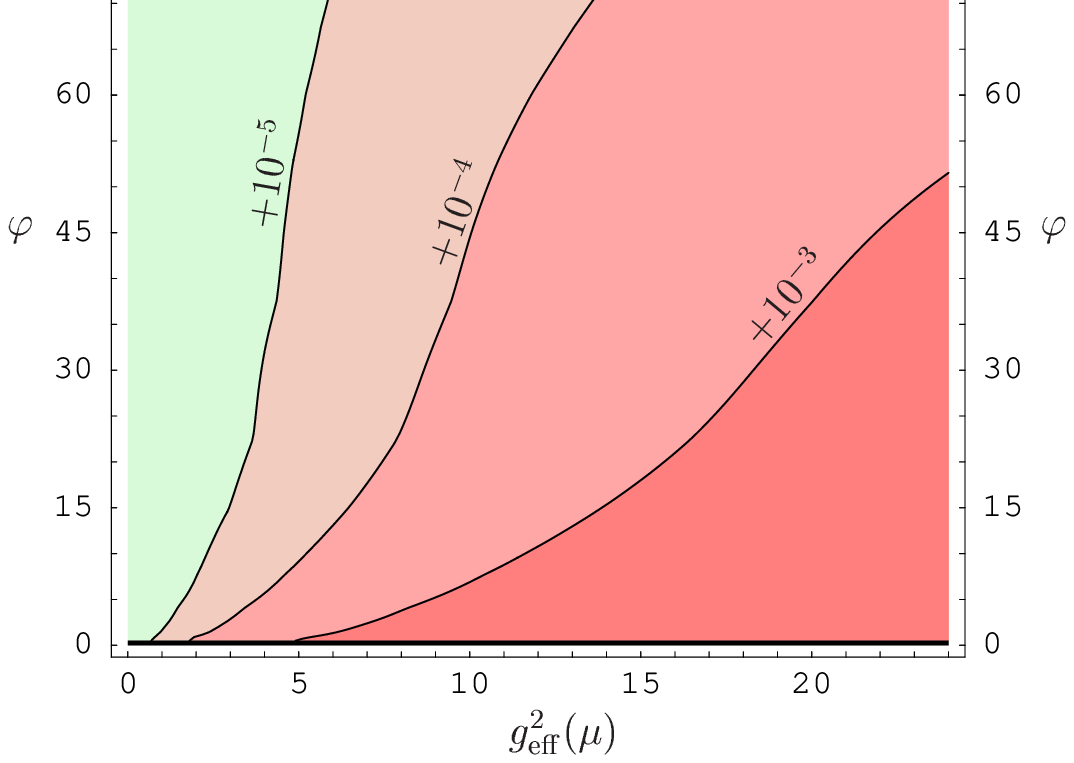}} \par}

\caption{Same as figure \ref{fig:contourFACm} except that the scale is now
FAC-g instead of FAC-m, that is applying fastest apparent convergence
to \protect\( g_{E}^{2}\protect \). Contrary to the FAC-m scale,
the error is now always positive, meaning that dimensional reduction
at this renormalization scale always overestimates the exact result.
Apart from that the error of the FAC-g scale is comparable to the
one of the FAC-m scale. \label{fig:contourFACg}}
\end{figure}
Another natural choice for scale might be to apply FAC to \( g_{E}^{2} \).
This produces in the \( N_{f}\rightarrow \infty  \) limit\begin{equation}
\label{FACg}
\muMS (\FACg )=\pi T\exp \left[ -\gamma _{E}-\frac{1}{2}\mathcal{D}(\frac{\mu }{\pi T})\right] 
\end{equation}
 which curiously is just a factor \( 1/\sqrt{e} \) smaller than the
\( \FACm  \) scale but shows otherwise exactly the same functional
dependence on \( \mu  \) and \( T \). 

Figure \ref{fig:contourFACg} shows the difference of the pressures
now for the FAC-g scale. It is conspicuous that this time there is
no sign change in the plot. Indeed, contrary to what we saw in figure
\ref{fig:contourFACm} for the FAC-m scale, the difference between
\( P_{\DR } \) and \( P_{\rm exact}=P_{\NLO } \) is now always positive,
meaning that dimensional reduction at this renormalization scale always
overestimates the exact result. But apart from that, the absolute
value of the errors of the FAC-g scale plot is comparable to the one
of the FAC-m scale. One should not be irritated by the horizontal
line in figure \ref{fig:contourFACm} at \( \phi \approx 25^{\circ } \)
where the range of small error is extended to larger values of the
coupling \( \geff ^{2} \): This merely indicates the change of sign
of the error.

\begin{figure}
{\centering \resizebox*{12cm}{!}{\includegraphics{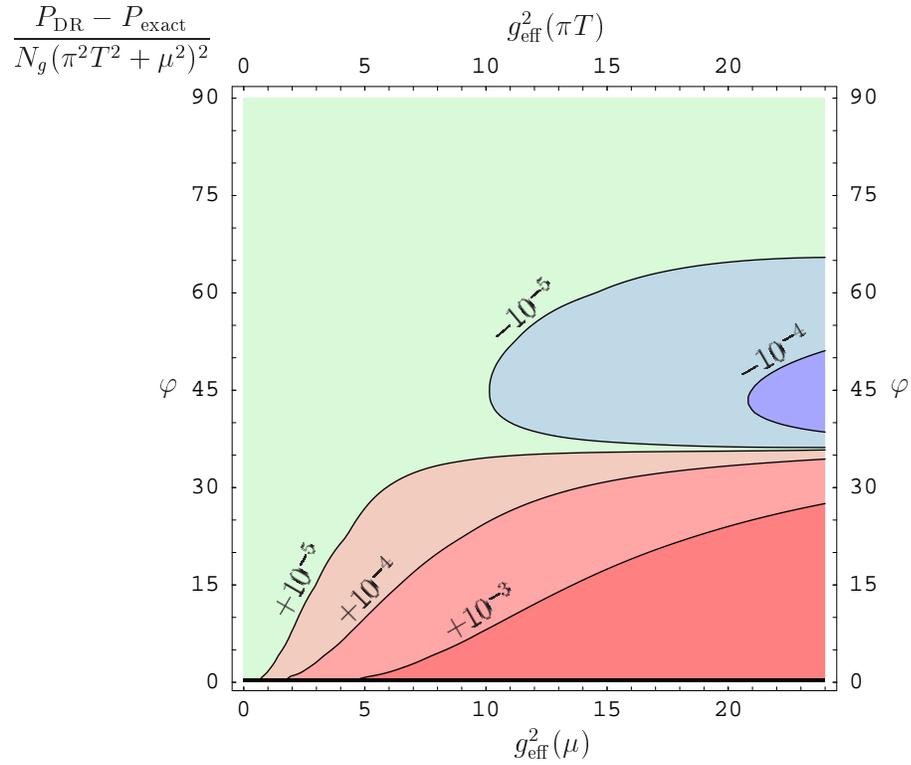}} \par}

\caption{Same as figures \ref{fig:contourFACm} and \ref{fig:contourFACg}
except that the scale is now chosen as the arithmetic mean of the
two scales FAC-m and FAC-g. We see remarkable agreement for \protect\( \phi \gtrsim 65^{\circ }\protect \)
over the whole range of couplings down to \protect\( \phi \gtrsim 35^{\circ }\protect \)
for \protect\( \geff ^{2}\lesssim 10\protect \). Below \protect\( \phi \lesssim 30^{\circ }\protect \)
the spreading of errors does not differ much from that of the scales
FAC-m and FAC-g separately. \label{fig:contourFAC0803265}}
\end{figure}
Having such good agreement for a wide range of \( \phi  \) and \( \geff ^{2} \),
it seems tempting to try to improve the result by averaging over the
two scales: Taking the arithmetic mean of the two scales FAC-m and
FAC-g we obtain a scale that is a factor \( (1+1/\sqrt{e})/2\approx 0.803265.. \)
smaller than the FAC-m scale. Remember that the FAC-m and FAC-g scale
are directly proportional to each other with \( \muMS (\FACg )=\muMS (\FACm )/\sqrt{e}\approx 0.606531\muMS (\FACm ) \)
so that the arithmetic mean scale is also proportional to them. 

Figure \ref{fig:contourFAC0803265} shows the remarkable agreement
that follows from this choice of scale: The thermal pressure of dimensional
reduction agrees to the exact large \( N_{f} \) result for \( \phi \gtrsim 65^{\circ } \)
over the whole range of couplings. If we look at smaller couplings
\( \geff ^{2}\lesssim 10 \), the two results agree perfectly down
to \( \phi \gtrsim 35^{\circ } \). Only below \( \phi \lesssim 30^{\circ } \)
the spreading of errors does not differ much from that of the scales
FAC-m and FAC-g separately. One should note however, that this might
be a mere coincidence in the large \( N_{f} \) limit and that no
real conclusion can be inferred from this result for finite, smaller
\( N_{f} \). After all, for finite \( N_{f} \) FAC-m and FAC-g are
not necessarily proportional to each other (except for \( \mu =0 \)
where both are simply proportional to the temperature \( T \)).

\subsection{Pressure at zero chemical potential\label{section_pressurezeromu}}

\begin{figure}
\selectlanguage{english}
{\centering \includegraphics{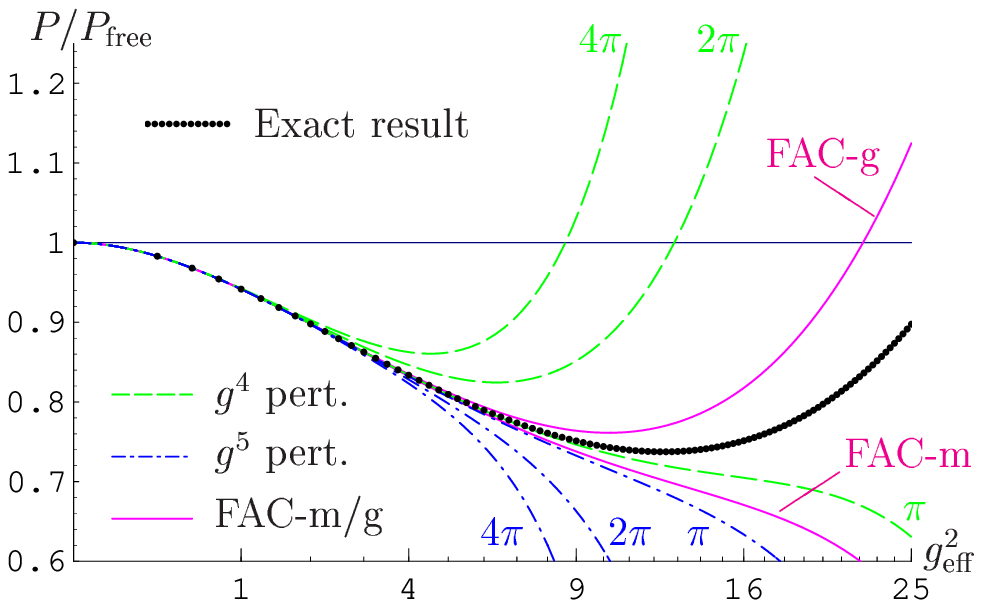} \par}

\caption{Exact result for \protect\( P_{\NLO}/P_{\textrm{free}}\protect \)
as a function of \(g^{2}_{\textrm{eff}}(\bar{\mu}=\pi e^{-\gamma_{E}}T)\),
rendered with an abscissa linear in \(g_{\textrm{eff}}\), in comparison
with the perturbative results through order \protect\( g^{4}\protect \) and \protect\( g^{5}\protect \) with a renormalization point chosen within a power of 2
of \(2\pi T\). The line marked ``FAC-m'' corresponds to \(\bar{\mu}=\pi e^{{1\over2}-\gamma_{E}}T\) where the perturbative result to order \(g^{4}\) coincides with the one to order \(g^{5}\). The
line marked {}``FAC-g'' corresponds to the renormalization point
\protect\( \bar{\mu }=\pi e^{-\gamma _{E}}T\protect \) and shows
the result through order \protect\( g^{5}\protect \). \label{figurePnlo}}
\selectlanguage{american}
\end{figure}

\begin{figure}
\selectlanguage{english}
{\centering \includegraphics{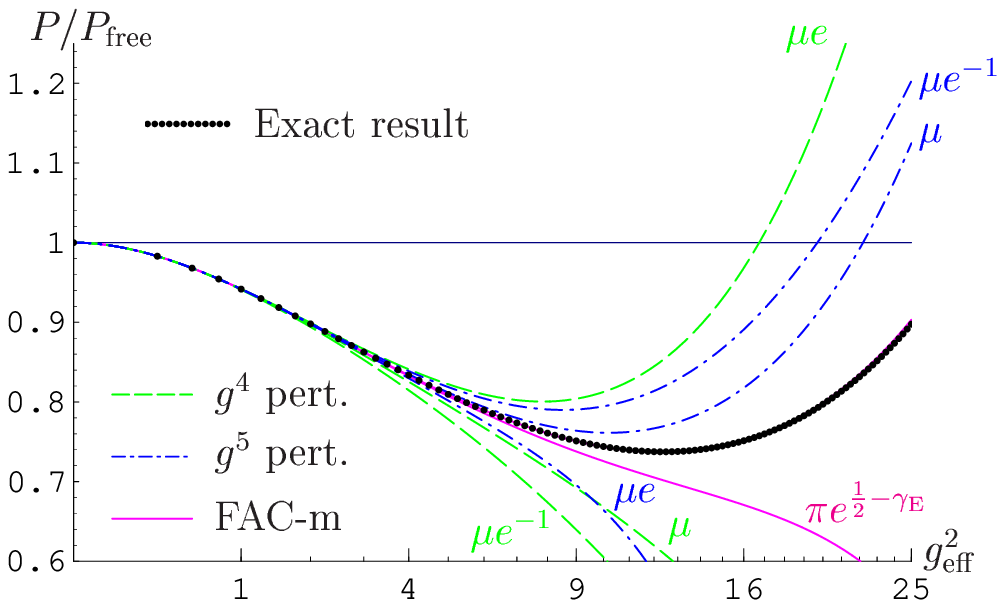} \par}

\caption{Same as figure \ref{figurePnlo} only with renormalization point
chosen within a power of \(e\) of \(\mu=\pi e^{-\gamma_{E}}T\). The
FAC-g result is not explicitly marked, but coincides with the scale
choice of \protect\( \mu \protect \) in this plot. Therefore the
central line of the \protect\( g^{5}\protect \) result shows the
FAC-g up to order \protect\( g^{5}\protect \), and the central line
of the \protect\( g^{4}\protect \) result shows the FAC-g result
up to order \protect\( g^{4}\protect \). Note that unlike the FAC-m
result, the FAC-g result differs considerably from order \protect\( g^{4}\protect \)
to order \protect\( g^{5}\protect \). Plotted in this graph is also
the arithmetic mean of the two FAC scales as described in the text.
This line is almost indistinguishable from the exact result. \label{figurePnlo2}}
\selectlanguage{american}
\end{figure}

At zero chemical potential we shall study in more detail the convergence
properties and renormalization scale dependences of perturbation theory
on the one hand, and the Landau pole ambiguity of our large \( N_{f} \)
result on the other hand. In figure \ref{figurePnlo} we give the
numerical result of the thermal pressure as a function of \( g^{2}_{\textrm{eff}}(\bar{\mu }=\pi e^{-\gamma _{E}}T) \).
The perturbative results depend on the value of the renormalization
point \( \bar{\mu } \), which we vary between \( \pi T \) and \( 4\pi T \).
In order to compare the different perturbative results with the exact
large \( N_{f} \) result, we use the exact running coupling (\ref{gscal})
to rescale everything to \( \bar{\mu }=\pi e^{-\gamma _{E}}T \).
We see that the perturbative result through order \( g^{5} \) gives
reliable results only for effective couplings \( \geff ^{2}\lesssim 4 \).
If the perturbative result to order \( g^{5} \) is optimized by fastest
apparent convergence of \( m_{E}^{2} \) (FAC-m) from equation (\ref{FACm})
which for \( \mu =0 \) means the scale \( \bar{\mu }=\pi e^{{1\over2 }-\gamma _{E}}T \),
the agreement with perturbation theory is improved and extends to
\( g^{2}_{\textrm{eff}}\approx 7 \). Figure \ref{figurePnlo} shows
for comparison also the FAC-g scale. 

Figure \ref{figurePnlo2} shows the same thermal pressure as figure
\ref{figurePnlo}, but with the perturbative results varied between
\( \mu /e \) and \( \mu e \) where \( \mu =\pi e^{-\gamma _{E}}T \)
which is in fact the FAC-g scale. The central line of the \( g^{5} \)
result is just the FAC-g line from the previous plot. The central
line of the \( g^{4} \) result also corresponds to the FAC-g scale.
Different from the FAC-m scale, the FAC-g scale stays not the same
from order \( g^{4} \) to order \( g^{5} \), but instead changes
noticeably. Remember that the FAC-m scale was just chosen such that
the \( g^{4} \) and the \( g^{5} \) results lie on top of each other.
\foreignlanguage{english}{Figure} \ref{figurePnlo2} \foreignlanguage{english}{also
includes the arithmetic mean of the two FAC scales which is \( \frac{1}{2}(\FACm +\FACg )=\frac{1}{2}(1+1/\sqrt{e})\FACm =\frac{1}{2}(\sqrt{e}+1)\FACg  \).
It turns out that this line is almost indistinguishable from the exact
result. Only for larger values of \( \geff ^{2} \) can one see a
tiny deviation from the exact result.}

\begin{figure}
\selectlanguage{english}
{\centering \includegraphics{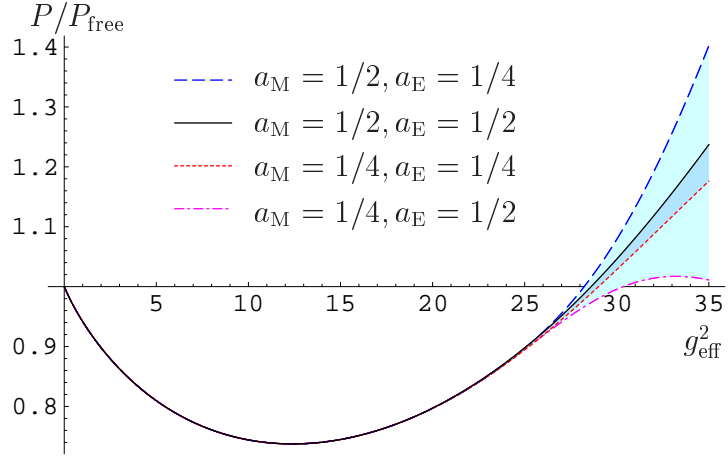} \par}

\caption{The result for \protect\( P_{NLO}/P_{\textrm{free}}\protect \)
up to \(g_{\textrm{eff}}^{2}=35\) and 
with different cutoffs.\label{figurePland}}
\selectlanguage{american}
\end{figure}
The exact large \( N_{f} \) result flattens out for higher values
of \( g^{2}_{\textrm{eff}} \) and reaches a minimum at \( g^{2}_{\textrm{eff}}\approx 12 \).
For this coupling, \( \Lambda _{\textrm{Landau}}\simeq 480T \) and
the ambiguity introduced by the Landau singularity is completely negligible.
For still higher values the pressure rises and starts to exceed the
free theory pressure at \( g_{\textrm{eff}}^{2}>28 \), as can be
seen in figure \ref{figurePland}. This occurs at a coupling for which
\( \Lambda _{\textrm{Landau}}/T<34 \). While this still seems to
be a reasonably large number, the numerical result starts to become
sensitive to the cutoff just where the pressure approaches the free
one. The four curves displayed in figure \ref{figurePland} show the
result of varying the parameter \( a \) in the UV cutoff \( \sqrt{a}\Lambda _{\textrm{Landau}} \)
in the Minkowski and Euclidean parts of the calculation (\( a_{M} \)
and \( a_{E} \) respectively) from \( a=1/4 \) to \( a=1/2 \).
The numerical result is rather insensitive to this below \( g^{2}_{\textrm{eff}}\approx 25 \),
but very sensitive in the region where the pressure starts to exceed
the free one.

\begin{figure}
\selectlanguage{english}
{\centering \includegraphics{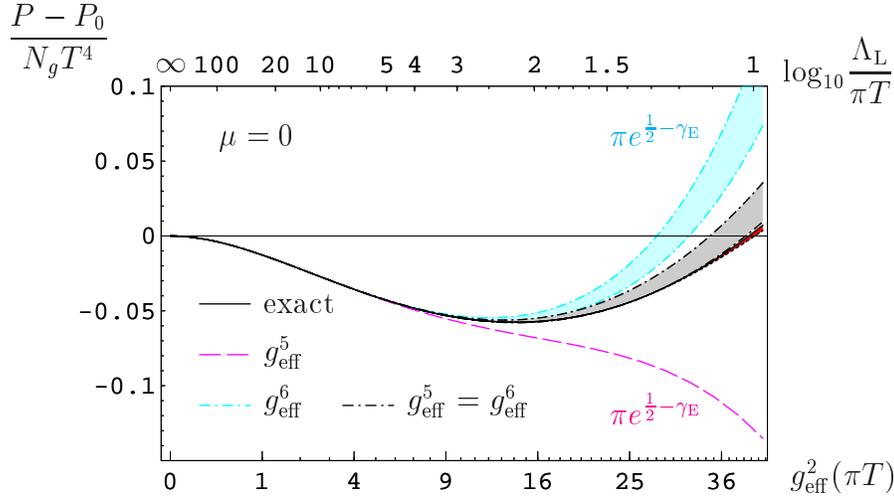} \par}

\caption{Exact result for the interaction pressure at zero
chemical potential as in references \cite{Moore:2002md,Ipp:2003zr}
but as a function of \(\g^{2}(\bar{\mu}_{\textrm{MS}}=\pi T)\) or, alternatively,
\(\log_{10}(\Lambda_{\textrm{L}}/\pi T)\). The \rem{purple }dashed line is
the perturbative result when the latter is evaluated
with renormalization scale
\(\bar{\mu}_{\textrm{MS}}=\bar{\mu}_{\textrm{FAC}}\equiv\pi e^{1/2-\gamma}T\);
the \rem{blue }dash-dotted lines include the numerically determined
coefficient to order \(\g^{6}\) (with its estimated error)
at the same renormalization scale.
The result marked ``\(\g^{5}=\g^{6}\)'' 
corresponds to choosing \(\bar{\mu}_{\textrm{MS}}\) such
that the order-\(\g^{6}\) coefficient vanishes and retaining
all higher-order terms contained in the plasmon term \(\propto m_{E}^{3}\).
In this and the following plots the (tiny) red band appearing
around the exact result at large coupling displays the
effect of varying the cut-off from 50\% to 70\% of the
Landau scale \(\Lambda_{\textrm{L}}\).\label{fig:zeromu}}
\selectlanguage{american}
\end{figure}

In figure \ref{fig:zeromu} we show the pressure at \( \mu =0 \)
and compare with even higher orders of perturbation theory. Here the
result is presented as a function of \( \g ^{2}(\bar{\mu }_{\textrm{MS}}=\pi T) \)
and not divided by the free NLO contribution from the pressure, but
with the ideal-gas limit from equation (\ref{largenfP0}) subtracted.
In the plots we also give the corresponding values of the Landau pole
\( \log _{10}(\Lambda _{\textrm{L}}/\pi T) \) to different values
of the effective coupling \( \geff ^{2} \). This plot also contains
the result up to \( \geff ^{6} \) which is not known analytically
yet, but obtained from our numerical results at small coupling. For
small coupling the agreement of our numerical results with perturbation
theory is sufficiently accurate so that we can numerically extract
coefficients to this order as follows: From the pressure in dimensional
reduction (\ref{pressureDRorder5}) we can infer by applying the exact
running coupling of large \( N_{f} \) (\ref{gscal}) that the \( \g ^{6} \)-term
in the pressure at \( \mu =0 \) has the form \begin{eqnarray}
{1\0N _{g}}P\Big |_{\g ^{6},\mu =0} & \! \! \! \! = & \! \! \! {\left( \frac{\g }{4\pi }\right) }^{6}T^{4}\left[ C_{6}+10\ln ^{2}\frac{\bar{\mu }_{\textrm{MS}}}{\pi T}\right. \label{pressurecoefficient6} \\
 & \! \! \! - & \! \! \! \! \! \left. \frac{16{\pi }^{2}}{81}\left( 1+12\gamma -\frac{464\ln 2}{5}-8\frac{\zeta '(-3)}{\zeta (-3)}+16\frac{\zeta '(-1)}{\zeta (-1)}\right) \ln \frac{\bar{\mu }_{\textrm{MS}}}{\pi T}\right] .\nonumber 
\end{eqnarray}
 Notice the appearance of the (as yet) unknown constant \( C_{6} \).
For small values of \( \geff ^{2} \), perturbation theory is expected
to give excellent agreement with the exact result, so that we can
actually use our (numerically obtained) exact result to determine
unknown coefficients in the perturbative series. By least-square fitting
we obtain numerically the estimate \( C_{6}=+20(2) \).

In real, finite-\( N_{f} \) QCD this result corresponds to the \( N_{f}^{3}g^{6}T^{4} \)
coefficient of the pressure. As we saw in figure \ref{fig:perturbativeQCD},the
purely gluonic contribution \( \propto N_{f}^{0}g^{6}T^{4} \) of
the same coupling order \( g^{6} \) is completely nonperturbative.

In the figures \ref{figurePnlo} and \ref{figurePnlo2} the agreement
between exact result and perturbative result up to order \( \geff ^{5} \)
was limited to a range of about \( \geff ^{2}\lesssim 9 \). In figure
\ref{fig:zeromu} we show the improved result that remains accurate
up to \( \g ^{2}\sim 16 \) by including our numerical estimate of
the \( \g ^{6} \)-coefficient and using \( \bar{\mu }_{\textrm{FAC}} \).
The agreement with the exact result can even be further improved by
fixing the renormalization point such that the \( \g ^{6} \) coefficient
vanishes and keeping all orders of the odd terms in \( \g  \) by
leaving the plasmon term \( \propto m_{E}^{3} \) unexpanded in \( \g ^{2} \).
The result of this procedure is indicated by the gray area in figure
\ref{fig:zeromu}. Obviously the result now gets very close to the
exact result. Of course, this procedure does by no means guarantee
that an analogous scheme for real QCD with finite \( N_{f} \) will
produce similarly convincing results, but we can take it as another
indication of the observation that keeping the parameters of the dimensionally
reduced theory unexpanded can improve the convergence of thermal perturbation
theory \cite{Blaizot:2003iq}.

\begin{figure}
{\centering \resizebox*{0.8\textwidth}{!}{\includegraphics{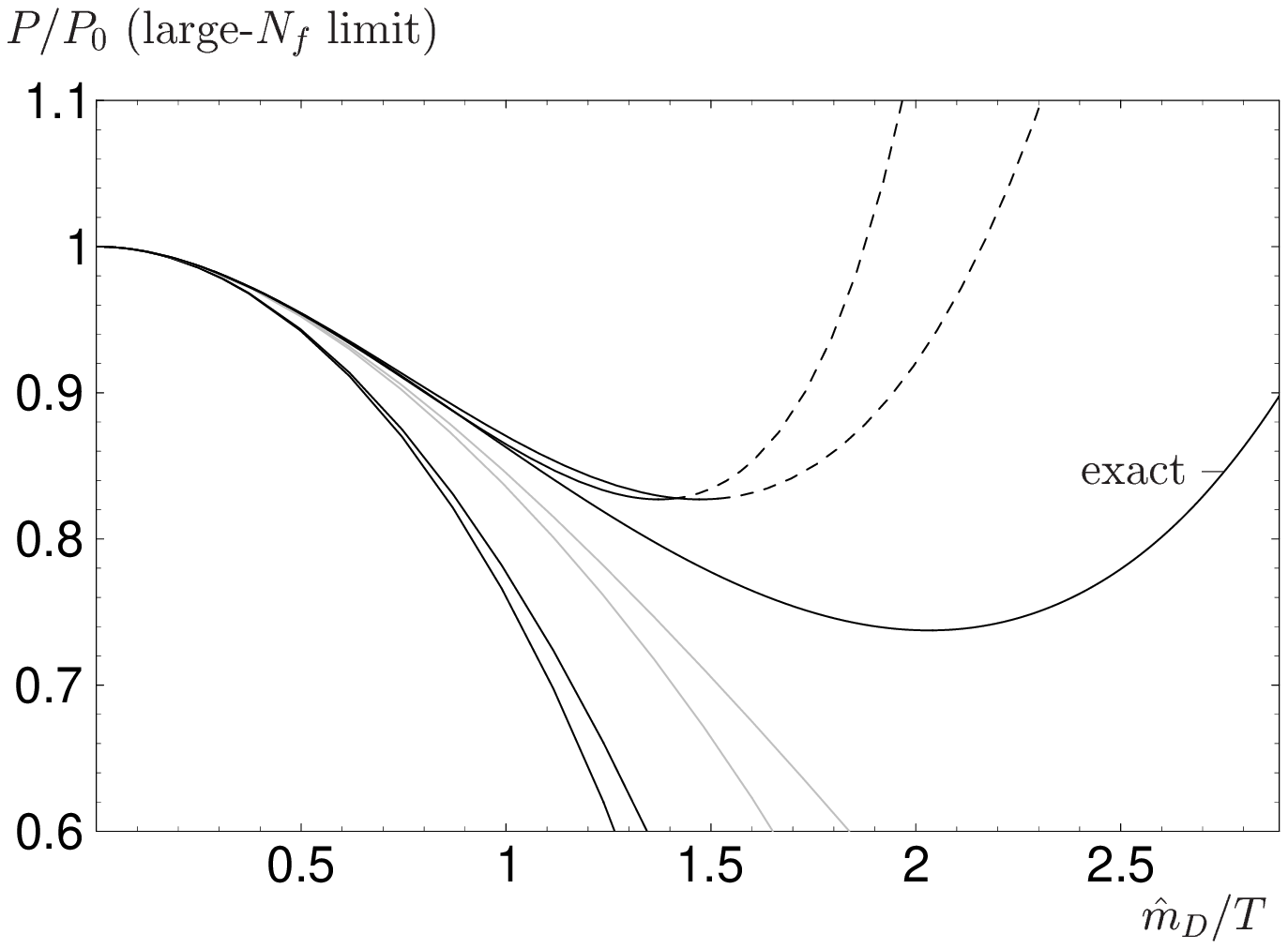}} \par}

\caption{Comparison (taken from reference \cite{Rebhan:2003fj}) of the \protect\( \Phi \protect \)-derivable
2-loop result in the HTL approximation (full lines) and in the next-to-leading
approximation (full lines ending in dashed lines), for \protect\( \bar{\mu }=T\protect \)
and \protect\( 4\pi T\protect \), and the exact result for the pressure
in the limit of large \protect\( N_{f}\protect \). The gray lines
denote the next-to-leading approximation with quadratic fermionic
gap equation considered in reference \cite{Blaizot:2000fc}, but which
Rebhan argues in reference \cite{Rebhan:2003fj} needs to be replaced
by equation (16) therein. The Debye mass appearing on the lower axis
of the plot is related to the effective coupling via \protect\( \hat{m}_{D}/T\equiv \geff (\mu _{\textrm{DR}})/\sqrt{3}\protect \).
\label{fig:largenfphiderivable}}
\end{figure}
Figure \ref{fig:largenfphiderivable} shows a comparison of the \( \Phi  \)-derivable
2-loop result in the HTL approximation (full lines) and in the next-to-leading
approximation (full lines ending in dashed lines) as performed by
Rebhan \cite{Rebhan:2003fj}. In the \( \Phi  \)-derivable approach
of reference \cite{Blaizot:2000fc}, a fermionic gap equation had
been assumed that turned out not to be compatible with the limit of
large \( N_{f} \). As is shown in \cite{Rebhan:2003fj}, the originally
quadratic gap equation for the fermions does not comply with the Casimir
scaling in the large \( N_{f} \) limit which would favor a linear
dependence on the asymptotic mass \( \bar{M}_{\infty } \). The new
gap equation (16) in reference \cite{Rebhan:2003fj} reproduces qualitatively
the nonmonotonic behavior in \( \geff ^{2} \) of the exact result.
By a curious coincidence, for $N=3$ and $N_f=3$ the revised gap equation
has exactly the same solutions as the uncoupled quadratic gap equations
that have been previously in use. Only for $N_f>3$ there exists a
coupling where the fermionic mass ceases to grow monotonicly with
$g$. Because of this coincidence, the numerical changes in the previous
results of \cite{Blaizot:2000fc} are almost completely negligible.

\begin{figure}
{\centering \resizebox*{0.8\textwidth}{!}{\includegraphics{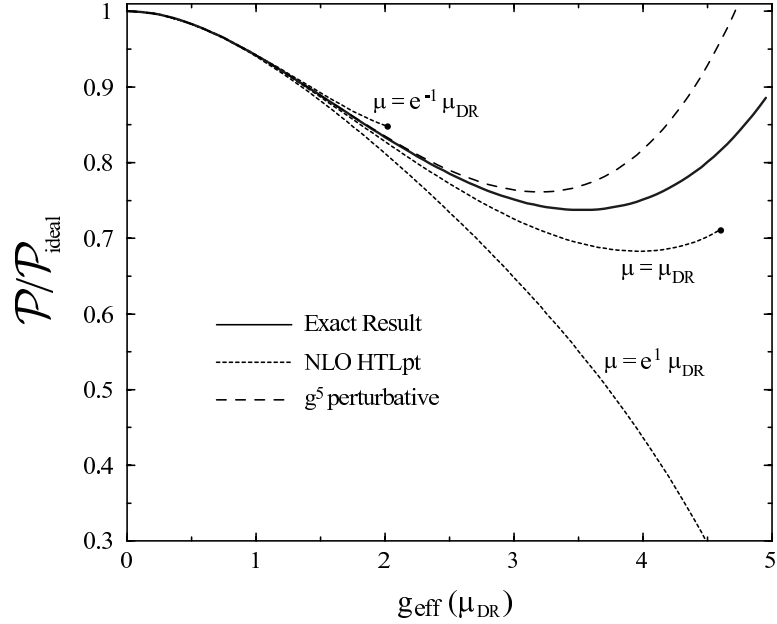}} \par}

\caption{Comparison (taken from reference \cite{Andersen:2003zk}) of the
NLO HTLpt prediction for the $O(N_f^0)$ contribution to the free energy,
the exact numerical in the large \protect\( N_{f}\protect \) limit,
and the perturbative prediction accurate to $g^5$ as a function of
$g_{\rm eff}(\mu_{\rm DR}) = \sqrt{s_f} g(\mu_{\rm DR}) = 2 \pi \sqrt{s_f \alpha_s(\mu_{\rm DR})} $
where $\mu_{\rm DR} = \pi e^{-\gamma} T$. Dots indicate the point
at which there is no longer a real-valued solution to the gap equation
for $m_D$. The renormalization scale $\mu$ is varied by a factor of
$e$ around $\mu_{\rm DR}$ and the perturbative $g^5$ result is evaluated
at the central scale.\label{fig_largenfhtlpta}}
\end{figure}

\begin{figure}
{\centering \resizebox*{0.8\textwidth}{!}{\includegraphics{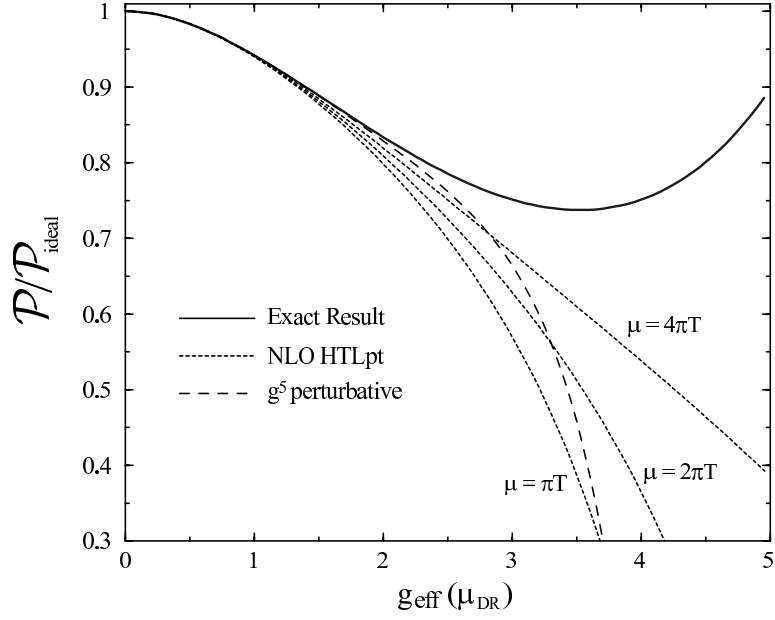}} \par}

\caption{Same as figure \ref{fig_largenfhtlpta} only with the renormalization
scale $\mu$ varied by a factor of $2$ around $2\pi T$. The perturbative
$g^5$ result is again evaluated at the central scale. \cite{Andersen:2003zk}\label{fig_largenfhtlptb}}
\end{figure}
Figures \ref{fig_largenfhtlpta} and \ref{fig_largenfhtlptb} show
the HTLpt (Hard Thermal Loop - perturbation theory) comparison to
the large \( N_{f} \) limit by Andersen et al.~\cite{Andersen:2003zk}.
The plots show the NLO HTLpt prediction for the $O(N_f^0)$ contribution
to the free energy, the perturbative prediction accurate to $g^5$,
and the exact result up to NLO in the large \( N_{f} \) limit. It
should be noted, however, that both the HTLpt and the 2-loop \( \Phi  \)-derivable
results are perturbatively accurate only up to and including order
\( \geff ^{3} \), where the perturbative result is rather ill-behaved.
As noticed in \cite{Andersen:2003zk}, the HTLpt predictions for both
the free energy and the Debye mass (which was also discussed in \cite{Andersen:2003zk})
seemed to diverge from the exact result around $g_{\rm eff} \sim 2$
regardless of the scale that was chosen; however, for both quantities,
choosing the scale to be $\mu = \mu_{\rm DR} = \pi e^{-\gamma} T$
seemed to reasonably reproduce the exact results. The authors of \cite{Andersen:2003zk}
concluded that their result is comparable to the performance of the
$\Phi$-derivable approach in the large $N_f$ limit in figure \ref{fig:largenfphiderivable}.

\subsection{Quark number susceptibilities\label{section_quarknumbersusceptibilities}}

\subsubsection{Linear quark number susceptibility}

The (linear) quark number susceptibility is defined as the first derivate
of the quark number density \( \mathcal{N}\equiv N/V \) with respect
to chemical potential (see also appendix \ref{section_derivativerelations}),
\begin{equation}
\chi ={\6 \mathcal{N}\0 \6 \mu }={\6 ^{2}P\0 \6 \mu ^{2}}.
\end{equation}

\begin{figure}
\selectlanguage{english}
{\centering \includegraphics{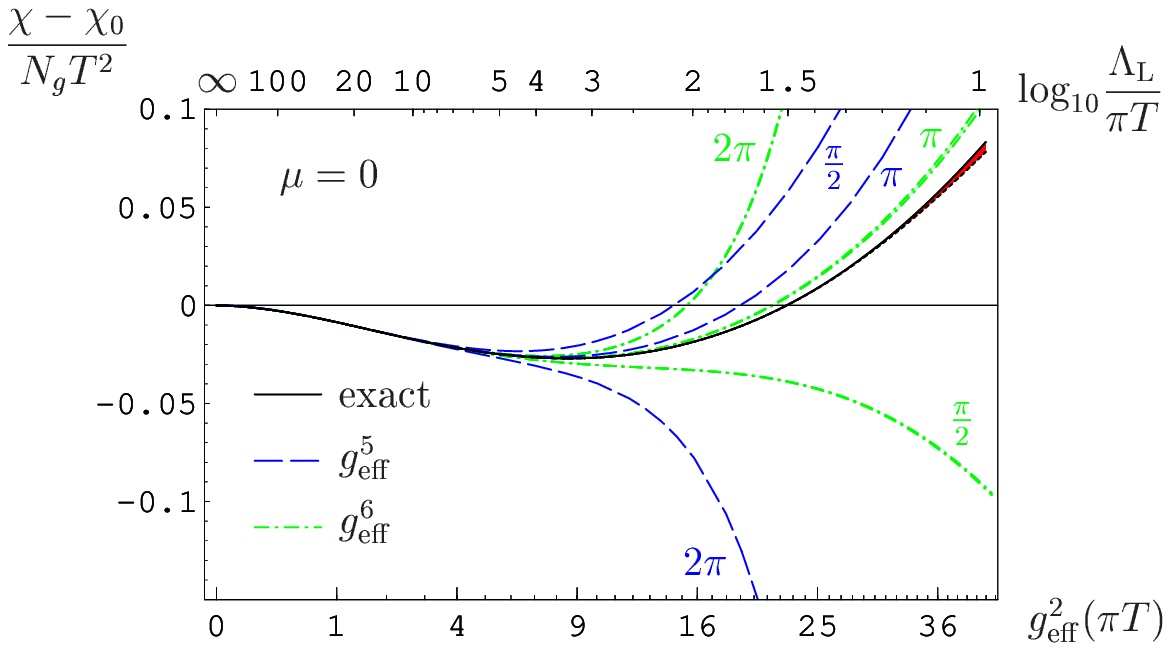} \par}

\caption{The interaction part of the quark number susceptibility at \(\mu=0\)
compared with strict perturbation theory to order \(\g^{5}\) and
\(\g^{6}\), respectively, with renormalization scale varied about
\(\pi T\) by a factor of 2. \label{fig:chi}}
\selectlanguage{american}
\end{figure}

Figure \ref{fig:chi} displays the exact large-\( N_{f} \) result
for the interaction part of \( \chi  \) at zero chemical potential
as a function of \( \g  \) (or alternatively \( \log _{10}(\Lambda _{\textrm{L}}/\pi T) \)).
Similar to the thermal pressure, the result is nonmonotonic, but the
minimum already occurs at \( \g ^{2}(\pi T)\approx 8.6 \), and the
free-theory value is recovered at \( \g ^{2}(\pi T)\approx 22.5 \),
where the Landau ambiguity is still well under control since \( \Lambda _{\textrm{L}}/T\approx 100 \)
at that coupling.

The perturbative (dimensional reduction) result can be read from the
linear term in \( \mu ^{2} \) of (\ref{pintdr}) and gives \begin{eqnarray}
{\chi -\chi _{0}\0N _{g}T^{2}} & = & 2{\6 \0 \6 (\mu ^{2})}{P-P_{0}\0N _{g}}\Big |_{\mu =0}=-{\g ^{2}\08 \pi ^{2}}+{\g ^{3}\04 \pi ^{3}\sqrt{3}}\nonumber \\
 &  & +{\g ^{4}\048 \pi ^{4}}\left[ \ln {\bar{\mu }_{\textrm{MS}}\04 \pi e^{-\gamma }T}-{13\012 }-{4\03 }\ln2 \right] \nonumber \\
 &  & +{\g ^{5}\016 \pi ^{5}\sqrt{3}}\left[ -\ln {\bar{\mu }_{\textrm{MS}}\0e ^{1/2-\gamma }\pi T}+{7\018 }\, \zeta (3)\right] .
\end{eqnarray}
 The coefficient of \( \g ^{4} \) has only recently been obtained
in \cite{Vuorinen:2002ue} in a three-loop calculation. We can confirm
its closed-form value by a numerical fit, which gives agreement with
an accuracy of \( 2\times 10^{-4} \), thus providing a good check
on both our numerics and the analytical calculations of \cite{Vuorinen:2002ue}.
This level of accuracy allows us to also extract the order-\( g^{6} \)
term as (for \( \mu _{\textrm{MS}}=\pi T \)) \begin{equation}
\label{quarksusceptibilitycoefficient6}
{\chi |_{\g ^{6}}\0N _{g}T^{2}}=-4.55(9)\times \left( {\g \04 \pi }\right) ^{6}.
\end{equation}

In Fig.~\ref{fig:chi} we show the perturbative results to order
\( \g ^{5} \) and \( \g ^{6} \), varying the renormalization scale
about \( \pi T \) by a factor of 2 (now without the improvement of
keeping effective-theory parameters unexpanded). The value \( \bar{\mu }_{\textrm{MS}}=\pi T \)
is in fact close to \( \bar{\mu }_{\textrm{FAC}} \) where it makes
no difference whether \( m_{E}^{2} \) is kept unexpanded or not.
We find that the quality of the perturbative result for the susceptibility
is comparable to that observed in the pressure, with \( \bar{\mu }_{\textrm{MS}}=\pi T \)
being close to the optimal choice.

\subsubsection{Higher-order quark susceptibility}

We also computed explicitly the higher-order susceptibility \( \6 ^{4}P/\6 \mu ^{4}|_{\mu =0} \)
(which has been investigated in lattice QCD with \( N_{f}=2 \) in
reference \cite{Gavai:2003mf}).
\begin{figure}
\selectlanguage{english}
{\centering \includegraphics{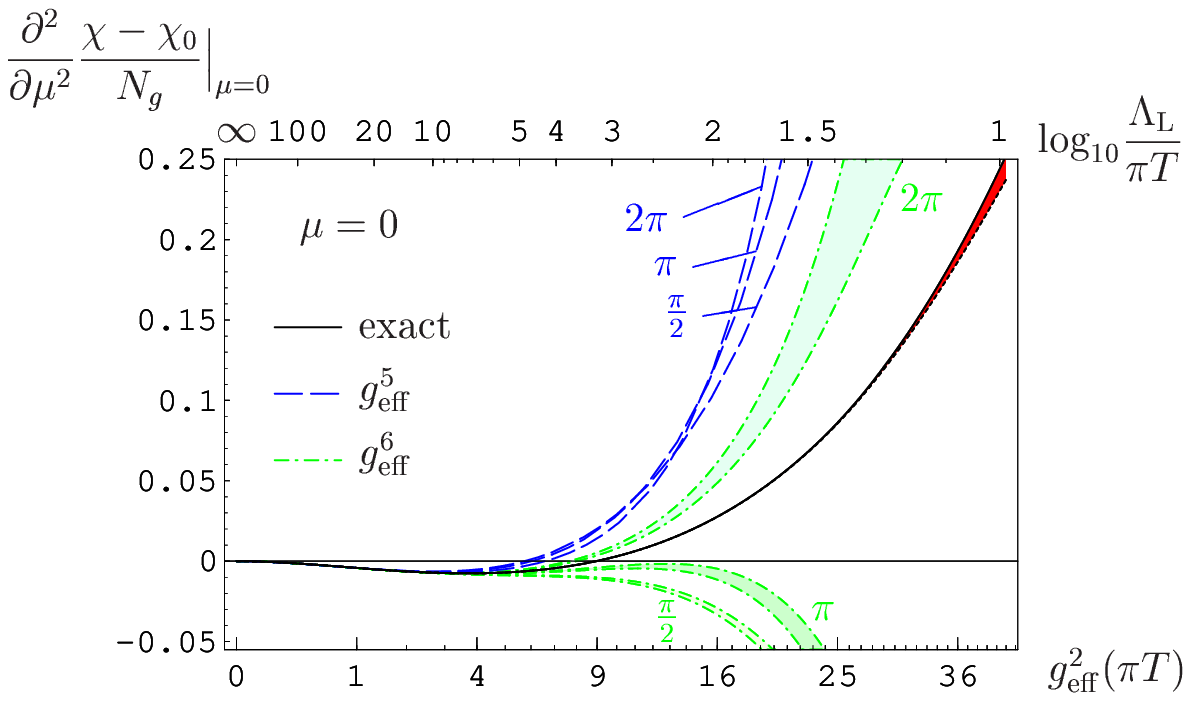} \par}

\caption{The interaction part of the higher-order quark number susceptibility 
\(\6_{\mu}^{2}(\chi-\chi_{0})\) at \(\mu=0\)
compared with strict perturbation theory to order \(\g^{5}\) and
\(\g^{6}\), respectively, with renormalization scale varied about
\(\pi T\) by a factor of 2. The coloured bands of the \(\g^{6}\)-results
cover the estimated
error of the numerically extracted perturbative coefficients.\label{fig:chi4}}
\selectlanguage{american}
\end{figure}

Our exact result in the large-\( N_{f} \) limit is shown in figure
\ref{fig:chi4}. In this quantity, we find that the nonmonotonic behavior
observed above in the pressure and the linear susceptibility is much
more pronounced. The minimum now occurs at \( \g ^{2}\approx 3.7 \),
where perturbation theory is still in good shape, and the free-theory
value is exceeded for \( \g ^{2}\gtrapprox 9 \). Using (\ref{pintdr})
we find to order \( g^{5} \)\begin{eqnarray}
{\6 ^{2}\0 \6 \mu ^{2}}{\chi -\chi _{0}\0N _{g}}\Big |_{\mu =0} & \! \! \! \! = & \! \! \! {\6 ^{4}\0 \6 \mu ^{4}}{P-P_{0}\0N _{g}}\Big |_{\mu =0}=12{\6 ^{2}\0 (\6 \mu ^{2})^{2}}{P-P_{0}\0N _{g}}\Big |_{\mu =0}\label{chinonlinpert} \\
 & \! \! \! \! = & \! \! \! -{3\g ^{2}\04 \pi ^{4}}+{3\sqrt{3}\g ^{3}\04 \pi ^{5}}+{\g ^{4}\08 \pi ^{6}}\left[ \ln {\bar{\mu }_{\textrm{MS}}\04 \pi T}+\gamma +C_{4}\right] \nonumber \\
 & \! \! \! \!  & \! \! \! \! \! +{3\sqrt{3}\g ^{5}\016 \pi ^{7}}\left[ -\ln {\bar{\mu }_{\textrm{MS}}\0e ^{1/2-\gamma }\pi T}+{7\03 }\, \zeta (3)-{31\054 }\, \zeta (5)\right] +O(\g ^{6}).\nonumber 
\end{eqnarray}

In the original publication of these results \cite{Ipp:2003jy} the
coefficient appearing at order \( \g ^{4} \) has been numerically
extracted as \( C_{4}=-7.02(3) \). In the meantime, the complete
\( \mu  \) dependence of the dimensional reduction result to order
\( \g ^{4} \) has been worked out in reference \cite{Vuorinen:2003fs}
from where one can obtain the exact result \begin{equation}
\label{C4exact}
C_{4}=-\frac{1}{12}-12\ln2 +\frac{7}{6}\zeta (3)=-6.9986997796998\ldots 
\end{equation}
 The complete agreement with reference \cite{Vuorinen:2003fs} provided
on the one hand an independent check on the correctness of the 3-loop
calculations of \cite{Vuorinen:2003fs} and on the other hand a check
on the accuracy of our numerical analysis.

Using (\ref{C4exact}) we can extract the term of order \( \g ^{6} \)
in equation (\ref{chinonlinpert}) as \( -39(1)\g ^{6}/(128\pi ^{8}) \)
for \( \bar{\mu }=\pi T \). The perturbative results to order \( \g ^{5} \)
and to order \( \g ^{6} \) are compared with the exact result in
figure \ref{fig:chi4}. This shows that the accuracy of the perturbative
result again improves by going from order \( \g ^{5} \) to order
\( \g ^{6} \), but the renormalization scale dependence increases
sharply at large coupling.

\subsubsection{Pressure at larger chemical potential from susceptibilities}

With regard to the attempts to explore QCD at finite chemical potential
by means of lattice gauge theory \cite{Fodor:2002km,deForcrand:2002ci,Allton:2002zi,D'Elia:2002gd},
it is of interest how well the pressure at larger chemical potential
can be approximated by the first few terms of a Taylor series in \( \mu ^{2} \).

In reference \cite{Fodor:2002km} it has been observed that the ratio
of \( \Delta P=P(T,\mu )-P(T,\mu =0) \) over the corresponding free-theory
quantity \( \Delta P_{0} \) is practically independent of \( \mu  \)
for the range of chemical potentials explored. This is also realized
when quasi-particle models are used for a phenomenological extrapolation
of lattice data \cite{Szabo:2003kg,Rebhan:2003wn} in a method introduced
by Peshier et al. \cite{Peshier:1999ww}.
\begin{figure}
\selectlanguage{english}
{\centering \includegraphics{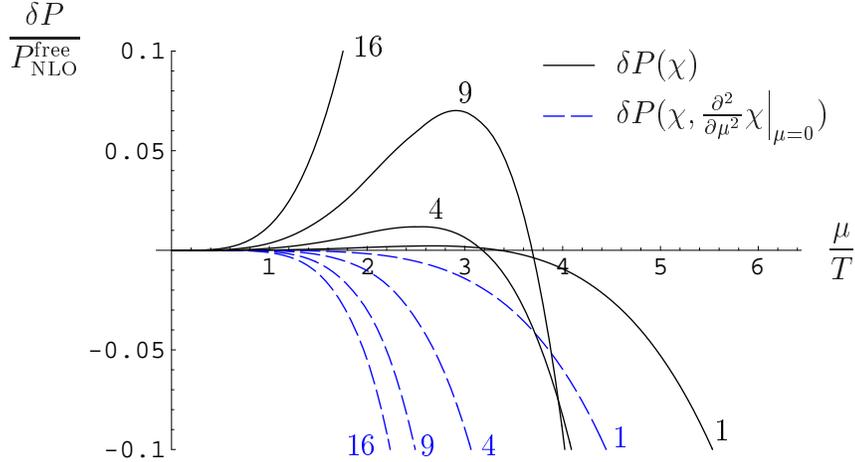} \par}

\caption{Deviation from the scaling observed in Ref.~\cite{Fodor:2002km}
in lattice QCD for small chemical potential
in the quantity
\(\delta P=P(T,\mu)-P(T,0)-{1\02}\chi|_{\mu=0} 
(\mu^{2}+\mu^{4}/(2\pi^{2}
T^{2}))\) (full lines) and
in \(\delta P=P(T,\mu)-P(T,0)
-{\mu^{2}\02}\chi|_{\mu=0}  - {\mu^{4}\04!}{\6^{4}P\0\6\mu^{4}}|_{\mu=0}\) 
(dashed lines),
both
normalized to \(P_{\textrm{NLO}}^{\textrm{free}}=N_{g} \pi^{2} T^{4}/45\),
for \(\g^{2}(\pi T)=1,4,9,16\).\label{fig:deltap}}
\selectlanguage{american}
\end{figure}
In figure \ref{fig:deltap} we show the deviation from this {}``scaling''
at higher values of \( \mu /T \) by considering the quantity \( \delta P=P(T,\mu )-P(T,0)-{1\02 }\chi |_{\mu =0}(\mu ^{2}+\mu ^{4}/(2\pi ^{2}T^{2})) \)
divided by \( P_{\textrm{NLO}}^{\textrm{free}}=N_{g}\pi ^{2}T^{4}/45 \).
The combination \( (\mu ^{2}+\mu ^{4}/(2\pi ^{2}T^{2})) \) appearing
therein is such that a replacement of \( P \) and \( \chi  \) by
their interaction-free values \( P_{0} \) and \( \chi _{0} \) makes
\( \delta P \) vanish identically. (As can be seen from the above
perturbative results, \( \delta P \) also vanishes for the leading-order
interaction parts \( \propto \g ^{2} \).) In the exact large-\( N_{f} \)
results of figure \ref{fig:deltap} we observe that for coupling \( \g ^{2}\lesssim 4 \)
the deviation \( \delta P \) is at most a few percent of \( P_{\textrm{NLO}}^{\textrm{free}} \)
for \( \mu /T\lesssim \pi  \), but it rapidly grows for \( \mu /T\gtrsim \pi  \).
This is in fact also nicely illustrated by the 3-dimensional plot
of the pressure in figure \ref{fig:3d}, which has a rather conspicuous
kink at \( \phi =45^{\circ } \) corresponding to \( \mu =\pi T \).

It turns out that including the exact higher-order susceptibility
at \( \mu =0 \) does not lead to a better approximation of the pressure
at larger chemical potential. The dashed lines in Fig.~\ref{fig:deltap}
correspond to \( \delta P=P(T,\mu )-P(T,0)-{\mu ^{2}\02 }\chi |_{\mu =0}-{\mu ^{4}\04 !}{\6 ^{4}P\0 \6 \mu ^{4}}|_{\mu =0} \).
While this slightly improves matters at small \( \mu /T \), it results
into even quicker deviations for larger \( \mu /T \).

It is of course impossible to say whether this behavior would also
appear in real QCD, but since it occurs already at comparatively small
\( \g  \) in the large-\( N_{f} \) limit, where the peculiar nonmonotonic
behavior of the pressure as a function of \( \g  \) does not yet
arise (the minimum in the normalized interaction pressure occurs at
\( \g ^{2}(\pi T)\approx 14 \)), it may be taken as an indication
that extrapolations of lattice data on the equation of state from
small chemical potential to large \( \mu /T \) are generally problematic.
If anything, real QCD should be more complicated because of the existence
of phase transitions which are absent at NLO in the large-\( N_{f} \)
limit.

\subsection{Pressure at zero temperature}

\begin{figure}
\selectlanguage{english}
{\centering \includegraphics{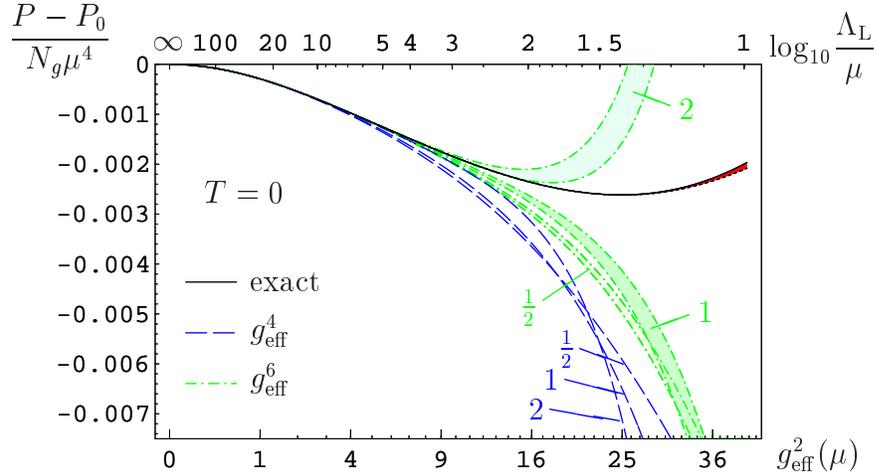} \par}

\caption{The interaction part of the pressure at zero temperature and
finite chemical potential as a function of \(\g^{2}(\bar{\mu}_{\textrm{MS}}=\mu)\)
or, alternatively, \(\log_{10}(\Lambda_{\textrm{L}}/\mu)\),
compared with the perturbative result of Freedman and McLerran
\cite{Freedman:1977dm,Baluni:1978ms}
to order \(\g^{4}\), and our numerically extracted order-\(\g^{6}\)
result, both with renormalization scale in the perturbative
results varied around \(\bar{\mu}_{\textrm{MS}}=\mu\) by a factor of 2.
The coloured bands of the \(\g^{6}\)-results
cover the error of the numerically extracted perturbative coefficients.\label{fig:zerot}}
\selectlanguage{american}
\end{figure}

Our exact result for the thermal pressure at zero temperature and
finite chemical potential is given in figure \ref{fig:zerot} as a
function of \( \g ^{2}(\bar{\mu }_{\textrm{MS}}=\mu ) \). In contrast
to the pressure at zero chemical potential and finite temperature,
the interaction pressure divided by \( \mu ^{4} \) is monotonically
decreasing essentially all the way up to the point where the Landau
ambiguity becomes noticeable.

The thermal pressure at zero temperature and large chemical potential
for QED and QCD has been obtained to order \( g^{4} \) long ago by
Freedman and McLerran \cite{Freedman:1977dm,Baluni:1978ms}. At this
order, there is a non-analytic zero-temperature plasmon term \( \propto g^{4}\ln (g) \),
whose prefactor is known exactly, but the constant under the logarithm
only numerically. The transposition of their result, which has been
obtained in a particular momentum-subtraction scheme, to the gauge-independent
\( \overline{\hbox {MS}} \) scheme can be found in references \cite{Blaizot:2000fc,Fraga:2001id}.
The large-\( N_{f} \) limit of this result reads \begin{equation}
\label{PFMcL}
{P-P_{0}\0N _{g}\, \mu ^{4}}\Big |_{T=0}=-{\g ^{2}\032 \pi ^{4}}-\left[ \ln {\g ^{2}\02 \pi ^{2}}-{2\03 }\ln {\bar{\mu }_{\textrm{MS}}\0 \mu }-\tilde{C}_{4}\right] {\g ^{4}\0128 \pi ^{6}}+O(\g ^{6}\ln \g ^{\phantom {6}})
\end{equation}
 and involves one of the numerical constants computed in reference
\cite{Freedman:1977dm}, \begin{equation}
\tilde{C}_{4}=\frac{79}{18}-\frac{{\pi }^{2}}{3}-\frac{7\, \log (2)}{3}-\frac{2\, b}{3}\approx 0.536,
\end{equation}
 where \( b \) has an integral representation, given in equation
(II.3.25)%
\footnote{\selectlanguage{english}
Note that there is a typo in the equation mentioned: Comparison with
equation (II.3.24) shows that there is a missing exponent 2 after
the second set of large round parenthesis in equation (II.3.25) of
\cite{Freedman:1977dm} as pointed out by \cite{Ipp:2003jy}.
\selectlanguage{american}
} of reference \cite{Freedman:1977dm}, that apparently cannot be evaluated
in closed form. With better computer equipment, \( b \) can however
be evaluated numerically to higher accuracy than that given in \cite{Freedman:1977dm}
as \( b=-1.581231511\ldots  \), which leads to \( \tilde{C}_{4}=0.5358316747\ldots \,  \).

The accuracy of our numerical results is sufficiently high to confirm
the correctness of the result for \( \tilde{C}_{4} \) with an accuracy
of \( \sim 2\times 10^{-4} \). With the knowledge of the exact value
of \( \tilde{C}_{4} \) we can also extract, with lower precision,
the next coefficients at order \( \g ^{6} \), which again involve
a logarithmic term: \begin{eqnarray}
{P-P_{0}\0N _{g}\, \mu ^{4}}\Big |_{T=0} & \! \! \! \! = & \! \! \! -{\g ^{2}\032 \pi ^{4}}-\left[ \ln {\g ^{2}\02 \pi ^{2}}-{2\03 }\ln {\bar{\mu }_{\textrm{MS}}\0 \mu }-\tilde{C}_{4}\right] {\g ^{4}\0128 \pi ^{6}}\label{pressurezerotempcoefficients6} \\
 &  & \! \! \! \qquad -\biggl [\left( 3.18(5)-{16\03 }\ln {\bar{\mu }_{\textrm{MS}}\0 \mu }\right) \ln {\g ^{2}\02 \pi ^{2}}\nonumber \\
 &  & \! \! \! +{16\09 }\ln ^{2}{\bar{\mu }_{\textrm{MS}}\0 \mu }+{16\03 }\left( \tilde{C}_{4}-{1\02 }\right) \ln {\bar{\mu }_{\textrm{MS}}\0 \mu }-3.4(3)\biggr ]{\g ^{6}\02048 \pi ^{8}}+\ldots \nonumber 
\end{eqnarray}

In figure \ref{fig:zerot} we also study the renormalization scale
dependence and apparent convergence of the perturbative result. We
have varied \( \bar{\mu }_{\textrm{MS}} \) about \( \mu  \) by a
factor of 2, and it emerges that the larger values are somewhat favored.

At low temperature \( T\ll \mu  \), dimensional reduction does not
occur. If one nevertheless considers the effective-field-theory parameter
\( m_{E}^{2} \) of (\ref{mE2}) in this limit, one finds that the
function \( {\mathcal{D}}(x) \) therein approaches \( -2\, (\ln 2x+\gamma ) \),
so that the \( T\rightarrow0  \) limit of \( m_{E}^{2} \) exists
and reads \begin{equation}
m_{E}^{2}\rightarrow \mu ^{2}{\g ^{2}\0 \pi ^{2}}\left\{ 1-{\g ^{2}\06 \pi ^{2}}\left[ \ln {\bar{\mu }_{\textrm{MS}}\02 \mu }-\2 \right] \right\} +O(\g ^{6}).
\end{equation}
 Fastest apparent convergence applied to this quantity would suggest
\( \bar{\mu }_{\textrm{MS}}=2e^{\2 }\mu \approx 3.3\mu  \). This
turns out to be not as good as the choice of \( 2\mu  \), though
slightly better than \( \bar{\mu }_{\textrm{MS}}=\mu  \).

The region of low temperature \( T\ll \mu  \) might be readily explored
in the large \( N_{f} \) limit, where it gives rise to anomalous
contributions of the pressure in the context of non-Fermi liquid.
We will study this region in the next chapter. 

\tmpbibtex

\rem{ ========= Non-Fermi Liquid CHAPTER ========== }

\chapter{Non-Fermi Liquid}

\section{Introduction}

\subsection{From ideal gas to non-Fermi liquid}

What is a non-Fermi liquid? The straightforward answer is to start
from the ideal gas and extend the model layer by layer: The ideal
gas is a statistical description of non-interacting point-like particles.
If we go to low temperatures, we cannot neglect quantum mechanics
and we have the choice of describing either bosons, which leads to
the Bose-gas and to Bose-Einstein condensation at very low temperatures,
or fermions, called (ideal) Fermi gas. Still, the Fermi gas is an
accumulation of non-interacting fermions. If we turn on short-range
interaction we get to the description of a (Landau-) Fermi-liquid.
Thermodynamic quantities will mostly show the same order of dependence
on the temperature \( T \) as in a Fermi gas, but their magnitude
might change drastically. Finally, if the interactions we introduce
change the dependence on \( T \) qualitatively, which is the case
for long-range quasi-static interactions, the system cannot be described
by the quasi-particle picture of the Landau-Fermi-liquid theory, and
is therefore called a non-Fermi-liquid. Thus, a non-Fermi liquid is
a cold gas/liquid of fermions with long-range interactions that change
the behavior of a Landau-Fermi liquid qualitatively.

The ideal gas of not too dense, not too cold, and {}``non-interacting''
(actually only interacting by elastic impact for thermalization) particles
obeys the ideal gas law \( pV=Nk_{B}T \). While simple and universal,
this law is not sufficient to calculate all thermodynamic quantities
like energy or entropy of the system. To calculate the energy for
a fixed number of particles \( E(S,V) \), we need an additional quantity,
for example the specific heat \( C_{V}\equiv T\left( \frac{\partial S}{\partial T}\right) _{V} \),
which depends on the inner degrees of the ideal gas. Each degree of
freedom in every particle of a system will contain the same energy
on average so that the specific heat is given by \begin{equation}
\label{specificheat_idealgas}
C_{V}(T)=\frac{f}{2}Nk_{B}\qquad (\textrm{ideal gas})
\end{equation}
for \( f \) degrees of freedom per particle and the Boltzmann constant
\( k_{B} \). For pointlike particles in three dimensions \( f=3 \),
for molecules of two atoms for example there is one additional degree
of freedom for rotation and one for the change of bond length so that
\( f=5 \). The specific heat is an extensive quantity, but we can
divide by the volume \( \mathcal{C}_{V}\equiv C_{V}/V=\frac{f}{2}nk_{B} \)
which now depends on the particle density \( n\equiv N/V \). Note
that the specific heat for an ideal gas is independent of temperature.
The entropy of the system can be calculated for example by integrating
\( \left( \frac{\partial S}{\partial T}\right) _{V}=C_{V}/T \) and
the Maxwell relation \( \left( \frac{\partial S}{\partial V}\right) _{T}=\left( \frac{\partial p}{\partial T}\right) _{V}=Nk_{B}/V \)
which for the ideal gas gives \( S(T,V)=C_{V}\ln (T/T_{0})+Nk_{B}\ln (V/V_{0})+S(T_{0},V_{0}) \).
Statistical distribution of quantities like the mean velocity is given
by the Maxwell-Boltzmann distribution function, which is not valid
anymore for small temperatures where quantum mechanics starts to play
a role. Here we have to use Bose-Einstein statistics for bosons or
Fermi-Dirac statistics for fermions.

The ideal Fermi gas is a description of cold, non-interacting fermions
\cite{Landau:1985aa,Reif:1965aa}. Examples include electrons in metals
and semiconductors (when the Coulomb interaction between them is neglected),
as well as neutrons in a neutron star (again when neglecting the interaction).
Fermions obey the Pauli principle: no two fermions with the same quantum
numbers can be in the same energy state. As a consequence, fermions
will fill up the Fermi sphere in momentum space - if one possible
momentum is occupied by a fermion, the next fermion has to go to the
next higher momentum. (If they differ by a quantum number, e.g. the
spin, then they might occupy the same energy level.) The particle
density corresponding to a completely filled Fermi sphere of radius
\( p_{F} \) in momentum space for a spin 1/2 fermion gas is given
by \( n=N/V=p_{F}^{3}/(3\pi ^{2}\hbar ^{3}) \) at zero temperature.
The number-density distribution for \( T=0 \) is a step-function:
1 for energy states below the Fermi energy, and 0 for energy states
above. For non-zero temperature, the number-density distribution is
given by the Fermi-Dirac distribution \( n=(e^{(\epsilon -\mu )/(k_{B}T)}+1)^{-1} \),
where the chemical potential is just the energy on the Fermi surface
at zero temperature \( \mu |_{T=0}=\varepsilon _{F}\equiv \varepsilon (p_{F}) \).
The specific heat can be calculated from the energy change due to
the Fermi-Dirac distribution at non-zero temperature. It is given
by

\begin{eqnarray}
C_{V}(T) & \approx  & \frac{\pi ^{2}}{3}k_{B}^{2}D(\epsilon _{F})T+O(T^{3})\qquad (\textrm{ideal Fermi gas})\nonumber \\
 & = & V\frac{mp_{F}}{3\hbar ^{3}}k_{B}^{2}T+O(T^{3})\nonumber \\
 & = & \frac{\pi ^{2}}{2}Nk_{B}\frac{T}{T_{F}}+O(T^{3})\label{specificheat_idealFermiliquid} 
\end{eqnarray}
with a density of states factor \( D(\epsilon )=\frac{\sqrt{2}}{\pi ^{2}}V(m/\hbar ^{2})^{3/2}\epsilon ^{1/2} \)
and the Fermi energy \( \epsilon _{F}=\frac{\hbar ^{2}}{2m}(3\pi ^{2}N/V)^{2/3}=p_{F}/(2m)=k_{B}T_{F} \).
We see that the specific heat of the ideal Fermi gas grows linearly
in \( T \) for small temperatures. This is in contrast to the specific
heat of an ideal gas (\ref{specificheat_idealgas}) that is independent
of temperature. Another striking difference to the ideal gas is the
pressure: Since the fermions fill up the Fermi sphere, only a small
fraction of fermions can be in the ground state. As a consequence,
the pressure of a Fermi gas is nonzero even at zero temperature. This
is in contrast to the pressure of an ideal gas that would be zero
(\( pV=Nk_{B}T \)). This so-called degeneracy pressure stabilizes
a neutron star (Fermi gas of neutrons) or a White Dwarf star (Fermi
gas of electrons) against the inward pull of gravity.  

If we include (short-range) interactions between the fermions, we
call the system a Fermi liquid. A description of Fermi liquid was
first given in 1956 by Landau \cite{Landau:1956aa,Lifshitz:1980aa}
and is thus commonly dubbed Landau-Fermi-liquid theory. It can be
applied to liquid \( {}_{2}^{3}\textrm{He} \) (which contains 2 protons,
1 neutron and 2 electrons - an odd number of fermions per atom - such
that the atom itself is a fermion%
\footnote{Note that liquid \( {}_{2}^{4}\textrm{He} \) with an even number
of fermions per atom forms a boson gas and is a superfluid below 2.17
K. At this temperature, \( {}_{2}^{3}\textrm{He} \) behaves like
a Landau-Fermi liquid. Only at much smaller temperatures, around 2.7
mK, two atoms of \( {}_{2}^{3}\textrm{He} \) align themselves to
give an overall spin \( s=1 \) and angular momentum \( l=1 \) and
also form a superfluid. This effect was found in 1971 by Osheroff,
Lee, and Richardson who in 1996 received the Nobel prize for their
work. \rem{ 1971, Doug Osheroff, David Lee, and Robert
Richardson}
}), as well as to the electrons in a normal metal. Landau-Fermi-liquid
theory is qualitatively very similar to the theory of an ideal Fermi
gas in which quasi-particles take the role of the non-interacting
particles of the Fermi gas. Quasi-particles are collective excitations
of the macroscopic system with certain energy \( \epsilon  \) and
momentum \( p \). The properties of quasiparticles are mainly characterized
by the dispersion relation \( \epsilon (p) \). Quasi-particles might
have finite lifetime which gets sufficiently long in the vicinity
of the Fermi surface to allow for a description similar to the particles
in a Fermi gas. Strictly speaking, only certain kinds of interaction
lead to the Landau-Fermi liquid, for example quasi-particles should
still be fermions. (If there are attractive forces that favor a pairing
of the fermions into bosons, the description of the Landau-Fermi-liquid
is not applicable anymore.) Since the quasi-particles are fermions,
they also fill up the Fermi sphere in momentum space. Quasi-particles
always have a spin 1/2 spectrum. If the underlying particles had spins
different from 1/2, this would lead to a degeneracy of the energies
of the quasi-particles so that each branch corresponds to a spin 1/2
quasiparticle. The velocity of a quasiparticle is defined as \( v_{F}=(\partial \epsilon /\partial p)|_{p=p_{F}} \).
In the non-relativistic case one can introduce the effective mass
of the quasiparticle as \( m^{*}=p_{F}/v_{F} \). With this definition,
the specific heat can be derived as\begin{equation}
C_{V}(T)\approx V\frac{m^{*}p_{F}}{3\hbar ^{3}}k_{B}^{2}T+O(T^{3}\ln T).\qquad (\text {Landau-Fermi\textrm{ }liquid})
\end{equation}
The only difference to the formula (\ref{specificheat_idealFermiliquid})
in the leading order contribution is that the particle mass \( m \)
is replaced by the effective mass \( m^{*} \) of the quasi-particle.
We still have the same linear dependence on temperature%
\footnote{For a Bose liquid the specific heat grows proportional to the cube
of the temperature \( C_{V}=V2\pi ^{2}T^{3}/(15(\hbar u)^{3}) \)
\rem{(\( k_{B}=1 \))}with the sound velocity \( u=\epsilon /p \)
\cite{Lifshitz:1980aa}. 
} \cite{Luttinger:1960aa}. Note also that the next order contribution
changes from \( O(T^{3}) \) for the Fermi gas to \( O(T^{3}\ln T) \)
for the Landau-Fermi liquid \cite{Carneiro:1975}. Not only the specific
heat, but also other thermodynamic quantities like compressibility
or spin-susceptibility show the same qualitative temperature dependence
as in the Fermi gas, but might have different coefficients. The classical
Landau Fermi liquid theory was extended to relativistic Fermi systems,
enabling the study of high density matter with weak interaction via
scalar and vector meson exchange \cite{Baym:1976va}. As mentioned,
not every kind of interaction leads to a Landau-Fermi liquid: If the
interaction favors a pairing of the fermions, we effectively deal
with a Bose gas. (In fact, in a Fermi gas arbitrarily small attractive
forces lead to Cooper pairing of fermions.) If we have long-range,
quasi-static interactions, the quasi-particle description is also
not valid any more, and we obtain non-Fermi-liquid behavior.

Anomalous contributions to the specific heat were first calculated
by Holstein et al. \cite{Holstein:1973} who found a deviation from
the linear dependence on \( T \) by a term proportional to \( T\ln T^{-1} \).
For very small temperatures \( T \), the temperature dependence will
actually be dominated by the \( T\ln T^{-1} \) term. For QCD, quasi-static
transverse gauge boson interactions further lead to a series with
fractional powers of \( T \). The specific heat of this non-Fermi
liquid is given by (now with the usual quantum-mechanical natural
units of \( k_{B}=\hbar =1 \)\rem{apple}) \begin{eqnarray}
\mathcal{C}_{V}\equiv C_{V}/V & \approx  & \mu _{q}^{2}T\frac{NN_{f}}{3}+..T\ln ..T^{-1}\qquad \qquad (\text {\textrm{non}-\textrm{Fermi liquid}})\nonumber \\
 &  & +..T^{5/3}+..T^{7/3}+O(T^{3}\ln T)
\end{eqnarray}
where the coefficients were first calculated in 2003 \cite{Ipp:2003cj}
and will be presented in the following sections.

\subsection{Non-Fermi liquid}

A {}``non-Fermi liquid'' is in principle any thermodynamic system
of fermions that goes beyond the classical Landau-Fermi liquid description
(including for example Cooper pairing of fermions and color superconductivity).
Here we will narrow its usage to a Landau-Fermi liquid that is changed
qualitatively by the introduction of long-range interactions. It was
noticed by Holstein et al. \cite{Holstein:1973} in 1973 when studying
the de Haas-van Alphen effect%
\footnote{The de Haas-van Alphen effect states that magnetization shows an oscillatory
dependence on the inverse magnetic field if the magnetic field is
strong enough, \( T\lesssim \mu _{B}B\ll \mu  \) with the Bohr magneton
\( \mu _{B}=e\hbar /(2m_{e}) \) times the magnetic induction \( B \)
still smaller than the chemical potential \( \mu  \) \cite{Lifshitz:1980aa,Ashcroft:1985aa}.
A determination of oscillations in the magnetization \( M \) as a
function of the inverse magnetic field \( 1/H \) (the de Haas-van
Alphen effect) for different orientations of the field has been the
most successful method for mapping out the shape of the Fermi surface
of metals \cite{Ashcroft:1985aa}.
} that long-range interactions lead to a deviation from the linear
temperature dependence of the specific heat for small \( T \). In
order to explain the special de Haas-van Alphen oscillations in some
noble metals in a slight modification of the original model, a current-current
interaction, which in contrast to the Coulomb potential is not screened
at zero frequency, leads to an anomalous contribution to the specific
heat of the form \( T\ln T^{-1} \). The coefficient of this term
is of the order \( \sim \alpha (v_{F}/c)(m^{*}/m)^{2}\sim 10^{-5} \)
smaller than that of the dominant term linear in \( T \) for non-relativistic
QED applications, so originally it was argued that this anomalous
contribution cannot be the cause for measurable effects. (A factor
of 4 that was missing in the original calculation was corrected by
\cite{Chakravarty:1995} so that the leading log correction to the
specific heat is \( \mathcal{C}_{V}=(g_{0}^{2}p_{F}^{2}/36\pi ^{2})T\ln T^{-1} \)).
While this contribution implies that the entropy of the non-Fermi
liquid exceeds the entropy of a Landau-Fermi liquid below some temperature,
it is not possible to infer at which temperature this happens without
complete knowledge of the argument of the logarithm: As long as we
take the logarithm of a dimensionful quantity \( \ln T \), we are
missing the information on the relevant scale for the appearance of
the anomalous behavior.

Long-range interactions are generally screened in the presence of
a large Fermi sea. The reason why this does not happen for transverse
gauge bosons like photons is that gauge invariance prevents them from
acquiring a mass, unless gauge invariance is spontaneously broken
(as in superconductors). An electron gas interacting via transverse
gauge bosons shows non-Fermi liquid behavior depending on the number
of space dimensions \( D \). The specific heat is proportional to
\( \mathcal{C}_{V}\sim T\ln T^{-1} \) only in \( D=3 \) dimensions.
For \( D<3 \) dimensions, \( \mathcal{C}_{V}\sim T^{D/3} \) while
for \( D>3 \) the system behaves like an ordinary Landau-Fermi liquid
\cite{Chakravarty:1995,Gan:1993}. In this sense \( D=3 \) corresponds
to a quantum critical point with respect to variation of the dimension.

A recent calculation of non-Fermi-liquid corrections using renormalization
group resummation techniques in reference \cite{Boyanovsky:2000zj}
suggests a different leading nonanalytic behavior of the specific
heat proportional to \( T^{3}\ln T \), which actually is of the same
order as ordinary Landau-Fermi liquid corrections coming for example
from electron-phonon interactions \cite{Carneiro:1975,Reizer:1989,Coffey:1988aa,Eliashberg:1962}.
\rem{However, reference \cite{Boyanovsky:2000zj} omitted essential
contributions \( \propto \alpha T \) in the course of calculation
that were considered to be free of nonanalytic terms.}However, the
starting point of reference \cite{Boyanovsky:2000zj} already neglected
diagrammatic contributions which would give \( T\ln T^{-1} \) corrections.

It is sometimes assumed that the anomalous contributions to entropy
or specific heat will hardly play any role at all, since color superconductivity
(CSC) will dominate dense quark matter long before non-Fermi liquid
behavior becomes effective \cite{Son:1998uk,Schaefer:2003yh}. The
argument is based on simple dimensional analysis of the energy scale,
and compares the critical temperature of color superconductivity to
the scale where the leading log correction becomes as large as the
leading contribution, that is where non-Fermi liquid becomes \emph{nonperturbative}.
Using the known formulae for the CSC critical temperature and our
complete leading logarithm contribution presented in the following,
we can calculate the \emph{perturbative} correction stemming from
non-Fermi liquid effects. As we will see in figure \ref{fig:cscgapNFLcorrection}
of section \ref{section_csceffects}, perturbative NFL corrections
to the specific heat are of the order of 10\%-20\% at the CSC critical
temperature for \( g\sim O(1) \) and of the order of \( g/(3\sqrt{2}) \)
for small \( g \). Moreover, quark components which do not participate
in the formation of diquark condensates will produce non-Fermi-liquid
behavior even in the color superconducting phase.

Non-Fermi liquid behavior may therefore occur in astrophysical situations,
for example the cooling rate of proto-neutron stars \cite{Iwamoto:1980eb}
if they involve a normal (non-superconducting) degenerate quark matter
component.

In solid state physics experiments, non-Fermi-liquid behavior has
been measured in so-called heavy-fermion metals \cite{Coleman:2001aa},
for example in the specific heat of the recently analyzed \( \textrm{YbRh}_{2}\textrm{Si}_{2} \)
crystal \cite{Custers:2003aa,Trovarelli:1999aa}. There, \( T\ln T^{-1} \)
has been demonstrated experimentally over more than an order of magnitude
in the vicinity of the quantum critical point for non-Fermi liquid
behavior. At ambient pressure there are only a few undoped compounds
that show non-Fermi-liquid behavior, like \( \rm {UBe_{13}} \) \cite{Steglich:1997aa},
\( \rm {CeNi_{2}Ge_{2}} \) \cite{Grosche:1998aa}, or \( \rm {CeCu_{2}Si_{2}} \)
\cite{Gegenwart:1998aa}. Other heavy fermion metals can be tuned
to a quantum critical point by varying for example doping\rem{ (changing
chemical pressure)}, pressure, or magnetic field. 

In the following we will show how the specific heat for ultradegenerate
QED and QCD can be calculated from the pressure integral we used for
calculating the large \( N_{f} \) limit. Non-analytic contributions
in \( T \) come from transverse gauge boson contributions. We will
complete the leading logarithmic contribution \( T\ln T^{-1} \) in
that we calculate the argument of the logarithm. Beyond this contribution,
dynamical screening gives rise to anomalous fractional powers \( T^{(3+2n)/3} \),
for which we calculate the coefficients up to and including order
\( T^{7/3} \).

\section{Entropy at small temperatures}

We want to study thermodynamic quantities in the region of small temperature
\( T\ll \mu  \). The anomalous contributions that we would like to
calculate, can be located in the entropy density \begin{equation}
\mathcal{S}\equiv S/V=\left( {\6 P\0 \6 T}\right) _{\mu }\, ,
\end{equation}
 from which the specific heat can be derived. One might expect that
the {}``\( n_{b} \)'' contributions containing the Bose-Einstein
distribution in the pressure (\ref{PNLO}) are negligible compared
to the {}``non-\( n_{b} \)'' contributions, because \( n_{b}(\omega )=1/(e^{\omega /T}-1) \)
will vanish exponentially with small \( T \) for each (fixed) \( \omega  \).
However, we will see that {}``\( n_{b} \)'' contributions actually
cannot be neglected for a region with \( \omega \lesssim T \) and
that they are the source of anomalous terms in the low-temperature
series. 

Looking at the {}``non-\( n_{b} \)'' contributions first (which
are those parts of the calculation that we have to integrate in Euclidean
space and which require the introduction of a cutoff due to the Landau
pole), the corresponding part of the entropy is given by\begin{eqnarray}
\frac{\mathcal{S}_{\mbox {\scriptsize non-}n_{b}}}{N_{g}} & = & \mu ^{2}T\left\{ -{\g ^{2}\08 \pi ^{2}}+{\g ^{4}\032 \pi ^{4}}\left[ {2\03 }\ln {\bar{\mu }_{\textrm{MS}}\0 \mu }\right. \right. \label{Tsigma} \\
 &  & \left. \left. -0.328(1)\times \ln {\g ^{2}\02 \pi ^{2}}+0.462(5)\right] +O(\g ^{6}\ln \g )\right\} +O(T^{3}).\nonumber 
\end{eqnarray}
 We see that the coefficient at order \( \g ^{2} \) is the same as
the \( \g ^{2} \) part of the strictly perturbative result for the
pressure. The latter is also known as the exchange term \cite{Kap:FTFT}
and coincides with the \( \g ^{2} \) part of (\ref{pintdr}). Here
we have also numerically extracted the order-\( g^{4}\ln (g) \) corrections
from the exact large \( N_{f} \) result of \( \mathcal{S}_{\mbox {\scriptsize non-}n_{b}} \)
at small coupling. 

\begin{figure}
{\centering \includegraphics{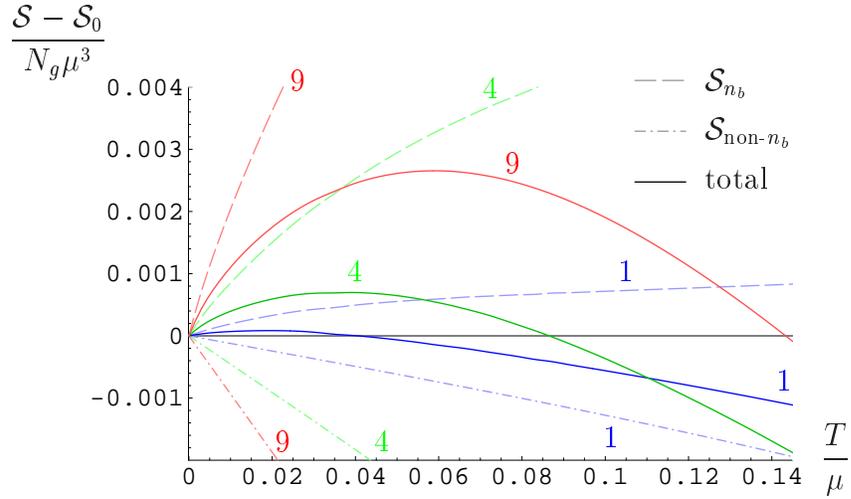} \par}

\caption{The interaction part of the entropy at small \(T/\mu\) for \(\g^{2}(\bar{\mu}_{\textrm{MS}}\!=\!\mu)=1\), \(4\), and \(9\).
The ``non-\(n_{b}\)'' contributions (dash-dotted lines) are negative
and approximately linear in \(T\) with a coefficient agreeing with
the exchange term \(\propto \g^{2}\) in the pressure at small coupling; the
``\(n_{b}\)'' contributions (dashed lines) are positive and nonlinear
in \(T\) such that the total entropy exceeds the free-theory value
at sufficiently small \(T/\mu\). Transverse gauge boson modes dominate the
anomalous contribution in the ``non-\(n_{b}\)'' part. \label{fig:sst}}
\end{figure}

The dash-dotted lines in figure~\ref{fig:sst} show the exact (NLO
large \( N_{f} \)) result for \( \mathcal{S}_{\mbox {\scriptsize non-}n_{b}} \)
for different couplings \( \g ^{2}(\bar{\mu }_{\textrm{MS}}\! =\! \mu )=1 \),
\( 4 \), and \( 9 \) and small temperatures \( 0<T/\mu <0.15 \).
In this range of temperatures, \( \mathcal{S}_{\mbox {\scriptsize non-}n_{b}} \)
is well approximated by the linear term (\ref{Tsigma}).
The dashed lines in figure~\ref{fig:sst} show the {}``\( n_{b} \)''
contributions to the specific heat in the large \( N_{f} \) limit
evaluated numerically. For small \( T \) we see that these parts
cannot be neglected. For sufficiently small \( T/\mu  \), \( \mathcal{S}_{n_{b}} \)
is positive and even dominates so that the total result for the entropy
turns out to exceed its free-theory value for a certain range of \( T/\mu  \).
The size of the range where the the anomalous contribution dominates
in the entropy gets larger with increasing \( \g ^{2} \). The largest
part of the positive and nonlinear contributions at small \( T/\mu  \)
in fact comes from the transverse vector-boson modes. These are only
weakly dynamically screened at small frequencies and completely unscreened
in the static limit because of gauge invariance. We will see that
it is the transverse part of the {}``non-\( n_{b} \)'' term that
will fully contain the temperature-dependent anomalous contribution.
The large \( N_{f} \) limit will allow us to calculate the anomalous
contributions in a very straightforward manner, so let us start by
looking more closely at the transverse contribution.

\section{Transverse contribution}

We started our large \( N_{f} \) calculation from equation (\ref{PNLO})
which gives the pressure associated with the gauge boson loop with
a resummed fermion loop insertion. All interesting non-Fermi liquid
(\( T\ln T \), \( T^{5/3} \), and \( T^{7/3} \)) behavior to the
specific heat arise from the thermal part, the {}``non-\( n_{b} \)''
part, of the transverse propagator. In the following, we will calculate
the contribution to the pressure coming from the transverse part\begin{equation}
\label{pressuretransverse}
\frac{P_{T,n_{b}}}{N_{g}}=-\int \frac{d^{3}q}{(2\pi )^{3}}\int _{0}^{\infty }\frac{dq_{0}}{\pi }2n_{b}\textrm{Im}\ln (q^{2}-q_{0}^{2}+\Pi _{T}+\Pi _{\vac }).
\end{equation}
We will start with a motivating calculation for the leading logarithmic
contribution to the pressure and systematically extend this calculation
to finally obtain the anomalous Fermi liquid contributions up to and
including order \( T^{7/3} \).

\subsection{Leading log contribution}

To calculate the pressure at small \( T \), we start from equation
(\ref{pressuretransverse}) where we can write the argument function
as \( \im \ln x=\arctan (\im \, x/\re \, x) \) as long as \( \re \, x \)
is positive. Keeping this in mind, we use the following low-temperature
approximations: We first neglect the real part of \( \textrm{Re}(\Pi _{T}+\Pi _{\vac })\simeq 0+O(q_{0}^{2}) \)
so that the main contribution comes from \( q^{2} \). For the imaginary
part we only go through order \( O(q) \) (the complete results for
real and imaginary parts of \( \Pi _{T} \) are given in appendix
\ref{chapter_zerotemperaturelimit}) so that we use the following
expression\begin{equation}
\label{ImlnPiTapprox}
\textrm{Im}\ln (q^{2}-q_{0}^{2}+\Pi _{T}+\Pi _{\vac })\simeq \arctan \frac{-\geff ^{2}4\mu ^{2}q_{0}\theta (\mu -\frac{q}{2})/(16\pi q)}{q^{2}}.
\end{equation}
We first apply the \( q \)-integration of the form \begin{eqnarray}
\int _{0}^{q_{m}}dq\, q^{2}\arctan \frac{q_{0}}{q^{3}} & = & \frac{1}{6}\left( \pi q_{m}^{3}-2q_{m}^{3}\arctan \frac{q_{m}^{3}}{q_{0}}+q_{0}\ln \left( 1+\frac{q_{m}^{6}}{q_{0}^{2}}\right) \right) \nonumber \\
 & \simeq  & \frac{1}{3}q_{0}\left( 1+\ln \frac{q_{m}^{3}}{q_{0}}\right) +O(q^{3}_{0})
\end{eqnarray}
for a maximum upper bound \( q_{m} \). From the imaginary part we
see that \( q_{m}=2\mu  \). We are left with the following \( q_{0} \)
integration which gives\begin{eqnarray}
\frac{P_{T,n_{b}}}{N_{g}} & = & \frac{4\pi }{8\pi ^{3}}\int _{0}^{\infty }\frac{dq_{0}}{\pi }\frac{2}{e^{q_{0}/T}-1}\frac{\geff ^{2}\mu ^{2}}{12\pi }q_{0}\left( 1+\ln \frac{4\pi q^{3}_{m}}{\geff ^{2}\mu ^{2}q_{0}}\right) \nonumber \\
 & = & \frac{\geff ^{2}\mu ^{2}T^{2}}{72\pi ^{2}}\left( \ln \frac{4\pi q^{3}_{m}}{\geff ^{2}\mu ^{2}T}+\gamma _{E}-\frac{6}{\pi ^{2}}\zeta '(2)\right) \nonumber \\
 & = & \frac{\geff ^{2}\mu ^{2}T^{2}}{72\pi ^{2}}\left( \ln \frac{32\pi \mu }{\geff ^{2}T}+\gamma _{E}-\frac{6}{\pi ^{2}}\zeta '(2)\right) \label{nonfermiPressureT1} 
\end{eqnarray}
with \( \gamma _{E}-\frac{6}{\pi ^{2}}\zeta '(2)\simeq 1.14718 \).
One does not have to worry about first expanding in small \( q_{0} \)
and then integrating \( q_{0} \) from \( 0 \) to \( \infty  \):
The bosonic distribution function \( n_{b}=1/(\exp (q_{0}/T)-1) \)
ensures that for small \( T \) only small values of \( q_{0}\lesssim T \)
are sampled in the integrand - larger \( q_{0} \) contributions are
suppressed exponentially. We find the contribution to the entropy
density \( \mathcal{S}\equiv S/V=(\partial P/\partial T)_{V,\mu } \)
as \begin{equation}
\frac{\mathcal{S}_{T,n_{b}}}{N_{g}}=\frac{\geff ^{2}\mu ^{2}T}{36\pi ^{2}}\left( \ln \frac{32\pi \mu }{\geff ^{2}T}-\frac{1}{2}+\gamma _{E}-\frac{6}{\pi ^{2}}\zeta '(2)\right) .
\end{equation}
We can also form the specific heat \( \mathcal{C}_{V}\equiv C_{V}/V \)
at constant volume and number density which for unit-volume is given
by \cite{LL:VCv}\begin{equation}
\mathcal{C}_{V}=T\left\{ \left( \frac{\partial \mathcal{S}}{\partial T}\right) _{\mu }-{\left( \frac{\partial \mathcal{N}}{\partial T}\right) ^{2}_{\mu }}{\left( \frac{\partial \mathcal{N}}{\partial \mu }\right) ^{-1}_{T}}\right\} .
\end{equation}
 where \( \mathcal{N} \) is the number density \( \mathcal{N}\equiv N/V=(\partial P/\partial \mu )_{T,V} \).
The application of this formula including the leading-order contribution
is given in equation (\ref{specificheatNLOgeneral}) from appendix
\ref{chapter_thermodynamics}. We find that to leading order the coefficient
of the logarithm remains the same for entropy and specific heat. There
is only a shift by \( -1 \) in the sublogarithmic term \begin{equation}
\frac{\mathcal{C}_{V,T,n_{b}}}{N_{g}}=\frac{\geff ^{2}\mu ^{2}T}{36\pi ^{2}}\left( \ln \frac{32\pi \mu }{\geff ^{2}T}-\frac{3}{2}+\gamma _{E}-\frac{6}{\pi ^{2}}\zeta '(2)\right) .
\end{equation}
This first version of the result already includes the correct pre-factor
of the leading logarithmic contribution of the specific heat as given
in the literature. It confirms the calculation of reference \cite{Chakravarty:1995}
who also found the leading log contribution \( C_{V}=(g_{0}^{2}p_{F}^{2}/36\pi ^{2})T\ln T^{-1} \)
as the specific heat correction, and even goes beyond this known result
in that dimensional dependence and the constants {}``under'' the
logarithm are determined. But the term proportional to \( T \) (the
argument of the logarithm) still calls for improved approximations
and should not be trusted yet: It will change by inclusion of higher
order terms as we will see in the next section.

\subsection{Straightforward improvements\label{sectionstraightforwardimprovements}}

It is possible to retain another term and replace \( 4\mu ^{2} \)
by \( (4\mu ^{2}+q^{2}) \) in (\ref{ImlnPiTapprox}). This will enable
us to obtain the correct leading contribution to the \( q \)-integration
even for larger \( q \). The integral over the arc tangent involves
roots of some cubic equation, but it is still solvable. After performing
the small-\( q_{0} \) expansion, the result is almost the same, up
to an additional term.\begin{equation}
\label{integralimprovement1}
\int _{0}^{q_{m}}dq\, q^{2}\arctan \frac{q_{0}(4\mu ^{2}+q^{2})}{q^{3}}\simeq \frac{4\mu ^{2}}{3}q_{0}\left( 1+\ln \frac{q_{m}^{3}}{q_{0}4\mu ^{2}}+\frac{3}{2}\frac{q_{m}^{2}}{4\mu ^{2}}\right) +O(q^{5/3}_{0}).
\end{equation}
Using \( q_{m}=2\mu  \), this term only contributes by a shift of
\( +3/2 \) inside the braces. The second integral is straightforward
and leads to the following pressure\begin{equation}
\label{nonfermiPressureT2}
\frac{P_{T,n_{b}}}{N_{g}}=\frac{\geff ^{2}\mu ^{2}T^{2}}{72\pi ^{2}}\left( \ln \frac{32\pi \mu }{\geff ^{2}T}+\frac{3}{2}+\gamma _{E}-\frac{6}{\pi ^{2}}\zeta '(2)\right) 
\end{equation}
and its corresponding entropy \begin{equation}
\frac{\mathcal{S}_{T,n_{b}}}{N_{g}}=\frac{\geff ^{2}\mu ^{2}T}{36\pi ^{2}}\left( \ln \frac{32\pi \mu }{\geff ^{2}T}+1+\gamma _{E}-\frac{6}{\pi ^{2}}\zeta '(2)\right) .
\end{equation}

It is also possible to add the vacuum term which also contains a part
proportional to \( q^{2} \): Instead of \( q^{2} \) in the denominator
of (\ref{ImlnPiTapprox}) we could start with \begin{equation}
q^{2}-q_{0}^{2}+\Pi _{T}+\Pi _{\vac }\simeq q^{2}+q^{2}\frac{\geff ^{2}}{6\pi ^{2}}\ln \frac{\muMS }{2\mu }+O(q_{0}^{2})+O(q^{3}).
\end{equation}
The calculation will be almost unaffected, only in the final step
we need to change \( \geff ^{2}\rightarrow \geff ^{2}/(1+\frac{\geff ^{2}}{6\pi ^{2}}\ln \frac{\muMS }{2\mu }) \).
It is clear that this only gives corrections subleading in \( \geff ^{2} \).
In fact, this substitution is nothing else than a change in scale
as given by the exact scale dependence of \( \geff  \) in (\ref{gscal}).

Another extension that we should consider is the \( T \)-dependence
of the self-energies. Up to now we tacitly assumed constant self-energies
\( \Pi _{T}(T)=\Pi _{T}(0) \) at small \( T \), but this is a simplification
which might affect the next-to-leading order in temperature.  We find
the following temperature dependencies:\begin{eqnarray}
\textrm{Im}\Pi _{T}(q_{0},q,T) & = & \textrm{Im}\Pi _{T}(q_{0},q,0)-T^{2}\frac{\geff ^{2}\pi }{12q}q_{0}\theta (2\mu -q)+O(q_{0}^{2}),\qquad \quad \label{selfenergytempimPiT} \\
\textrm{Re}\Pi _{T}(q_{0},q,T) & = & \textrm{Re}\Pi _{T}(q_{0},q,0)+T^{2}\frac{\geff ^{2}q^{2}}{36\mu ^{2}}+O(q_{0}^{2})+O(q^{3}).\label{selfenergytemprePiT} 
\end{eqnarray}
The leading order contributions turn out to have the same order as
the \( T=0 \) terms, so that our calculation does not change qualitatively.
Introducing new variables\begin{eqnarray}
X & = & 1+\frac{\pi ^{2}T^{2}}{3\mu ^{2}},\\
Y & = & 1+\frac{\geff ^{2}T^{2}}{36\mu ^{2}}-\frac{\geff ^{2}}{6\pi ^{2}}\ln \frac{2\mu }{\muMS }
\end{eqnarray}
we can write the resulting pressure as \begin{equation}
\frac{P_{T,n_{b}}}{N_{g}}=\frac{\geff ^{2}\mu ^{2}T^{2}}{72\pi ^{2}}\left( \ln \frac{32\pi \mu }{\geff ^{2}T}+\ln \frac{Y}{X}+\gamma _{E}-\frac{6}{\pi ^{2}}\zeta '(2)+\frac{3}{2X}\right) \frac{X}{Y}.
\end{equation}
effectively adding \( T^{4} \) contributions to the pressure (or
equivalently \( T^{3} \) contributions to the entropy or the specific
heat). This expression reduces to (\ref{nonfermiPressureT2}) if we
set \( T\rightarrow 0 \) in \( X \) and \( Y \). 

We should also consider consistently including higher order terms
of \( q \) and \( q_{0} \) in the arctan arguments of (\ref{ImlnPiTapprox})
or (\ref{integralimprovement1}). In the next section we will see
how this will eventually lead to an integral with a tenth degree polynomial
in the denominator that cannot be readily solved. We have to introduce
approximations in order to factorize the polynomial to obtain doable
integrals, and we shall carefully check that we don't omit any vital
contributions while applying our approximations. \rem{We will shortly
see that this leads to an integral that even in principle cannot be
solved in closed form (unless one uses hypergeometric, modular, elliptic,
Siegel functions or the like), because it contains a tenth degree
polynomial in the denominator. But let us start systematically from
the beginning.}

\subsection{Approximations to order \protect\( T^{3}\protect \) }

\begin{figure}
{\centering \includegraphics{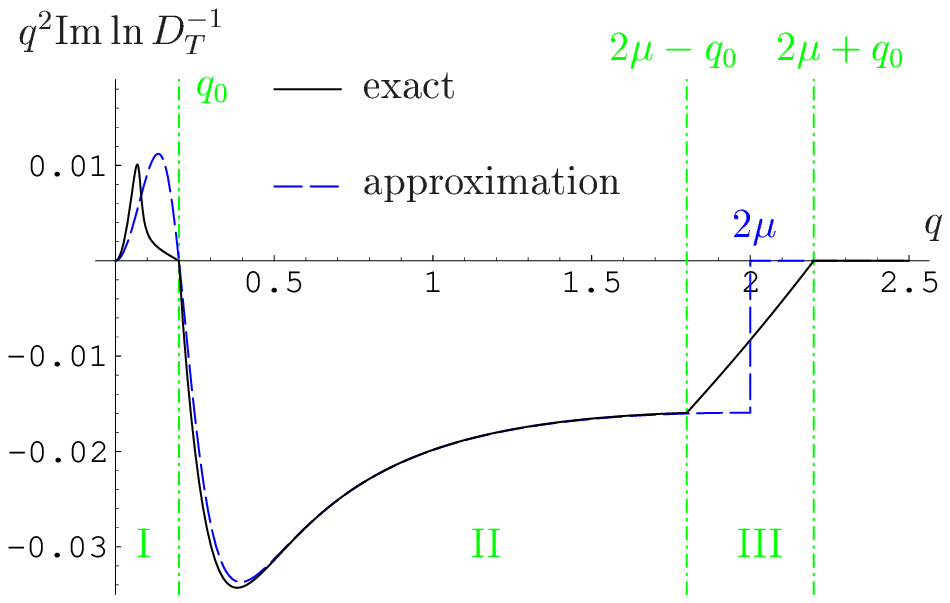} \par}

\caption{Integrand for the \protect\( q\protect \)-integration \protect\( q^{2}\im \ln (q^{2}-q_{0}^{2}+\Pi _{T}+\Pi _{\vac })\protect \)
for \protect\( \mu =\muMS /2=1\protect \), \protect\( q_{0}=0.2\protect \),
\protect\( \geff ^{2}=1\protect \). The solid line shows the exact
result that follows from the full \protect\( T=0\protect \) self
energy expressions, the dashed line shows the result with \protect\( \re \, \Pi _{T}\protect \)
expanded through order \protect\( O(q^{2}_{0})\protect \) and \protect\( O(q^{2})\protect \)
and \protect\( \theta (2\mu -|q\pm q_{0}|)\protect \) replaced by
\protect\( \theta (2\mu -q)\protect \). The parameter \protect\( q_{0}=0.2\protect \)
is chosen this large as to clearly show the three different ranges.
We will see that the error introduced by changing regions I and III
by our approximation only contribute at order \protect\( O(q_{0}^{3})\protect \)
or higher, whereas the main contribution for lower orders only comes
from region II, if the dashed line is integrated from \protect\( q_{0}\protect \)
to \protect\( 2\mu \protect \). \label{figintegrand}}
\end{figure}
We want to carefully examine the full integrand of (\ref{pressuretransverse})
and study which kind of approximations can be applied consistently
up to order \( T^{3} \). For \( T=0 \) an exact solution of the
real and imaginary parts of the gluon self energy \( \Pi _{T} \)
and \( \Pi _{L} \) can be given (see appendix \ref{chapter_zerotemperaturelimit},
equations (\ref{zerotPiL}), (\ref{zerotPiT}), (\ref{zeroPiLRe}),
and (\ref{zeroPiTRe})). These solutions contain expressions like
\( \ln |q-q_{0}| \) or \( \theta (2\mu -|q_{0}\pm q|) \) so that
the \( q \)-integration naturally splits into three regions. The
full line in figure \ref{figintegrand} shows the exact integrand.
The dashed line in the figure shows the integrand that we want to
use: We resolve all absolute values in the integral in a region \( q_{0}<q<2\mu  \),
and expand the real part \( \re (\Pi _{T}+\Pi _{\vac }) \) through
order \( O(q_{0}^{2}) \) and \( O(q^{2}) \). Also, as before, we
replace \( \theta (2\mu -|q_{0}\pm q|) \) by \( \theta (2\mu -q) \)
so that our inverse transverse propagator \( D_{T}^{-1}=q^{2}-q_{0}^{2}+\Pi _{T}+\Pi _{\vac } \)
for \( T=0 \) reads\begin{eqnarray}
\re \, D_{T}^{-1} & = & \left( q^{2}-q^{2}\frac{\geff ^{2}}{6\pi ^{2}}\ln \frac{2\mu }{\muMS }+O(q^{4})\right) \nonumber \\
 &  & +\left( \frac{\geff ^{2}\mu ^{2}}{\pi ^{2}q^{2}}-1+\frac{\geff ^{2}}{6\pi ^{2}}\left[ \frac{1}{2}+\ln \frac{2\mu }{\muMS }\right] +\frac{\geff ^{2}q^{2}}{80\pi ^{2}\mu ^{2}}+O(q^{4})\right) q_{0}^{2}\nonumber \\
 &  & +O(q_{0}^{4}),\label{approxReDT} \\
\im \, D_{T}^{-1} & = & -\frac{\geff ^{2}q_{0}}{48\pi q^{3}}\left( q^{2}-q_{0}^{2}\right) \left( 12\mu ^{2}+3q^{2}+q_{0}^{2}\right) \theta (2\mu -q).\label{approxImDT} 
\end{eqnarray}
All powers of higher order in \( q \) or \( q_{0} \) in (\ref{approxImDT})
are suppressed at least by a factor of \( \geff ^{2} \). The expansion
in the first line of (\ref{approxImDT}) in \( q \) is a priori only
justifiable for anomalous contributions that arise from a region of
small \( q_{0} \) and small \( q \). Actually, the \( q \) integral
gives contributions all the way up to \( 0<q<2\mu  \) and higher
order contributions in \( q \) contribute to the argument of the
leading logarithm \( T\ln T^{-1} \). These corrections will be suppressed
by \( \geff ^{2} \) at least so that we neglect them for the moment.

Both real and imaginary part can be extended from region II between
\( q_{0} \) and \( 2\mu -q_{0} \) to the integration region \( q_{0} \)
to \( 2\mu  \), and we should verify that this procedure does not
introduce an error to the order of interest. In the following we will
explicitly examine to what order the regions I and III will contribute.

In region I we can \rem{simply apply the worst case scenario}argue
by the maximum upper bound of the integrand: The arc tangent function
limits the integrand to \( \pi /2 \) so that the integral can maximally
contribute with \( \int _{0}^{q_{0}}q^{2}\frac{\pi }{2}dq=\pi q_{0}^{3}/6 \).
This is a result of order \( q_{0}^{3} \) that can be neglected if
we expect results of the order \( q_{0}\ln q_{0} \), \( q_{0}^{5/3} \),
or \( q_{0}^{7/3} \).

In region III the argumentation is more subtle. If we expand the exact
self energy functions in a region \( 2\mu -q_{0}<q<2\mu +q \) and
integrate them in this interval, we obtain as a leading order contribution
\( -72\pi \geff ^{2}\pi \mu ^{2}q_{0}^{2}/(4\mu (72\pi ^{2}+13\geff ^{2}-12\geff ^{2}\ln \frac{4\mu }{\muMS }))+O(q^{3}_{0}) \).
This result is quadratic in the leading order of \( q_{0} \) and
could therefore not be neglected as such. However, it turns out that
this contribution exactly matches the quadratic contribution that
one would obtain from extending the approximating formula from region
II to region III and integrating it between \( 2\mu -q_{0} \) and
\( 2\mu  \). This is also suggested by figure \ref{figintegrand}
where the triangle in region III of the exact solution is replaced
by a rectangle of half the width of our dashed approximate expression.
So by integrating our region II formula from \( q_{0} \) to \( 2\mu  \),
we actually reduce the error to order \( O(q_{0}^{3}) \). In the
following we can expect to obtain correct expansion terms below third
order in \( q_{0} \) by integrating out our approximating functions
(\ref{approxReDT}) and (\ref{approxImDT}) from \( q_{0} \) to \( 2\mu  \).

\subsection{Momentum integration}

We will integrate (\ref{pressuretransverse}) using our approximating
functions (\ref{approxReDT}) and (\ref{approxImDT}), abbreviated
conveniently as \( \re  \) (containing only the explicit terms given
in (\ref{approxReDT}) omitting higher order contributions) and \( \im  \),
using integration by parts. \begin{equation}
\label{arctanintegrationparts}
\int _{q_{0}}^{2\mu }q^{2}\arctan \frac{\im }{\re }dq=\left. \frac{q^{3}}{3}\arctan \frac{\im }{\re }\right| _{q_{0}}^{2\mu }-\int _{q_{0}}^{2\mu }\frac{q^{3}}{3}\frac{\re \, \im' -\im \, \re' }{\re ^{2}+\im ^{2}}dq.
\end{equation}
The first part of this integral poses no problem, and it is straightforward
to expand the result in terms of small \( q_{0} \): \begin{equation}
\label{arctancontribution}
\left. \frac{q^{3}}{3}\arctan \frac{\im }{\re }\right| _{q_{0}}^{2\mu }=-\frac{\geff ^{2}\mu ^{2}q_{0}}{6\pi }+O(q_{0}^{3}).
\end{equation}
As it should be, this result is independent of whether we start the
integration from \( q_{0} \) or from \( 0 \). The second part of
the integral is more demanding as we get a bulky polynomial in the
denominator. (To simplify the following expressions, we will set \( \muMS =2\mu , \)
thereby getting rid of logarithmic constants, and reintroduce them
only in the final result):\begin{eqnarray}
57\lefteqn {600\mu ^{4}\pi ^{4}q^{6}\left( \re ^{2}+\im ^{2}\right) } &  & \nonumber \\
 & = & \! \! q^{10}\left( 57600\mu ^{4}\pi ^{4}+1440\geff ^{2}\mu ^{2}\pi ^{2}q_{0}^{2}+9\geff ^{2}q_{0}^{4}\right) \nonumber \\
 &  & \! \! +q^{8}\left( 75\mu ^{4}\pi ^{2}\left[ 128\geff ^{2}\! +\! 3\geff ^{4}\! -\! 1536\pi ^{2}\right] q_{0}^{2}+120\geff ^{2}\mu ^{2}\left[ \geff ^{2}\! -\! 12\pi ^{2}\right] q_{0}^{4}\right) \nonumber \\
 &  & \! \! +q^{6}\left( 1800\geff ^{2}\left[ 64+\geff ^{2}\right] \mu ^{6}\pi ^{2}q_{0}^{2}\right. \nonumber \\
 &  & \! \! \qquad \left. -20\mu ^{4}\left[ 480\geff ^{2}\pi ^{2}-2880\pi ^{4}+\geff ^{4}(-92+15\pi ^{2})\right] q_{0}^{4}\right) \nonumber \\
 &  & \! \! +q^{4}\left( 3600\geff ^{4}\mu ^{8}\pi ^{2}q_{0}^{2}-600\geff ^{2}\mu ^{6}\left[ 192\pi ^{2}+\geff ^{2}(-16+5\pi ^{2})\right] q_{0}^{4}\right. \nonumber \\
 &  & \! \! \qquad \left. -50\geff ^{4}\mu ^{4}\pi ^{2}q_{0}^{8}\right) \nonumber \\
 &  & \! \! +q^{2}\left( -7200\geff ^{4}\mu ^{8}\left[ -8+\pi ^{2}\right] q_{0}^{4}+600\geff ^{4}\mu ^{6}\pi ^{2}q_{0}^{6}+100\geff ^{4}\mu ^{4}\pi ^{2}q_{0}^{8}\right) \nonumber \\
 &  & \! \! +\left( 3600\geff ^{4}\mu ^{8}q_{0}^{6}+600\geff ^{4}\mu ^{6}\pi ^{2}q_{0}^{8}+25\geff ^{4}\mu ^{4}\pi ^{2}q_{0}^{10}\right) .\label{fulldenominator} 
\end{eqnarray}
Bearing in mind that this is just the denominator of the integrand,
a direct integration seems impossible: This is a polynomial of tenth
degree in \( q \), or given the fact that only even powers of \( q \)
appear, we have to cope with a quintic at least. Performing rational
integrals requires knowledge of the roots of the denominator (see
appendix \ref{section_puiseuxseries}, equation (\ref{generalintegralsquarefree})).
But for a quintic, solutions in form of root expressions can not be
given anymore in general. However, all we need is an expansion of
the final integral in terms of \( q_{0} \), that is for small \( q_{0} \).
Can we therefore locate those terms of this polynomial that are important
to low order terms of the \( q_{0} \) expansion? A first naive approach
of simply expanding the denominator in terms of small \( q_{0} \)
will certainly fail, because also our integration variable \( q \)
can become small at the same time. A possible solution might be found
in the pole structure of this expression for small \( q_{0} \) as
plotted in figure \ref{figtenpoles}. The location of the poles reveals
some symmetry, and presumably only a few terms from the polynomial
(\ref{fulldenominator}) determine this structure. In the following
we will see how to extract these terms.

\subsubsection{Pole structure}

\begin{figure}
{\centering \includegraphics{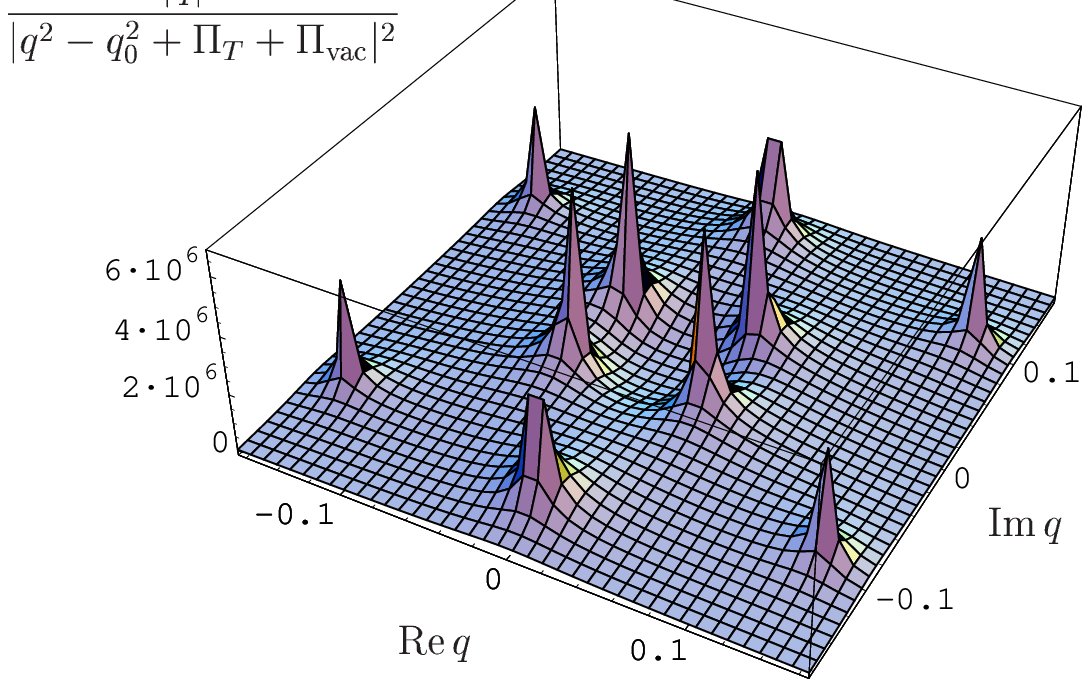} \par}

\caption{Pole structure of the \emph{truncated} absolute squared transverse
propagator \protect\( \re ^{2}+\im ^{2}\protect \) at \protect\( \mu =\muMS /2=1\protect \),
\protect\( q_{0}=0.05\protect \), \protect\( \geff ^{2}=1\protect \)
in the complex \protect\( q\protect \)-plane. This 10-pole structure
arises from the special form of truncation that we chose in (\ref{approxReDT}).
For small \protect\( q_{0}\protect \) the 10-pole structure naturally
decomposes into a rectangular structure with the four poles in the
middle at the order of \protect\( |q|\approx q_{0}\protect \), while
the six surrounding poles, prominently arranged in the shape of a
honeycomb, stay at the order \protect\( |q|\approx (\geff ^{2}\mu ^{2}q_{0}/4\pi )^{1/3}\protect \).
While the inner four poles basically contribute to the leading logarithmic
contribution, it is these outer six poles that give rise to anomalous
Fermi-liquid behavior of order \protect\( T^{5/3}\protect \) and
\protect\( T^{7/3}\protect \). Corrections to this simple picture
will be discussed in the text. The plot is multiplied by \protect\( |q|^{-3}\protect \)
which does not change the pole structure since the propagator contains
an overall factor of \protect\( q^{6}\protect \), but merely makes
the poles have similar heights in this plot.\label{figtenpoles}}
\end{figure}
We obtain (what we will call) the {}``low-order front'' of the polynomial
by first retaining only the leading order in \( q_{0} \) for each
order of \( q \) separately, and then keeping only the leading order
terms in \( q \) for each order of \( q_{0} \). Applying this procedure
to our polynomial (\ref{fulldenominator}), we first get terms of
the orders \( q^{10} \), \( q^{8}q_{0}^{2} \), \( q^{6}q_{0}^{2} \),
\( q^{4}q_{0}^{2} \), \( q^{2}q_{0}^{4} \), and \( q_{0}^{6} \)
which are the relevant terms for small \( q_{0} \). From the three
terms proportional to \( q_{0}^{2} \) we can omit \( q^{8}q_{0}^{2} \)
and \( q^{6}q_{0}^{2} \) for small \( q \) and \( q_{0} \) so that
we are left with only four terms that fully determine the leading
order behavior of the integrand denominator for small \( q \) and
\( q_{0} \):\begin{equation}
\label{loworderfront1}
\re ^{2}+\im ^{2}\approx \frac{1}{q^{6}}\left( q^{10}+\frac{\geff ^{4}\mu ^{4}}{16\pi ^{2}}q^{4}q_{0}^{2}-\frac{\geff ^{4}\mu ^{4}(\pi ^{2}-8)}{8\pi ^{4}}q^{2}q_{0}^{4}+\frac{\geff ^{4}\mu ^{4}}{16\pi ^{2}}q_{0}^{6}\right) .
\end{equation}
\rem{It is crucial to notice that this structure of the {}``low-order
front'' does not change even if we include higher-order terms of
\( q \) or \( q_{0} \) from the beginning of the calculation in
(\ref{approxReDT}): higher order contributions in \( q_{0} \) are
shielded by the \( q_{0}^{6} \) term, higher order terms in \( q \)
are shielded by the \( q^{10} \) term (for small \( q \) where the
small \( q_{0} \) pole structure appears) and all other power mixtures
in between are shielded by this low-order front of exactly four terms:
They uniquely characterize the small \( q_{0} \) and small \( q \)
pole structure. Only higher contributions of the order \( q \) }This
{}``low-order front'' uniquely characterize the small \( q_{0} \)
and small \( q \) pole structure. The first three terms will not
change even if we include higher-order terms of \( q \) or \( q_{0} \)
from the beginning of the calculation in (\ref{approxReDT}): higher
order contributions in \( q_{0} \) are shielded by the \( q_{0}^{6} \)
term, higher order terms in \( q \) are shielded by the \( q^{10} \)
term (for small \( q \) where the small \( q_{0} \) pole structure
appears) and all other power mixtures in between are shielded by this
low-order front of four terms. It should be noted however that in
the expansion of (\ref{approxReDT}) also negative powers of \( q \)
appear for higher order \( q_{0} \) corrections that add terms of
the order \( (q_{0}/q)^{2n} \) with \( 2n\geq 6 \). Their omission
turns out to be negligible since we restricted our integral to the
region \( q_{0}\leq q \) and the coefficients of this series quickly
get smaller.\rem{actually, here is the point to improve the calculation! but
the effect would be small, if any.} 

Still, we are left with a quintic whose general solution cannot be
given in form of root expressions. Let us see, if we can reproduce
the pole structure suggested by figure \ref{figtenpoles}. As a first
guess we would start with \begin{equation}
\re ^{2}+\im ^{2}\approx \frac{1}{q^{6}}\left( A+q^{6}\right) \left( F+q^{4}\right) 
\end{equation}
which would give four poles arranged in a square at a distance \( F^{1/4} \)
and six poles arranged in the shape of a honeycomb at a distance \( A^{1/6} \).
Expanded, this term gives \( AF+Aq^{4}+Fq^{6}+q^{10} \). We can read
off \( A \) from the \( q^{4} \) term \( A=\geff ^{4}\mu ^{4}q_{0}^{2}/(16\pi ^{2}) \)
which determines \( F \) from the constant term as \( F=q_{0}^{4} \).
This already explains the main features of the pole structures. Unfortunately,
our low-order front is not correctly reproduced yet. This is not dramatic
for the additional \( Fq^{6} \)-term, because it is sub-leading to
the low-order front, but the term proportional to \( q^{2} \) is
not reproduced at all. Therefore we have to extend our guess slightly:\begin{equation}
\label{factorizationattempt1}
\re ^{2}+\im ^{2}\approx \frac{1}{q^{6}}\left( A+q^{6}\right) \left( F+Gq^{2}+q^{4}\right) \equiv D_{0}
\end{equation}
where we can determine \( G=-2(\pi ^{2}-8)q_{0}^{2}/\pi ^{2} \).
In this way we change the inner pole structure from a square to a
rectangle. We can now calculate the integral. Since we know how to
factor our denominator, we can apply partial fraction decomposition
and reduce the integral (\ref{arctanintegrationparts}) to doable
simpler integrals\begin{equation}
\int \frac{N}{D}dq\approx \int \frac{N}{D_{0}}dq=\int \frac{N_{\alpha }}{(q^{6}+..)}dq+\int \frac{N_{\beta }}{(q^{4}+..)}dq
\end{equation}
where we abbreviated the numerator as \( N=q^{3}(\re \, \im' -\im \, \re' )/3 \),
the original denominator as \( D=\re ^{2}+\im ^{2} \), and its first
factorization as \( D_{0} \) from (\ref{factorizationattempt1}).
Applying this procedure, we obtain the following result: \rem{scratch033y\_cvexact.nb,
{} resicool4list; Ergebnisse:
specificheat12b\_strange\_diss.nb, scratch074a\_dissertationconstanten.nb}
\begin{eqnarray}
-\int _{q_{0}}^{2\mu }\frac{N}{D_{0}}dq & \! \! =\! \!  & -\frac{\geff ^{2}\mu ^{2}q_{0}}{12\pi }\left( \frac{1}{2}+\ln \frac{32\pi \mu }{\geff ^{2}q_{0}}\right) +\frac{2^{2/3}(\geff \mu )^{4/3}q_{0}^{5/3}}{9\sqrt{3}\pi ^{2/3}}\left( 1+\frac{\geff ^{2}}{64}\right) \nonumber \\
 & \! \! \! \!  & +\frac{8\, 2^{1/3}(\geff \mu )^{2/3}q_{0}^{7/3}}{3\sqrt{3}\pi ^{4/3}}\left( 1+\frac{\geff ^{2}}{72}\right) +O(q_{0}^{3}\ln q_{0}).
\end{eqnarray}
The result looks insofar good as we obtain the same leading order
logarithm as from our first approach (\ref{integralimprovement1})
if we take (\ref{arctancontribution}) into account, and we obtain
anomalous \( q_{0}^{5/3} \) and \( q_{0}^{7/3} \) contributions.
However, those are not complete yet, as we will see soon. This integral
is only the first of a series that we can form: If we denote the terms
we neglected in the denominator as \( \delta D_{0}=D-D_{0} \), we
can write an expansion series that should contain the complete result:\begin{equation}
\int \frac{N}{D}dq=\int \frac{N}{D_{0}+\delta D_{0}}dq=\int \frac{N}{D_{0}}dq-\int \frac{N\, \delta D_{0}}{D_{0}^{2}}dq+\int \frac{N\, \delta D_{0}^{2}}{D_{0}^{3}}dq-+...
\end{equation}
Again, as before, by partial fraction decomposition, all of the integrals
on the right hand side are doable. Looking at the second term in this
series, we again find contributions of the order \( q_{0}^{5/3} \)
and \( q_{0}^{7/3} \):\rem{scratch038a\_dilaton\_orders.nb; NEUE
Ergebnisse: specificheat12b\_strange\_diss.nb, scratch074a\_dissertationconstanten.nb}\begin{eqnarray}
\int \frac{N\, \delta D_{0}}{D_{0}^{2}}dq & = & \frac{2\, 2^{2/3}(\geff \mu )^{4/3}q_{0}^{5/3}}{9\sqrt{3}\pi ^{2/3}}\left( 1+\frac{\geff ^{2}}{64}\right) \\
 &  & +\frac{64\, 2^{1/3}(\geff \mu )^{2/3}q_{0}^{7/3}}{27\sqrt{3}\pi ^{4/3}}\left( 1-\frac{3\geff ^{2}}{256}-\frac{5\geff ^{4}}{16384}\right) +(q_{0}^{3}\ln q_{0}).\nonumber 
\end{eqnarray}
Although we expect to calculate a correction contribution, the \( q_{0}^{5/3} \)
term turns out to be twice as big as the first one. Apparently the
series does not converge fast enough. The third term in the series
\( \int N\, \delta D_{0}^{2}/D_{0}^{3}dq \) again seems promising
for it starts at order \( O(q_{0}^{7/3}) \) and does not contain
a \( q_{0}^{5/3} \) term anymore, but still we would like to find
a more controlled kind of approximation.

\subsubsection{Completing the hypercube}

The problem of the series lies in the correction \( \delta D_{0} \)
which is of the order \( q_{0}^{2} \). This order is not enough to
compensate negative orders of \( q_{0} \) that are introduced by
powers of the denominator \( D_{0} \). We could ask if it is possible
to shuffle the \( q_{0}^{2} \) terms from \( \delta D_{0} \) into
\( D_{0} \) in such a way that \( D_{0} \) remains a product of
a fourth-order and a sixth-order polynomial. In this way we could
form a new denominator \( D_{1} \) with a corresponding \( \delta D_{1}=D-D_{1}=O(q_{0}^{4}) \)
and we could expect a faster converging series in terms of \( q_{0} \).
It turns out that it is indeed possible to complete the square - or
actually the 10-dimensional hypercube - up to a given order of \( q_{0} \)
so that corrections are higher-order in \( q_{0} \) by the following
procedure: We start from \begin{eqnarray}
\re ^{2}+\im ^{2} & = & \frac{1}{q^{6}}\left\{ a+bq^{2}+cq^{4}+dq^{6}+eq^{8}+fq^{10}\right\} \nonumber \\
 & \approx  & \frac{1}{q^{6}}M\left( A+Bq^{2}+Cq^{4}+q^{6}\right) \left( F+Gq^{2}+q^{4}\right) \equiv D_{1}\nonumber \\
 & = & \frac{1}{q^{6}}M\left\{ AF+(AG+BF)q^{2}+(A+BG+CF)q^{4}\right. \nonumber \\
 &  & \qquad \quad \left. +(B+CG+F)q^{6}+(C+G)q^{8}+q^{10}\right\} \label{factorizationattempt2} 
\end{eqnarray}
where \( a \), \( b \), \( c \), \( d \), \( e \), and \( f \)
can be read off from (\ref{fulldenominator}). We already determined
\( A \), \( F \), and \( G \) before. \( \delta D_{0} \) contains
\( q_{0}^{2} \) contributions for \( q^{10} \), \( q^{8} \), and
\( q^{6} \), and we have three variables left: \( B \), \( C \),
and \( M \). This should not be too difficult, since the mixing terms
\( BF \), \( BG \), \( CF \), ... naturally turn out to be of higher
order in \( q_{0} \). We can summarize the procedure as follows:\begin{eqnarray}
M & \rightarrow  & f\sim 1+O(q_{0}^{2}),\nonumber \\
A & \rightarrow  & \frac{c}{M}\sim O(q_{0}^{2}),\nonumber \\
F & \rightarrow  & \frac{a}{AM}\sim O(q_{0}^{4}),\nonumber \\
G & \rightarrow  & \frac{b}{AM}\sim O(q_{0}^{2}),\nonumber \\
C & \rightarrow  & \frac{e-MG}{M}\sim O(q_{0}^{2}),\nonumber \\
B & \rightarrow  & \frac{d}{M}\sim O(q_{0}^{2})
\end{eqnarray}
where the coefficients are series expanded in \( q_{0} \) and truncated
to form polynomials. In our case we get the following assignments\begin{eqnarray}
A & \rightarrow  & \frac{\geff ^{4}\mu ^{4}}{16\pi ^{2}}q_{0}^{2},\nonumber \\
B & \rightarrow  & \frac{\geff ^{2}\mu ^{2}(64+\geff ^{2})}{32\pi ^{2}}q_{0}^{2},\nonumber \\
C & \rightarrow  & \frac{3\geff ^{4}+128\geff ^{2}-12288}{768\pi ^{2}}q_{0}^{2},\nonumber \\
F & \rightarrow  & q_{0}^{4},\nonumber \\
G & \rightarrow  & -2\frac{\pi ^{2}-8}{\pi ^{2}}q_{0}^{2},\nonumber \\
M & \rightarrow  & 1+\frac{\geff ^{2}}{40\pi ^{2}\mu ^{2}}q_{0}^{2}
\end{eqnarray}
Using these in (\ref{factorizationattempt2}) we obtain a correction
term \( \delta D_{1} \) of the order \( O(q_{0}^{4}) \). It is possible
to pursue the same strategy again to get rid of the \( O(q_{0}^{4}) \)
term in the \( \delta D_{1} \) correction to form a \( D_{2} \)
result by the very same procedure as above, including the next order
term in each variable. \( \delta D_{2} \) would then be a correction
of order \( O(q_{0}^{6}) \).

Let us see how this new denominator \( D_{1} \) changes the result.
There are many terms involved in this calculation, so it is best to
let a computer do all integrals each of which is doable just as before.
Here is the final result: \rem{scratch045d\_extractcoefficients.nb,
NEUE Koeffizienten: scratch074a\_dissertationconstanten.nb}\begin{eqnarray}
-\int _{q_{0}}^{2\mu }\frac{N}{D_{1}}dq & \! \! =\! \!  & \! \! -\frac{\geff ^{2}\mu ^{2}q_{0}}{12\pi }\left( \frac{1}{2}+\ln \frac{32\pi \mu }{\geff ^{2}q_{0}}\right) +\frac{2^{2/3}(\geff \mu )^{4/3}q_{0}^{5/3}}{3\sqrt{3}\pi ^{2/3}}\left( 1+\frac{\geff ^{2}}{64}\right) \, \, \, \, \, \nonumber \\
 & \! \! -\! \!  & \! \! \! \frac{8\, 2^{1/3}(\geff \mu )^{2/3}q_{0}^{7/3}}{9\sqrt{3}\pi ^{4/3}}\left( 1-\frac{\geff ^{2}}{32}-\frac{\geff ^{4}}{2048}\right) +O(q_{0}^{3}\ln q_{0}).\label{integral_NoD1} 
\end{eqnarray}
Another calculation shows that the correction to this result \( \int N\, \delta D_{1}/D_{1}^{2}dq \)
is already of order \( O(q_{0}^{13/3}) \).\rem{scratch046e\_dilaton\_extractcoefficient73.nb}
Also, using the second correction \( \int N/D_{2}dq \) gives the
same result up to \( O(q_{0}^{3}) \) so that we can now trust the
\( q_{0}^{5/3} \) and \( q_{0}^{7/3} \) coefficients. \rem{scratch045d\_extractcoefficients.nb,
\char`\"{}Beta-Correction\char`\"{}}

Just to give an impression of how the series would continue, we give
the \( q_{0}^{3} \) coefficient of the expansion (\ref{integral_NoD1})
\begin{eqnarray}
O(q_{0}^{3}\ln q_{0}) & \rightarrow  & \frac{q_{0}^{3}}{12\pi ^{3}}\left\{ \frac{896}{3}-8\pi ^{2}+\frac{10\geff ^{2}}{3}-\frac{123\geff ^{4}}{320}-\frac{67\geff ^{6}}{12288}\right. \nonumber \\
 &  & \qquad \quad -2(\pi ^{2}-16)\sqrt{\pi ^{2}-4}\left( \pi -2\arctan \frac{2}{\sqrt{\pi ^{2}-4}}\right) \nonumber \\
 &  & \qquad \quad +\left( 32\pi ^{2}-\frac{512}{3}-\frac{\geff ^{2}\pi ^{2}}{12}+\frac{\geff ^{4}}{30}+\frac{\geff ^{6}}{3072}\right) \ln \frac{32\pi \mu }{\geff ^{2}q_{0}}\nonumber \\
 &  & \qquad \quad \left. -8(3\pi ^{2}-16)\ln \frac{1024\pi }{\geff ^{4}}\right\} +O(q_{0}^{11/3}).\label{orderq3lnqpreview} 
\end{eqnarray}
Note that this expansion results from the terms explicitly given in
equations (\ref{approxReDT}) and (\ref{approxImDT}). Including higher
order contributions of \( q_{0} \) and \( q \) in (\ref{approxReDT})\rem{
and (\ref{approxImDT})} will change the higher order coefficients
of \( \geff  \) given here.

\subsection{Pressure, entropy and specific heat}

As before, we calculate pressure by integrating over \( q_{0} \)
and entropy and specific heat from derivatives thereof. We introduce
the following abbreviations \begin{eqnarray}
\tilde{c}_{\ln } & = & \frac{3}{2}+\gamma _{E}-\frac{6}{\pi ^{2}}\zeta '(2),\\
\tilde{c}_{8/3} & = & -\frac{2^{2/3}(\geff \mu )^{4/3}\Gamma (\frac{8}{3})\zeta (\frac{8}{3})}{3\sqrt{3}\pi ^{11/3}}(1+\frac{\geff ^{2}}{64}),\\
\tilde{c}_{10/3} & = & \frac{8\, 2^{1/3}(\geff \mu )^{2/3}\Gamma (\frac{10}{3})\zeta (\frac{10}{3})}{9\sqrt{3}\pi ^{13/3}}(1-\frac{\geff ^{2}}{32}-\frac{\geff ^{4}}{2048}).
\end{eqnarray}
Note that inclusion of higher order terms in \( q_{0} \) and \( q \)
in equation (\ref{approxReDT}) may change the higher order coefficients
in \( \geff  \) given here. The leading order in \( \geff  \) for
each coefficient is complete though. The results are\begin{eqnarray}
\frac{P_{T,n_{b}}}{N_{g}} & = & \frac{\geff ^{2}\mu ^{2}T^{2}}{72\pi ^{2}}\left( \ln \frac{32\pi \mu }{\geff ^{2}T}+\tilde{c}_{\ln }\right) \label{nonfermiPressureT3} \\
 &  & +\tilde{c}_{8/3}T^{8/3}+\tilde{c}_{10/3}T^{10/3}+O(T^{4}\ln T),\nonumber 
\end{eqnarray}
\begin{eqnarray}
\frac{\mathcal{S}_{T,n_{b}}}{N_{g}} & = & \frac{\geff ^{2}\mu ^{2}T}{36\pi ^{2}}\left( \ln \frac{32\pi \mu }{\geff ^{2}T}+\tilde{c}_{\ln }-\frac{1}{2}\right) \label{nonfermiEntropyT3} \\
 &  & +\frac{8}{3}\tilde{c}_{8/3}T^{5/3}+\frac{10}{3}\tilde{c}_{10/3}T^{7/3}+O(T^{3}\ln T),\nonumber 
\end{eqnarray}
\begin{eqnarray}
\frac{\mathcal{C}_{V,T,n_{b}}}{N_{g}} & = & \frac{\geff ^{2}\mu ^{2}T}{36\pi ^{2}}\left( \ln \frac{32\pi \mu }{\geff ^{2}T}+\tilde{c}_{\ln }-\frac{3}{2}\right) \label{nonfermiSpecificHeatT3} \\
 &  & +\frac{40}{9}\tilde{c}_{8/3}T^{5/3}+\frac{70}{9}\tilde{c}_{10/3}T^{7/3}+O(T^{3}\ln T).\nonumber 
\end{eqnarray}

\begin{figure}
{\centering \includegraphics{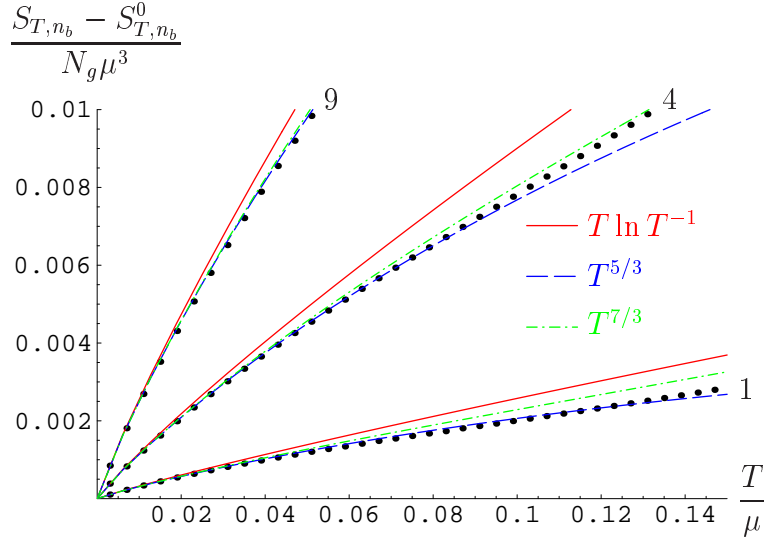} \par}

\caption{Transverse \protect\( n_{b}\protect \)-contribution to the interaction
part of the low-temperature entropy density in the large-\protect\( N_{f}\protect \)
limit for the three values \protect\( \g ^{2}=1,4,9\protect \). The
heavy dots give the exact numerical results; the full, dashed, and
dash-dotted lines correspond to our perturbative result up to and
including the \protect\( T\ln T^{-1}\protect \), \protect\( T^{5/3}\protect \),
and \protect\( T^{7/3}\protect \) contributions.\label{figST}\label{figEntropyComparison}}
\end{figure}

Figure \ref{figEntropyComparison} shows a comparison between a full
numerical result and the series expansion for small \( T \) for the
transverse \( n_{b} \)-part of the entropy in the large \( N_{f} \)
limit. Clearly, the leading order contribution of the curves is accurately
reproduced by the small-\( T \) expansion for different \( \geff ^{2} \)
of 1, 4, or 9 in a region \( T/\mu \lesssim \g /(2\pi ^{2}) \). This
is also the region where the complete large-\( N_{f} \) result for
the low-temperature entropy exceeds the ideal-gas value in figure
\ref{fig:sst}.

\subsection{Higher order corrections}

We can now apply full temperature-dependence of the self energies
and keep the renormalization scale \( \muMS  \) explicitly in our
formulae as in section \ref{sectionstraightforwardimprovements}.
Our final result for the pressure reads as follows: \rem{scratch048a\_photon\_tempextracted.nb}
Using the following dimensionless quantities \begin{eqnarray}
X & = & 1+\frac{\pi ^{2}T^{2}}{3\mu ^{2}},\\
Y & = & 1+\frac{\geff ^{2}T^{2}}{36\mu ^{2}}-\frac{\geff ^{2}}{6\pi ^{2}}\ln \frac{2\mu }{\muMS }
\end{eqnarray}
and the following abbreviations \begin{eqnarray}
\hat{c}_{\ln } & \! \! =\! \!  & \frac{3}{2X}+\gamma _{E}-\frac{6}{\pi ^{2}}\zeta '(2)+\ln \frac{Y}{X},\\
\hat{c}_{8/3} & \! \! =\! \!  & -\frac{2^{2/3}(\geff \mu )^{4/3}\Gamma (\frac{8}{3})\zeta (\frac{8}{3})}{3\sqrt{3}\pi ^{11/3}}(1+\frac{\geff ^{2}}{64Y})\left( \frac{X}{Y}\right) ^{2/3},\\
\hat{c}_{10/3} & \! \! =\! \!  & \frac{8\, 2^{1/3}(\geff \mu )^{2/3}\Gamma (\frac{10}{3})\zeta (\frac{10}{3})}{9\sqrt{3}\pi ^{13/3}}(1\! -\! \frac{\geff ^{2}}{32Y}\! -\! \frac{\geff ^{4}}{2048Y^{2}})\left( \frac{X}{Y}\right) ^{1/3}\qquad \quad 
\end{eqnarray}
we can write the resulting pressure as\begin{eqnarray}
\frac{P_{T,n_{b}}}{N_{g}} & = & \frac{\geff ^{2}\mu ^{2}T^{2}}{72\pi ^{2}}\left( \ln \frac{32\pi \mu }{\geff ^{2}T}+\hat{c}_{\ln }\right) \frac{X}{Y}\nonumber \\
 &  & +\hat{c}_{8/3}T^{8/3}+\hat{c}_{10/3}T^{10/3}+O(T^{4}\ln T).
\end{eqnarray}
 The temperature corrections of \( X \) and \( Y \) only contribute
to order \( O(T^{4}) \) so that in a strict expansion sense, this
expression reduces to (\ref{nonfermiPressureT3}) up to \( \ln \muMS  \)
terms.

Let us finally note that with the tools presented here, it seems a
straightforward but tedious task to calculate the complete \( O(T^{4}) \)
contribution to the pressure (or the \( O(T^{3}) \) contribution
to entropy or specific heat). First, from the region II integral we
expect a \( T^{3}\ln T \) contribution to the entropy as is suggested
by the (still incomplete) contribution in equation (\ref{orderq3lnqpreview}).
Then we have to take a careful look at regions I and III again, which
could start contributing at this order. Finally, the temperature dependencies
of the self-energies in equations (\ref{selfenergytempimPiT}) and
(\ref{selfenergytemprePiT}) will also start contributing at this
order via higher corrections similar to the \( X \) and \( Y \)
functions above. In a full \( O(T^{3}\ln T) \) calculation of the
entropy, all of these terms would have to be taken into account.

\section{Longitudinal contribution}

For the longitudinal {}``\( n_{b} \)''-part from our starting equation
(\ref{PNLO}) we cannot neglect the real part of the self-energy \( \Pi _{L} \)
compared to the vacuum self energy as we did in equation (\ref{ImlnPiTapprox}).
We have to add a term \begin{equation}
\textrm{Re}\Pi _{L}(q_{0},q)=\frac{\geff ^{2}\mu ^{2}}{\pi ^{2}}+O(q_{0}^{2})+O(q^{2})
\end{equation}
which is absent in the transverse part. This contribution does not
vanish for small \( q_{0} \) and \( q \) and actually simplifies
the calculation. This time we can expand the arc tangent for small
argument \( \arctan x\simeq x \) so that we can write\begin{equation}
\textrm{Im}\ln (q^{2}-q_{0}^{2}+\Pi _{L}+\Pi _{\vac })\simeq \frac{\geff ^{2}(4\mu ^{2}-q^{2})q_{0}\theta (\mu -\frac{q}{2})/(8\pi q)}{q^{2}+(\geff ^{2}\mu ^{2})/\pi ^{2}}.
\end{equation}
Performing the \( q \)-integration as before we get\begin{eqnarray}
Q & \equiv  & \int _{0}^{2\mu }dqq^{2}\frac{\geff ^{2}(4\mu ^{2}-q^{2})q_{0}/(8\pi q)}{q^{2}+(\geff ^{2}\mu ^{2})/\pi ^{2}}\\
 & = & \frac{\geff ^{2}\mu ^{2}}{4\pi }q_{0}\left\{ \left( 1+\frac{\geff ^{2}}{4\pi ^{2}}\right) \ln (1+\frac{4\pi ^{2}}{\geff ^{2}})-1\right\} .
\end{eqnarray}
This integral can be performed exactly, so here and in the following
step there is no need for any truncation so far. From this integral
\( Q \) we get the pressure as \begin{eqnarray}
\frac{P_{L,n_{b}}}{N_{g}} & = & -\frac{4\pi }{8\pi ^{3}}\int _{0}^{\infty }\frac{dq_{0}}{\pi }\frac{1}{e^{q_{0}/T}-1}Q\nonumber \\
 & = & -\frac{\geff ^{2}\mu ^{2}T^{2}}{48\pi ^{2}}\left\{ \left( 1+\frac{\geff ^{2}}{4\pi ^{2}}\right) \ln (1+\frac{4\pi ^{2}}{\geff ^{2}})-1\right\} 
\end{eqnarray}
and the entropy density as\begin{equation}
\frac{\mathcal{S}_{L,n_{b}}}{N_{g}}=-\frac{\geff ^{2}\mu ^{2}T}{24\pi ^{2}}\left\{ \left( 1+\frac{\geff ^{2}}{4\pi ^{2}}\right) \ln (1+\frac{4\pi ^{2}}{\geff ^{2}})-1\right\} .
\end{equation}
The logarithm can be expanded giving\begin{equation}
\label{entropyL}
\frac{\mathcal{S}_{L,n_{b}}}{N_{g}}=\frac{\geff ^{2}\mu ^{2}T}{24\pi ^{2}}\left( \ln \frac{\geff ^{2}}{4\pi ^{2}}+1\right) +O(\geff ^{4})+O(T^{3}).
\end{equation}

\section{Full result}

Let us take together all the results we obtained so far. We have three
main contributions to the entropy correction \begin{equation}
\mathcal{S}-\mathcal{S}_{0}=\mathcal{S}_{\non n_{b}}+\mathcal{S}_{T,n_{b}}+\mathcal{S}_{L,n_{b}}
\end{equation}
 with \( \mathcal{S}_{\mbox {\scriptsize non-}n_{b}} \) from equation
(\ref{Tsigma}), \( \mathcal{S}_{T,n_{b}} \) from (\ref{nonfermiEntropyT3}),
and \( \mathcal{S}_{L,n_{b}} \) from (\ref{entropyL}). \( \mathcal{S}_{0} \)
is the ideal-gas value per unit volume obtained from (\ref{largenfP0})
and is given by \[
\mathcal{S}_{0}=\left( \frac{\partial P_{0}}{\partial T}\right) _{V,\mu }=NN_{f}\left( \frac{\mu ^{2}}{3}T+\frac{7\pi ^{2}}{45}T^{3}\right) +N_{g}\frac{4\pi ^{2}}{15}T^{3}.\]
Using the following abbreviations \begin{eqnarray}
c_{5/3} & = & -\frac{8\; 2^{2/3}\Gamma (\frac{8}{3})\zeta (\frac{8}{3})}{9\sqrt{3}\pi ^{11/3}}(\geff \mu )^{4/3},\label{c53} \\
c_{7/3} & = & \frac{80\; 2^{1/3}\Gamma (\frac{10}{3})\zeta (\frac{10}{3})}{27\sqrt{3}\pi ^{13/3}}(\geff \mu )^{2/3}\label{c73} 
\end{eqnarray}
we can write the final result to the entropy correction as \begin{eqnarray}
{\mathcal{S}-\mathcal{S}_{0}\0N _{g}} & = & \frac{\geff ^{2}\mu ^{2}T}{36\pi ^{2}}\left( \ln {4\g \mu \0 \pi ^{2}T}-2+\gamma _{E}-\frac{6}{\pi ^{2}}\zeta '(2)\right) \nonumber \label{Sseries} \\
 &  & +\, c_{5/3}T^{5/3}+c_{7/3}T^{7/3}+O(T^{3}\ln T)\label{entropyNLO} 
\end{eqnarray}
and for the specific heat as \begin{eqnarray}
\frac{\mathcal{C}_{V}-\mathcal{C}_{V}^{0}}{N_{g}} & = & \frac{\geff ^{2}\mu ^{2}T}{36\pi ^{2}}\left( \ln \frac{4\geff \mu }{\pi ^{2}T}-3+\gamma _{E}-\frac{6}{\pi ^{2}}\zeta '(2)\right) \nonumber \\
 & + & \! {5\03 }c_{5/3}T^{5/3}+{7\03 }c_{7/3}T^{7/3}+O(T^{3}\ln T)\label{specificheatNLO} 
\end{eqnarray}
with the ideal-gas contribution of the specific heat as calculated
in appendix \ref{chapter_thermodynamics}\begin{equation}
\mathcal{C}_{V}^{0}=NN_{f}\left( \frac{\mu ^{2}T}{3}+\frac{7\pi ^{2}T^{3}}{15}-\frac{4\mu ^{2}T^{3}}{9\left( \frac{T^{2}}{3}+\frac{\mu ^{2}}{\pi ^{2}}\right) }\right) +N_{g}\frac{4\pi ^{2}T^{3}}{15}.
\end{equation}

\subsection{Extension to full QED and QCD}

Having calculated the large \( N_{f} \) result of entropy and specific
heat, the natural question is how these results translate back to
the full theory of QED or QCD. It turns out that for entropy and specific
heat of QED and QCD all we calculated so far in the first few orders
of \( T\ll \mu  \) is actually all there is. 

In fact, our NLO expression for the pressure at large \( N_{f} \)
(\ref{PNLO}) can be seen as the starting point for an expansion of
small \( g \) but finite, smaller \( N_{f} \) in the regime \( T/\mu \ll g \)
with an error of order \( g^{4} \). This is because bosonic loop
insertions that we could omit in the large \( N_{f} \) limit are
also negligible in this regime to the order of interest. The pressure
would then be given by \begin{eqnarray}
{P} & = & NN_{f}\left( {\mu ^{4}\012 \pi ^{2}}+{\mu ^{2}T^{2}\06 }+{7\pi ^{2}T^{4}\0180 }\right) \nonumber \\
 &  & -N_{g}\int \frac{d^{3}q}{(2\pi )^{3}}\int _{0}^{\infty }\frac{dq_{0}}{\pi }\nonumber \\
 & \times  & \! \biggl [2\left( [n_{b}+\frac{1}{2}]\textrm{Im}\ln D^{-1}_{T}-\frac{1}{2}\textrm{Im}\ln D^{-1}_{\textrm{vac}}\right) \nonumber \\
 & + & \! \! \left( [n_{b}+\frac{1}{2}]\textrm{Im}\ln \frac{D^{-1}_{L}}{q^{2}-q_{0}^{2}}-\frac{1}{2}\textrm{Im}\ln \frac{D^{-1}_{\textrm{vac}}}{q^{2}-q_{0}^{2}}\right) \biggr ]\nonumber \\
 &  & +O(g^{4}\mu ^{4}),\qquad (T/\mu \ll g)\label{pressurecomplete} 
\end{eqnarray}
 where \( N=3 \), \( N_{g}=8 \) for QCD, and both equal to one for
QED. The temperature \( T \) is assumed to be the smallest mass scale
in the problem. \( D_{T} \) and \( D_{L} \) are the spatially transverse
and longitudinal gauge boson propagators at finite temperature \( T \)
and (electron or quark) chemical potential \( \mu  \) obtained by
Dyson-resumming one-loop fermion loops, and \( D_{\textrm{vac}} \)
is the corresponding quantity at zero temperature and chemical potential,
just as in equation (\ref{PNLO}). This expression is sufficient to
obtain results up to (but not including) order \( T^{4} \) in the
pressure or to order \( T^{3} \) in the entropy for the regime \( T/\mu \ll g \).
Since the fractional powers \( T^{5/3} \) and \( T^{7/3} \) are
included in these bounds they indeed give the anomalous behavior also
to full QED and QCD.

In order to obtain the anomalous terms through order \( T^{7/3} \)
one actually only needs the following terms of the real transverse
part of the propagator (\ref{approxReDT}) where we only keep the
lowest orders in \( q_{0}/q \) and \( q_{0}/\mu  \) \begin{eqnarray}
\re \, D_{T}^{-1} & = & q^{2}\left( 1+O(\g ^{2})\right) \nonumber \\
 & + & \! \! \left( \frac{\geff ^{2}\mu ^{2}}{\pi ^{2}q^{2}}-1+O(\g ^{2}q^{0})+O(\g ^{2}q^{2}/\mu ^{2})\right) q_{0}^{2}\nonumber \\
 & + & \! \! O(\g ^{2}q_{0}^{4}).\label{approxReDTpaper} 
\end{eqnarray}
These are the necessary terms to give the same {}``low order front''
as presented in equation (\ref{loworderfront1}). 

For \( T/\mu \ll g \) the anomalous contributions we calculated for
the \emph{pressure} are actually negligible compared to zero-temperature
contributions \( \sim g^{4}\mu ^{4} \) which we omitted in our starting
formula (\ref{pressurecomplete}) and which can be found for QED and
QCD to order \( g^{4} \) in references \cite{Freedman:1977dm,Baluni:1978ms}.
However, when calculating the \emph{entropy density} \( \mathcal{S}=(\partial P/\partial T)_{V,\mu } \)
or the \emph{specific heat}, these terms drop out and we obtain the
correct low temperature series expansions for QED and QCD by the results
calculated above in equations (\ref{entropyNLO}) and (\ref{specificheatNLO}). 

A useful way of understanding the organization of the series expansion
is to make explicit that \( T/\mu  \) is the smallest scale by writing
\( T/\mu \sim \g ^{1+\delta } \) with \( \delta >0 \). The terms
in the expansion(\ref{entropyNLO}) then correspond to the orders
\( \g ^{3+\delta }\ln (c/\g ) \), \( \g ^{3+(5/3)\delta } \), and
\( \g ^{3+(7/3)\delta } \), respectively, with a truncation error
of the order \( \g ^{3+3\delta } \). The expansion parameter in this
low-temperature series is \( T/(\g \mu ) \), which is also the scaleless
parameter appearing in the argument of the leading logarithm of (\ref{entropyNLO})
and (\ref{specificheatNLO}), but interestingly only after the transverse
and the longitudinal contributions have been added together. The combination
\( \g \mu  \) is the scale of the Debye mass at high chemical potential,
whose leading-order value is \( m_{D}=\g \mu /\pi  \). In fact, the
calculation of the coefficients in (\ref{entropyNLO}) required keeping
the leading-order {}``hard-dense-loop'' (HDL) part of the gauge
boson propagator \cite{Braaten:1990mz,Altherr:1992mf}, in particular
the dynamic screening in (\ref{approxImDT}), but also a HDL correction
to the real part of the transverse self energy in (\ref{approxReDTpaper}).
The above calculation is therefore in a certain sense another application
of HDL resummation \cite{Altherr:1992mf}, which thus turns out to
be necessary also for a perturbative treatment of the low-temperature
regime \( T/\mu \ll g \).

Summarizing, we have found that for the entropy and the specific heat
(but not for the pressure) the expressions given in (\ref{entropyNLO})
and (\ref{specificheatNLO}) already give the leading order contribution
for full QED and QCD in a regime of \( T/\mu \ll g \) with \( \geff ^{2}=g^{2}N_{f} \)
for QED and \( \geff ^{2}=g^{2}N_{f}/2 \) for QCD.

\section{Discussion}

\subsection{Specific heat}

\begin{figure}
{\centering \includegraphics{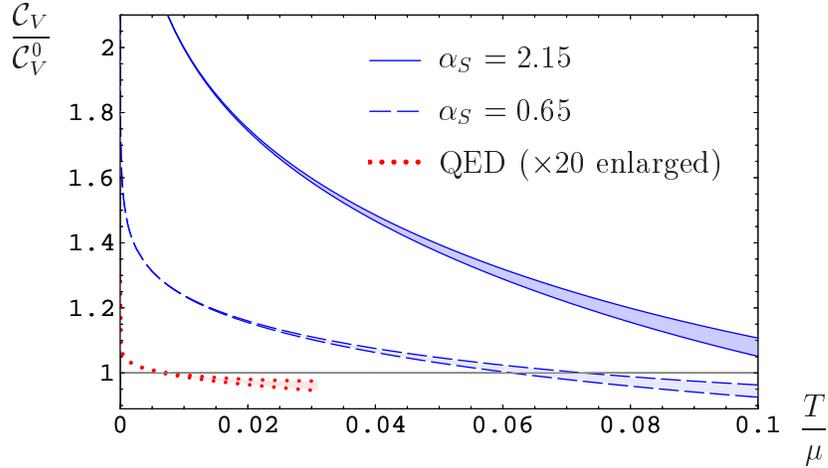} \par}

\caption{The perturbative result for the specific heat, normalized to the
ideal-gas value, to order \protect\( T^{5/3}\protect \) and \protect\( T^{7/3}\protect \)
(lower and upper curves, respectively) for two particular values of
\protect\( \alpha _{s}\protect \) in two-flavor QCD (chosen for comparability
to reference \cite{Boyanovsky:2000bc}) and \protect\( \g \approx 0.303\protect \)
for QED. The deviation of the QED result from the ideal-gas value
is enlarged by a factor of 20, and the plot terminates where the expansion
parameter \protect\( (\pi ^{2}T)/(\g \mu )\approx 1\protect \).\label{figspecificheat}}
\end{figure}

The specific heat \( \mathcal{C}_{V}\equiv C_{V}/V \) is an important
quantity for potential phenomenological applications in astrophysical
systems. Figure \ref{figspecificheat} shows the ratio of \( \mathcal{C}_{V} \)
as given by (\ref{specificheatNLO}) to the ideal-gas value \( \mathcal{C}_{V}^{0} \)
for QCD with two massless quark flavors. We compute the ratio for
two values for \( \alpha _{s} \) which have been used also in reference
\cite{Boyanovsky:2000bc} and which correspond to one-loop running
couplings with renormalization point \( 0.5 \) GeV (full line) and
1 GeV (dashed line). The upper limit of the shaded bands give the
result to order \( T^{5/3} \) and the lower band to order \( T^{7/3} \).
Alternatively we can think of the two QCD bands as roughly corresponding
to QCD with a quark chemical potential of \( 0.5 \) GeV and the total
variation corresponding to different renormalization schemes with
minimal subtraction scale varied between \( \mu  \) and \( 2\mu  \).
The critical temperature for the color superconducting phase transition
may be anywhere between 6 and 60 MeV \cite{Rischke:2003mt}, so the
range \( T/\mu \geq 0.012 \) in figure \ref{figspecificheat} might
correspond to normal quark matter. We will explore this region in
more detail in the next section, looking at the anomalous correction
at the critical temperature of color superconductivity. While it is
certainly questionable to apply perturbative results for \( \alpha _{s}\gtrsim 0.65 \),
figure \ref{figspecificheat} suggests that the anomalous feature
of an excess of the specific heat over its ideal-gas value may possibly
come into play in astrophysical situations, in particular in the cooling
of (proto-)neutron stars \cite{Iwamoto:1980eb,Carter:2000xf,Wong:2003mw}.
The anomalous correction \( \mathcal{C}_{V}/\mathcal{C}_{V}^{0}>1 \)
should be compared to the ordinary perturbative estimate for \( \mathcal{C}_{V}/\mathcal{C}_{V}^{0} \)
based on the well-known \cite{Kap:FTFT} exchange term \( \propto g^{2} \)
(which requires \( T/\mu \gg g \)). The latter would predict \( \mathcal{C}_{V}/\mathcal{C}_{V}^{0}\lesssim 0.6 \)
for \( \alpha _{s}\gtrsim 0.65 \).

In figure \ref{figspecificheat} we also show the effect on QED, where
\( \g =\sqrt{4\pi \alpha }\approx \sqrt{4\pi /137} \) \( \approx  \)
\( 0.303 \). Here the range of temperature, where the specific heat
exceeds the ideal-gas value, and the deviations from the latter, are
much smaller (the deviations from the ideal-gas value have been enlarged
by a factor of 20 in figure \ref{figspecificheat} to make them more
visible). As mentioned, the effect of the anomalous contribution to
the specific heat remains small in QED, but it might play a noticeable
role in the thermodynamics of a normal quark matter component of neutron
or proto-neutron stars.

\subsection{Effect near the CSC critical temperature\label{section_csceffects}}

It has been argued recently that color superconductivity (CSC) will
dominate dense quark matter long before non-Fermi liquid behavior
becomes effective \cite{Son:1998uk,Schaefer:2003yh}. The argument
is based on simple dimensional analysis of the energy scale, which
for the 2SC color superconducting gap is given by \cite{Son:1998uk,Schafer:1999jg,Pisarski:1999bf,Hong:1999fh,Brown:2000eh,Schaefer:2003jn,Schmitt:2002sc}
\begin{equation}
\label{energyscale_superconductor}
E=bb_{0}'\mu g^{-5}\exp (-c/g),
\end{equation}
\begin{equation}
b=512\pi ^{4}\left( \frac{2}{N_{f}}\right) ^{5/2},\quad b_{0}'=\exp (-\frac{4+\pi ^{2}}{8}),\quad c=\sqrt{\frac{6N_{c}}{N_{c}+1}}\pi ^{2}
\end{equation}
 and for non-Fermi liquid effects is given by \begin{equation}
\label{energyscale_nonfermiliquid}
\bar{E}=\bar{b}\mu g\exp (-x\bar{c}/g^{2}),
\end{equation}
\begin{equation}
\bar{b}=\frac{4}{\pi ^{2}}\sqrt{\frac{N_{f}}{2}}\exp (-3+\gamma _{E}-\frac{6}{\pi ^{2}}\zeta '(2)),\quad \bar{c}=\frac{24\pi ^{2}N_{c}}{N_{g}}.
\end{equation}
 This NFL result is obtained from equating \( x \) times the coefficient
of the leading order \( T \) contribution to the \( T\ln T^{-1} \)
coefficient from the NFL entropy correction (\ref{entropyNLO}) (for
the specific heat (\ref{specificheatNLO}), \( \bar{b} \) gets another
factor \( e \)). The parameter \( x \) gives the kind of NFL correction:
for \( x<1 \), \( \bar{E} \) gives the energy scale for corrections
of the order \( x \); for \( x\sim 1 \) we get into non-perturbative
NFL correction regime. Note that the pre-exponential factor \( b_{0}' \)
in the color superconducting energy gap (\ref{energyscale_superconductor})
stems from non-Fermi liquid behavior through the quark-selfenergy
\cite{Brown:1999aq,Brown:2000eh,Wang:2001aq}.

\begin{figure}
{\centering \includegraphics{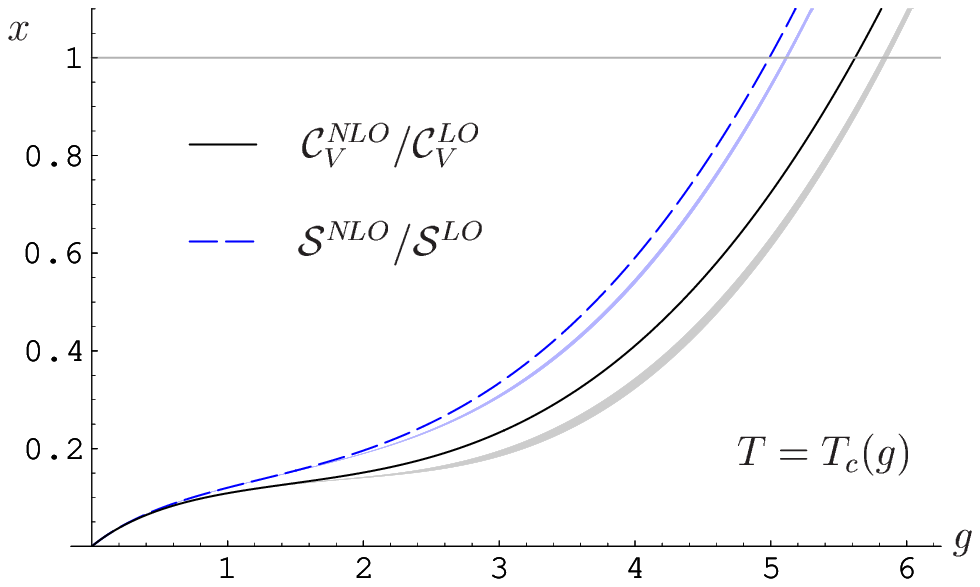} \par}

\caption{\rem{/home/ipp/andi02/vuorinen/label/csccorr02b\_label.tex}Non-Fermi
liquid correction at the critical temperature \protect\( T_{c}\protect \)
of the color superconductor as a function of the coupling \protect\( g\protect \)
for entropy and specific heat. \rem{The factor \protect\( x\protect \)
relates leading order (LO) contribution \protect\( \propto T\protect \)
to next-to-leading order (NLO) contribution \protect\( \propto T\ln T^{-1}\protect \)
via \protect\( x\times [\LO ]=[\NLO ]\protect \).}The factor \protect\( x\protect \)
is the rate of the next-to-leading order contribution \protect\( \propto T\ln T^{-1}\protect \)
to the leading order (Stefan-Boltzmann) contribution \protect\( \propto T\protect \).
The shaded bands give the result including \protect\( T^{5/3}\protect \)
(lower limit of band) and \protect\( T^{7/3}\protect \) (upper limit
of band) contributions in the NLO expression. For \protect\( g\lesssim 5\protect \),
non-Fermi liquid corrections are perturbative (\protect\( x<1\protect \))
and the size of correction is given by \protect\( x\protect \) for
the specific heat and the entropy. For \protect\( g\gtrsim 5\protect \)
nonperturbative NFL contributions cannot be neglected. For asymptotically
small \protect\( g\rightarrow 0\protect \) the correction is linear
\protect\( x=g/(3\sqrt{2})\protect \). \label{fig:cscgapNFLcorrection}\protect \\
\rem{\textbackslash{}vspace{*}\{-1.3cm\}}}
\end{figure}
In references \cite{Son:1998uk,Schaefer:2003yh}, the energy scale
for color superconductivity is related to the energy scale where NFL
effects get nonperturbative. For a 2SC system, the energy scale of
the critical temperature of the color superconductor \( T_{c} \)
is related to the gap energy (\ref{energyscale_superconductor}) by
\( T_{c}=Ee^{\gamma _{E}}/\pi  \) \cite{Schmitt:2002sc}. It is clear
that for arbitrarily small couplings \( g \) the critical energy
\( T_{c}\propto E \) will be larger than the NFL energy \( \bar{E} \).
Thus, nonperturbative NFL effects will not play a role because color
superconductivity will set in already at higher temperatures.

On the other hand, if we are interested in when \emph{perturbative}
corrections cannot be neglected, it boils down to a choice of the
size of corrections we wish to regard, entering via the parameter
\( x \): Indeed, if we want to study NFL corrections of the order
of \( g \) by setting \( x=g \) in (\ref{energyscale_nonfermiliquid}),
we find the same parametrical energy dependence \( \propto \exp (-\bar{c}/g) \)
for \( T_{c} \) and \( \bar{E} \). Here of course the magnitudes
of \( c \), \( \bar{c} \), and the large exponential prefactor \( b \)
start to play a role and one has to be more careful. Solving for the
correction in \( T_{c}=\bar{E} \) gives\begin{equation}
x=g\frac{c}{\bar{c}}+\frac{g^{2}}{\bar{c}}\ln \frac{\bar{b}g^{6}}{bb_{0}'e^{\gamma _{E}}/\pi }.
\end{equation}
From another point of view, this is just the correction \( x=\mathcal{C}_{V}^{\NLO }(T_{c})/\mathcal{C}_{V}^{\LO }(T_{c}) \)
evaluated at the critical temperature \( T_{c}(g)=T_{c}(E(g)) \)
for each coupling \( g \) for the NLO contribution \( \propto T\ln T^{-1} \).
Figure \ref{fig:cscgapNFLcorrection} shows the dependence of the
size of the correction \( x \) on the coupling \( g \) for the specific
heat and the entropy. According to this plot, below \( g\lesssim 5 \)
we are in the perturbative NFL regime where our formulae (\ref{entropyNLO})
and (\ref{specificheatNLO}) are well applicable, as can be seen from
the shaded bands in the plots which show the result including \( T^{5/3} \)
and \( T^{7/3} \) contributions. These are the biggest corrections
that we expect to see from perturbative non-Fermi liquids because
for higher temperatures \( T>T_{c} \) the NLO corrections get smaller,
and for smaller temperatures \( T<T_{c} \) we are in the color superconducting
phase. Since the color superconductor gap equation (\ref{energyscale_superconductor})
might not be valid in this coupling range anymore, this plot should
be treated with care. The only purpose of this plot is to show that
from the gap energy given above, one can not readily derive that perturbative
NFL effects could be neglected. Indeed, the energy gap formula might
break down for \( g \) as low as \( g\gtrsim 0.8 \) \cite{Rajagopal:2000rs}. 

For small \( g \) the correction decreases linearly as \( x=g/(3\sqrt{2}) \).
This is larger than naively expected from the perturbative NLO NFL
term \( \propto g^{2}\ln g \) and an effect of the exponential decrease
of the critical temperature \( T_{c} \) with \( \exp (-c/g) \). 

\tmpbibtex

\rem{ ========= Summary ========== }

\chapter{Summary and Outlook}

\setcounter{section}{1}

In this thesis we calculated pressure, entropy, and specific heat
of the high-temperature and high-density quark-gluon plasma in the
limit of a large number of quark flavors. The theoretical understanding
of the quark-gluon plasma is of current interest, as existing heavy-ion
colliders start to reach energy densities that allow for observation
of this new state of matter. In the low-temperature but high-density
region, the quark-gluon plasma features non-Fermi-liquid behavior
with anomalous contributions to thermodynamic quantities. In this
thesis we derived for the first time in a detailed calculation the
low-temperature series beyond leading-log accuracy and including anomalous
fractional powers.

We have focused on large \( N_{f} \) as the theoretical limit in
which we let the number of flavors go to \( N_{f}\rightarrow \infty  \)
and the coupling \( g\rightarrow 0 \) such that the effective coupling
\( \geff \propto g^{2}N_{f} \) stays of order \( O(1) \). We argued
that this theory is of particular interest because it is exactly solvable
up to next-to-leading order (NLO) in the \( 1/N_{f} \) expansion
and thus provides an ideal test for resummation schemes that try to
overcome the poor convergence properties of strict perturbative expansions
of thermodynamic quantities like the pressure. The NLO large \( N_{f} \)
limit technically corresponds to resumming a boson loop with any number
of fermion loop insertions which can be resummed by the Schwinger-Dyson
method. Although we cannot give a closed form of the NLO pressure,
we can numerically evaluate the result with a small error introduced
by a cutoff due to a Landau singularity in the coupling. Given that
the renormalization scale dependence is exact to the order of interest,
we have an exact result for all couplings and renormalization scales.
We calculated the pressure for the whole range of temperature and
chemical potentials where the error introduced by the cutoff is negligible
numerically. For small values of the coupling, our numerics turned
out to be sufficiently accurate to allow comparison to perturbation
theory and to extract coefficients in the perturbative expansion of
the pressure that had not been calculated analytically before. This
table shows a summary of our predictions:

\medskip{}
{\centering \begin{tabular}{c|c|c|c}
\centering Coefficient&
Predicted value&
Equation&
Confirmed\\
\hline
\hline 
\( P \): \( C_{6} \)&
+20(2)&
(\ref{pressurecoefficient6})&
not yet\\
\( \chi  \): \( C_{6} \)&
-4.55(9)&
(\ref{quarksusceptibilitycoefficient6})&
not yet\\
\( \partial ^{2}\chi /\partial \mu ^{2} \): \( C_{4} \)&
-7.02(3)&
(\ref{chinonlinpert})&
-6.9986997..\cite{Vuorinen:2003fs}\\
\( P_{T=0} \): \( C_{6,\ln } \)&
+3.18(5)&
(\ref{pressurezerotempcoefficients6})&
not yet\\
\( P_{T=0} \): \( C_{6} \)&
-3.4(3)&
(\ref{pressurezerotempcoefficients6})&
not yet\\
\end{tabular}\par}
\medskip{}

\noindent Since the original publication of the predicted values in
\cite{Ipp:2003jy} so far only one has been confirmed by an analytic
calculation. The other four constants still need to be confirmed by
means of an analytic calculation. 

\noindent For larger values of the coupling, we studied the convergence
properties of perturbative series including renormalization scale
dependence and the FAC (fastest apparent convergence) scale applied
to \( m_{E}^{2} \) (FAC-m) and to \( g_{E}^{2} \) (FAC-g) at zero
chemical potential. The latter choice leads to quite accurate results
up to \( \geff ^{2}\sim 9 \). We further studied the scaling from
pressure at small chemical potentials to higher chemical potentials
and noticed an unexpected breakdown at about \( \mu \approx \pi T \).
This might indicate a generic obstruction for extrapolating lattice
data from \( \mu \ll T \) to \( \mu \gg T \). 

In the second part of the thesis we examined the region of small \( T \)
and large \( \mu  \) more closely and found non-Fermi-liquid behavior
in an anomalous series expansion of the entropy and specific heat
for small temperatures involving a leading \( T\ln T^{-1} \) term
and fractional powers \( T^{(2n+3)/3} \). We gave a brief overview
of the theoretical picture from the ideal gas to the non-Fermi liquid.
While the classical ideal gas shows no temperature dependence in the
specific heat, the ideal Fermi gas and the Landau-Fermi liquid show
a linear dependence on temperature in the specific heat. Only the
introduction of long-range interactions changes this behavior qualitatively:
The leading \( T\ln T^{-1} \) contribution in the entropy or the
specific heat cannot be explained by classical Landau-Fermi liquid
theory anymore and the system is therefore called a non-Fermi liquid.
In our case long-range interactions are introduced by transverse gauge
boson interactions which are only weakly screened at low frequencies
\( q_{0} \) and not screened at all in the static limit of \( q_{0}\rightarrow 0 \).
We showed that the anomalous leading logarithmic contribution leads
to a range where the entropy exceeds the ideal gas value of the entropy
for small temperatures. We calculated the leading log coefficient
in a straightforward way, confirming a result by reference \cite{Chakravarty:1995}.
We further completed the argument of the leading logarithm and went
beyond the leading order to find anomalous contributions to order
\( T^{5/3} \) and \( T^{7/3} \) in the entropy and the specific
heat. These contributions indeed come from transverse {}``non-\( n_{b} \)''
parts (i.e. contributions in the original expression of the pressure
that do not contain a factor of the bosonic distribution function
\( n_{b} \)) and we calculated them by spotting a {}``low order
front'' in the denominator of the pressure integral which we can
factor into a sixth order and a fourth order polynomial, hiding higher
order terms in the frequency \( q_{0} \) in a correction term. The
final result for the entropy and the specific heat turns out to be
readily applicable to full QED and full QCD in a range where \( T/\mu \ll \geff  \).
For the entropy we found the result\begin{eqnarray}
\mathcal{S} & = & N_{c}N_{f}\frac{\mu ^{2}}{3}T+\frac{\geff ^{2}N_{g}\mu ^{2}T}{36\pi ^{2}}\left( \ln \frac{4\geff \mu }{\pi ^{2}T}-2+\gamma _{E}-\frac{6}{\pi ^{2}}\zeta '(2)\right) \nonumber \label{Sseries} \\
 &  & +N_{g}c_{5/3}T^{5/3}+N_{g}c_{7/3}T^{7/3}+O(T^{3}\ln T)
\end{eqnarray}
 from which the specific heat can be derived as \begin{eqnarray}
\mathcal{C}_{V} & = & N_{c}N_{f}\frac{\mu ^{2}T}{3}+\frac{\geff ^{2}N_{g}\mu ^{2}T}{36\pi ^{2}}\left( \ln \frac{4\geff \mu }{\pi ^{2}T}-3+\gamma _{E}-\frac{6}{\pi ^{2}}\zeta '(2)\right) \nonumber \\
 &  & +N_{g}{5\03 }c_{5/3}T^{5/3}+N_{g}{7\03 }c_{7/3}T^{7/3}+O(T^{3}\ln T)
\end{eqnarray}
with\begin{eqnarray}
c_{5/3} & = & -\frac{8\; 2^{2/3}\Gamma (\frac{8}{3})\zeta (\frac{8}{3})}{9\sqrt{3}\pi ^{11/3}}(\geff \mu )^{4/3},\label{summary_c53} \\
c_{7/3} & = & \frac{80\; 2^{1/3}\Gamma (\frac{10}{3})\zeta (\frac{10}{3})}{27\sqrt{3}\pi ^{13/3}}(\geff \mu )^{2/3}.\label{summary_c73} 
\end{eqnarray}
We pointed out a possible implication in astrophysical situations,
namely the cooling rate of proto-neutron stars, and calculated the
effect of the non-Fermi-liquid contributions near the critical temperature
of color-super\-conduct\-ivity. We found that while \emph{non-perturbative}
non-Fermi-liquid behavior might play a role for \( g\gtrsim 5 \),
there are non-negligible \emph{perturbative} non-Fermi-liquid effects
for \( g\sim O(1) \) of the order of \( 10\%-20\% \) compared to
the leading order contribution, and that perturbative non-Fermi-liquid
effects are of the order of \( g/(3\sqrt{2}) \) for small \( g \). 

In the appendix we calculated the Feynman rules trying to match different
conventions in the literature, gave a derivation of the specific heat
at next-to-leading order and presented the zero-temperature limits
of the boson self-energy with one fermion insertion. Finally, we provided
a \emph{Mathematica} extension that is necessary in order to correctly
calculate the fractional power series. This calculation involves a
series expansion that depends crucially on assumptions about the variables
involved that standard symbolic manipulation programs implicitly ignore.

Possible extensions to the work presented here first of all include
improvement of the non-Fermi-liquid series expansion. Within the large
\( N_{f} \) limit, the \( T^{3} \) contribution might be of special
interest. With the tools presented here it seems straightforward to
calculate it, but one has to take into account that disproportionately
more contributions have to be considered at this order. Apart from
higher orders in \( T \), the series expansion could also be pushed
towards higher orders in the coupling \( \geff  \). 

The analytic treatment of non-Fermi-liquid behavior presented in this
work paves the way for various possible applications: One could try
and repeat the calculation of neutron star cooling rates with the
improved series expansion presented here; one could study the interplay
of the effects of the non-Fermi-liquid contributions with color superconductivity;
finally, one could extend the analysis to the nonrelativistic case
(i.e.~include fermion masses) which may be of interest in view of
the recent experimental findings \cite{Custers:2003aa} in condensed
matter physics.

\newpage

\appendix

\chapter{Abbreviations}

\begin{lyxlist}{00.00.0000}
\item [2PI]2 Particle Irreducible
\item [2SC]2-Flavor Color Superconductor
\item [CFL]Color-Flavor Locked
\item [CSC]Color Superconductivity
\item [DR]Dimensional Reduction
\item [FAC]Fastest Apparent Convergence
\item [HDL]Hard Dense Loop
\item [HTL]Hard Thermal Loop
\item [ITF]Imaginary Time Formalism
\item [Large~\( N_{f} \)]Large number of Quark Flavors
\item [LHC]Large Hadron Collider
\item [LO]Leading Order
\item [NLO]Next-to-Leading Order
\item [PMS]Principle of Minimal Sensitivity
\item [QCD]Quantum Chromodynamics
\item [QED]Quantum Electrodynamics
\item [QGP]Quark Gluon Plasma
\item [RHIC]Relativistic Heavy Ion Collider
\item [RTF]Real Time Formalism
\item [SPS]Super Proton Synchrotron
\end{lyxlist}
\rem{DRSPT Dimensionally Reduced Screened Perturbation Theory, HTLpt
HTL-screened perturbation theory, VLHC Very Large Hadron Collider}


\chapter{QCD Feynman rules\label{section_qcdfeynmanrules}}

\renewcommand{\braket}[2]{\left\langle #1 \right| \! \! \left. #2 \right\rangle }

\section{Motivation for a unifying approach}

\rem{REDO: Fr, 25. Juli 2003}The theoretical description of the quark-gluon
plasma is based on quantum field theory (QFT). Its application to
a 'Many-Body Problem' using tools from statistical mechanics started
in the late 1950s. In the mean-time the theoretical tools improved
dramatically, for example by reorganized perturbation theory in the
1990's, and it is no wonder that a plethora of conventions and notations
have been invented so far. The drawback of this kind of scientific
independence is that it renders comparison between different authors
often a tedious task. In this chapter we will try to find a formal
{}``common denominator'' for modern literature on thermal field
theory, whilst providing a brief introduction to Feynman rules and
some basic diagrams we will be using later on.

We will use the Imaginary Time Formalism (ITF) for the statistical
description of the quantum field. This formalism is based on the observation
that the statistical density matrix for a system \( \rho (\beta )=\exp (-\beta H) \)
with inverse temperature \( \beta =1/T \) looks similar to a quantum
mechanical time evolution operator \( \exp (iHt) \) if one uses formally
imaginary time \( t\rightarrow i\beta  \). It is therefore tempting
to simply rotate QFT formulae from zero temperature, as taught in
modern textbooks like Peskin and Schroeder \cite{Peskin:1995ev},
in the complex time plane to finite temperature in ITF. Unfortunately,
naive comparisons quickly result in missing factors of \( -1 \) or
\( i \), so that it seems worth to invest some effort in locating
the sources of discrepancies. In the end we will see that only two
factors determine the sign discrepancies for Feynman rules between
various conventions, namely the sign in the generating functional,
which we will call \( \widetilde{s} \), and a factor in the covariant
derivative \( \widetilde{g} \).

Let us start by looking at the gluon propagator, which basically resembles
the photon propagator up to group factors. It can be found in various
definitions in the literature: \begin{equation}
\label{introPropPS}
\widetilde{D}_{F}^{\mu \nu }(k)=\frac{-i}{k^{2}+i\epsilon }\left( g^{\mu \nu }-(1-\xi )\frac{k^{\mu }k^{\nu }}{k^{2}}\right) ,
\end{equation}

\begin{eqnarray}
G_{\mu \nu }^{0}(k) & = & -g_{\mu \nu }\Delta (k)+(\lambda -1)k_{\mu }k_{\nu }\Delta ^{2}(k)\label{introPropBI} \\
 & = & -g_{\mu \nu }\frac{-1}{k^{2}}+(\lambda -1)k_{\mu }k_{\nu }\frac{+1}{k^{4}},\nonumber 
\end{eqnarray}

\begin{equation}
\label{introPropLB}
D^{F}_{\mu \nu }=\frac{1}{Q^{2}}\left( \delta _{\mu \nu }-(1-\xi )\frac{Q_{\mu }Q_{\nu }}{Q^{2}}\right) .
\end{equation}

Equation (\ref{introPropPS}) for \( T=0 \) can be found in Peskin
and Schroeder \cite{Peskin:1995ev}, p.~297, equation (\ref{introPropBI})
applies to the ITF and is the definition of Blaizot and Iancu \cite{Blaizot:2001nr},
p.~176, while equation (\ref{introPropLB}) is the ITF definition
according to Le Bellac \cite{LeBellac:1996aa}, p.~107. The first
two equations make use of the Minkowski metric \( g_{\mu \nu }=\diag (1,-1,-1,-1) \)
while the third equation does not. We can unify the first two equations,
using the following {}``generic'' equation:\begin{equation}
\label{introPropGeneric}
\widetilde{G}^{\mu \nu }(k)=-\widetilde{\chi }\left( g^{\mu \nu }\frac{1}{k^{2}}-(1-\xi )\frac{k^{\mu }k^{\nu }}{k^{4}}\right) 
\end{equation}
where we introduced a {}``generic constant'' \( \widetilde{\chi } \)
whose value resembles different definitions. Here we can identify
\( \widetilde{\chi }=i \) for the \( T=0 \) case in equation~(\ref{introPropPS})
and \( \widetilde{\chi }=-1 \) for the ITF in equation~(\ref{introPropBI}).
We cannot easily identify equation~(\ref{introPropLB}) with this
method, though, because it uses Euclidean metric. We will see later
how we can convert results obtained in this work from Minkowski metric
to Euclidean metric.

Another example of using different definitions is the covariant derivative.
In \cite{Peskin:1995ev}, p.~490, it is defined as \( D_{\mu }=\partial _{\mu }-igA_{\mu }^{a}t^{a} \),
while \cite{Blaizot:2001nr}, p.~171, defines the covariant derivative
with a positive sign \( D_{\mu }=\partial _{\mu }+igA_{\mu }^{a}t^{a} \).
Again these two definitions can be easily unified, using the generic
equation:\begin{equation}
\label{introCovariantGeneric}
D_{\mu }\equiv \partial _{\mu }+\widetilde{g}gA_{\mu }^{a}t^{a}
\end{equation}
with \( \widetilde{g}=-i \) in the first case and \( \widetilde{g}=+i \)
in the second. Itzykson and Zuber \cite{Itzykson:1980rh}, p.~584,
use yet another convention \( \widetilde{g}=-1 \). 

Yet another striking difference concerns the definition of the generating
functional. While \cite{Peskin:1995ev}, p.~290, defines the generating
functional for \( T=0 \) as \begin{equation}
\label{introGenFunctPS}
Z[J]=\int \cD \phi \exp \left[ i\int d^{4}x\left[ \cL +J(x)\phi (x)\right] \right] 
\end{equation}
we can find the following definition for ITF in \cite{Blaizot:2001nr},
p.~28,\begin{equation}
\label{introGenFunctBI}
Z[j]=\cN \int _{\phi (0)=\phi (\beta )}\cD (\phi )\exp \left\{ -\int _{0}^{\beta }\dd \tau \int \dd ^{3}x\left( \cL _{E}(x)-j(x)\phi (x)\right) \right\} 
\end{equation}
where the Euclidean Lagrangian \( \cL _{E} \) basically equals the
negative Minkowski Lagrangian\( -\cL  \) in the common Minkowski
metric. Defining a {}``generic'' position-integral as \begin{equation}
\label{introPosIntegralGeneric}
\int d^{4}\widetilde{x}\equiv \left\{ \begin{array}{ll}
\int d^{4}x & \text {for\, }T=0\\
\int ^{\beta }_{0}d\tau \int d^{3}\vec{x} & \text {for\, ITF}
\end{array}\right. 
\end{equation}
we can again combine the two equations to form the generic equation\begin{equation}
\label{introGenFunctGeneric}
Z[j]\equiv \cN \int \cD \phi \exp \left[ \widetilde{s}\int d^{4}\widetilde{x}\left( \cL +\widetilde{j}j(x)\phi (x)\right) \right] 
\end{equation}
with \( \widetilde{s}=i \), \( \widetilde{j}=1 \) for \( T=0 \),
and \( \widetilde{s}=1 \), \( \widetilde{j}=1 \) for ITF.

What we shall find in the following is that the Feynman rules can
be derived basically from these three generic definitions presented
here, namely the definition of the propagator (\ref{introPropGeneric}),
the definition of the covariant derivative (\ref{introCovariantGeneric})
and the definition of the generating functional (\ref{introGenFunctGeneric}).
From these definitions we can derive the QCD Feynman rules. In the
end the Feynman rules will in a simple way depend on the constants
\( \widetilde{\chi } \), \( \widetilde{g} \), and \( \widetilde{s} \).

\section{Derivation of Feynman rules}

\rem{Mi, 3. April 2002}

In the following we will present the key equations that define constants
like \( \widetilde{\chi } \), \( \widetilde{g} \), or \( \widetilde{s} \).
The aim is to unify the Feynman rules for \( T=0 \) and the Imaginary
Time Formalism (ITF). We use natural units \( \hbar =c=1 \).

\subsection*{Metric}

Equations in both formalisms can be formulated using the Minkowski
metric \begin{equation}
\label{sumMetric}
g_{\mu \nu }=g^{\mu \nu }=\diag (1,-1,-1,-1).
\end{equation}

In the case of the imaginary-time formalism we adopt the definition
of Blaizot and Iancu \cite{Blaizot:2001nr} and write \( x^{\mu }=(x_{0},\bx )=(t_{0}-i\tau ,\bx ) \)
with \( 0\leq \tau \leq \beta  \), with the inverse temperature \( \beta =T^{-1} \),
and \( k^{\mu }=(k_{0},\bk )=(i\omega _{n},\bk ) \) with \( \omega _{n}=2n\pi T \)
for bosonic fields or \( \omega _{n}=(2n+1)\pi T \) for fermionic
fields. This definition has the advantage that the scalar product
of position and momentum vector still has the form of the Minkowski
metric:\begin{equation}
\label{sumFourVector}
k_{\mu }x^{\mu }=k_{0}x_{0}-\bk \bx =\omega _{n}\tau -\bk \bx .
\end{equation}

\rem{Fr, 5. April 2002}

We define the generic position integral to be\begin{equation}
\label{sumPosIntegralGeneric}
\int d^{4}\widetilde{x}\equiv \left\{ \begin{array}{ll}
\int d^{4}x & \text {for\, }T=0\\
\int ^{\beta }_{0}d\tau \int d^{3}\vec{x} & \text {for\, ITF}.
\end{array}\right. 
\end{equation}

Analogous {}``generic'' definitions can be given for the position
integral and the Dirac \( \delta  \)-function.

\subsection*{Lagrangian}

In all textbooks we compared, the Lagrangian takes the following form:

\begin{equation}
\label{sumLagrangian}
\cL =-\frac{1}{4}F^{a}_{\mu \nu }F^{\mu \nu a}+\overline{\psi }_{i}\left( i\slD -m\right) _{ij}\psi _{j}.
\end{equation}

The covariant derivative \( \, \slD =\gamma ^{\mu }D_{\mu } \) is
defined as \begin{equation}
\label{sumCovariantGeneric}
D_{\mu }\equiv \partial _{\mu }+\widetilde{g}gA_{\mu }^{a}t^{a}
\end{equation}
with \( \widetilde{g}=i \), \( -1 \), or \( -i \). We use the commutation
relation of the generators \( t_{a} \) of the Lie algebra of the
gauge group in the fundamental representation\begin{equation}
\label{sumComGenerators}
[t^{a},t^{b}]=if^{abc}t^{c}
\end{equation}
 where \( f^{abc} \) are the structure constants of the group. The
commutator of the covariant derivative \begin{equation}
\label{sumComDer}
[D_{\mu },D_{\nu }]=\widetilde{g}g\left( \partial _{\mu }A_{\nu }^{a}-\partial _{\nu }A_{\mu }^{a}+i\widetilde{g}gf^{abc}A_{\mu }^{b}A_{\nu }^{c}\right) t^{a}\equiv \widetilde{g}gF_{\mu \nu }^{a}t^{a}
\end{equation}
 defines the field strength tensor \( F_{\mu \nu }^{a} \). The adjoint
representation is given by \( (t_{G}^{b})_{ac}=if^{abc} \). In this
representation the covariant derivative is given by\begin{equation}
\label{sumCovAdjoint}
(D_{\mu })_{ac}=\partial _{\mu }\delta _{ac}+i\widetilde{g}gA_{\mu }^{b}f^{abc}.
\end{equation}
The generating functional for gluons can be written as\begin{equation}
\label{sumGenFunctGluonGeneric}
Z[j]\equiv \cN \int [\cD A^{\mu }]\exp \left[ \widetilde{s}\int \dd ^{4}\widetilde{x}\left( \cL +\widetilde{j}j^{\nu a}(x)A^{a}_{\nu }(x)\right) \right] 
\end{equation}
with \( \widetilde{s}=i \) for \( T=0 \) and \( \widetilde{s}=+1 \)
for imaginary time formalism. We usually find \( \widetilde{j}=1 \)
in both cases. For the quarks which are fermions we have to consider
two anticommuting fields and the generating functional takes the form\begin{equation}
\label{sumGenFunctQuarkGeneric}
Z[\bar{\eta },\eta ]\equiv \cN \int \cD \bar{\psi }\cD \psi \exp \left[ \widetilde{s}\int \dd ^{4}\widetilde{x}\left( \cL +\widetilde{\eta }\bar{\eta }_{i}(x)\psi _{i}(x)+\widetilde{\eta }\bar{\psi }_{i}(x)\eta _{i}(x)\right) \right] 
\end{equation}
with \( \widetilde{\eta }=1 \). For ghost fields there is a similar
generating functional.

\subsection*{Propagators\label{sectionSummaryPropagators}}

\rem{Mo, 8. April 2002}

Starting from the generating functional (\ref{sumGenFunctGluonGeneric})
we can write the two-point correlation function as\begin{eqnarray}
<0|TA_{\mu }^{a}(x_{1})A_{\nu }^{b}(x_{2})|0> & \equiv  & \frac{\cN }{Z[0]}\int [\cD A^{\alpha }]A_{\mu }^{a}(x_{1})A_{\nu }^{b}(x_{2})\exp \left[ \widetilde{s}\int \dd ^{4}\widetilde{x}\cL \right] \nonumber \\
 & = & \frac{1}{Z[0]}\left. \frac{\delta ^{2}Z[j]}{(\widetilde{s}\widetilde{j}\delta j^{\mu a}(x_{1}))(\widetilde{s}\widetilde{j}\delta j^{\nu b}(x_{2}))}\right| _{j=0}\label{sum2PtCorGeneric} 
\end{eqnarray}

We can rewrite the generating functional by a shift of the gauge field
\( A_{\mu }^{a}(x)=A'{}^{a}_{\mu }(x)+(-\widetilde{j}/\widetilde{\chi })\int \dd ^{4}\widetilde{y}G_{\mu \nu }^{ab}(x-y)j^{\nu b}(y) \)
where the Green's function \( G_{\mu \nu }^{ab}(x-y) \) satisfies
the generic equation\begin{equation}
\label{sumGreenPosGluonGeneric}
\left( \partial ^{2}g_{\mu \nu }-\left( 1-\frac{1}{\xi }\right) \partial _{\mu }\partial _{\nu }\right) G^{\nu \rho ab}(x-y)\equiv \widetilde{\chi }\delta _{\mu }^{\rho }\delta ^{ab}\delta ^{(4)}(x-y)
\end{equation}
with \( \widetilde{\chi }=i \) for \( T=0 \) and \( \widetilde{\chi }=-1 \)
for ITF. Here we already added the Faddeev-Popov term \( -(\partial ^{\mu }A_{\mu }^{a})^{2}/2\xi  \)
to the Lagrangian. The shift of \( A_{\mu }^{a}(x) \) does not affect
the result of the integral and we can rewrite%
\footnote{In the literature this procedure is often referred to as {}``completing
the square''.
} the generating functional (\ref{sumGenFunctGluonGeneric}) as\begin{equation}
\label{sumGenFunctGreenGluonGeneric}
Z[j]=Z[0]\exp \left[ -\frac{\widetilde{s}\widetilde{j}^{2}}{2\widetilde{\chi }}\int \dd ^{4}\widetilde{x}\int \dd ^{4}\widetilde{y}j^{\mu a}G_{\mu \nu }^{ab}(x-y)j^{\nu b}(y)\right] .
\end{equation}
Using this result, we can calculate the two-point correlation function
(\ref{sum2PtCorGeneric}) as\begin{equation}
\label{sum2PtCorGreen}
\bra{0}TA_{\mu }^{a}(x_{1})A_{\nu }^{b}(x_{2})\ket{0}=\frac{-1}{\widetilde{s}\widetilde{\chi }}G_{\mu \nu }^{ab}(x_{1}-x_{2}).
\end{equation}
As expected, this expression is independent of the constant \( \widetilde{j} \).
We can calculate the Green's function for the momentum space by Fourier
transforming equation (\ref{sumGreenPosGluonGeneric}):\begin{equation}
\label{sumGreenMomGluonGeneric}
\widetilde{G}_{\mu \nu }^{ab}(k)=-\widetilde{\chi }\left( g_{\mu \nu }\frac{1}{k^{2}}+(\xi -1)\frac{k_{\mu }k_{\nu }}{k^{4}}\right) \delta ^{ab}.
\end{equation}
This is the form that we already introduced in equation (\ref{introPropGeneric}).
We can now write the Fourier transform of the two-point correlation
function

\begin{equation}
\label{sum2PtCorMom}
\begin{array}{ccccc}
\bra{0}TA_{\mu }^{a}(x_{1})A_{\nu }^{b}(x_{2})\ket{0} & \! \! \! =\! \! \!  & \underbrace{\int \frac{\dd ^{4}\widetilde{k}}{(2\pi )^{4}}} & \underbrace{\begin{array}{c}
\! \\
\! 
\end{array}\! \! \! \! e^{-ik(x_{1}-x_{2})}} & \underbrace{\frac{-1}{\widetilde{s}\widetilde{\chi }}\widetilde{G}_{\mu \nu }^{ab}(k)}.\\
 & \! \! \!  & \begin{array}{c}
\! \! \! \textrm{momentum}\! \! \! \\
\textrm{integral}
\end{array} & \textrm{external fields} & \textrm{propagator}
\end{array}
\end{equation}
Written in this way, the connection to the Feynman rules is clear
and we can readily read off the propagator that will be used in calculating
Feynman diagrams:

\begin{picture}(100,00)(80,70)
\begin{feynartspicture}(300,100)(1.,1.)
\FADiagram{} 
\FAProp(1.,10.)(20.,10.)(0.,){/Sine}{0}
\FALabel(0,11)[b]{\( a,\mu  \)} 
\FALabel(20,11)[b]{\( b,\nu  \)} 
\FAProp(6,8)(15,8)(0,){/Straight}{0}
\FAProp(14.9,8)(15,8)(0,){/Straight}{1}
\FALabel(10,7)[t]{\( k \)} 
\end{feynartspicture}
\end{picture}\begin{eqnarray}
\qquad \qquad \qquad \qquad  & = & \frac{-1}{\widetilde{s}\widetilde{\chi }}\widetilde{G}_{\mu \nu }^{ab}(k)\label{sumPropGluon} \\
 & = & \frac{1}{\widetilde{s}}\left( g_{\mu \nu }\frac{1}{k^{2}}+(\xi -1)\frac{k_{\mu }k_{\nu }}{k^{4}}\right) \delta ^{ab}.\nonumber 
\end{eqnarray}

It is interesting to see that the last line is actually independent
of \( \widetilde{\chi } \) and thus independent of our choice of
defining the propagator \( G_{\mu \nu }^{ab}(x) \). It merely depends
on the definition of the generating functional (\ref{sumGenFunctGluonGeneric}).

For fermions we can derive the propagator in a similar way starting
from the generating functional equation~(\ref{sumGenFunctQuarkGeneric}).
Again, we can shift the fermion field by introducing the fermionic
Green's function \( S_{F}^{ij}(x-y) \) which satisfies\begin{equation}
\label{sumGreenPosQuarkGeneric}
\left( i\slpd -m\right) S_{F}^{ij}(x-y)\equiv \widetilde{\sigma }\delta ^{(4)}(x-y)\delta ^{ij}\dsOne _{4\times 4}
\end{equation}
with \( \dsOne _{4\times 4} \) being the unit matrix in the space
of the anticommuting \( \gamma  \)-matrices and \( \widetilde{\sigma } \)
a generic constant. Similarly as above, we can write the two-point
correlation function of two fermionic fields as\begin{equation}
\label{sum2PtCorQuark}
\bra{0}T\psi ^{i}(x_{1})\bar{\psi }^{j}(x_{2})\ket{0}=\frac{-1}{\widetilde{s}\widetilde{\sigma }}S_{F}^{ij}(x-y)
\end{equation}
 and we find for the propagator used in the Feynman rules

\begin{picture}(100,0)(80,70)
\begin{feynartspicture}(300,100)(1.,1.)
\FADiagram{} 
\FAProp(1.,10.)(20.,10.)(0.,){/Straight}{1}
\FALabel(0,11)[b]{\( i \)} 
\FALabel(20,11)[b]{\( j \)} 
\FAProp(6,8)(15,8)(0,){/Straight}{0}
\FAProp(14.9,8)(15,8)(0,){/Straight}{1}
\FALabel(10,7)[t]{\( p \)} 
\end{feynartspicture}
\end{picture}
\begin{eqnarray}
\qquad \qquad \qquad \qquad \qquad  & = & \frac{-1}{\widetilde{s}\widetilde{\sigma }}\widetilde{S}_{F}^{ij}(p)\label{sumPropQuark} \\
 & = & \frac{-1}{\widetilde{s}}\frac{1}{\, \slp -m}\delta ^{ij}=\frac{-1}{\widetilde{s}}\frac{\, \slp +m}{p^{2}-m^{2}+i\varepsilon }\delta ^{ij}.\nonumber 
\end{eqnarray}

Finally we have similar results for the ghost field. The ghost Green's
function satisfies\begin{equation}
\label{sumGreenPosGhostGeneric}
\left( -\partial ^{2}\delta ^{ab}\right) G^{bc}_{gh}(x-y)\equiv \widetilde{c}\delta ^{ac}\delta ^{(4)}(x-y)
\end{equation}
with the generic constant \( \widetilde{c} \). This Green's function
gives the two-point correlation function\begin{equation}
\label{sum2PtCorGhost}
\bra{0}Tc^{a}(x_{1})\bar{c}^{b}(x_{2})\ket{0}=\frac{-1}{\widetilde{s}\widetilde{c}}G_{gh}^{ab}(x_{1}-x_{2})
\end{equation}
 and the ghost propagator for the Feynman rules

\begin{picture}(100,0)(80,70)
\begin{feynartspicture}(300,100)(1.,1.)
\FADiagram{} 
\FAProp(1.,10.)(20.,10.)(0.,){/GhostDash}{1}
\FALabel(0,11)[b]{\( a \)} 
\FALabel(20,11)[b]{\( b \)} 
\FAProp(6,8)(15,8)(0,){/Straight}{0}
\FAProp(14.9,8)(15,8)(0,){/Straight}{1}
\FALabel(10,7)[t]{\( k \)} 
\end{feynartspicture}
\end{picture}\begin{equation}
\label{sumPropGhost}
\qquad \qquad =\, \, \frac{-1}{\widetilde{s}\widetilde{c}}\widetilde{G}_{gh}^{ab}(k)=\frac{-1}{\widetilde{s}}\frac{\delta ^{ab}}{k^{2}}.
\end{equation}

~

\subsection*{Vertices}

We use the field strength tensor from equation (\ref{sumComDer})
\begin{equation}
\label{sumFieldStrengthTensor}
F_{\mu \nu }^{a}=\partial _{\mu }A_{\nu }^{a}-\partial _{\nu }A_{\mu }^{a}+i\widetilde{g}gf^{abc}A_{\mu }^{b}A_{\nu }^{c}
\end{equation}
to calculate the expression we need for the Lagrangian (\ref{sumLagrangian})\begin{eqnarray}
\cL _{gluon}\, \, =\, \, -\frac{1}{4}F^{a}_{\mu \nu }F^{\mu \nu a} & = & -\frac{1}{2}\left( \partial _{\mu }A_{\nu }^{a}\partial ^{\mu }A^{\nu a}-\partial _{\mu }A_{\nu }^{a}\partial ^{\nu }A^{\mu a}\right) \}\qquad \label{sumFmunu} \\
 &  & -i\partial _{\mu }A_{\nu }^{a}\widetilde{g}gf^{abc}A^{\mu b}A^{\nu c}\nonumber \\
 &  & +\frac{1}{4}\widetilde{g}^{2}g^{2}f^{abc}f^{ade}A_{\mu }^{b}A_{\nu }^{c}A^{\mu d}A^{\nu e}.\nonumber 
\end{eqnarray}

The first line in equation (\ref{sumFmunu}) gave the gluon propagator
(\ref{sumGreenPosGluonGeneric}). The following two lines can be treated
as small perturbations with the order parameter \( g \) and contribute
to the Feynman rules as 3-boson vertex (second line) and 4-boson vertex
(third line). We can calculate the 3-boson vertex for example by calculating
the 3-point function \begin{eqnarray}
\bra{\Omega }TA_{\mu }^{a}(x)A_{\nu }^{b}(y)A_{\rho }^{c}(z)\ket{\Omega } & \! \! \! =\! \! \!  & \bra{0}AAA\exp (\widetilde{s}\int \dd ^{4}\widetilde{x}(\partial A\widetilde{g}gfAA)\ket{0}\label{sum3PtFunctGluon} \\
 & \! \! \! \! \! \! \! \! \! =\! \! \! \! \!  & \! \! \bra{0}AAA\ket{0}\! +\widetilde{s}\widetilde{g}gf\bra{0}AAA\! \int \! (\partial A)AA\ket{0}+...\nonumber 
\end{eqnarray}
and applying the Wick theorem. After some careful calculation we get
for the 3-boson vertex

\begin{samepage}~

\nopagebreak

\begin{picture}(100,10)(80,90)
\begin{feynartspicture}(300,100)(1.,1.)
\FADiagram{} 
\FAProp(1.,1.)(10.,8.)(0.,){/Sine}{0}
\FAProp(19.,1.)(10.,8.)(0.,){/Sine}{0}
\FAProp(10.,19.)(10.,8.)(0.,){/Sine}{0}
\FAVert(10.,8.){0}
\FALabel(0,0)[t]{\( c,\rho  \)} 
\FALabel(20,0)[t]{\( b,\nu  \)} 
\FALabel(10,20)[b]{\( a,\mu  \)} 
\FAProp(11.5,17)(11.5,10)(0,){/Straight}{0}
\FAProp(11.5,10.1)(11.5,10)(0,){/Straight}{1}
\FALabel(12,13.5)[l]{\( k \)} 
\FAProp(16,1.5)(11,5.5)(0,){/Straight}{0}
\FAProp(11.1,5.42)(11,5.5)(0,){/Straight}{1}
\FALabel(13,3)[tr]{\( p \)} 
\FAProp(2,4)(7,8)(0,){/Straight}{0}
\FAProp(6.9,7.92)(7,8)(0,){/Straight}{1}
\FALabel(3.5,6.5)[br]{\( q \)} 
\end{feynartspicture} 
\end{picture}
\nopagebreak \begin{eqnarray}
\qquad \qquad \qquad \qquad  & =\, \, (\widetilde{s}\widetilde{g})gf^{abc} & \left( g^{\mu \nu }(k-p)^{\rho }\right. \label{sumVertex3Boson} \\
 &  & \left. +g^{\nu \rho }(p-q)^{\mu }\right. \nonumber \\
 &  & \left. +g^{\rho \mu }(q-k)^{\nu }\right) \nonumber 
\end{eqnarray}

\nopagebreak ~

\nopagebreak ~

\nopagebreak ~

\end{samepage}

\noindent where we use the short notation \( (k-p)^{\rho }\equiv k^{\rho }-p^{\rho } \).
For the 4-boson vertex we can start similarly from a 4-point correlation
function and obtain

\begin{samepage}

\nopagebreak ~

\begin{picture}(100,10)(80,90)
\begin{feynartspicture}(300,100)(1.,1.)
\FADiagram{} 
\FAProp(1.,1.)(10.,10.)(0.,){/Sine}{0}
\FAProp(19.,1.)(10.,10.)(0.,){/Sine}{0}
\FAProp(19.,19.)(10.,10.)(0.,){/Sine}{0}
\FAProp(1.,19.)(10.,10.)(0.,){/Sine}{0}
\FAVert(10.,10.){0}
\FALabel(0,0)[t]{\( d,\sigma  \)} 
\FALabel(20,0)[t]{\( c,\rho  \)} 
\FALabel(20,20)[b]{\( b,\nu  \)} 
\FALabel(0,20)[b]{\( a,\mu  \)} 
\FAProp(4,18)(9,13)(0,){/Straight}{0}
\FAProp(8.9,13.1)(9,13)(0,){/Straight}{1}
\FALabel(6.5,16.5)[bl]{\( k \)} 
\FAProp(18,16)(13,11)(0,){/Straight}{0}
\FAProp(13.1,11.1)(13,11)(0,){/Straight}{1}
\FALabel(16.5,13.5)[tl]{\( p \)} 
\FAProp(16,2)(11,7)(0,){/Straight}{0}
\FAProp(11.1,6.9)(11,7)(0,){/Straight}{1}
\FALabel(13.5,3.5)[tr]{\( q \)} 
\FAProp(2,4)(7,9)(0,){/Straight}{0}
\FAProp(6.9,8.9)(7,9)(0,){/Straight}{1}
\FALabel(3.5,6.5)[br]{\( r \)} 
\end{feynartspicture}
\end{picture}
\nopagebreak \begin{eqnarray}
\qquad \qquad \qquad \qquad \qquad  & \! =\, (\widetilde{s}\widetilde{g}^{2})g^{2} & \! \! \! \! \! \! \left( f^{abe}f^{cde}(g^{\mu \rho }g^{\nu \sigma }\! -\! g^{\mu \sigma }g^{\nu \rho })\right. \label{sumVertex4Boson} \\
 &  & \! \! \! \! \! \! \left. +f^{ace}f^{bde}(g^{\mu \nu }g^{\rho \sigma }\! -\! g^{\mu \sigma }g^{\nu \rho })\right. \nonumber \\
 &  & \! \! \! \! \! \! \left. +f^{ade}f^{bce}(g^{\mu \nu }g^{\rho \sigma }\! -\! g^{\mu \rho }g^{\nu \sigma })\right) .\nonumber 
\end{eqnarray}

\nopagebreak ~

\nopagebreak ~

\nopagebreak ~

\end{samepage}

The part of the Lagrangian (\ref{sumLagrangian}) that describes the
fermions is given by \begin{equation}
\label{sumFermionLagrangian}
\cL _{quark}=\overline{\psi }_{i}\left( i\slD -m\right) _{ij}\psi _{j}=\overline{\psi }_{i}\left( i\slpd -m\right) \delta _{ij}\psi _{j}+i\widetilde{g}g\overline{\psi }_{i}\gamma ^{\mu }A^{a}_{\mu }t^{a}_{ij}\psi _{j}
\end{equation}
where we used the covariant derivative \( i\slD =i\gamma ^{\mu }\partial _{\mu }+i\widetilde{g}g\gamma ^{\mu }A^{a}_{\mu }t^{a} \)
from equation (\ref{sumCovariantGeneric}) . The first part of equation
(\ref{sumFermionLagrangian}) gave the fermion propagator (\ref{sumPropQuark})
while the second part gives the quark-gluon vertex

\begin{samepage} ~

\nopagebreak ~

\begin{picture}(100,10)(80,80)
\begin{feynartspicture}(300,100)(1.,1.)
\FADiagram{} 
\FAProp(1.,1.)(10.,8.)(0.,){/Straight}{-1}
\FAProp(19.,1.)(10.,8.)(0.,){/Straight}{1}
\FAProp(10.,19.)(10.,8.)(0.,){/Sine}{0}
\FAVert(10.,8.){0}
\FALabel(0,0)[t]{\( i \)} 
\FALabel(20,0)[t]{\( j \)} 
\FALabel(10,20)[b]{\( a,\mu  \)} 
\FAProp(11.5,17)(11.5,10)(0,){/Straight}{0}
\FAProp(11.5,10.1)(11.5,10)(0,){/Straight}{1}
\FALabel(12,13.5)[l]{\( k \)} 
\FAProp(16,1.5)(11,5.5)(0,){/Straight}{0}
\FAProp(11.1,5.42)(11,5.5)(0,){/Straight}{1}
\FALabel(13,3)[tr]{\( p \)} 
\FAProp(2,4)(7,8)(0,){/Straight}{0}
\FAProp(2.1,4.08)(2,4)(0,){/Straight}{1}
\FALabel(3.5,6.5)[br]{\( q \)} 
\end{feynartspicture}
\end{picture}\\
\nopagebreak \begin{equation}
\label{sumVertexQuarkGluon}
=\, (i\widetilde{s}\widetilde{g})g\gamma ^{\mu }t^{a}_{ij}.
\end{equation}

\nopagebreak ~

\nopagebreak ~

\nopagebreak ~

\nopagebreak ~

\nopagebreak ~

\end{samepage}

\rem{Di, 9. April 2002 - Zahnarzttermin.. oje =)}

Similarly we can write down the Lagrangian part for the ghost fields\begin{equation}
\label{sumGhostLagrangian}
\cL _{ghost}=\bar{c}^{a}\left( -\partial ^{\mu }D_{\mu }^{ac}\right) c^{c}=\bar{c}^{a}\left( -\partial ^{\mu }\partial _{\mu }\delta ^{ac}\right) c^{c}+\bar{c}^{a}\left( -i\widetilde{g}gf^{abc}\partial ^{\mu }A^{b}_{\mu }\right) c^{c}
\end{equation}
where we used the adjoint representation of the covariant derivative
(\ref{sumCovAdjoint}). Note that the derivative in the second part
acts on both the gauge field \( A_{\mu }^{b} \) and the ghost field
\( c^{c} \). The first part determined the ghost propagator (\ref{sumPropGhost})
while the second part gives the following Feynman rule:

\begin{samepage} ~

\nopagebreak ~

\begin{picture}(100,10)(80,80)
\begin{feynartspicture}(300,100)(1.,1.)
\FADiagram{} 
\FAProp(1.,1.)(10.,8.)(0.,){/GhostDash}{-1}
\FAProp(19.,1.)(10.,8.)(0.,){/GhostDash}{1}
\FAProp(10.,19.)(10.,8.)(0.,){/Sine}{0}
\FAVert(10.,8.){0}
\FALabel(0,0)[t]{\( a \)} 
\FALabel(20,0)[t]{\( c \)} 
\FALabel(10,20)[b]{\( b,\mu  \)} 
\FAProp(11.5,17)(11.5,10)(0,){/Straight}{0}
\FAProp(11.5,10.1)(11.5,10)(0,){/Straight}{1}
\FALabel(12,13.5)[l]{\( k \)} 
\FAProp(16,1.5)(11,5.5)(0,){/Straight}{0}
\FAProp(11.1,5.42)(11,5.5)(0,){/Straight}{1}
\FALabel(13,3)[tr]{\( p \)} 
\FAProp(2,4)(7,8)(0,){/Straight}{0}
\FAProp(2.1,4.08)(2,4)(0,){/Straight}{1}
\FALabel(3.5,6.5)[br]{\( q \)} 
\end{feynartspicture}
\end{picture} 
\nopagebreak \begin{equation}
\label{sumVertexGhostGluon}
\qquad =\, (-\widetilde{s}\widetilde{g})gq^{\mu }f^{abc}.
\end{equation}

\nopagebreak ~

\nopagebreak ~

\nopagebreak ~

\nopagebreak ~

\nopagebreak ~

\end{samepage}

In this Feynman rule \( q^{\mu } \) denotes the momentum of the outgoing
ghost.

\subsection*{Euclidean metric\label{sectionSumEuclideanMetric}}

\rem{10. April 2002}

We can now transform our results into Euclidean metric. In deriving
the Feynman rules we did not make use of the metric \( g_{\mu \nu } \)
other than contracting momentum vectors like \( k^{\mu }k_{\mu } \)
or momentum and position vectors as in the Fourier transform \( k^{\mu }x_{\mu } \).
The latter is taken care of by our generic definitions of the integrals
(\ref{sumPosIntegralGeneric}) and delta functions. The former just
needs {}``undoing'' the contractions. Since we started with lower
indices at the beginning (\ref{sum3PtFunctGluon}), resulting in propagators
with lower indices and vertices with upper indices, we now have to
{}``pull down'' all indices. That is\begin{eqnarray}
k^{\mu } & \rightarrow  & g^{\mu \mu '}k_{\mu '}\, ,\label{sumMetricTransform} \\
k^{2}=k^{\mu }k_{\mu } & \rightarrow  & g^{\mu \mu '}k_{\mu '}k_{\mu }\nonumber 
\end{eqnarray}
 such that in our formula there are only vectors \( k_{\mu },\, p_{\mu },\, q_{\mu },\, ... \)
with lower indices. 

\rem{Do, 11. April 2002}

We could have done the complete calculation by using the Euclidean
metric this way, so we can simply replace \begin{equation}
\label{sumMetricConvert}
g_{\mu \nu }=g^{\mu \nu }\rightarrow -\delta _{\mu \nu }
\end{equation}
and write \( k^{2}=g^{\mu \nu }k_{\mu }k_{\nu }\rightarrow -\delta _{\mu \nu }K_{\mu }K_{\nu }=-K^{2} \),
where we use capital letters to denote vectors \( K^{2}=K_{\mu }K_{\mu }=K_{4}^{2}+K_{1}^{2}+K_{2}^{2}+K_{3}^{2} \)
in the Euclidean metric. 

For example the gluon propagator (\ref{sumGreenMomGluonGeneric})
becomes\begin{eqnarray}
\widetilde{G}_{\mu \nu }^{ab}(k) & = & -\widetilde{\chi }\left( g_{\mu \nu }\frac{1}{g^{\alpha \beta }k_{\alpha }k_{\beta }}+(\xi -1)\frac{k_{\mu }k_{\nu }}{g^{\alpha \beta }g^{\gamma \delta }k_{\alpha }k_{\beta }k_{\gamma }k_{\delta }}\right) \delta ^{ab}\qquad \label{sumMetricPropGluon} \\
 & \rightarrow  & -\widetilde{\chi }\left( -\delta _{\mu \nu }\frac{1}{-K^{2}}+(\xi -1)\frac{K_{\mu }K_{\nu }}{+K^{4}}\right) \delta ^{ab}.\nonumber 
\end{eqnarray}

For vertices we have to pull down all indices, so for example the
3-boson vertex (\ref{sumVertex3Boson}) changes to \begin{eqnarray}
 & (\widetilde{s}\widetilde{g})gf^{abc}\left( g^{\mu \nu }g^{\rho \rho '}(k-p)_{\rho '}+g^{\nu \rho }g^{\mu \mu '}(p-q)_{\mu '}+g^{\rho \mu }g^{\nu \nu '}(q-k)_{\nu '}\right)  & \nonumber \\
 & \rightarrow \, \, (\widetilde{s}\widetilde{g})gf^{abc}\left( +\delta _{\mu \nu }(K-P)_{\rho }+\delta _{\nu \rho }(P-Q)_{\rho }+\delta _{\rho \mu }(Q-K)_{\nu }\right) . & \qquad \label{sumMetricVertex3Boson} 
\end{eqnarray}

In the case of slashed quantities \( \, \slp  \) we apply the same
transformation and write\begin{equation}
\label{sumTransSlash}
\, \qquad \qquad \qquad \, \slp =\gamma ^{\mu }p_{\mu }=g^{\mu \mu '}\gamma _{\mu }p_{\mu '}\rightarrow -\delta _{\mu \mu '}\gamma _{\mu }P_{\mu '}=-\slP .
\end{equation}

The fermion Green's function (\ref{sumPropQuark}) then becomes\begin{equation}
\label{sumTransQuark}
\widetilde{S_{F}}^{ij}(p)=\widetilde{\sigma }\frac{1}{\, \slp -m}\delta ^{ab}\rightarrow -\widetilde{\sigma }\frac{1}{\, \slP +m}\delta ^{ab}.
\end{equation}

\section{Comparison with literature}

\subsubsection{Values for constants\label{sectionComparisonValues}}

\newcommand{\TableComparison}{

\vspace{0.3cm}
{\centering \begin{tabular}{|l|@{}c@{}|@{}c@{}||@{}c@{}|@{}c@{}||@{}c@{}|@{}c@{}|@{}c@{}|}
\hline 
\selectlanguage{english}
~~~~~Reference
\selectlanguage{american}&
\selectlanguage{english}
M/E
\selectlanguage{american}&
\selectlanguage{english}
T
\selectlanguage{american}&
\selectlanguage{english}
\( \widetilde{s} \)
\selectlanguage{american}&
\selectlanguage{english}
\( \widetilde{g} \)
\selectlanguage{american}&
\selectlanguage{english}
\( \widetilde{\chi } \)
\selectlanguage{american}&
\selectlanguage{english}
\( \widetilde{\sigma } \)
\selectlanguage{american}&
\selectlanguage{english}
\( \widetilde{c} \)
\selectlanguage{american}\\
\hline
\hline 
\selectlanguage{english}
Peskin \& Schroeder \cite{Peskin:1995ev}
\selectlanguage{american}&
\selectlanguage{english}
M
\selectlanguage{american}&
\selectlanguage{english}
\( T_{0} \)
\selectlanguage{american}&
\selectlanguage{english}
\( i \)
\selectlanguage{american}&
\selectlanguage{english}
\( -i \)
\selectlanguage{american}&
\selectlanguage{english}
\( i \)
\selectlanguage{american}&
\selectlanguage{english}
\( i \)
\selectlanguage{american}&
\\
\hline 
\selectlanguage{english}
Itzykson \& Zuber \cite{Itzykson:1980rh}
\selectlanguage{american}&
\selectlanguage{english}
M
\selectlanguage{american}&
\selectlanguage{english}
\( T_{0} \)
\selectlanguage{american}&
\selectlanguage{english}
\( i \)
\selectlanguage{american}&
\selectlanguage{english}
\( -1 \)
\selectlanguage{american}&
\selectlanguage{english}
\( i \)
\selectlanguage{american}&
&
\\
\hline 
\selectlanguage{english}
Le Bellac, QSFT \cite{LeBellac:1991cq}
\selectlanguage{american}&
\selectlanguage{english}
M
\selectlanguage{american}&
\selectlanguage{english}
\( T_{0} \)
\selectlanguage{american}&
\selectlanguage{english}
\( i \)
\selectlanguage{american}&
\selectlanguage{english}
\( +i \)
\selectlanguage{american}&
\selectlanguage{english}
\( i \)
\selectlanguage{american}&
\selectlanguage{english}
\( i \)
\selectlanguage{american}&
\\
\hline 
\selectlanguage{english}
Huang \cite{Huang:1982ik}
\selectlanguage{american}&
\selectlanguage{english}
M
\selectlanguage{american}&
\selectlanguage{english}
\( T_{0} \)
\selectlanguage{american}&
\selectlanguage{english}
\( i \)
\selectlanguage{american}&
\selectlanguage{english}
\( +i \)
\selectlanguage{american}&
\selectlanguage{english}
\( i \)
\selectlanguage{american}&
&
\\
\hline 
\selectlanguage{english}
Le Bellac, TFT-M \cite{LeBellac:1996aa}
\selectlanguage{american}&
\selectlanguage{english}
M
\selectlanguage{american}&
\selectlanguage{english}
RTF
\selectlanguage{american}&
\selectlanguage{english}
\( i \)
\selectlanguage{american}&
\selectlanguage{english}
\( +i \)
\selectlanguage{american}&
\selectlanguage{english}
\( i \)
\selectlanguage{american}&
\selectlanguage{english}
\( i \)
\selectlanguage{american}&
\\
\hline 
\selectlanguage{english}
Blaizot \& Iancu \cite{Blaizot:2001nr}
\selectlanguage{american}&
\selectlanguage{english}
~~~M~~~
\selectlanguage{american}&
\selectlanguage{english}
~~ITF~~
\selectlanguage{american}&
\selectlanguage{english}
\( \, \, +1\, \,  \)
\selectlanguage{american}&
\selectlanguage{english}
\( \, \, +i\, \,  \)
\selectlanguage{american}&
\selectlanguage{english}
\( \, -1\,  \)
\selectlanguage{american}&
\selectlanguage{english}
\( \, -1\,  \)
\selectlanguage{american}&
\selectlanguage{english}
\( \, -1\,  \)
\selectlanguage{american}\\
\hline 
\selectlanguage{english}
Kapusta \cite{Kapusta:1989tk}
\selectlanguage{american}&
\selectlanguage{english}
M
\selectlanguage{american}&
\selectlanguage{english}
ITF
\selectlanguage{american}&
\selectlanguage{english}
\( +1 \)
\selectlanguage{american}&
\selectlanguage{english}
\( +i \)
\selectlanguage{american}&
\selectlanguage{english}
\( -1 \)
\selectlanguage{american}&
\selectlanguage{english}
\( +1 \)
\selectlanguage{american}&
\selectlanguage{english}
\( +1 \)
\selectlanguage{american}\\
\hline 
\selectlanguage{english}
Le Bellac, TFT-E \cite{LeBellac:1996aa}
\selectlanguage{american}&
\selectlanguage{english}
E
\selectlanguage{american}&
\selectlanguage{english}
ITF
\selectlanguage{american}&
\selectlanguage{english}
\( +1 \)
\selectlanguage{american}&
\selectlanguage{english}
\( +i \)
\selectlanguage{american}&
\selectlanguage{english}
\( -1 \)
\selectlanguage{american}&
\selectlanguage{english}
\( -1 \)
\selectlanguage{american}&
\\
\hline
\end{tabular}\par}
\vspace{0.3cm}

}
\rem{

Annotations:

The following formulae were used to adapt the constants.

Peskin \& Schroeder: Feynman rules: p801, \( \widetilde{\chi } \):
p297 (9.58). \( \widetilde{\sigma } \): p63 (3.120). 

Itzykson \& Zuber: Feynman rules: p698, p583. \( \widetilde{\chi } \):
p272 (6-28), p319 (7-1).

Le Bellac, QSFT: Feynman rules: p581, p509. \( \widetilde{\chi } \):
p439 (12.1.11). p429 (11.4.17). \( \widetilde{\sigma } \): p439 (12.1.7a).

Huang: Feynman rules: p254, Fig. 12.1. \( \widetilde{\chi } \): p262,
(12.39). 

Blaizot \& Iancu: \( \widetilde{\chi } \): p176 (B.4). \( \widetilde{\sigma } \):
p176 (B.3). \( \widetilde{c} \): p185, before (B.53).

Kapusta: Feynman rules: p124. \( \widetilde{g} \): p122 (8.4). \( \widetilde{\chi } \),
\( \widetilde{\sigma } \), \( \widetilde{c} \): p124, \( G \),
\( D \), \( W \).

Le Bellac, TFT-M: Feynman rules: p242. \( \widetilde{\chi } \): p104,
p242. \( \widetilde{\sigma } \): p242.

Le Bellac, TFT-E: Feynman rules: p241. \( \widetilde{\chi } \): p107,
p240. \( \widetilde{\sigma } \): p240.

}

\begin{table}
\TableComparison

\caption{Comparison of definitions given in literature. {}``M/E'' means
the metric (Minkowski or Euclidean), T stands for temperature \protect\( T=0\protect \)
(\protect\( T_{0}\protect \)), Imaginary Time Formalism (ITF) or
Real Time Formalism (RTF). Le Bellac \cite{LeBellac:1996aa} gives
two sets of Feynman rules which are compared independently. Only \protect\( \widetilde{s}\protect \)
and \protect\( \widetilde{g}\protect \) are relevant for the sign
changes in the Feynman rules. \label{tableComparisonConstants}}
\end{table}

We want to compare our {}``generic'' results to various results
of Feynman rules in the literature. The results are summarized in
table \ref{tableComparisonConstants}. The first constant \( \widetilde{s} \)
describes the sign in the Lagrangian of the generating functional
\( Z=\exp (\widetilde{s}\int \cL ) \) and can be obtained by comparing
equations in the literature to equation (\ref{introGenFunctGeneric}).
Another way is to compare the propagators of the Feynman rules as
in equations (\ref{sumPropGluon}), (\ref{sumPropQuark}), and (\ref{sumPropGhost}).
These are independent of the sign of the Green's function (\( \widetilde{\chi } \),
\( \widetilde{\sigma } \), \( \widetilde{c} \)) and are usually
given in a table of Feynman rules. A first consistency check can be
applied here to see whether the definitions of the three kinds of
propagators (gluon, quark, and ghost) are consistent with each other.

The next constant \( \widetilde{g} \) appears in the definition of
the covariant derivative \( D_{\mu }\equiv \partial _{\mu }+\widetilde{g}gA_{\mu }^{a}t^{a} \)
(\ref{sumCovariantGeneric}). The next three constants \( \widetilde{\chi } \),
\( \widetilde{\sigma } \), \( \widetilde{c} \) for the gluon, quark,
and ghost Green's functions respectively are not used in all books.
They are not essential for the Feynman rules, though, as they do not
enter the equations of the Feynman vertices and Feynman propagators.

\subsubsection{Vertices\label{sectionComparisonVertices}}

\newcommand{\TableVertices}{

{\centering \begin{tabular}{|l|@{}c@{}|c|@{}c@{}|c|}
\hline 
\selectlanguage{english}
~~~~~Reference
\selectlanguage{american}&
\selectlanguage{english}
3-Gluon
\selectlanguage{american}&
\selectlanguage{english}
4-Gluon
\selectlanguage{american}&
\selectlanguage{english}
Quark
\selectlanguage{american}&
\selectlanguage{english}
Ghost
\selectlanguage{american}\\
\hline
\hline 
\selectlanguage{english}
~~~Generic - Minkowski
\selectlanguage{american}&
\selectlanguage{english}
\( \widetilde{s}\widetilde{g} \)
\selectlanguage{american}&
\selectlanguage{english}
\( \widetilde{s}\widetilde{g}^{2} \)
\selectlanguage{american}&
\selectlanguage{english}
\( i\widetilde{s}\widetilde{g} \)
\selectlanguage{american}&
\selectlanguage{english}
\( -\widetilde{s}\widetilde{g} \)
\selectlanguage{american}\\
\hline
\hline 
\selectlanguage{english}
\( \!  \)Peskin \& Schroeder \cite{Peskin:1995ev}\( \! \!  \)
\selectlanguage{american}&
\selectlanguage{english}
\( 1 \)
\selectlanguage{american}&
\selectlanguage{english}
\( -i \)
\selectlanguage{american}&
\selectlanguage{english}
\( i\, (^{\ddagger }) \)
\selectlanguage{american}&
\selectlanguage{english}
\( -1 \) ({*})
\selectlanguage{american}\\
\hline 
\selectlanguage{english}
\( \!  \)Itzykson \& Zuber \cite{Itzykson:1980rh}
\selectlanguage{american}&
\selectlanguage{english}
\( -i \)
\selectlanguage{american}&
\selectlanguage{english}
\( i \)
\selectlanguage{american}&
\selectlanguage{english}
\( 1 \)
\selectlanguage{american}&
\selectlanguage{english}
\( +i \)
\selectlanguage{american}\\
\hline 
\selectlanguage{english}
\( \!  \)Le Bellac, QSFT \cite{LeBellac:1991cq}
\selectlanguage{american}&
\selectlanguage{english}
\( -1 \)
\selectlanguage{american}&
\selectlanguage{english}
\( -i \)
\selectlanguage{american}&
\selectlanguage{english}
\( -i \)
\selectlanguage{american}&
\selectlanguage{english}
\( 1 \)
\selectlanguage{american}\\
\hline 
\selectlanguage{english}
\( \!  \)Huang \cite{Huang:1982ik}
\selectlanguage{american}&
\selectlanguage{english}
\( -1\neq -i\bad  \)
\selectlanguage{american}&
\selectlanguage{english}
\( \!  \)\( \! -i\neq i\bad \!  \)\( \!  \)
\selectlanguage{american}&
\selectlanguage{english}
\( -i\, (^{\ddagger }) \)
\selectlanguage{american}&
\selectlanguage{english}
\( \! 1\neq \, ?\bad \!  \)
\selectlanguage{american}\\
\hline 
\selectlanguage{english}
\( \!  \)Le Bellac, TFT-M \cite{LeBellac:1996aa}\( \! \!  \)
\selectlanguage{american}&
\selectlanguage{english}
\( -1 \)
\selectlanguage{american}&
\selectlanguage{english}
\( -i \)
\selectlanguage{american}&
\selectlanguage{english}
\( -i \)
\selectlanguage{american}&
\selectlanguage{english}
\( 1 \)
\selectlanguage{american}\\
\hline 
\selectlanguage{english}
\( \!  \)Blaizot \& Iancu \cite{Blaizot:2001nr}
\selectlanguage{american}&
\selectlanguage{english}
\( i\, (^{\dagger }) \)
\selectlanguage{american}&
\selectlanguage{english}
\( -1\, (^{\dagger }) \)
\selectlanguage{american}&
\selectlanguage{english}
\( -1\, (^{\dagger }) \)
\selectlanguage{american}&
\selectlanguage{english}
\( -i\, (^{\dagger }) \)
\selectlanguage{american}\\
\hline 
\selectlanguage{english}
\( \!  \)Kapusta \cite{Kapusta:1989tk}
\selectlanguage{american}&
\selectlanguage{english}
\( i\neq -i\bad  \)
\selectlanguage{american}&
\selectlanguage{english}
\( -1 \)
\selectlanguage{american}&
\selectlanguage{english}
\( \!  \)\( -1\neq +1\bad  \)\( \!  \)
\selectlanguage{american}&
\selectlanguage{english}
\( -i \)
\selectlanguage{american}\\
\hline
\multicolumn{5}{l}{\selectlanguage{english}
{\tiny ~}
\selectlanguage{american}}\\
\hline 
\selectlanguage{english}
~~~Generic - Euclidean
\selectlanguage{american}&
\selectlanguage{english}
\( \widetilde{s}\widetilde{g} \)
\selectlanguage{american}&
\selectlanguage{english}
\( \widetilde{s}\widetilde{g}^{2} \)
\selectlanguage{american}&
\selectlanguage{english}
\( -i\widetilde{s}\widetilde{g} \)
\selectlanguage{american}&
\selectlanguage{english}
\( +\widetilde{s}\widetilde{g} \)
\selectlanguage{american}\\
\hline
\hline 
\selectlanguage{english}
\( \!  \)Le Bellac, TFT-E \cite{LeBellac:1996aa}\( \! \!  \)
\selectlanguage{american}&
\selectlanguage{english}
\( i \)
\selectlanguage{american}&
\selectlanguage{english}
\( -1 \)
\selectlanguage{american}&
\selectlanguage{english}
\( 1 \)
\selectlanguage{american}&
\selectlanguage{english}
\( +i \)
\selectlanguage{american}\\
\hline
\end{tabular}\par}

}

{
\begin{table}
{\raggedright \TableVertices\par}

\caption{Prefactors of Feynman rule vertices. The vertex equations are given
in (\ref{sumVertex3Boson}), (\ref{sumVertex4Boson}), (\ref{sumVertexQuarkGluon}),
and (\ref{sumVertexGhostGluon}) for the 3-gluon, 4-gluon, quark-gluon,
and ghost-gluon vertices respectively. If results differ from literature
(marked by \protect\( \bad \protect \)) then the l.h.s.~of \protect\( \neq \protect \)
denotes the generic value and the r.h.s.~denotes the value from literature.
The ghost vertex of Peskin and Schroeder ({*}) is wrong in the book
\protect\( (+1)\protect \), but corrected in the online list of errata
\cite{Peskin:1995web}. Blaizot and Iancu \protect\( (^{\dagger })\protect \)
do not give the Feynman rules explicitly, so the values in the table
are the generic ones. Peskin and Schroeder and Huang \protect\( (^{\ddagger })\protect \)
do not give the quark indices in the graphs, resulting in a possible
\protect\( \pm 1\protect \) factor from the antisymmetric structure
constants. \label{tablePrefactors} }
\end{table}
\par}

Table \ref{tablePrefactors} shows the prefactors of the Feynman rule
vertices of the 3-gluon (\ref{sumVertex3Boson}), 4-gluon (\ref{sumVertex4Boson}),
quark-gluon (\ref{sumVertexQuarkGluon}), and ghost-gluon (\ref{sumVertexGhostGluon})
interactions. Before results from literature can be compared to our
generic results, some obvious transformations have to be made: These
include relabelling indices, using the antisymmetry property of the
structure constants \( f^{abc} \), or flipping signs because of different
orientation of momentum vectors. 

\rem{Fr, 12. April 2002}

Our results agree with most of the authors. At \( T=0 \) in the Minkowski
metric our results perfectly agree with Peskin and Schroeder \cite{Peskin:1995ev},
Itzykson \& Zuber \cite{Itzykson:1980rh} and Le Bellac \cite{LeBellac:1991cq}.
In the imaginary time formalism in Euclidean metric our results agree
with Le Bellac \cite{LeBellac:1996aa}. We calculated the prefactors
for Euclidean metric in table \ref{tablePrefactors} by applying the
transformation rules from Minkowski to Euclidean metric of section
\ref{sectionSumEuclideanMetric}.

The results of Itzykson \& Zuber \cite{Itzykson:1980rh} use structure
constants \( C_{abc}=if^{abc} \) and momentum vectors pointing outwards
instead of inwards. Taking care of these obvious changes, their results
are in accordance with our generic Feynman rules.

Our results do not agree entirely with Kapusta \cite{Kapusta:1989tk}
in imaginary time formalism with Minkowski metric. There are relative
sign errors \( -1 \) in the 3-gluon vertex and in the quark-gluon
vertex. His group generators are labelled \( G_{ij}^{a}=t_{ij}^{a} \).

In the case of Huang \cite{Huang:1982ik} Feynman rules disagree in
a weird way for the 3-gluon vertex, the 4-gluon vertex and the ghost
vertex. In the case of the ghost vertex, the momentum of the boson
enters the Feynman rule instead of the momentum of the anti-ghost
as in the rest of the literature cited.

Peskin and Schroeder \cite{Peskin:1995ev} do not provide quark indices
\( i \), \( j \) in their Feynman graphs. A swapping of the quark
indices results in a relative minus sign \( -1 \) stemming from the
antisymmetric nature of the representation matrices \( (t^{a})_{ij}=if^{iaj}=-(t^{a})_{ji} \).
It is assumed that incoming quarks correspond to the second matrix
index and outgoing quarks correspond to the first matrix index, matching
the indices in the fermion interaction Lagrangian (\ref{sumFermionLagrangian}).

\subsubsection{Conclusion}

\rem{The surprising result: }Only two generic constants determine
the prefactors of all Feynman rules: \( \widetilde{s} \) and \( \widetilde{g} \).
The factor \( \widetilde{s} \) is determined by the generating functional
(\ref{introGenFunctGeneric}). It is \( i \) in the case of \( T=0 \)
or the Real Time Formalism (RTF). In the case of Imaginary Time Formalism
(ITF) it takes the value \( +1 \), corresponding to the change of
sign in the time-component from \( x_{0}\rightarrow -i\tau  \) and
\( \dd x_{0}\rightarrow -i\dd \tau _{0} \) in the exponent of the
generating functional. 

The second essential generic constant is \( \widetilde{g} \). Its
choice is less motivated physically, but merely a matter of taste.

Together, these two constants (\( \widetilde{s}=i \) or \( +1 \),
\( \widetilde{g}=-i,\, -1, \) or \( +i \)) explain the prefactors
of propagators and vertices used in Feynman rules.



\section{Summary of rules}

The constants \( \widetilde{s} \) and \( \widetilde{g} \) can be
determined by comparison with the following formulae. In standard
literature \( \widetilde{g} \) assumes the values \( -i \) \cite{Peskin:1995ev},
\( -1 \) \cite{Itzykson:1980rh}, or \( +i \) \cite{Kapusta:1989bd,LeBellac:1996aa,LeBellac:1991cq}.
The constant \( \widetilde{s} \) takes the value \( i \) for zero
temperature field theory or real time formalism and \( +1 \) for
imaginary time formalism. An overview of these constants can be found
in the tables \ref{tableComparisonConstants} and \ref{tablePrefactors}
of section \ref{sectionComparisonValues}.

\subsubsection{\noindent Lagrangian}

\noindent \begin{eqnarray}
\cL  & = & -\frac{1}{4}F^{a}_{\mu \nu }F^{\mu \nu a}+\overline{\psi }_{i}\left( i\slD -m\right) _{ij}\psi _{j}-\frac{1}{2\xi }(\partial ^{\mu }A_{\mu }^{a})^{2}+\bar{c}^{a}\left( -\partial ^{\mu }D_{\mu }^{ac}\right) c^{c}\quad \quad \nonumber \\
 & = & -\frac{1}{2}\left( \partial _{\mu }A_{\nu }^{a}\partial ^{\mu }A^{\nu a}-\partial _{\mu }A_{\nu }^{a}\partial ^{\nu }A^{\mu a}\right) -\frac{1}{2\xi }(\partial ^{\mu }A_{\mu }^{a})^{2}\nonumber \\
 &  & -i\partial _{\mu }A_{\nu }^{a}\widetilde{g}gf^{abc}A^{\mu b}A^{\nu c}+\frac{1}{4}\widetilde{g}^{2}g^{2}f^{abc}f^{ade}A_{\mu }^{b}A_{\nu }^{c}A^{\mu d}A^{\nu e}\nonumber \\
 &  & +\overline{\psi }_{i}\left( i\slpd -m\right) \delta _{ij}\psi _{j}+i\widetilde{g}g\overline{\psi }_{i}\gamma ^{\mu }A^{a}_{\mu }t^{a}_{ij}\psi _{j}\nonumber \\
 &  & +\bar{c}^{a}\left( -\partial ^{\mu }\partial _{\mu }\delta ^{ac}\right) c^{c}-i\widetilde{g}gf^{abc}\bar{c}^{a}\partial ^{\mu }(A^{b}_{\mu }c^{c})
\end{eqnarray}

\subsubsection{\noindent Generic integral and Generating functional}

\begin{equation}
\label{quickPosIntegralGeneric}
\int d^{4}\widetilde{x}\equiv \left\{ \begin{array}{ll}
\int d^{4}x & \text {for\, }T=0\\
\int ^{\beta }_{0}d\tau \int d^{3}\vec{x} & \text {for\, ITF}
\end{array}\right. \quad \longleftrightarrow \quad Z\equiv \int D..\exp \left( \widetilde{s}\int \dd ^{4}\widetilde{x}\cL \right) 
\end{equation}

\subsubsection{\noindent Covariant derivative and Field strength tensor}

\begin{equation}
\label{quickCovariantGeneric}
D_{\mu }\equiv \partial _{\mu }+\widetilde{g}gA_{\mu }^{a}t^{a}\quad \longleftrightarrow \quad F_{\mu \nu }^{a}=\partial _{\mu }A_{\nu }^{a}-\partial _{\nu }A_{\mu }^{a}+i\widetilde{g}gf^{abc}A_{\mu }^{b}A_{\nu }^{c}
\end{equation}

\section*{\noindent Feynman propagators }

\noindent In the following, the arrow means conversion from Minkowski
space result to Euclidean space result (Minkowski \( \rightarrow  \)
Euclidean).

\begin{picture}(100,00)(80,70)
\begin{feynartspicture}(300,100)(1.,1.)
\FADiagram{} 
\FAProp(1.,10.)(20.,10.)(0.,){/Sine}{0}
\FALabel(0,11)[b]{\( a,\mu  \)} 
\FALabel(20,11)[b]{\( b,\nu  \)} 
\FAProp(6,8)(15,8)(0,){/Straight}{0}
\FAProp(14.9,8)(15,8)(0,){/Straight}{1}
\FALabel(10,7)[t]{\( k \)} 
\end{feynartspicture}
\end{picture}\begin{eqnarray}
\qquad \qquad \qquad \qquad  & = & \frac{-1}{\widetilde{s}\widetilde{\chi }}\widetilde{G}_{\mu \nu }^{ab}(k)\label{quickPropGluon} \\
 & = & \frac{1}{\widetilde{s}}\left( g_{\mu \nu }\frac{1}{k^{2}}+(\xi -1)\frac{k_{\mu }k_{\nu }}{k^{4}}\right) \delta ^{ab}\nonumber \\
 & \rightarrow  & \frac{1}{\widetilde{s}}\left( g_{\mu \nu }\frac{1}{K^{2}}+(\xi -1)\frac{K_{\mu }K_{\nu }}{K^{4}}\right) \delta ^{ab}\nonumber 
\end{eqnarray}

\begin{picture}(100,0)(80,70)
\begin{feynartspicture}(300,100)(1.,1.)
\FADiagram{} 
\FAProp(1.,10.)(20.,10.)(0.,){/Straight}{1}
\FALabel(0,11)[b]{\( i \)} 
\FALabel(20,11)[b]{\( j \)} 
\FAProp(6,8)(15,8)(0,){/Straight}{0}
\FAProp(14.9,8)(15,8)(0,){/Straight}{1}
\FALabel(10,7)[t]{\( p \)} 
\end{feynartspicture}
\end{picture}
\begin{eqnarray}
\qquad \qquad \qquad \qquad \qquad  & = & \frac{-1}{\widetilde{s}\widetilde{\sigma }}\widetilde{S}_{F}^{ij}(p)\label{quickPropQuark} \\
 & = & \frac{-1}{\widetilde{s}}\frac{1}{\, \slp -m}\delta ^{ij}=\frac{-1}{\widetilde{s}}\frac{\, \slp +m}{p^{2}-m^{2}+i\varepsilon }\delta ^{ij}\nonumber \\
 & \rightarrow  & \frac{1}{\widetilde{s}}\frac{1}{\, \slP +m}\delta ^{ij}=\frac{1}{\widetilde{s}}\frac{\, \slP -m}{P^{2}+m^{2}}\delta ^{ij}\nonumber 
\end{eqnarray}

\begin{picture}(100,0)(80,70)
\begin{feynartspicture}(300,100)(1.,1.)
\FADiagram{} 
\FAProp(1.,10.)(20.,10.)(0.,){/GhostDash}{1}
\FALabel(0,11)[b]{\( a \)} 
\FALabel(20,11)[b]{\( b \)} 
\FAProp(6,8)(15,8)(0,){/Straight}{0}
\FAProp(14.9,8)(15,8)(0,){/Straight}{1}
\FALabel(10,7)[t]{\( k \)} 
\end{feynartspicture}
\end{picture}\begin{equation}
\label{quickPropGhost}
\qquad \qquad \qquad \qquad \qquad \qquad \quad \, \, =\, \, \frac{-1}{\widetilde{s}\widetilde{c}}\widetilde{G}_{gh}^{ab}(k)=\frac{-1}{\widetilde{s}}\frac{\delta ^{ab}}{k^{2}}\rightarrow \frac{1}{\widetilde{s}}\frac{\delta ^{ab}}{K^{2}}
\end{equation}

~

\section*{\noindent Vertices }

\noindent (Minkowski \( \rightarrow  \) Euclidean)

\begin{samepage}

\begin{picture}(100,10)(80,110)
\begin{feynartspicture}(300,100)(1.,1.)
\FADiagram{} 
\FAProp(1.,1.)(10.,8.)(0.,){/Sine}{0}
\FAProp(19.,1.)(10.,8.)(0.,){/Sine}{0}
\FAProp(10.,19.)(10.,8.)(0.,){/Sine}{0}
\FAVert(10.,8.){0}
\FALabel(0,0)[t]{\( c,\rho  \)} 
\FALabel(20,0)[t]{\( b,\nu  \)} 
\FALabel(10,20)[b]{\( a,\mu  \)} 
\FAProp(11.5,17)(11.5,10)(0,){/Straight}{0}
\FAProp(11.5,10.1)(11.5,10)(0,){/Straight}{1}
\FALabel(12,13.5)[l]{\( k \)} 
\FAProp(16,1.5)(11,5.5)(0,){/Straight}{0}
\FAProp(11.1,5.42)(11,5.5)(0,){/Straight}{1}
\FALabel(13,3)[tr]{\( p \)} 
\FAProp(2,4)(7,8)(0,){/Straight}{0}
\FAProp(6.9,7.92)(7,8)(0,){/Straight}{1}
\FALabel(3.5,6.5)[br]{\( q \)} 
\end{feynartspicture} 
\end{picture} \begin{eqnarray}
\qquad \qquad \qquad \qquad  & =\, \, (\widetilde{s}\widetilde{g})gf^{abc} & \left( g^{\mu \nu }(k-p)^{\rho }\right. \label{quickVertex3Boson} \\
 &  & \left. +g^{\nu \rho }(p-q)^{\mu }\right. \nonumber \\
 &  & \left. +g^{\rho \mu }(q-k)^{\nu }\right) \nonumber \\
 &  & \nonumber \\
 & \rightarrow \, \, (\widetilde{s}\widetilde{g})gf^{abc} & \left( \delta _{\mu \nu }(K-P)_{\rho }\right. \nonumber \\
 &  & \left. +\delta _{\nu \rho }(P-Q)_{\mu }\right. \nonumber \\
 &  & \left. +\delta _{\rho \mu }(Q-K)_{\nu }\right) \nonumber 
\end{eqnarray}

~

\end{samepage}


\begin{picture}(100,10)(80,110)
\begin{feynartspicture}(300,100)(1.,1.)
\FADiagram{} 
\FAProp(1.,1.)(10.,10.)(0.,){/Sine}{0}
\FAProp(19.,1.)(10.,10.)(0.,){/Sine}{0}
\FAProp(19.,19.)(10.,10.)(0.,){/Sine}{0}
\FAProp(1.,19.)(10.,10.)(0.,){/Sine}{0}
\FAVert(10.,10.){0}
\FALabel(0,0)[t]{\( d,\sigma  \)} 
\FALabel(20,0)[t]{\( c,\rho  \)} 
\FALabel(20,20)[b]{\( b,\nu  \)} 
\FALabel(0,20)[b]{\( a,\mu  \)} 
\FAProp(4,18)(9,13)(0,){/Straight}{0}
\FAProp(8.9,13.1)(9,13)(0,){/Straight}{1}
\FALabel(6.5,16.5)[bl]{\( k \)} 
\FAProp(18,16)(13,11)(0,){/Straight}{0}
\FAProp(13.1,11.1)(13,11)(0,){/Straight}{1}
\FALabel(16.5,13.5)[tl]{\( p \)} 
\FAProp(16,2)(11,7)(0,){/Straight}{0}
\FAProp(11.1,6.9)(11,7)(0,){/Straight}{1}
\FALabel(13.5,3.5)[tr]{\( q \)} 
\FAProp(2,4)(7,9)(0,){/Straight}{0}
\FAProp(6.9,8.9)(7,9)(0,){/Straight}{1}
\FALabel(3.5,6.5)[br]{\( r \)} 
\end{feynartspicture}
\end{picture}\begin{eqnarray}
\qquad \qquad \qquad \qquad \qquad \quad \, \,  & \! \! \! \! =(\widetilde{s}\widetilde{g}^{2})g^{2} & \! \! \! \! \! \! \left( f^{abe}f^{cde}(g^{\mu \rho }g^{\nu \sigma }\! -\! g^{\mu \sigma }g^{\nu \rho })\right. \label{quickVertex4Boson} \\
 &  & \! \! \! \! \! \left. +f^{ace}f^{bde}(g^{\mu \nu }g^{\rho \sigma }\! -\! g^{\mu \sigma }g^{\nu \rho })\right. \nonumber \\
 &  & \! \! \! \! \! \left. +f^{ade}f^{bce}(g^{\mu \nu }g^{\rho \sigma }\! -\! g^{\mu \rho }g^{\nu \sigma })\right) \nonumber \\
 &  & \nonumber \\
 & \! \! \rightarrow (\widetilde{s}\widetilde{g}^{2})g^{2} & \! \! \! \! \! \! \left( f^{abe}f^{cde}(\delta _{\mu \rho }\delta _{\nu \sigma }\! -\! \delta _{\mu \sigma }\delta _{\nu \rho })\right. \nonumber \\
 &  & \! \! \! \! \! \left. +f^{ace}f^{bde}(\delta _{\mu \nu }\delta _{\rho \sigma }\! -\! \delta _{\mu \sigma }\delta _{\nu \rho })\right. \nonumber \\
 &  & \! \! \! \! \! \left. +f^{ade}f^{bce}(\delta _{\mu \nu }\delta _{\rho \sigma }\! -\! \delta _{\mu \rho }\delta _{\nu \sigma })\right) \nonumber 
\end{eqnarray}

~

~

~

\begin{picture}(100,10)(80,80)
\begin{feynartspicture}(300,100)(1.,1.)
\FADiagram{} 
\FAProp(1.,1.)(10.,8.)(0.,){/Straight}{-1}
\FAProp(19.,1.)(10.,8.)(0.,){/Straight}{1}
\FAProp(10.,19.)(10.,8.)(0.,){/Sine}{0}
\FAVert(10.,8.){0}
\FALabel(0,0)[t]{\( i \)} 
\FALabel(20,0)[t]{\( j \)} 
\FALabel(10,20)[b]{\( a,\mu  \)} 
\FAProp(11.5,17)(11.5,10)(0,){/Straight}{0}
\FAProp(11.5,10.1)(11.5,10)(0,){/Straight}{1}
\FALabel(12,13.5)[l]{\( k \)} 
\FAProp(16,1.5)(11,5.5)(0,){/Straight}{0}
\FAProp(11.1,5.42)(11,5.5)(0,){/Straight}{1}
\FALabel(13,3)[tr]{\( p \)} 
\FAProp(2,4)(7,8)(0,){/Straight}{0}
\FAProp(2.1,4.08)(2,4)(0,){/Straight}{1}
\FALabel(3.5,6.5)[br]{\( q \)} 
\end{feynartspicture}
\end{picture}\begin{eqnarray}
 & = & (i\widetilde{s}\widetilde{g})g\gamma ^{\mu }t^{a}_{ij}\label{quickVertexQuarkGluon} \\
 &  & \nonumber \\
 & \rightarrow  & (-i\widetilde{s}\widetilde{g})g\gamma _{\mu }t^{a}_{ij}\nonumber 
\end{eqnarray}

~

~

~

~

~

\begin{picture}(100,10)(80,80)
\begin{feynartspicture}(300,100)(1.,1.)
\FADiagram{} 
\FAProp(1.,1.)(10.,8.)(0.,){/GhostDash}{-1}
\FAProp(19.,1.)(10.,8.)(0.,){/GhostDash}{1}
\FAProp(10.,19.)(10.,8.)(0.,){/Sine}{0}
\FAVert(10.,8.){0}
\FALabel(0,0)[t]{\( a \)} 
\FALabel(20,0)[t]{\( c \)} 
\FALabel(10,20)[b]{\( b,\mu  \)} 
\FAProp(11.5,17)(11.5,10)(0,){/Straight}{0}
\FAProp(11.5,10.1)(11.5,10)(0,){/Straight}{1}
\FALabel(12,13.5)[l]{\( k \)} 
\FAProp(16,1.5)(11,5.5)(0,){/Straight}{0}
\FAProp(11.1,5.42)(11,5.5)(0,){/Straight}{1}
\FALabel(13,3)[tr]{\( p \)} 
\FAProp(2,4)(7,8)(0,){/Straight}{0}
\FAProp(2.1,4.08)(2,4)(0,){/Straight}{1}
\FALabel(3.5,6.5)[br]{\( q \)} 
\end{feynartspicture}
\end{picture}\begin{eqnarray}
\qquad  & = & (-\widetilde{s}\widetilde{g})gq^{\mu }f^{abc}\label{quickVertexGhostGluon} \\
 &  & \nonumber \\
 & \rightarrow  & (\widetilde{s}\widetilde{g})gQ_{\mu }f^{abc}\nonumber 
\end{eqnarray}

\section*{~}

\section*{\noindent Green's functions}

\begin{equation}
\left( \partial ^{2}g_{\mu \nu }-\left( 1-\frac{1}{\xi }\right) \partial _{\mu }\partial _{\nu }\right) G^{\nu \rho ab}(x-y)\equiv \widetilde{\chi }\delta _{\mu }^{\rho }\delta ^{ab}\delta ^{(4)}(x-y)
\end{equation}

\begin{equation}
\widetilde{G}_{\mu \nu }^{ab}(k)\! =\! -\widetilde{\chi }\left( g_{\mu \nu }\frac{1}{k^{2}}\! +(\xi \! -\! 1)\frac{k_{\mu }k_{\nu }}{k^{4}}\right) \delta ^{ab}\! \rightarrow -\widetilde{\chi }\left( g_{\mu \nu }\frac{1}{K^{2}}\! +\! (\xi \! -\! 1)\frac{K_{\mu }K_{\nu }}{K^{4}}\right) \delta ^{ab}
\end{equation}

\begin{equation}
\left( i\slpd -m\right) S_{F}^{ij}(x-y)\equiv \widetilde{\sigma }\delta ^{(4)}(x-y)\delta ^{ij}\dsOne _{4\times 4}
\end{equation}

\begin{equation}
\widetilde{S}_{F}^{ij}(p)=\widetilde{\sigma }\frac{1}{\, \slp -m}\delta ^{ij}=\widetilde{\sigma }\frac{\, \slp +m}{p^{2}-m^{2}+i\varepsilon }\delta ^{ij}\rightarrow -\widetilde{\sigma }\frac{1}{\, \slP +m}\delta ^{ij}
\end{equation}

\begin{equation}
\left( -\partial ^{2}\delta ^{ab}\right) G^{bc}_{gh}(x-y)\equiv \widetilde{c}\delta ^{ac}\delta ^{(4)}(x-y)
\end{equation}
\begin{equation}
\widetilde{G}_{gh}^{ab}(k)=\widetilde{c}\frac{\delta ^{ab}}{k^{2}}\rightarrow \frac{1}{\widetilde{s}}-\widetilde{c}\frac{\delta ^{ab}}{K^{2}}
\end{equation}

~\tmpbibtex


\chapter{Thermodynamic relations\label{chapter_thermodynamics}}

\section{Specific heat at NLO}

The specific heat can be calculated from the entropy by a transformation
of thermodynamic quantities which can be found in \cite{LL:VCv}\begin{equation}
\mathcal{C}_{V}\equiv C_{V}/V=T\left\{ \left( \frac{\partial \mathcal{S}}{\partial T}\right) _{\mu }-\frac{\left( \frac{\partial \mathcal{N}}{\partial T}\right) ^{2}_{\mu }}{\left( \frac{\partial \mathcal{N}}{\partial \mu }\right) _{T}}\right\} 
\end{equation}
where \( \mathcal{S} \) is the entropy density\[
\mathcal{S}\equiv S/V=\left( \frac{\partial P}{\partial T}\right) _{\mu ,V}\]
 and \( \mathcal{N} \) is the particle number density \begin{equation}
\mathcal{N}\equiv N/V=\left( \frac{\partial P}{\partial \mu }\right) _{T,V}.
\end{equation}
In order to calculate the next-to-leading order (NLO) specific heat
in large-\( N_{f} \) we have to expand this equation around the leading
order contribution (\ref{largenfP0}\rem{??LO contribution??}). The
denominator takes the form\begin{equation}
\left( \frac{\partial \mathcal{N}}{\partial \mu }\right) _{T}=\left( \frac{\partial ^{2}P_{0}}{\partial \mu ^{2}}\right) _{T}=NN_{f}\left( \frac{T^{2}}{3}+\frac{\mu ^{2}}{\pi ^{2}}\right) 
\end{equation}
and the specific heat can then be written as \[
\mathcal{C}_{V}=\mathcal{C}_{V}^{\LO }+\mathcal{C}_{V}^{\NLO }\]
 with the leading order (LO) contribution\begin{equation}
\mathcal{C}_{V}^{\LO }=NN_{f}\left( \frac{7\pi ^{2}T^{3}}{15}+\frac{\mu ^{2}T}{3}-\frac{4\mu ^{2}T^{3}}{9\left( \frac{T^{2}}{3}+\frac{\mu ^{2}}{\pi ^{2}}\right) }\right) 
\end{equation}
and the next-to-leading order (NLO) contribution\begin{equation}
\label{specificheatNLOgeneral}
\mathcal{C}^{\NLO }_{V}=T\left\{ \left( \frac{\partial \mathcal{S}}{\partial T}\right) _{\mu }-\frac{4\frac{\mu T}{3}\left( \frac{\partial ^{2}P}{\partial \mu \partial T}\right) }{\frac{T^{2}}{3}+\frac{\mu ^{2}}{\pi ^{2}}}+\frac{\left( \frac{2\mu T}{3}\right) ^{2}\left( \frac{\partial ^{2}P}{\partial \mu ^{2}}\right) _{T}}{\left( \frac{T^{2}}{3}+\frac{\mu ^{2}}{\pi ^{2}}\right) ^{2}}\right\} +O(\frac{P^{2}_{\NLO }}{P_{0}^{2}})
\end{equation}
with NLO-contributions of \( \mathcal{S} \) and \( P \). The ideal-gas
limit of the specific heat contains, apart from the LO-contribution,
a term proportional to \( N_{f}^{0} \) and is given by \begin{equation}
\mathcal{C}_{V}^{0}=NN_{f}\left( \frac{7\pi ^{2}T^{3}}{15}+\frac{\mu ^{2}T}{3}-\frac{4\mu ^{2}T^{3}}{9\left( \frac{T^{2}}{3}+\frac{\mu ^{2}}{\pi ^{2}}\right) }\right) +N_{g}\frac{4\pi ^{2}T^{3}}{15}.
\end{equation}

\section{Derivative relations\label{section_derivativerelations}}

For the calculation of quark susceptibilities from the pressure in
section \ref{section_quarknumbersusceptibilities}, we evaluated the
derivative with respect to a squared quantity. The derivation with
respect to a squared quantity can be obtained as follows, using \( \mu ^{2}=t \)
\begin{equation}
\frac{\partial f(\mu )}{\partial \mu ^{2}}=\frac{\partial f(\sqrt{t})}{\partial t}=f'(\sqrt{t})\frac{1}{2\sqrt{t}}=f'(\mu )\frac{1}{2\mu }.
\end{equation}
The chain rule then works as follows\begin{equation}
\frac{\partial g(f(\mu ))}{\partial \mu ^{2}}=\frac{\partial g(f(\sqrt{t}))}{\partial t}=g'(f(\sqrt{t}))\frac{\partial f(\sqrt{t})}{\partial t}=g'(f(\mu ))\frac{\partial f(\mu )}{\partial \mu ^{2}}.
\end{equation}
Similarly for the second derivative we obtain\begin{equation}
\frac{\partial ^{2}}{(\partial \mu ^{2})^{2}}g(f(\mu ))=g''(f(\mu ))\left( \frac{\partial f(\mu )}{\partial \mu ^{2}}\right) ^{2}+g'(f(\mu ))\frac{\partial ^{2}f(\mu )}{(\partial \mu ^{2})^{2}}.
\end{equation}
In the case of a quadratic lowest order term of the series expansion
of \( f(\mu ) \) we can further write\begin{equation}
\left. \frac{\partial ^{2}}{\partial \mu ^{2}}f(\mu )\right| _{\mu =0}=\left. 2\frac{\partial }{\partial \mu ^{2}}f(\mu )\right| _{\mu =0}.
\end{equation}
This is however only valid if \( f(\mu )=O(\mu ^{2}) \). If \( f(\mu ) \)
contains a linear term, the l.h.s. vanishes, while the r.h.s. diverges.
Similarly, for the fourth derivative we see that \( \partial ^{4}\mu ^{4}/\partial \mu ^{4}=24 \)
but \( \partial ^{2}(\mu ^{2})^{2}/(\partial \mu ^{2})^{2}=\partial ^{2}t^{2}/\partial t^{2}=2 \)
so that we can write the following relationship if \( f(\mu ) \)
only contains positive even orders in \( \mu  \) up to \( O(\mu ^{4}) \),
i.e. \( f(\mu )=c\mu ^{2}+O(\mu ^{4}) \): \begin{equation}
\left. \frac{\partial ^{4}}{\partial \mu ^{4}}f(\mu )\right| _{\mu =0}=\left. 12\frac{\partial ^{2}}{(\partial \mu ^{2})^{2}}f(\mu )\right| _{\mu =0}.
\end{equation}

\chapter{Gauge boson self-energy\label{chapter_zerotemperaturelimit}}

\section{Exact result at \protect\( T=0\protect \)}

In this section we will extend the calculation of the leading order
gauge boson self-energy from section \ref{section_gaugebosonselfenergy}
to the limit of finite chemical potential \( \mu  \) and zero temperature
\( T=0 \). In this region, the fermionic distribution function from
equation (\ref{fermionicdistributionfunction}) is given by the step-function\begin{equation}
\label{fermionicdistributionfunctionzeroT}
n_{f}(k,T=0,\mu )=\frac{1}{2}\left( \theta (\mu -k)+\theta (-\mu -k)\right) 
\end{equation}
and we can explicitly calculate the real and imaginary parts of the
bosonic self energy for Minkowski space.

\subsubsection{Imaginary part}

In the equations (\ref{functionsF123finiteT}) we can calculate the
functions \( F_{i} \) using the \( T\rightarrow 0 \) limit of the
fermionic distribution function (\ref{fermionicdistributionfunctionzeroT})
which give \begin{eqnarray}
F_{1}(x) & \equiv  & \int _{x}^{\infty }\, n_{f}(k)dk=\frac{\mu -x}{2}\theta (\mu -x),\\
F_{2}(x) & \equiv  & \int _{x}^{\infty }k\, n_{f}(k)dk=\frac{\mu ^{2}-x^{2}}{4}\theta (\mu -x),\\
F_{3}(x) & \equiv  & \int _{x}^{\infty }k^{2}\, n_{f}(k)dk=\frac{\mu ^{3}-x^{3}}{6}\theta (\mu -x).
\end{eqnarray}
The resulting functions can be further simplified. Using the following
definitions\begin{eqnarray}
U(q_{0},q) & = & (2\mu -|q_{0}+q|)\theta (2\mu -|q_{0}+q|),\nonumber \\
V(q_{0},q) & = & (2\mu -q-q_{0})(2\mu +2q-q_{0}),\nonumber \\
W(q_{0},q) & = & \theta (q_{0}+q)V(q_{0},q)+\theta (-q_{0}-q)V(-q_{0},-q),\label{zerotuvw} 
\end{eqnarray}
we get\begin{eqnarray}
\textrm{Im}G(q_{0},q) & \! \! \! =\! \! \!  & \frac{1}{16\pi q}(q^{2}-q_{0}^{2})\left\{ U(|q_{0}|,q)-U(-|q_{0}|,q)\right\} ,\label{zerotG} \\
\textrm{Im}H(q_{0},q) & \! \! \! =\! \! \!  & -\frac{1}{96\pi q}\left\{ W(|q_{0}|,q)U(|q_{0}|,q)\! -\! W(-|q_{0}|,q)U(-|q_{0}|,q)\right\} .\qquad \quad \label{zerotH} 
\end{eqnarray}
From these we get the self-energies\begin{eqnarray}
\textrm{Im}\Pi _{L}(q_{0},q) & = & \frac{-\geff ^{2}(q^{2}-q_{0}^{2})}{48\pi q^{3}}\left\{ W(|q_{0}|,q)U(|q_{0}|,q)\right. \nonumber \\
 &  & \qquad \qquad \qquad \, \, \left. -W(-|q_{0}|,q)U(-|q_{0}|,q)\right\} ,\label{zerotPiL} \\
\textrm{Im}\Pi _{T}(q_{0},q) & = & \frac{\geff ^{2}(q^{2}-q_{0}^{2})}{96\pi q^{3}}\left\{ \left[ W(|q_{0}|,q)+6q^{2}\right] U(|q_{0}|,q)\right. \label{zerotPiT} \\
 &  & \qquad \qquad \qquad \, \, \left. -\left[ W(-|q_{0}|,q)+6q^{2}\right] U(-|q_{0}|,q)\right\} .\nonumber 
\end{eqnarray}

\subsubsection{Real part}

Also for the real part we can give the exact solution in the limit
\( T\rightarrow 0 \). Starting from (\ref{WeldonRealG}) and (\ref{WeldonRealH})
the momentum integration can be analytically done since the distribution
function reduces again to a step function for zero temperature. Rearranging
the result, we can write\begin{eqnarray}
R(q_{0},q) & = & (q_{0}+q)\ln |q_{0}+q|-(2\mu +q_{0}+q)\ln |2\mu +q_{0}+q|,\qquad \\
S(q_{0},q) & = & \left( -12\mu ^{2}q_{0}+(2q-q_{0})(q+q_{0})^{2}\right) \ln |q_{0}+q|\\
 &  & +\left( 2\mu -2q+q_{0}\right) \left( 2\mu +q+q_{0}\right) ^{2}\ln |2\mu +q+q_{0}|.\nonumber 
\end{eqnarray}
We use the following abbreviation to keep a lot of terms in dense
notation\begin{equation}
R^{\pm }_{\pm }(q_{0},q)\equiv R(q_{0},q)-R(q_{0},-q)+R(-q_{0},q)-R(-q_{0},-q)
\end{equation}
and similarly for \( S^{\pm }_{\pm } \) (just replace \( R \) by
\( S \)). Then the real part of \( G \) and \( H \) can be written
as\begin{eqnarray}
\textrm{Re}G(q_{0},q) & = & \frac{1}{2\pi ^{2}}\left( \mu ^{2}+\frac{q^{2}-q_{0}^{2}}{8q}R^{\pm }_{\pm }(q_{0},q)\right) ,\\
\textrm{Re}H(q_{0},q) & = & \frac{1}{96\pi ^{2}}\left( 32\mu ^{2}+\frac{1}{q}S^{\pm }_{\pm }(q_{0},q)\right) .
\end{eqnarray}
Therefore the exact solution for the real part of the self-energies
in Minkowski space is given by\begin{eqnarray}
\textrm{Re}\Pi _{L}(q_{0},q) & = & \frac{\geff ^{2}(q^{2}-q_{0}^{2})}{48\pi ^{2}q^{2}}\left( 32\mu ^{2}+\frac{1}{q}S^{\pm }_{\pm }(q_{0},q)\right) ,\label{zeroPiLRe} \\
\textrm{Re}\Pi _{T}(q_{0},q) & = & \frac{\geff ^{2}}{96\pi ^{2}q^{3}}\left\{ 16\mu ^{2}q(q^{2}+2q_{0}^{2})\right. \label{zeroPiTRe} \\
 &  & \qquad \qquad \left. +(q^{2}-q_{0}^{2})\left( 6q^{2}R^{\pm }_{\pm }(q_{0},q)-S^{\pm }_{\pm }(q_{0},q)\right) \right\} .\nonumber 
\end{eqnarray}

\section{Approximations to second order in frequency}

We can calculate the following \( T\rightarrow 0 \) limits in Minkowski
space by using our exact formulae for the imaginary part of the self-energies
(\ref{zerotPiL}) and (\ref{zerotPiT}). In the following we will
use these equations in a regime where \( 0\leq q_{0}\ll q,\mu  \).
This is a reasonable assumption, as the bosonic distribution function
\( n_{b} \) will probe smaller and smaller \( q_{0} \) if we let
\( T\rightarrow 0 \). In this limit we can simplify our functions
(\ref{zerotuvw}) to \( U(q_{0},q)\rightarrow (2\mu -q_{0}-q)\theta (2\mu -q) \)
and \( W(q_{0},q)\rightarrow V(q_{0},q) \). (Actually in \( U(q_{0},q) \)
we could have used \( \theta (2\mu -q_{0}-q) \) which would make
the following expressions slightly more complicated, but it naturally
reduces to \( \theta (2\mu -q) \) as soon as we expand \( q_{0} \)
around \( 0 \).) The functions \( G \) and \( H \) from (\ref{zerotG})
and (\ref{zerotH}) then become \begin{equation}
\textrm{Im}G(q_{0},q)=\frac{q_{0}(q_{0}^{2}-q^{2})}{8\pi q}\theta (2\mu -q),
\end{equation}
\begin{equation}
\textrm{Im}H(q_{0},q)=\frac{q_{0}(12\mu ^{2}+q_{0}^{2}-3q^{2})}{48\pi q}\theta (2\mu -q),
\end{equation}
where the \( \theta  \)-functions come from the step-function shape
of the fermionic distribution function at zero temperature. These
are exact solutions in the limit of \( T\rightarrow 0 \) and \( q_{0}\rightarrow 0 \).
Furthermore, for the real parts of \( G \) and \( H \) we can provide
the following expansions in small \( q_{0} \): \begin{eqnarray}
\textrm{Re}G(q_{0},q) & = & \frac{1}{8\pi ^{2}}\left\{ 4\mu ^{2}+q\left[ (2\mu -q)\ln |2\mu -q|+2q\ln q\right. \right. \nonumber \\
 &  & \left. \left. -(2\mu +q)\ln |2\mu +q|\right] \right\} +O(q_{0}^{2}),
\end{eqnarray}
\begin{eqnarray}
\textrm{Re}H(q_{0},q) & \! \! =\! \!  & \frac{1}{24\pi ^{2}q}\left\{ 8\mu ^{2}q-(2\mu -q)^{2}(\mu +q)\ln |2\mu -q|+2q^{3}\ln q\right. \qquad \nonumber \\
 & \! \! \! \!  & \left. +(\mu -q)(2\mu +q)^{2}\ln |2\mu +q|\right\} +O(q_{0}^{2}),
\end{eqnarray}
with the chemical potential \( \mu  \). 

We can combine the functions \( G \) and \( H \) to form the self-energies
\( \Pi _{L} \) and \( \Pi _{T} \) according to (\ref{zerotPiL})
and (\ref{zerotPiT}). Again, for the imaginary part we get nice analytic
expressions in the limit of small \( T \) and \( q_{0} \):\begin{equation}
\textrm{Im}\Pi _{L}(q_{0},q)=-\geff ^{2}\frac{q_{0}(-q_{0}^{2}+q^{2})(-12\mu ^{2}+3q^{2}-q_{0}^{2})}{24\pi q^{3}}\theta (2\mu -q),
\end{equation}
\begin{equation}
\textrm{Im}\Pi _{T}(q_{0},q)=-\geff ^{2}\frac{q_{0}(-q_{0}^{2}+q^{2})(12\mu ^{2}+3q^{2}+q_{0}^{2})}{48\pi q^{3}}\theta (2\mu -q).
\end{equation}
Since this is valid for small \( q_{0} \), it makes sense to expand
these expressions for small \( q_{0} \). The first coefficient in
the \( q_{0} \)-expansion reads\begin{equation}
\textrm{Im}\Pi _{L}(q_{0},q)=-\geff ^{2}\frac{(-4\mu ^{2}+q^{2})}{8\pi q}q_{0}\theta (2\mu -q)+O(q_{0}^{2}),
\end{equation}
\begin{equation}
\textrm{Im}\Pi _{T}(q_{0},q)=-\geff ^{2}\frac{(4\mu ^{2}+q^{2})}{16\pi q}q_{0}\theta (2\mu -q)+O(q_{0}^{2}),
\end{equation}
and for the real parts\begin{eqnarray}
\textrm{Re}\Pi _{L}(q_{0},q) & \! \! =\! \!  & \frac{\geff ^{2}}{12\pi ^{2}q}\left\{ 8\mu ^{2}q-(2\mu -q)^{2}(\mu +q)\ln |2\mu -q|+2q^{3}\ln q\right. \qquad \quad \\
 &  & \left. +(\mu -q)(2\mu +q)^{2}\ln |2\mu +q|\right\} +O(q_{0}^{2}),\nonumber 
\end{eqnarray}
\begin{eqnarray}
\textrm{Re}\Pi _{T}(q_{0},q) & \! \! =\! \!  & \frac{\geff ^{2}}{24\pi ^{2}q}\left\{ 4\mu ^{2}q+(2\mu \! -\! q)(2\mu ^{2}\! +\mu q+2q^{2})\ln |2\mu -q|\right. \qquad \qquad \\
 &  & \left. +4q^{3}\ln q-(2\mu +q)(2\mu ^{2}-\mu q+2q^{2})\ln |2\mu +q|\right\} +O(q_{0}^{2}).\nonumber 
\end{eqnarray}
For the temperature-independent vacuum contribution the imaginary
part vanishes for small \( q_{0} \), \( \textrm{Im}\Pi _{\vac }(q_{0},q)=O(q_{0}^{2}) \)
while the real part takes the value\begin{equation}
\textrm{Re}\Pi _{\vac }(q_{0},q)=-\geff ^{2}\frac{q^{2}}{12\pi ^{2}}\left( \ln \frac{q^{2}}{\bar{\mu }^{2}}-\frac{5}{3}\right) +O(q_{0}^{2})
\end{equation}
 with the renormalization scale \( \bar{\mu } \). 

~

\rem{ ========= Mathematica CHAPTER ========== }

\chapter{Fractional power expansion}

\section{Introduction to Puiseux series\label{section_puiseuxseries}}

The first time fractional powers appear in (\ref{integralimprovement1}).
In this section we want to analyze where fractional powers come from
and which terms determine their coefficients. 

Series involving fractional powers are also known as Puiseux series
\cite{allouche99b}. They appear because one expands around a singular
point that does not have a Taylor or Laurent series with integer powers
at this point. A simple expansion where the singular point is inherent
is \begin{equation}
\exp (\sqrt{x})\approx 1+\sqrt{x}+\frac{x}{2}+\frac{x^{3/2}}{6}+O(x^{2})
\end{equation}
which can be easily derived by variable substitution \( x=y^{2} \).
Another kind of example is\begin{equation}
f(x)=\left( x+x^{2}\right) ^{1/3}\approx x^{1/3}+\frac{x^{4/3}}{3}-\frac{x^{7/3}}{9}+O(x^{10/3}).
\end{equation}
In this case the famous Taylor series formula\begin{equation}
f(x)\approx f(0)+f'(0)x+f''(0)\frac{x^{2}}{2}+...
\end{equation}
 would give \( 0+\infty x-\infty x^{2}/2+... \) and clearly fails
to work. One has to pull out the singular part of the function and
perform the Taylor series on the rest:\begin{equation}
\left( x+x^{2}\right) ^{1/3}=x^{1/3}\left( 1+x\right) ^{1/3}=x^{1/3}\left( 1+\frac{x}{3}-\frac{x^{2}}{9}+O(x^{3})\right) 
\end{equation}
which gives the result above. In this way, one can obtain fractional
power series also from more complicated function compositions.

The simplest example of fractional powers of the kind we encountered
appear in the following expression (for positive \( A \), \( M \),
\( B \), and \( C \)) \rem{specificheat11b_alpha.nb}\-\begin{eqnarray}
\int _{q_{0}}^{\qmax }\! \! \frac{Aq^{5}q_{0}}{Mq^{6}+Bq_{0}^{2}+Cq^{2}q_{0}^{2}}dq & \! \! \! =\! \! \!  & \frac{Aq_{0}}{6M}\ln \frac{M\qmax ^{6}}{Bq_{0}^{2}}\nonumber \\
 & - & \! \! \! \! \! \frac{AC\pi }{9\sqrt{3}B^{2/3}M^{4/3}}q_{0}^{5/3}\! +\frac{AC^{2}\pi }{27\sqrt{3}B^{4/3}M^{5/3}}q_{0}^{7/3}\nonumber \\
 & - & \! \! \! \! \! \frac{A\left( 6B^{3}+9B^{2}C\qmax ^{2}-C^{3}\qmax ^{6}\right) }{36B^{2}M^{2}}q_{0}^{3}\nonumber \\
 & + & \! \! \! \! \! \frac{5AC^{4}\pi q_{0}^{11/3}}{3^{6}\sqrt{3}B^{8/3}M^{7/3}}-\frac{7AC^{5}\pi q_{0}^{13/3}}{3^{7}\sqrt{3}B^{10/3}M^{8/3}}\nonumber \\
 & + & \! \! \! \! \! O(q_{0}^{5})\label{generictrinityintegral01} 
\end{eqnarray}
Without the \( q^{2}q_{0}^{2} \) term in the denominator (that is
\( C=0 \)) we would just get the leading logarithmic contribution
and integer powers of \( q_{0} \) (as one can also see from the series
by setting \( C=0 \)). It is this additional term \( Cq^{2}q_{0}^{2} \)
that determines additional fractional power terms. The integral can
be calculated using the following formula \cite{Bronstein:1997aa}
for square-free \( g(x) \) and \( \deg (f)<\deg (g) \):\begin{equation}
\label{generalintegralsquarefree}
\int \frac{f(x)}{g(x)}dx=\sum _{\alpha }\frac{f(\alpha )}{g'(\alpha )}\ln (x-\alpha )
\end{equation}
where \( \alpha  \) is the set of all roots of \( g(\alpha )=0 \).
In our case we get a sum over roots of a cubic equation which first
have to be expanded for small \( q_{0} \). We can motivate the exponents
by \begin{equation}
x\ln \left( x+x^{1/3}\right) =\frac{x}{3}\ln x+x^{5/3}-\frac{x^{7/3}}{2}+\frac{x^{3}}{3}-\frac{x^{11/9}}{4}+O(x^{4})
\end{equation}
 but we cannot explain factors of \( \pi /\sqrt{3} \) appearing in
the coefficients of (\ref{generictrinityintegral01}): They only appear
together with pure fractional coefficients like \( T^{5/3} \) and
\( T^{7/3} \), but not in the \( T^{3} \) coefficient. We can give
an idea of their appearance by this simple sum over roots\begin{equation}
\sum _{\alpha \in (x|x^{3}+1=0)}(\alpha ^{2}+\alpha )\ln \alpha =-\frac{2\pi }{\sqrt{3}}
\end{equation}
but to fully calculate the complete coefficients of the series in
(\ref{generictrinityintegral01}) it seems unavoidable to plug in
the roots of the denominator into (\ref{generalintegralsquarefree}). 

In the course of the \( q_{0} \) expansion, terms from the roots
of the denominator of (\ref{generictrinityintegral01}) contain expressions
like \( -9B\sqrt{M}+\sqrt{3}\sqrt{27B^{2}M+4C^{3}q_{0}^{2}} \) which
upon symbolic series expansion reduce to \( -9B\sqrt{M}+9\sqrt{B^{2}M} \).
\rem{, which seems to reduce to \( 0 \). Since such terms also appear
in the denominator of the \( q_{0} \) expansion, we have to use a
trick to obtain useful results: We replace \( -9B\sqrt{M}+9\sqrt{B^{2}M}\rightarrow 2\times 9B\sqrt{M}\Phi  \)
where we introduce a phase variable \( \Phi =\exp \varphi  \). In
the end of the calculation, there is a unique way to determine \( \Phi  \)
such that all terms appearing in the series only involve real quantities.
For an integral like (\ref{generictrinityintegral01}) the phase might
be \( \varphi =-5i\pi  \), implying \( \Phi =-1 \), \( \ln \Phi =-5i\pi  \),
\( \Phi ^{1/3}=(-1)^{1/3} \), and so on. The validity of this procedure
can be tested by applying simple definite values to the variables
\( M \), \( A \), \( B \),.. which might lead to simpler expansions
without this kind of ambiguity, or by numerical integration.} \rem{Since
this expression reduces to \( 0 \) upon inserting positive \( B \)
and \( M \), but also appears in the denominator of the \( q_{0} \)
expansion, effectively giving infinite or indeterminate expressions,
the series expansion obtained in this way cannot be correct.} This
expression is zero for positive \( B \) and non-zero for negative
\( B \) which would result in two completely different series. It
turns out that from the beginning of the expansion, \( B \) and \( M \)
have to be assumed to have a certain sign, in our case to be positive
quantities, in order to get the correct series. Unfortunately, symbolic
manipulation programs do not readily allow for this choice (in the
course of an expansion, \( B \) is not considered equal to \( \sqrt{B^{2}} \)
which in our case means the implicit assumption that \( B \) is negative.)
A possible way to accomplish the series expansion also for positive
\( B \) in \emph{Mathematica} is presented in the next section\rem{
\ref{math_assumepositive}}.

For the sake of completeness we also give the other integrals with
the same denominator as (\ref{generictrinityintegral01}) but different
odd powers of \( q \) in the numerator:\begin{eqnarray}
\int _{q_{0}}^{\qmax }\frac{Aqq_{0}^{3}}{Mq^{6}+Bq_{0}^{2}+Cq^{2}q_{0}^{2}}dq & \! \! =\! \!  & \frac{A\pi }{3\sqrt{3}B^{2/3}M^{1/3}}q_{0}^{5/3}\! -\frac{AC\pi }{9\sqrt{3}B^{4/3}M^{2/3}}q_{0}^{7/3}\nonumber \\
 &  & -\frac{A\left( 3B^{3}+6B^{2}M\qmax ^{4}-C^{2}\qmax ^{4}\right) }{12B^{2}M\qmax ^{4}}q_{0}^{3}\nonumber \\
 &  & -\frac{5AC^{3}\pi q_{0}^{11/3}}{3^{5}\sqrt{3}B^{8/3}M^{4/3}}+\frac{7AC^{5}\pi q_{0}^{13/3}}{3^{6}\sqrt{3}B^{10/3}M^{5/3}}\nonumber \\
 &  & +O(q_{0}^{5})\label{generictrinityintegral02} 
\end{eqnarray}
\begin{eqnarray}
\int _{q_{0}}^{\qmax }\frac{Aq^{3}q_{0}^{3}}{Mq^{6}+Bq_{0}^{2}+Cq^{2}q_{0}^{2}}dq & = & \frac{A\pi }{3\sqrt{3}B^{1/3}M^{2/3}}q_{0}^{7/3}\nonumber \\
 &  & -\frac{A\left( 3B+C^{2}\qmax ^{2}\right) }{6BM\qmax ^{2}}q_{0}^{3}\nonumber \\
 &  & +\frac{AC^{2}\pi q_{0}^{11/3}}{3^{3}\sqrt{3}B^{5/3}M^{4/3}}-\frac{4AC^{3}\pi q_{0}^{13/3}}{3^{5}\sqrt{3}B^{7/3}M^{5/3}}\nonumber \\
 &  & +O(q_{0}^{5})\label{generictrinityintegral03} 
\end{eqnarray}
Note that equation (\ref{generictrinityintegral02}) contains the
same fractional power coefficients as equation (\ref{generictrinityintegral01})
up to a factor \( -C/(3M) \). This is because the combination\begin{equation}
\int \frac{6Mq^{5}+2Cqq_{0}^{2}}{Mq^{6}+Bq_{0}^{2}+Cq^{2}q_{0}^{2}}dq=\ln (Mq^{6}+Bq_{0}^{2}+Cq^{2}q_{0}^{2})
\end{equation}
is a series in integer powers of \( q_{0} \).

\section{Workaround for \emph{Mathematica}}

\subsubsection{Assume positive\label{math_assumepositive}}

The calculation of fractional powers crucially depends on a series
expansion which in the current version of \emph{Mathematica} \cite{Mathematica:40,Mathematica:50}
is not straightforward to calculate. As a simple example, let us calculate
the series of\begin{equation}
s=\frac{-A+\sqrt{A^{2}+x}}{-A+\sqrt{A^{2}+2x}}.
\end{equation}
Expressions of the form of the denominator or numerator naturally
appear when solving quadratic, cubic, or quartic equations. The series
of \( s \) delicately depends on whether \( A \) is positive or
negative. We get the following series expansion for \( |x|<A^{2}/2 \):
\begin{equation}
\label{math_seriesexample}
s\simeq \left\{ \begin{array}{cc}
\displaystyle \ds \frac{1}{2}+\frac{x}{8A^{2}}-\frac{x^{2}}{8A^{4}}+\frac{19x^{3}}{128A^{6}}-\frac{25x^{4}}{128A^{8}}+O(x^{5}) & \textrm{for }A>0,\\
 & \\
\ds 1-\frac{x}{4A^{2}}+\frac{5x^{2}}{16A^{4}}-\frac{7x^{3}}{16A^{6}}+\frac{167x^{4}}{256A^{8}}+O(x^{5}) & \textrm{for }A<0.
\end{array}\right. 
\end{equation}
Trying to calculate the series symbolically results in subexpressions
of the form \( -A+\sqrt{A^{2}} \) which is not automatically reduced
to \( 0 \). Therefore \emph{Mathematica} does not recognize that
for positive \( A \) l'Hospital's rule has to be applied to calculate
the series elements as \( x\rightarrow 0 \). Unfortunately, even
in the most recent version of \emph{Mathematica} (which by the time
of writing is \emph{Mathematica} 5.0 \cite{Mathematica:50}) commands
like Assuming{[}\{A>0\}, ...{]} or options like Limit{[}s ,x\( \rightarrow  \)0,
Assumptions\( \rightarrow  \)\{A>0\}{]} are applied only before or
after a series calculation and not in the course of the calculation,
and therefore only provides a solution which is valid for \( -A+\sqrt{A^{2}}\neq 0 \)
which is satisfied by negative \( A \). Trying to plug in positive
\( A \) in the resulting series unfortunately gives the wrong result,
as can be checked numerically or by plotting the functions. We get
a complete nonsense result by the simple \emph{Mathematica} command
s+O{[}x{]}\textasciicircum{}3//PowerExpand \( \rightarrow  \) 1+ComplexInfinity
x+Indeterminate x\textasciicircum{}2+O{[}x{]}\textasciicircum{}3 since
series expansion (implicitly) assumes \( A<0 \) while PowerExpand
assumes \( A>0 \). 

In order to get the correct result, we have to modify the standard
behavior of the internal Power{[}..{]} function: Whenever we find
an expression \( (x^{a})^{b} \) with positive \( x \), we immediately
expand it to \( x^{ab} \). Our rules also include \( (x^{a}y)^{b}\rightarrow x^{ab}y^{b} \)
for \( x>0 \) and its standard-case for \( a=1 \). We use Length{[}Cases{[}ListOfPositives,x{]}{]}
> 0 to check whether x is part of varlist of positive variables. The
attribute HoldRest is set first, so that expressions are only evaluated
after the power rules have been modified. After calculation, the power
rules are reset to their original state, and the new result is returned.
The following routine also expands the Log{[}..{]} function accordingly.

\begin{lyxcode}
SetAttributes{[}AssumePositive,~HoldRest{]};

AssumePositive{[}varlist\_,~expression\_{]}~:=~

~~Block{[}\{ListOfPositives~=~varlist\},~

~~({*}Andreas~Ipp,~Sept~12,~2003{*})~

~~({*}First~we~change~Power~Rules{*})~

~~Unprotect{[}Power,~Log{]};~

~~Power{[}Power{[}x\_~/;~(Length{[}Cases{[}ListOfPositives,~x{]}{]}~

~~~~~~~~>~0),~n\_.{]}{*}y\_,~m\_{]}~:=~Power{[}x,~n{*}m{]}{*}Power{[}y,~m{]};~

~~Power{[}Power{[}x\_~/;~(Length{[}Cases{[}ListOfPositives,~x{]}{]}~

~~~~~~~~>~0),~n\_{]},~m\_{]}~:=~Power{[}x,~n{*}m{]};

~~Log{[}Power{[}x\_~/;~(Length{[}Cases{[}ListOfPositives,~x{]}{]}~>~0),

~~~~~~n\_.{]}{*}y\_{]}~:=~n~Log{[}x{]}~+~Log{[}y{]};~

~~Log{[}Power{[}x\_~/;~(Length{[}Cases{[}ListOfPositives,~x{]}{]}~>~0),~

~~~~~~n\_{]}{]}~:=~n{*}Log{[}x{]};~Protect{[}Power,~Log{]};

~~({*}There~is~a~bug~if~we~try~to~use~FullSimplify{[}~..{]},~

~~~~so~we~turn~the~Error~Message~off~{*})~

~~Off{[}Pattern::\char`\"{}nodef\char`\"{}{]};~

~~({*}Then~we~evaluate~expression{*})~

~~\{expression,~

~~~On{[}Pattern::\char`\"{}nodef\char`\"{}{]};~

~~~({*}and~finally~we~change~rules~back~to~original~state{*})

~~~Unprotect{[}Power,~Log{]};~

~~~Power{[}Power{[}x\_~/;~(Length{[}Cases{[}ListOfPositives,~x{]}{]}~

~~~~~~~~~>~0),~n\_.{]}{*}y\_,~m\_{]}~=.;

~~~Power{[}~Power{[}x\_~/;~(Length{[}Cases{[}ListOfPositives,~x{]}{]}~

~~~~~~~~~>~0),~n\_{]},~m\_{]}~=.;~

~~~Log{[}Power{[}x\_~/;~(Length{[}Cases{[}ListOfPositives,~x{]}{]}~

~~~~~~~~~>~0),~n\_.{]}{*}y\_{]}~=.;~

~~~Log{[}Power{[}x\_~/;~(Length{[}Cases{[}ListOfPositives,~x{]}{]}~

~~~~~~~~~>~0),~n\_{]}{]}~=.;~

~~~Protect{[}Power,~Log{]};~

~~\}{[}{[}1{]}{]}({*}but~we~take~the~expression~from~before~-~

~~~in~this~way~we~don't~have~to~introduce~a~new~variable{*})~

{]}
\end{lyxcode}
Now it is straightforward to calculate the series above: Use AssumePositive{[}\{A\},
s + O{[}x{]}\textasciicircum{}5{]} for the upper series and AssumePositive{[}\{B\},
s + O{[}x{]}\textasciicircum{}5 /. A \( \rightarrow  \) -B{]} /.
B \( \rightarrow  \) -A for the lower series in (\ref{math_seriesexample}).
AssumePositive can take any list of variables, e.g. \( \textrm{AssumePositive}[\{a,b\},\sqrt{ab^{2}cd^{2}}]\rightarrow \sqrt{a}\, b\sqrt{cd^{2}} \).
Possible extensions to this routine might include \( \textrm{Abs}[x]\rightarrow x \)
or \( \textrm{UnitStep}[x]\rightarrow 1 \) for \( x>0 \).

\tmpbibtex

\newpage

\addcontentsline{toc}{chapter}{Bibliography}
\providecommand{\href}[2]{#2}\begingroup\raggedright\endgroup

%





%

\chapter*{Acknowledgements}

\addcontentsline{toc}{chapter}{Acknowledgements}

\renewcommand{\chaptermark}[1]{\markboth{\uppercase{#1}}{\uppercase{#1}}}

\chaptermark{Acknowledgements}

\rem{Oct 9, 2003}I would like to thank my supervisor Anton Rebhan
for excellently guiding me through my last two years as a doctoral
candidate at the Technical University of Vienna. I am indebted to
him for introducing Large \( N_{f} \) to me, a topic that during
the last year turned out to be remarkably fruitful with new unexpected
results, and let me rise to new challenges regarding my numerical
computing skills.

Anton Rebhan's current team of doctoral candidates has been a wonderful
environment for fruitful discussions: I want to thank Paul Romatschke
for expertise on 2PI, Robert Wimmer for discussions regarding the
foundations of quantum field theory, and Andreas Gerhold for insights
regarding non-Fermi liquid and color superconductors.

During the last two years I profited a lot from lectures by Manfred
Markytan who increased my interest in particle colliders, Dominik
Schwarz who introduced me to hydrodynamics and kept me in touch with
cosmology, and Albert Reiner who deepened my working knowledge on
\emph{Mathematica}.

From the Institute for Theoretical Physics I would like to single
out the head of the institute Wolfgang Kummer for providing an open
and lively working environment, and further Erwin Riegler, Herbert
Balasin, Daniel Grumiller, and Maximilian Kreuzer for permanent helpfulness
regarding computing and Linux support. Thanks also to \rem{Peter Fischer
for discussions and }Sebastian Guttenberg for his valuable experience
with \LyX{} which helped meeting the deadline. A big thank you also
to the administration team of the institute Elfriede Mössmer, Roswitha
Unden, and Franz Hochfellner who wonderfully help with all the important
work {}``behind the scenes''.

I also want to thank for the warm hospitality that I experienced during
my stay at the SPhT, CEA-Saclay in France. Thanks to Jean-Paul Blaizot,
Edmond Iancu, and especially Urko Reinosa for discussions on Large-\( N_{f} \),
2PI, renormalization issues, and non-Fermi-liquids, amongst other
topics.

On the technical side I would like to thank Dirk Rischke and Aleksi
Vuorinen for correspondence.

I further would like to thank Daniel Hofmann and Christopher Summer
for encouraging me during the tough initial part of my studies, and
Mike Strickland for wonderful improvisational music sessions.

Last but not least I would like to thank my parents for their incessant
support, my grandmother, my brother and my sister for moral support,
and of course Sun Ying for lovingly accompanying me through the final
stages of my thesis. 

This work has been supported by the Austrian Science Foundation FWF,
project no. 14632-TPH.

\newpage
\chapter*{Curriculum Vitae}

\addcontentsline{toc}{chapter}{Curriculum Vitae}

\renewcommand{\chaptermark}[1]{\markboth{\uppercase{#1}}{\uppercase{#1}}}
\chaptermark{Curriculum Vitae}

{\centering \textbf{\LARGE Andreas Ipp}\LARGE \par}
\medskip{}

\begin{tabular}{ll}
\selectlanguage{english}
Address: 
\selectlanguage{american}&
\selectlanguage{english}
Krotenbachgasse 27
\selectlanguage{american}\\
&
\selectlanguage{english}
A-2345 Brunn am Gebirge
\selectlanguage{american}\\
&
\selectlanguage{english}
AUSTRIA / EUROPE 
\selectlanguage{american}\\
\selectlanguage{english}
Phone:
\selectlanguage{american}&
\selectlanguage{english}
+43-1-2236-378811
\selectlanguage{american}\\
\selectlanguage{english}
Email:
\selectlanguage{american}&
\selectlanguage{english}
ipp@hep.itp.tuwien.ac.at
\selectlanguage{american}\\
\selectlanguage{english}
Homepage:
\selectlanguage{american}&
\selectlanguage{english}
http://hep.itp.tuwien.ac.at/\textasciitilde{}ipp/
\selectlanguage{american}\\
\end{tabular}

\section*{Personal information}

Date and place of birth: July 12, 1975, Mödling, Austria.\\
Citizenship: Austria

\section*{Academic Career}

\begin{tabular}{ll}
\selectlanguage{english}
since 2002
\selectlanguage{american}&
\selectlanguage{english}
Ph.D.~student, Institute for Theoretical Physics, TU Vienna
\selectlanguage{american}\\
\selectlanguage{english}
March 2000
\selectlanguage{american}&
\selectlanguage{english}
Second Diploma Examination in physics (Master's degree)
\selectlanguage{american}\\
&
\selectlanguage{english}
passed with distinction
\selectlanguage{american}\\
&
\selectlanguage{english}
Diploma thesis: {}``Phase Transition of a Scalar Field Theory
\selectlanguage{american}\\
&
\selectlanguage{english}
\textcolor{white}{Diploma thesis: {}``}in the Early Universe{}``
\selectlanguage{american}\\
&
\selectlanguage{english}
Advisor: Dr. Anton Rebhan, Dr. Laurence Yaffe
\selectlanguage{american}\\
\selectlanguage{english}
1998 - 1999 
\selectlanguage{american}&
\selectlanguage{english}
Exchange student, University of Washington Seattle, USA
\selectlanguage{american}\\
&
\selectlanguage{english}
Advisor: Dr. Laurence Yaffe
\selectlanguage{american}\\
\selectlanguage{english}
June 1997
\selectlanguage{american}&
\selectlanguage{english}
University of Music and Performing Arts Vienna
\selectlanguage{american}\\
&
\selectlanguage{english}
First diploma examination in piano pedagogic
\selectlanguage{american}\\
\selectlanguage{english}
January 1996
\selectlanguage{american}&
\selectlanguage{english}
Vienna University of Technology
\selectlanguage{american}\\
&
\selectlanguage{english}
First diploma examination in physics passed with distinction
\selectlanguage{american}\\
\selectlanguage{english}
1985 - 1993
\selectlanguage{american}&
\selectlanguage{english}
BG/BRG Mödling - Keimgasse Mödling (Secondary education)
\selectlanguage{american}\\
&
\selectlanguage{english}
Final examination passed with distinction
\selectlanguage{american}\\
\end{tabular}

\section*{International Experience}

\begin{tabular}{ll}
\selectlanguage{english}
2000 - 2001
\selectlanguage{american}&
\selectlanguage{english}
Alternative service / social service in Qiqihar, China
\selectlanguage{american}\\
\selectlanguage{english}
1998 - 1999 
\selectlanguage{american}&
\selectlanguage{english}
Exchange student, University of Washington Seattle, USA
\selectlanguage{american}\\
&
\selectlanguage{english}
Advisor: Dr. Laurence Yaffe
\selectlanguage{american}\\
\end{tabular}

\section*{Talks given at Conferences and Workshops}

\begin{tabular}{@{} cl}
\selectlanguage{english}
Oct 2003
\selectlanguage{american}&
\selectlanguage{english}
FAKT 2003, Strobl am Wolfgangsee, Austria.
\selectlanguage{american}\\
&
\selectlanguage{english}
{}``Thermodynamics of Large \( N_{f} \) QCD
\selectlanguage{american}\\
&
\selectlanguage{english}
at Finite Chemical Potential''
\selectlanguage{american}\\
\selectlanguage{english}
Sept 2003
\selectlanguage{american}&
\selectlanguage{english}
Quantum fields in and out of equilibrium, Bielefeld, Germany.
\selectlanguage{american}\\
&
\selectlanguage{english}
{}``Thermodynamics of Large \( N_{f} \) QCD 
\selectlanguage{american}\\
&
\selectlanguage{english}
at Finite \( \mu  \) and Non-Fermi-Liquid Behavior''
\selectlanguage{american}\\
\selectlanguage{english}
June 2003
\selectlanguage{american}&
\selectlanguage{english}
Common Trends in Cosmology and Particle Physics, 
\selectlanguage{american}\\
&
\selectlanguage{english}
Balatonfüred, Hungary.
\selectlanguage{american}\\
&
\selectlanguage{english}
{}``Phase Transition of a Scalar Field Theory
\selectlanguage{american}\\
&
\selectlanguage{english}
in the Early Universe''
\selectlanguage{american}\\
\end{tabular}

\section*{Further Conferences and Workshops attended}

\begin{tabular}{ll}
\selectlanguage{english}
Feb 2003
\selectlanguage{american}&
\selectlanguage{english}
41. Internationale Universitätswochen für Theoretische
\selectlanguage{american}\\
&
\selectlanguage{english}
Physik, {}``Flavor Physics'', Schladming, Austria.
\selectlanguage{american}\\
\selectlanguage{english}
Oct 2002
\selectlanguage{american}&
\selectlanguage{english}
Strong and Electroweak Matter 2002, Heidelberg, Germany.
\selectlanguage{american}\\
\selectlanguage{english}
Sept 2002
\selectlanguage{american}&
\selectlanguage{english}
WE-Heraeus-Doktorandenschule \char`\"{}Grundlagen und neue
\selectlanguage{american}\\
&
\selectlanguage{english}
Methoden der theoretischen Physik\char`\"{}, Wolfersdorf, Germany.
\selectlanguage{american}\\
\selectlanguage{english}
Aug 1998
\selectlanguage{american}&
\selectlanguage{english}
5th International Workshop on Thermal Field Theories
\selectlanguage{american}\\
&
\selectlanguage{english}
and Their Applications, Regensburg, Germany.
\selectlanguage{american}\\
\end{tabular}

\section*{International cooperation}

\begin{tabular}{ll}
\selectlanguage{english}
May 2003
\selectlanguage{american}&
\selectlanguage{english}
Institute for Theoretical Physics, CEA Saclay, France,
\selectlanguage{american}\\
&
\selectlanguage{english}
with Jean-Paul Blaizot, Edmond Iancu, and Urko Reinosa.
\selectlanguage{american}\\
\end{tabular}

\section*{Professional Experience}

\begin{tabular}{ll}
\selectlanguage{english}
since 2002
\selectlanguage{american}&
\selectlanguage{english}
TU Vienna, FWF research project no. 14632-PHY.
\selectlanguage{american}\\
\selectlanguage{english}
2000
\selectlanguage{american}&
\selectlanguage{english}
Part time computer support for In-Vision GmbH
\selectlanguage{american}\\
\selectlanguage{english}
Oct 1999 - 2000
\selectlanguage{american}&
\selectlanguage{english}
Part time computer support for Docter Optics
\selectlanguage{american}\\
\selectlanguage{english}
July - Sept 1999
\selectlanguage{american}&
\selectlanguage{english}
Docter Optics
\selectlanguage{american}\\
\selectlanguage{english}
July-Aug 96, 97, 98
\selectlanguage{american}&
\selectlanguage{english}
Siemens PSE / Competence Basis
\selectlanguage{american}\\
\selectlanguage{english}
July 1995
\selectlanguage{american}&
\selectlanguage{english}
Philips iR3 Video International GmbH
\selectlanguage{american}\\
\selectlanguage{english}
July 1994
\selectlanguage{american}&
\selectlanguage{english}
RZB Wien
\selectlanguage{american}\\
\end{tabular}
\end{document}